\documentclass[12pt]{report}
\usepackage{mathptmx}
\usepackage{times}
\usepackage{graphicx}
\usepackage{fancyhdr}
\renewcommand{\headrulewidth}{0pt}
\usepackage{geometry}
\usepackage[subfigure]{tocloft}
\usepackage{import}
\usepackage{amssymb}
\usepackage{bm}
\usepackage{mathtools}
\usepackage[hidelinks]{hyperref}
\usepackage{wrapfig}
\usepackage[sorting=none,backend=bibtex]{biblatex}
\usepackage[utf8]{inputenc}  
\usepackage{tabularx}
\usepackage{lipsum}

\usepackage{microtype}

\usepackage[]{url} 
\usepackage[]{hyperref}
\usepackage{graphicx}

\usepackage{float}

\usepackage{amsmath}

\usepackage{gensymb}
\usepackage{textcomp}
\usepackage{multirow}
\usepackage{color}
\usepackage{mathptmx}
\usepackage{booktabs}
\usepackage{subfigure}
\usepackage{amsfonts}
\usepackage{stackengine}

\usepackage{algorithm} 
\usepackage{algpseudocode} 

\usepackage[OT1]{fontenc} 
\usepackage{xspace}
\newcommand{\descr}[1]{\smallskip\noindent\textbf{#1}}

\usepackage[font=small,labelfont=bf]{caption}
\usepackage{paralist}

\usepackage[OT2,OT1]{fontenc}
\newcommand\cyr
{
\renewcommand\rmdefault{wncyr}
\renewcommand\sfdefault{wncyss}
\renewcommand\encodingdefault{OT2}
\normalfont
\selectfont
}
\DeclareTextFontCommand{\textcyr}{\cyr}

\usepackage{titlesec}

\newif\ifcomment
\commenttrue
\ifcomment
\newcommand{\kp}[1]{{\textcolor{blue}{#1}}}
\newcommand{\ms}[1]{{\textcolor{red}{#1}}}
\newcommand{\revision}[1]{{\textcolor{black}{#1}}}
\newcommand{\revcomment}[1]{{\textcolor{red}{}}}
\else
\newcommand{\kp}[1]{}
\newcommand{\ms}[1]{}
\newcommand{\revision}[1]{}
\newcommand{\revcomment}[1]{}
\fi

\titleclass{\subsubsubsection}{straight}[\subsection]

\newcounter{subsubsubsection}[subsubsection]
\renewcommand\thesubsubsubsection{\thesubsubsection.\arabic{subsubsubsection}}

\titleformat{\subsubsubsection}
  {\normalfont\normalsize\bfseries}{\thesubsubsubsection}{1em}{}
\titlespacing*{\subsubsubsection}
{0pt}{3.25ex plus 1ex minus .2ex}{1.5ex plus .2ex}

\makeatletter
\renewcommand\paragraph{\@startsection{paragraph}{5}{\z@}%
  {3.25ex \@plus1ex \@minus.2ex}%
  {-1em}%
  {\normalfont\normalsize\bfseries}}
\renewcommand\subparagraph{\@startsection{subparagraph}{6}{\parindent}%
  {3.25ex \@plus1ex \@minus .2ex}%
  {-1em}%
  {\normalfont\normalsize\bfseries}}
\def\toclevel@subsubsubsection{4}
\def\toclevel@paragraph{5}
\def\toclevel@paragraph{6}
\def\l@subsubsubsection{\@dottedtocline{4}{7em}{4em}}
\def\l@paragraph{\@dottedtocline{5}{10em}{5em}}
\def\l@subparagraph{\@dottedtocline{6}{14em}{6em}}
\makeatother

\begin{document}

\thispagestyle{empty}
\vspace*{3\baselineskip}
\begin{center}
    \textbf{\LARGE{Characterizing Abhorrent, Misinformative, and Mistargeted Content on YouTube}}

    \vspace{2\baselineskip}
    \textit{\large{Kostantinos Papadamou}}
    
    \vspace{12\baselineskip}

    \small
    A thesis submitted in partial fulfillment\\of the requirements for the degree of\\\textbf{Doctor of Philosophy}\\of\\\textbf{Cyprus University of Technology}
    \vspace{1.5\baselineskip}

	Department of Electrical Engineering, Computer Engineering and Informatics\\
	Cyprus University of Technology
    
    \vfill
    May 16, 2021
\end{center}

\newpage
\thispagestyle{empty}

\textbf{\LARGE{Abstract}}
\vspace{1\baselineskip}

YouTube has revolutionized the way people discover and consume video content.
Although YouTube facilitates easy access to hundreds of well-produced educational, entertaining, and trustworthy news videos, abhorrent, misinformative and mistargeted content is also common.
The platform is plagued by various types of inappropriate content including: 1) disturbing videos targeting young children; 2) hateful and misogynistic content; and 3) pseudoscientific and conspiratorial content.
While YouTube's recommendation algorithm plays a vital role in increasing user engagement and YouTube's monetization, its role in unwittingly promoting problematic content is not entirely understood.

In this thesis, we shed some light on the degree of abhorrent, misinformative, and mistargeted content on YouTube and the role of the recommendation algorithm in the discovery and dissemination of such content.
Following a data-driven quantitative approach, we analyze thousands of videos posted on YouTube. 
Specifically, we devise various methodologies to detect problematic content, and we use them to simulate the behavior of users casually browsing YouTube to shed light on:
1) the risks of YouTube media consumption by young children;
2) the role of YouTube's recommendation algorithm in the dissemination of hateful and misogynistic content, by focusing on the Involuntary Celibates (Incels) community;
and 
3) user exposure to pseudoscientific misinformation on various parts of the platform and how this exposure changes based on the user's watch history.

In a nutshell, our analysis reveals that young children are likely to encounter disturbing content when they randomly browse the platform starting from benign videos relevant to their interests and that YouTube's currently deployed counter-measures are ineffective in terms of detecting them in a timely manner.
By analyzing the Incel community on YouTube, we find that not only Incel activity is increasing over time, but platforms may also play an active role in steering users towards extreme content.
Finally, when studying pseudoscientific misinformation, we find among other things that YouTube suggests more pseudoscientific content regarding traditional pseudoscientific topics (e.g., flat earth) than for emerging ones (like COVID-19), and that these recommendations are more common on the search results page than on a user’s homepage or the video recommendations (up-next) section.

\newpage
\thispagestyle{empty}

\textbf{\LARGE{Acknowledgments}}
\vspace{1\baselineskip}    

First and foremost, I am grateful to my advisor, Michael Sirivianos, for his continuous support and valuable feedback throughout my PhD journey.
His support and guidance was instrumental in turning me into an independent and competent researcher. 
He was there to guide me when the research seemed fuzzy and disheartening.
More importantly, he has shown me how to analyze an important and complex problem, and divide it in small manageable problems that can be addressed in a more practical way.

Second, I would like to thank Jeremy Blackburn, Emiliano De Cristofaro, Gianluca Stringhini, and Savvas Zannettou. 
Their expertise and valuable feedback complemented the one received by my advisor, hence helping me in further expanding my research and writing skills.

Also, I want to thank various colleagues from the Cyprus University of Technology, Telefonica Research, and Max Planck Institute. 
Their help and feedback was vital to undertaking the studies presented in this thesis.

Furthermore, I would like to thank my family for their support, patience, and encouragement, which ensured that I was mentally strong to overcome all the obstacles faced during my PhD journey.

Finally, I owe gratitude to the European Commission's Horizon 2020 program and the Cyprus University of Technology for funding my research.

\newpage    
\thispagestyle{empty}

\textbf{\LARGE{In Memoriam of Prof. Vassos Soteriou}}
\vspace{1\baselineskip}    

I would like to pay tribute to the memory of Professor Vassos Soteriou.
I am grateful for having him as a professor during my barchelor's and master's studies, and I am thankful for the valuable feedback he gave me as one of my PhD proposal examiners.
His pressing questions during my proposal examination have helped to improve this work and to make me a better
presenter and researcher.

\begingroup
\pagestyle{fancy}
\renewcommand{\headrulewidth}{0pt}
\fancyhf{}
\fancyhead[R]{\thepage}
\fancyfoot{}

\newpage
\renewcommand{\contentsname}{\LARGE{Contents}}
\tableofcontents

\newpage
\renewcommand{\listfigurename}{\LARGE{List of Figures}}
\listoffigures

\newpage
\renewcommand{\listtablename}{\LARGE{List of Tables}}
\listoftables

\newpage 
\thispagestyle{empty}
\begin{flushleft}
    \textbf{\LARGE{Abbreviations}}
    \vspace{1\baselineskip}
    
    \begin{tabular}{l l}
        API: & Application Programming Interface\\
        AUC: & Area under the ROC Curve\\
        BERT: & Bidirectional Encoder Representations\\
        CEO: & Chief Executive Officer\\
        CNN: & Convolutional Neural Networks\\
        CDF: & Cumulative Distribution Function\\
        COVID: & Coronavirus Disease\\
        DDNN: & Double Dense Neural Network\\
        DL: & Deep Learning\\
        LSTM: & Long Short-Term Memory\\
        MGTOW: & Men Going Their Own Way\\
        ML: & Machine Learning\\
        MPAA & Motion Picture Association of America\\
        MRA: & Men Rights Activists\\
        NC-17: & No Children Under 17\\
        NLP: & Natural Language Processing\\
        OSN: & Online Social Network \\
        PG: & Parental Guidance suggested\\
        PUA: & Pick Up Artists\\
        ReLU: & Rectified Linear Unit\\
        REST: & Representational State Transfer\\
        RNN: & Recurrent Neural Network\\
        ROC: & Receiver Operating Characteristic\\
        SMOTE: & Synthetic minority Over-sampling Technique\\
        SVM: & Support Vector Machines\\
        URL: & Uniform Resource Locator\\
    \end{tabular}    
\end{flushleft}

\endgroup

\pagestyle{fancy}
\renewcommand{\headrulewidth}{0pt}
\fancyhf{}
\fancyhead[C]{\small{\itshape\nouppercase{\rightmark}}}
\fancyhead[R]{\thepage}

\newpage
\chapter{Introduction}
\label{chapter:introduction}

Over the last decade, user{-}generated video platforms have exploded in popularity.
YouTube is the leading video-sharing platform on the Web and the second most visited website worldwide, with billions of monthly active users, surpassing cable TV in terms of popularity, especially among teenagers.
For many users, YouTube has also become one of the most important information sources for news, world events, and various other topics~\cite{reutersdigital,pewyoutubenews}.
However, while YouTube facilitates easy access to hundreds of well-produced educational, entertaining and credible news videos, the platform is also fertile ground for the spread of mistargeted and abhorrent content.

In the last few years, we have seen extensive anecdotal evidence suggesting that YouTube, and in particular its recommendation algorithm, promote offensive and potentially harmful content (i.e., conspiracy theories), and even helping radicalize users~\cite{rooseyoutuberadical,tufekci2018youtube,mozillaregrets,ribeiro2020auditing}.
Some examples include disturbing and harmful videos targeting young children evading YouTube's attempts to control them ~\cite{nytiyoutubekids,bbdisturbingyoutube}.
Alas, the platform also hosts hateful and misogynistic content acting as a medium for extremist ideologies to thrive~\cite{beccayoutuberadicalize,bbcrodgers,regner2014youtube}.
On top of this, the platform is also plagued by misinformation, pseudoscientific and conspiratorial content~\cite{carneyoutubeconspricacies}.
At the same time, the opaque nature of YouTube's recommendation algorithm aids the dissemination of such content to millions of viewers~\cite{springyoutubefalse}.
This can have dire real-world consequences, especially when pseudoscientific content is promoted to users at critical times, such as the COVID-19 pandemic~\cite{springyoutubefalse,li2020youtube}.

Despite numerous reports that have highlighted the alarming presence of problematic content, YouTube and other social media platforms have struggled to mitigate the harm from this type of content.
The difficulty is partly due to the sheer scale and also because of the deployment of recommendation algorithms~\cite{fastcompeny2019conspiracies}.
In particular, to detect inappropriate content, YouTube heavily relies on users reporting videos they consider inappropriate, which are then inspected by YouTube employees. 
However, since the process involves manual labor, the process does not scale to the number of videos that a platform like YouTube serves. 
On rare occasions, the platforms have also refused to moderate such content stating that it does not violate their community guidelines~\cite{kashmir2014elliot,fernanda2020antivaccine}.

\revision{
In an attempt to solve the problem, YouTube employs automated moderation using Machine Learning (ML) to detect and flag videos that violate their community guidelines. However, purely automated moderation tools have thus far been insufficient to moderate content~\cite{dave2017youtubedetect,shu2018youtubemldetect}, and human moderators had to be brought back into the loop~\cite{vincent2020humanmoderators,heilweil2020}.
Whereas the scientific community has repeatedly raised the need for effectively detecting and moderating inappropriate content, the various types of problematic content on the platform are relatively unstudied.
At the same time, the role played by YouTube's recommendation algorithm in unwittingly promoting such content is not entirely understood, while its opaque nature makes it difficult to audit.
}

\revision{
In this thesis, we aim at: 
1) investigating how we can classify the quality of content available on YouTube;
2) developing methodologies that allow us to quantify the influence of YouTube's recommendation algorithm in the dissemination of problematic content and to investigate whether the recommendation algorithm contributes to steering users towards hateful communities; and
3) assessing how likely it is for users to come across abhorrent and misleading content while casually browsing the platform.
We do this by focusing on three diverse types of misleading and abhorrent content, namely: 1) inappropriate videos targeting young children; 2) hateful and misogynistic content; and 3) pseudoscientific misinformation.
We delve into these types of content mainly because they are of great societal interest and because the dissemination of such content to users can have dire real-consequences.
We also argue that providing knowledge, methodologies, and tools to analyze various types of content on YouTube is a significant step towards understanding and mitigating inappropriate content,
and will help the research community shed additional light on the recommendation algorithm and its potential influence.
We elaborate on each type of abhorrent and misleading content below.
}

\descr{Characterizing and Detecting Inappropriate Videos Targeting Young Children.}
A plethora of the most subscribed YouTube channels target children of very young age. 
While most of these channels feature inoffensive and educational videos, recent reports have highlighted the trend of inappropriate content targeting this demographic referring to such videos as \textit{disturbing}~\cite{wired2018disturbed}.
Unfortunately, YouTube’s algorithmic recommendation system regrettably suggests inappropriate content because some of it mimics or is derived from otherwise appropriate content.
Despite the company's attempts to curb this phenomenon, disturbing videos still appear even in its ``Kids'' app~\cite{disturbing2018youtubekids}, due to the difficulty in identifying them.
Prior work studied YouTube videos with inappropriate content for children at a small scale, however, none of them focused on the characterization and detection of disturbing videos that explicitly target toddlers. 
Hence, the prevalence of the problem at a large scale is not entirely understood, neither is the role of YouTube's recommendation algorithm in the dissemination of such videos among young viewers.

\descr{Characterizing Hateful and Misogynistic Content on YouTube Through the Lens of the Incel Community.}
Despite YouTube's attempts to tackle hateful and misogynistic content on the platform, the problem still abounds~\cite{martineau2020youtubeincels}.
One fringe community that has often been linked to sharing and publishing such content is the so-called Involuntary Celibates, also known as \textit{Incels}~\cite{serena2019incel}.
This community is one of the most extreme in the Manosphere~\cite{bbc2018incel}, a larger collection of movements discussing men’s issues~\cite{ging2019alphas} (see Section~\ref{sec:incels_background}).
While the concepts underpinning the Incel ``ideology'' may seem absurd, they also have grievous real-world consequences~\cite{beauchamp2020toronto}.
Despite the important implications that Incel ideology can have on society, research on Incels is limited and mostly qualitative~\cite{hoffman2020assessing}.
At the same time, the research community has mostly studied the Incel community and the broader Manosphere on Reddit, 4chan, and other online discussion forums like Incels.me~\cite{farrell2019exploring,jaki2019online}.
However, the fact that YouTube has been repeatedly accused of user radicalization prompts the need to study the extent to which Incels are exploiting YouTube to spread their views.

\descr{Assessing the Effect of Watch History on YouTube’s Pseudoscientific Video Recommendations.}
YouTube provides an ideal environment for disseminating information to a vast number of people in a short period of time.
Yet, when exploited by bad actors, YouTube is often fertile ground for the spread of misleading and potentially harmful information like conspiracy theories.
Conspiracy theories are usually built on tenuous connections between various events, with little to no actual evidence to support them. 
On user-generated content platforms like YouTube, these are often presented as facts, regardless of whether they are supported by facts and even though they have been widely debunked.
For certain types of content (e.g., health-related topics), harmful videos can have devastating effects on society, especially during crises like the COVID-19 pandemic~\cite{springyoutubefalse}.
Recently, YouTube has been under scrutiny for suggesting pseudoscientific and conspiratorial content~\cite{carneyoutubeconspricacies,fernanda2020antivaccine,fastcompeny2019conspiracies}, mainly due to the difficulty in identifying such videos.
However, as a research community, we lack an understanding of how YouTube's recommendation algorithm contributes to the promotion of pseudoscience.
More importantly, we lack tools and methodologies for effectively auditing video recommendation algorithms.

\revision{
Motivated by the above aspects and the pressing need to suppress abhorrent and misleading content on YouTube, we set out to answer the following overarching research questions (RQs):
}
\begin{itemize}
    \item \revision{\textbf{RQ1:}
    Can we effectively detect and characterize abhorrent and misleading content on YouTube?}

    \item \revision{\textbf{RQ2:}
    Can we effectively quantify the influence of YouTube's recommendation algorithm in the dissemination of problematic content?}

    \item \revision{\textbf{RQ3:}
    Does YouTube's recommendation algorithm contribute to steering users towards hateful communities?}

    \item \revision{\textbf{RQ4:}
    How likely is it for a user to come across abhorrent and misleading content when casually browsing the platform?
    What is the proportion of such content on YouTube user's homepage, in search results, and in the video recommendations section of YouTube?}

\end{itemize}

\revision{
The above overarching research questions lead to several research sub-questions posed in the relevant chapters. 
In particular, sub-questions annotated with RQX.a are posed in Chapter~\ref{chapter:disturbed_youtube}, RQX.b in Chapter~\ref{chapter:incels_youtube}, and RQX.c in Chapter~\ref{chapter:pseudoscience_youtube}.}
To provide answers to these research questions, we follow a large-scale data-driven quantitative approach.
To do so, we first implement a data collection infrastructure that consists of various crawlers, which allow us to collect information from the Web.
By mainly leveraging the YouTube Data API~\cite{youtubedataapi}, we create large-scale datasets from videos uploaded on YouTube.
\revision{
Then, we use various deep learning techniques to detect abhorrent and misleading videos.
Finally, we use statistical analysis and other techniques, and we devise methodologies that allow us to quantify the influence of YouTube's recommendation algorithm in promoting problematic content and extract meaningful insights from the large-scale datasets.
}
Specifically, we use the following techniques:
\begin{itemize}
    \item \textbf{Neural Networks \& Deep Learning:} 
    \revision{We apply neural networks for various purposes. For instance, we use pre-trained neural networks to extract useful insights from images. Also, we use neural networks to build two custom deep learning classifiers that consider the various metadata of videos uploaded on YouTube (e.g., title, transcript, comments, etc.), and use them to detect various types of problematic content on YouTube (see Sections~\ref{sec:disturbed_detectionofdisturbingvideos} and~\ref{sec:pseudoscience_videos_detection}).}
    
    \item \textbf{Graph Analysis:} 
    We leverage several graph analysis and visualization techniques to analyze data that can be modeled with graphs.
    In the context of our work, we build directed graphs that are snapshots of YouTube's recommendation graph where nodes are videos and edges are recommendations between them.
    Then, we use graph analysis to extract meaningful conclusions (e.g., degree distribution) about the interplay between various types of benign and inappropriate content.
    
    \item \textbf{Random Walks:} 
    We perform "random walks" by traversing snapshots of YouTube's recommendation graph while selecting videos according to recommendations.
    This is particularly useful, as it allows us to simulate the behavior of various users who browse YouTube starting from a certain type of videos and select the videos to watch based on recommendations.
    This methodology allows us to understand what types of videos YouTube's recommendation algorithm suggests to various types of users for various topics.
    More details regarding how the random walks methodology is used in each one of our lines of work can be found in Sections~\ref{subsec:disturbed_random_walks_analysis},~\ref{subsec:incels_recommendations_analysis}, and~\ref{subsubsec:pseudoscience_random_walks}.

    \item \textbf{Clustering Algorithms:} 
    We use traditional clustering algorithms to create groups of similar information.
    More precisely, we use the $k$-means clustering algorithm~\cite{hartigan1979algorithm} to create clusters of similar topics from a list of keywords associated with young children (see Section~\ref{subsec:disturbed_random_walks_analysis}).
\end{itemize}

\section{Contributions}
\revision{This thesis makes several contributions towards classifying the quality of content on YouTube, understanding the role of YouTube's recommendation algorithm in promoting problematic content, as well as towards quantifying the degree of various types of such content on the platform.
More precisely, we make the following contributions:
}

\begin{enumerate}
    \item We provide the first large-scale characterization of inappropriate or disturbing videos targeted at toddlers.
    Using a set of seed keywords that cover a wide range of child-related content available on YouTube, we collect and analyze a large number of videos\footnote{We make publicly available the metadata of all the collected videos so that the research community can further investigate the problem~\cite{disturbeddataset}} finding $1.1\%$ of them to be inappropriate for young children \textbf{(RQ1)}.

    \item We propose and implement a deep learning classifier\footnote{We make the disturbing videos detection classifier publicly available~\cite{disturbedyoutubeclassifier}} that can discern disturbing videos that target toddlers with $0.84$ accuracy. We then perform a live simulation in which we mimic a toddler randomly clicking on YouTube’s suggested videos while using the developed classifier to classify the quality of all the selected videos. We find a $3.5\%$ chance that a toddler following YouTube’s recommendations will encounter an inappropriate video within ten recommendation hops if she starts from a video that appears among the top ten results of a toddler-appropriate keyword search (e.g., Peppa Pig) \textbf{(RQ1 and RQ4)}.

    \item \revision{We devise a text-based methodology for collecting and annotating videos related to the Incel community and use it to perform the first large-scale data-driven characterization of this community on YouTube\footnote{We make publicly available the metadata of all the collected Incel-derived and Control videos to assist researchers further investigate the problem~\cite{incelsdataset}}. We found a non-negligible growth in Incel-related activity on YouTube over the past few years, both in terms of published Incel-related videos and comments likely posted by Incels. This result suggests that users gravitating around the Incel community are increasingly using YouTube to disseminate their views \textbf{(RQ1)}.}
    
    \item We study how YouTube's recommendation algorithm behaves with respect to Incel-related videos.
    By performing random walks on YouTube's recommendation graph, we find a $6.3\%$ chance for a user who starts by watching non-Incel-related videos to be recommended Incel-related ones within five recommendation hops.
    At the same time, users who have seen two or three Incel-related videos at the start of their walk see recommendations that consist of $9.4\%$ and $11.4\%$ Incel-related videos, respectively. 
    Moreover, the portion of Incel-related recommendations increases substantially as the user watches an increasing number of consecutive Incel-related videos \textbf{(RQ3)}.
    
    \item \revision{We propose an additional deep learning classifier\footnote{We make the pseudoscientific videos detection classifier publicly available~\cite{pseudosciencerepository}}
    that can detect pseudoscientific content with $0.79$ accuracy.
    Although our classifier outperforms several baselines, ultimately, its accuracy reflects the subjective nature of pseudoscientific vs. scientific content classification on YouTube. It is also evidence of the hurdles in devising models that automatically discover pseudoscientific content. 
    Nonetheless, we argue that our classifier is only the first step in this direction and can be further improved; overall, it does provide a meaningful signal on whether a video is pseudoscientific \textbf{(RQ1)}.}

    \item \revision{To the best of our knowledge, we present the first study focusing on multiple health-related pseudoscientific topics on YouTube pertaining to the COVID-19 pandemic. Inspired by the literature, we develop a complete framework that allows us to assess the prevalence of pseudoscientific content on various parts of the YouTube platform (i.e., homepage, search results, video recommendations) while accounting for the effect of a user's watch history. Specifically, we build user profiles with distinct watch histories and use them to perform experiments that allow us to quantify the prevalence of pseudoscientific content on various parts of the YouTube platform. We find that the watch history of the user substantially affects what videos are suggested to the user and that it is more likely to encounter pseudoscientific videos in the search results (i.e., when searching for a specific topic) than in the video recommendations section or the homepage of a user. We also perform experiments using the YouTube Data API finding that the results using the API are similar to those of using a non-logged-in user profile with no watch history (using a browser); this indicates that recommendations returned using the API are not subject to personalization \textbf{(RQ2 and RQ4)}.}
    
    \item We provide a set of resources\footnote{We make the pseudoscientific videos ground-truth dataset~\cite{pseudosciencedataset}, source code, and crawlers of our experiments~\cite{pseudosciencerepository} publicly available} to help researchers shed additional light on YouTube's recommendation algorithm and its potential influence. In particular, the ability to run experiments while taking into account the users' viewing history will be beneficial to researchers focusing on demystifying YouTube's recommendation algorithm, irrespective of the topic of interest \textbf{(RQ2)}. 
\end{enumerate}

\section{Research Papers}
The work presented in this thesis is already published or is currently under review in peer-reviewed journals and conferences. Specifically, some aspects of our work (in collaboration with other researchers and academics) appear in the following research articles:
\begin{itemize}
    \item Papadamou, K., Papasavva, A., Zannettou, S., Blackburn, J., Kourtellis, N., Leontiadis, I., Stringhini, G., and Sirivianos, M., 2020, May. Disturbed YouTube for Kids: Characterizing and Detecting Inappropriate Videos Targeting Young Children. In Fourteenth International AAAI Conference on Web and Social Media.~\cite{papadamou2020disturbed} \textbf{(Spotlight Paper Award)}.
    
    \item Papadamou, K., Zannettou, S., Blackburn, J., De Cristofaro, E., Stringhini, G., and Sirivianos, M., 2021. "How over is it?" Understanding the Incel Community on YouTube. 
    Under major revision.~\cite{papadamou2020understanding}
    
    \item Papadamou, K., Zannettou, S., Blackburn, J., De Cristofaro, E., Stringhini, G., and Sirivianos, M., 2021. "It is just a flu": Assessing the Effect of Watch History on YouTube's Pseudoscientific Video Recommendations. 
    Under major revision.~\cite{papadamou2020just}
\end{itemize}

\section{Thesis Organization} 
The remainder of this thesis is organized as follows.
In Chapter~\ref{chapter:background} we provide essential background information, while Chapter~\ref{chapter:related} describes prior work on: 1) inappropriate content for children; 2) harmful and other malicious activity; 3) misinformation and pseudoscientific content; and 4) recommendation algorithms and their auditing. In Chapter~\ref{chapter:disturbed_youtube} we present our work on characterizing and detecting inappropriate videos targeting young children. Chapter~\ref{chapter:incels_youtube} describes our work on characterizing hateful and misogynistic content on YouTube through the lens of the Incel community. In Chapter~\ref{chapter:pseudoscience_youtube} we present our work on assessing the effect of watch history on YouTube's pseudoscientific video recommendations. %
Finally, we discuss our findings and conclude in Chapter~\ref{chapter:conclusions}.

\chapter{Background} 
\label{chapter:background}
In this chapter, we provide essential background information regarding YouTube and other Web communities from which we collect and analyze data.

\section{YouTube}
\label{subsec:youtube_background}
In this section, we briefly describe YouTube, its recommendation algorithm, and the methodology for collecting data from it.

\subsection{YouTube}

\descr{General.}
YouTube is the leading video-sharing platform on the Web\footnote{According to Statista as of 2019 YouTube has over 2B monthly active users worldwide (\url{http://bit.ly/statista-youtube-viewers})}, owned by Google. %
YouTube does not produce media content itself, instead, it focuses on user-generated content by allowing users or content creators to create and upload content on the platform and YouTube deals with the efficient distribution of this content to its billion of users.
According to Burgess et al.~\cite{burgess2018youtube} YouTube has also been seen as a spearhead of participatory culture by allowing everyone from industry experts, to governments, citizens' groups, or any other person to make their voices heard via YouTube.
When uploading a video on the platform, among other things, content creators can set a title, a description, define multiple tags, which are keywords that describe a particular video, and they can even upload a transcript for its audio.
Users and content creators can also create their YouTube channel where they upload their videos. 
Other users of the platform can subscribe to these channels and receive notifications when a new video has been uploaded.
Furthermore, anyone over 13 years old can search and watch videos on YouTube.
While watching videos, users can also like or dislike them, share them to other social networks, or post a comment below them.
Other users can also reply to those comments in a structured manner (e.g., reply to a specific comment or reply to a specific reply).

\descr{YouTube's Recommendation Algorithm.}
One of the main reasons for YouTube's success is its recommendation algorithm algorithm\footnote{According to YouTube's product chief, more than $70\%$ of what people watch on YouTube is determined by its recommendation algorithm (\url{https://www.cnet.com/news/youtube-ces-2018-neal-mohan/})}.
The recommendation algorithm is central to YouTube's user experience for three main reasons.
First, it determines what videos are presented to each user's homepage.
Second, it determines and ranks the videos when a user searches the platform.
Last,  while users watch videos on the platform, the recommendation system suggests ``up-next'' related videos appearing in the sidebar next to the currently playing video.

In its early stages, YouTube was suggesting videos based on their number of views (a.k.a. clicks), while in 2012 the platform announced an update to the discovery system designed to identify videos that keep users engaged based on the watch time (a.k.a. view duration).
While not being fully transparent on how the recommendation algorithm operates, YouTube released a paper describing how deep neural networks are used in the architecture of its recommendation system~\cite{covington2016deep}. At the same time, they state that YouTube recommends videos based on user engagement and perceived satisfaction.
\revision{
More precisely, YouTube determines the ranks of the videos recommended to users based on various user engagement (e.g., a user clicks, degree of engagement with recommended videos, etc.) and satisfaction metrics (e.g., likes, dislikes, etc.).
Aiming to increase the time that a user spends watching a particular video, the platform also considers various other user personalization factors, such as demographics, location, or the watch history of the user, etc.~\cite{zhao2019recommending}.
}

\descr{Monetization.}
YouTube established a very effective business model on top of its users and content creators by serving advertisements on videos.
At the same time, content creators receive a small percentage of the income received by YouTube for serving advertisements on their videos (revenue sharing).
YouTube chooses the advertisements to be shown on a specific video automatically based on context like the video's metadata and whether the content is advertiser-friendly. At the same time, content creators can enable or disable monetization for their videos at any time.

\descr{Moderation and De-monetization.}
YouTube moderates videos that violate the platform's community guidelines~\cite{youtubecommunityguidelines} by removing them.
According to the community guidelines, YouTube moderates the following types of content and activity: 
1) spam and deceptive practices like fake engagement, impersonation, etc.; 
2) sensitive content related to child safety, nudity, suicide, and self-injury, etc.;
3) violent and dangerous content;
4) hate speech; etc.
Due to the sheer scale, YouTube heavily relies on its users and machine learning to detect and flag problematic content, and on YouTube employees to manually inspect and remove them.
However, YouTube has been lately criticized for promoting borderline content.
Such videos do not quite violate YouTube's community guidelines but contain harmful or misleading content.
According to YouTube's CEO, the platform punishes videos with borderline content by either reducing exposure to them or de-monetizing such videos (removing advertisements)~\cite{youtubeceo2020changes}.

\descr{YouTube Data API.}
To collect data from YouTube we mainly use the official YouTube Data API~\cite{youtubedataapi} to create the datasets we analyze in this thesis.
While the YouTube Data API allows developers to incorporate YouTube features into their own web applications, it also enables obtaining the metadata of all videos publicly available on YouTube.
More precisely, we use the YouTube Data API to obtain the following metadata for each video: 1) title, description, and duration; 2) a set of tags defined by the uploader; 3) video category; 4) thumbnail information; 5) video statistics such as the number of views; 6) video transcript; 7) top related videos; and 7) top comments, defined by YouTube’s relevance metric, and their replies.

\subsection{YouTube Kids}

\descr{General.} 
In an attempt to offer a safer online experience for its young audience, YouTube launched the YouTube Kids application~\cite{youtubekids}, which equips parents with several controls over what their children are allowed to watch on YouTube. 
Such controls include setting content levels and disabling the search functionality.
YouTube Kids is designed specifically for young children aged 5 or younger and it has been promoted as "a world of learning and fun, made just for kids." The YouTube Kids app includes both popular children's videos and diverse new content, delivered in a way that is usable for children.
According to the company, videos marked as age-restricted are not available on the YouTube Kids app.
Age-restricted videos usually contain content that is not suitable for anyone under 18 years old but does not violate YouTube's community guidelines.

\descr{Data Collection.} 
We note that in this thesis we collect and analyze videos only from YouTube and not from YouTube Kids. 
This is because YouTube does not provide an open API for collecting videos that appear on YouTube Kids.

\section{Reddit}
\label{subsec:reddit_background}

In this section, we briefly describe Reddit and the data collection methodology we used.

\descr{General.}
Reddit is called the "front page of the Internet" and is a popular social news aggregator, web content rating, and discussion platform. 
Reddit can also be viewed as a collection of forums where users can share news and content.
Users can create discussion threads, also called "submissions", by posting a URL along with a title.
Other users can comment or reply below in a structured manner (e.g., reply to submission or reply to specific reply).
The popularity of content within the platform is determined via a voting system.
Users can up-vote or down-vote each comment or submission, hence a score can be calculated for each one. 
Submissions and comments with higher scores appear on top of submissions and comments with a lower score. 
In addition, a user-based score called "karma" summarizes the scores of all user comments and submissions.
Last, the community structure on Reddit is not defined by a friendship/follower relation like other social networks.
Instead, a Reddit user can list another user as a friend but this does not affect the structure or use of the platform.

\descr{Subreddits.}
Reddit consists of more than a million communities known as "subreddits".
Subreddits are created from users of the platform and this has lead to a plethora of communities discussing a wide variety of topics ranging from generic entertainment topics (e.g., video games, cartoons, etc.), to politics, men's issues, and even meta-communities that summarize interactions of users on other subreddits/social networks.
Reddit's administrators regularly monitor subreddits and they remove them when they share extremely inappropriate or offensive content.
Several subreddits have been removed/banned from the platform in the last few years.
For instance, several subreddits have been removed for bullying and harassment and for promoting rape and suicide (e.g., /r/Incels and /r/Braincels), or for promoting conspiracy theories (e.g., /r/pizzagate and /r/greatawakening for promoting the Pizzagate~\cite{reddit2016pizzagate} and QAnon~\cite{reddit2018qanon} conspiracy theories, respectively).

\descr{Data Collection.}
To obtain the submissions and comments of the various subreddits we consider in this thesis, we use Pushshift~\cite{baumgartner2020pushshift}. 
Pushshift is a six-year-old platform that collects all the submissions and comments posted on Reddit and exposes them to the research community via an open REST API.
Specifically, Pushshift provides all the Reddit data since the platform's beginning in separate monthly dumps that someone can download and process.
We use Pushshift instead of the official Reddit API because it is much easier to query and retrieve historical Reddit data. 
At the same time, it also provides extended functionality by providing full-text search against comments and submissions and has larger query limits.

\subsection{Remarks}
In this section, we presented the data sources we used in this thesis. We selected them for specific reasons.
First, we elect to focus our efforts on YouTube because it is the most popular video-sharing platform worldwide.
This enables us to analyze and understand various types of inappropriate content on a large scale.
Second, we select YouTube mainly because of anecdotal evidence suggesting that its recommendation algorithm promotes offensive and dangerous content, and even that it helps radicalize users~\cite{carneyoutubeconspricacies, mozillaregrets, ribeiro2020auditing}. 
Last, we avoid using other popular social networks and video-sharing platforms like Facebook and its video platforms mainly due to limits imposed on their APIs by the company itself, hence constituting the task of obtaining data non-straightforward.
For more details on why we use Reddit as a data source in some of our lines of work please see Section~\ref{subsec:disturbed_data_collection} and Section~\ref{subsec:incels_data_collection}.

\chapter{Literature Review}
\label{chapter:related}

In this chapter, we provide an extensive literature review of prior work that focuses on problematic content and activity on YouTube and other video-sharing platforms on the Web. We review the following lines of work: 
1) inappropriate content targeting young children;
2) harmful, hateful, and other malicious activity;
3) misinformation and pseudoscientific content;
4) recommendation algorithms and audits; and 
5) various other relevant studies.

\section{Inappropriate Content for Children}
\label{sec:literature_inappropriate_children}

\subsection{Characterizing Inappropriate Content for Children}
Previous work focuses on characterizing and extracting meaningful insights about inappropriate content for children by analyzing data obtained from various video-sharing platforms and other types of Online Social Networks (OSNs). Once the use of social media became one of the most common activities of children and adolescents, the research community started investigating the possible implications. O'Keeffe et al.~\cite{o2011impact} focus on the impact of online social networks and video-sharing platforms such as Facebook, Twitter, and YouTube on children, adolescents, and families in general.
They conclude that it is important for parents to become aware of the nature of such platforms, and highlight that parents must monitor their children while using these platforms for potential problems such as cyberbullying, sexting, and exposure to inappropriate content.
Elias et al.~\cite{elias2017youtube} aim at profiling toddlers who watch YouTube videos, based on the child, parent, and family-related characteristics.
They interview 289 parents of toddlers aged 18-36 months and they find that watching videos on YouTube has become normative behavior among toddlers.
They also find that online viewing is deeply integrated into the daily routine of parents and that YouTube and other video-sharing platforms have become a tool for them to fulfill a wide range of their child-rearing needs. Chen et al.~\cite{chen2013children} are the first to study the appropriateness of advertising content in mobile applications from a children’s online safety perspective.
Among other things, they find that free applications mainly designed for kids contain advertisements, and a substantial portion of those advertisements contain inappropriate content. They suggest that a maturity monitoring mechanism is needed to detect or filter inappropriate advertising in children's applications.

Next, Izci et al.~\cite{izci2019youtube} survey research examining young children’s use of YouTube and YouTube Kids, as well as experts and parental concerns about children's digital media use.
Burroughs~\cite{burroughs2017youtube} focus on the relationship between the YouTube Kids application and the everyday viewing patterns and lives of young children.
He claims that YouTube and YouTube Kids’ recommendation algorithms consider infants as consumers, and as such, they are labeled as “algorithmic infants.” For example, if they like watching toy car videos, similar videos including toy cars appear on their screen as a result of the algorithm.
Other scholars also focus on YouTube videos targeting young children, and mainly on videos that feature controversial and inappropriate themes.

Craig et al.~\cite{craig2017toy} examine toy unboxing videos in social media and the concerns around these videos.
Toy-unboxing videos include other children’s or adults’ reviews of a set of objects inside a box.
Craig et al. provide a brief history of this phenomenon and describe how these videos represent forms of creator labor and operate within the structural and material interests of social media entertainment.
They also provide a detailed discussion of YouTube toy unboxing videos from a media regulation and economics perspective.
Nicoll et al.~\cite{nicoll2018mimetic} also focus on toy unboxing videos on YouTube.
They perform a content analysis of 100 toy unboxing videos and analyze their genre features. %
They categorize those features into five categories: 1) genre; 2) product; 3) narration; 4) production; and 5) branding, to analyze variations of expertise, professionalism, and promotions across the genre.
They find that amateur child unboxers mimic the production and branding strategies of the "professional" channels that often produce a semblance of playful amateur authenticity.
Jaakkola~\cite{jaakkola2020vernacularized} investigates the toy review genre as present in YouTube videos targeting children aiming to understand the forms and functions of toy reviews in the contexts of YouTube's political economy, branding, commercialization, and regulation.
She analyzes the strategies of 180 videos across 35 toy review channels finding that these channels produce repetitive content, which they call "kidbait", and that they employ creative aims and strategies to convince adults of the benefits of watching.
She concludes that toy reviews are a complex hybrid genre mediating children's commodities and play culture and that more attention should be dedicated to the ethical principles of their production.

Ara{\'u}jo et al.~\cite{araujo2017characterizing} study the audience profiles and comments posted on YouTube videos in popular children-oriented channels, and conclude that children under the age of 13 use YouTube and are exposed to advertising, inappropriate content, and privacy threats.
Paolillo et al.~\cite{paolillo2020youtube} focus on child-oriented YouTube videos that feature violent, disturbing, or otherwise inappropriate content. They collect and characterize children's videos published on YouTube between 2016 and 2018.
They code the collected videos for a variety of production and content features and then perform a cluster analysis.
Interestingly, they find that the use of branded materials takes place more frequently in the context of toy play and unboxing, while for other types of content they note that they do not find videos with inappropriate content in the analyzed sample.

\subsection{Detection and Containment of Inappropriate Content for Children}
Detecting inappropriate or disturbing content tailored toward young viewers and preventing its spread is not a straightforward task.
This is because such videos usually feature popular cartoon characters, like Spiderman, Mickey Mouse, etc., and include an innocent thumbnail aiming at tricking the toddlers and their custodians. 
In Table~\ref{tab:detection_containment_inappropriate_children_studies_overview} we summarize the studies that aim to solve the problem by detecting and containing inappropriate content targeting young children on YouTube.
Most studies try to solve the problem with handcrafted features and conventional machine learning techniques.
Recently, leveraging recent advancements in deep learning, the research community used deep neural networks to detect inappropriate content for children. In addition, we report studies that proposed additional parental controls as a way to prevent children from accessing inappropriate content on YouTube and other OSNs. Finally, we also report a variety of other methodologies and techniques that have been proposed for the detection and containment of inappropriate content for children.

\begin{table}[t!]
\footnotesize
\centering
\begin{tabular}{crrr}
\toprule
\textbf{Platform} & \textbf{Machine Learning} & \textbf{Parental Control Tools} & \textbf{Other Methods/Algorithms}  \\
\midrule
\multirow{6}{*}{\begin{tabular}[c]{@{}l@{}}YouTube\end{tabular}} %
& Kaushal et al.~\cite{kaushal2016kidstube} & \multirow{6}{*}{\begin{tabular}[c]{@{}r@{}}Buzzi~\cite{buzzi2011children}\end{tabular}} & \multirow{6}{*}{\begin{tabular}[c]{@{}r@{}}Alshamrani~\cite{alshamrani2020hiding}\\Reddy et al.~\cite{reddy2021development}\\St{\"o}cker et al.~\cite{stocker2020riding}\end{tabular}} \\
& Ishikawa et al.~\cite{ishikawa2019combating} & & \\
& Singh et al.~\cite{singh2019kidsguard} & & \\
& Eickhoff et al.~\cite{eickhoff2010identifying} & & \\
& Han et al.~\cite{han2020discovery} & & \\
& Alshamrani~\cite{alshamrani2020detecting} & & \\
\midrule
\multirow{2}{*}{\begin{tabular}[c]{@{}l@{}}YouTube Kids\end{tabular}} 
& \multirow{2}{*}{\begin{tabular}[c]{@{}l@{}}Tahir et al.~\cite{tahir2019bringing}\end{tabular}} & \multirow{2}{*}{\begin{tabular}[c]{@{}l@{}}-\end{tabular}} & Alghowinem~\cite{alghowinem2018safer} \\
& & & Neumann et al.~\cite{neumann2020young}\\
\midrule
\multirow{4}{*}{\begin{tabular}[c]{@{}l@{}}Other\end{tabular}} & \multirow{4}{*}{\begin{tabular}[c]{@{}l@{}}Wehrmann~\cite{wehrmann2018adult}\end{tabular}} & \multirow{4}{*}{\begin{tabular}[c]{@{}l@{}}Thierer~\cite{thierer2009parental}\end{tabular}} & Tsirtsis et al.~\cite{tsirtsis2016cyber}\\
& & & Luo et al.~\cite{luo2020automatic}\\
& & & Charalambous et al.~\cite{charalambous2020privacy}\\
& & & Ybarra et al.~\cite{ybarra2009associations}\\
& & & Parmaxi et al.~\cite{parmaxi2017safety}\\
\bottomrule
\end{tabular}
\caption{Studies that focus on the detection and containment of inappropriate content for children. The reported studies are separated based on their main methodology, as well as the considered OSNs.}
\label{tab:detection_containment_inappropriate_children_studies_overview}
\end{table}

\descr{Machine Learning.}
Kaushal et al.~\cite{kaushal2016kidstube} focus on the characterization and detection of unsafe content for children and its promoters on YouTube.
For the detection of unsafe content on videos, they propose a machine learning classifier based on Convolutional Neural Networks (CNNs). The classifier considers video frames, while for the detection of promoters of unsafe content they propose another approach based on the supervised classification that considers a set of video-, user-, and comment-level features. The proposed approaches can detect unsafe content on YouTube videos targeted at children and the promoters of such content with $85.7\%$ accuracy. Furthermore, their analysis of child-oriented videos reveals that unsafe content promoters are less popular and less engaging compared with all other users. They also find that unsafe content turns up very close to safe content and unsafe content promoters form very tight-knit communities with other users.
Ishikawa et al.~\cite{ishikawa2019combating} study the Elsagate~\cite{elsagatephenomenon} phenomenon on YouTube and they propose a deep learning model for detecting Elsagate-related content on YouTube that is based on deep convolutional neural networks combined with static (raw video frames) and motion (MPEG motion vectors) video information.
They use transfer learning to extract features from videos and they train the proposed model on a dataset of 3K videos, finding that it can detect Elsagate-related videos with $92.6\%$ accuracy. 
They also note that the proposed solution is compatible with mobile platforms. 
Similarly, Singh et al.~\cite{singh2019kidsguard} focus on the detection of child unsafe content.
This work is motivated by the fact that malicious video uploaders typically limit the child's unsafe content to only a few frames in the video to evade moderation.
They propose an LSTM-based deep neural network that is based on auto-encoders and a VGG16 CNN~\cite{zhang2015accelerating} to build a fine-grained detection method named KidsGUARD. 
The goal of this method is to prevent kids from experiencing unsafe content by detecting sparsely present child unsafe content in a video. 

Tahir et al.~\cite{tahir2019bringing} focus on the problem of inappropriate content targeting young children specifically on YouTube Kids.
They collect, manually review, and analyze 5K videos available on YouTube Kids finding more than 1K of them to contain fake, explicit, or violent content. Then, using the curated dataset they develop a deep learning architecture that can detect inappropriate videos with $90\%$ accuracy.
They conclude by stating that the proposed system can be successfully applied to various types of animations, cartoons, and CGI videos to detect any type of inappropriate content within them. Eickhoff et al.~\cite{eickhoff2010identifying} also propose a binary classifier based on the non-audio-visual data of a video, for identifying suitable YouTube videos for children. In a similar context, Wehrmann~\cite{wehrmann2018adult} propose ACORDE, a deep learning architecture for adult content detection in videos that is based on convolutional neural networks and LSTM recurrent networks.

Han et al.~\cite{han2020discovery} focus on the Elsagate phenomenon with a method for detecting Elsagate-related inappropriate YouTube videos based on Sparse Linear Discrimination (LSD), which can detect the violent scenes included in videos.
Alshamrani~\cite{alshamrani2020detecting} focuses on detecting and measuring the exposure of young children to inappropriate comments posted under YouTube videos targeting this demographic. He builds a data collection and processing pipeline and uses it to collect 3.7 million comments posted on 10K videos targeting young children on YouTube. Then, he creates an ensemble of video deep neural networks that classify comments into toxic, obscene, threat, insult, and identity hate. Using this ensemble of classifiers, he measures the exposure of each comments category by different age groups finding that children between 13-17 comprise the age group that is the most exposed to the inappropriate comments, followed by the 6-8 age group.

\descr{Parental Control Tools.}
Several studies assess or propose parental controls to monitor the use of YouTube and other social networks by young children, thus preventing them from watching inappropriate content. Thierer~\cite{thierer2009parental} provides an extensive survey and collection of useful resources and parental control tools for the protection of minors on the Web. Buzzi~\cite{buzzi2011children} focuses on pornography and inappropriate content on YouTube targeted at young children. Using an example of a video that features a famous Disney cartoon but its original audio was substituted with pornographic audio, she analyzes the effectiveness of YouTube's user interface for reporting inappropriate content.
She concludes by suggesting the addition of extra parental controls and indicators on YouTube to prevent children from accessing inappropriate content, %
such as an indication of whether a given video is suitable for all or adults only, before allowing the user to access the video. 

In another study, Charalambous et al.~\cite{charalambous2020privacy} propose a privacy-preserving architecture for the protection of adolescents in OSNs, called Cybersafety Family Advice Suite (CFAS). The proposed system includes virtual guardian avatars that provide cybersafety advice to children while preserving their privacy towards their custodians and towards the advice tool itself. The tool aims to prevent children's exposure to numerous risks and dangers while using Facebook, Twitter, and YouTube on a browser. Last, Ybarra et al.~\cite{ybarra2009associations} provide an extensive study conducted with teens (10-17 years old) and one of their custodians, in which they investigate the consequences of using parental control tools and other software. They find that techniques for preventing access to unsolicited inappropriate content, such as filtering, blocking, and monitoring software, were associated with a considerable decrease in unwanted exposure to sexual material.

\descr{Other Methodologies, Algorithms, and Systems}
Tsirtsis et al.~\cite{tsirtsis2016cyber} review existing work on the internet activity and the motivation of use by young children and list the identified risks and threats for young children who use the Web and OSNs. They also present a systematic process for designing and developing modern and state of the art techniques to protect
minors against the identified risks and dangers.
Alghowinem~\cite{alghowinem2018safer} investigates the safety of the YouTube Kids application and proposes an advanced content filtering approach using automated video and audio analysis as an extra layer for kids' safety.
The proposed solution is based on the thin-slicing theory~\cite{slepian2014thin} for checking in real-time whether the videos in YouTube Kids have problematic images or sounds. Specifically, this filter randomly cuts the clip into several 1-s slices and then extracts images and audio texts from each slice. Motivated by the increased popularity of YouTube and YouTube Kids among young children, Neumann et al.~\cite{neumann2020young} discuss the concerns raised by YouTube viewing on early childhood learning and development. They also discuss the risks of children's exposure to commercial advertisements and disturbing videos, and they propose that the key factors that should be considered when selecting YouTube videos for kids are age-appropriateness, content quality, video design features, and potential for the video to support learning outcomes. They conclude with some practical strategies that parents should adopt.

Alshamrani~\cite{alshamrani2020hiding} investigates the exposure of young children to malicious content in comments posted on YouTube videos targeting this demographic. They collect 4M comments posted on children's YouTube videos and look for URLs embedded in these comments that contain inappropriate topics for children, as well as the interactions of the viewers with these comments.
They find an alarming number of inappropriate URLs embedded in comments available for children and young users, and a high chance of kids exposure to them since the average number of views on videos containing such URLs is 48M.
Their findings indicate that monitoring the URLs provided within the comments of videos targeting young children is of vital importance, and would limit children's exposure to problematic content.

Focusing on the same demographic, Luo et al.~\cite{luo2020automatic} propose an automatic content inspection and forensics framework to identify Android applications that contain content not suitable for children under 12 years old (such as violence, pornography, gambling, and drugs).
In addition, their proposed framework provides evidence to inform users on why the certain inspected application is judged as unsuitable.
They evaluate the proposed framework on 70 android applications for children showing that it can identify improper applications for children with $85.7\%$ accuracy, while they also find $40\%$ of evaluated applications to contain inappropriate content for children.
Reddy et al.~\cite{reddy2021development} propose a proactive approach for video-sharing platforms like YouTube to handle situations where a child might accidentally open a video that includes problematic content.
The proposed methodology incorporates a face unlock system for every video that is clicked on YouTube that is based on age detection and a sentiment analysis model. Once the age detection process is successful, the detected age is passed to the sentiment analysis model, which then calculates the percentage of positive and negative sentiments of the requested video and decides whether the given video should be viewed by the user or not.

St{\"o}cker et al.~\cite{stocker2020riding} focus on contextually inappropriate content on YouTube.
They define contextually inappropriate content as content that in a specific context or when targeted at a different audience may be entirely harmful to the viewer. They focus their efforts on the recommendation algorithm and they perform a simulation to assess whether the recommendation algorithm steers users towards problematic content. Their findings indicate that the autoplay feature of YouTube is a problematic feature.
They conclude that completely preventing inappropriate recommendations is technically infeasible in the current context, and they propose some measures that can mitigate the problem of inappropriate recommendations, such as improving the quantity and quality of the feedback users can provide.

Last, Parmaxi et al.~\cite{parmaxi2017safety} focus on the research development of safety and security in Web 2.0 learning environments, and provide a review of web-based tools and applications that attempt to address security and privacy issues in OSNs.

\subsection{Inappropriate Content for Children - Remarks}
\revision{
The main findings from the review of the prior work on inappropriate content for children are that such content appears frequently on the platform, and  reports on the presence of inappropriate videos on YouTube increases over the years.
At the same time, we infer that machine learning techniques can assist in identifying inappropriate content. 
However, none of the available research in the field has focused on inappropriate videos that explicitly targets toddlers, thus it is not clear whether any of the proposed classifiers can really detect such videos. 
Finally, none of the above studies analyzed the problem on a large scale on YouTube, neither they quantify how likely it is for an inappropriate video to be served to a toddler who casually browses the platform \textbf{(RQ1 and RQ4)}.
}

\section{Misogyny and Other Harmful Activity}
\label{sec:literature_harmful_malicious_activity}
Online misogyny and hate against women, as well as other forms of harmful activity, are on the rise.
Social media platforms became the number one place for psychological violence -- in the form of sexist and misogynistic content -- and many other types of harmful activity.
In this section we review prior work that aims at characterizing, detecting, and containing misogyny, hate, and other types of malicious activity on YouTube and other social media platforms.
Table~\ref{tab:misogyny_harmful_activity_studies_overview} reports the reviewed work for each type of harmful activity as well as the platform examined in each study.

\begin{table}[t!]
\footnotesize
\centering
\begin{tabular}{crrr}
\toprule
\textbf{Platform} & \textbf{Misogyny} & \textbf{Hateful \& Harmful Activity} & \textbf{Spam \& Malicious Activity}  \\
\midrule
\multirow{10}{*}{\begin{tabular}[c]{@{}c@{}}YouTube\end{tabular}} &
\multirow{10}{*}{\begin{tabular}[c]{@{}r@{}}Pratana~\cite{pratama2018identifying} \textbf{(D.)}\\Zahir et al.~\cite{zahir2020arabic} \textbf{(D.)}\\Wotanis et al.~\cite{wotanis2014performing} \textbf{(A.)}\\Tucker-McLaughlin~\cite{tucker2013youtube} \textbf{(A.)}\\D{\"o}ring et al.~\cite{doring2019male} \textbf{(A.)}\end{tabular}}
& Sureka et al.~\cite{sureka2010mining} \textbf{(D.)} 
& \multirow{10}{*}{\begin{tabular}[c]{@{}r@{}}Chowdury et al.~\cite{chowdury2013data} \textbf{(D.)}\\Sureka et al.~\cite{sureka2011mining} \textbf{(D.)}\\Bulakh et al.~\cite{bulakh2014identifying} \textbf{(D.)}\\Benevenuto et al.~\cite{benevenuto2011practical} \textbf{(D.)}\\Chaudhary et al.~\cite{chaudhary2013contextual} \textbf{(D.)}\\Ezpeleta et al.~\cite{ezpeleta2018mood} \textbf{(D.)}\\Zannettou et al.~\cite{zannettou2018good} \textbf{(D.)}\end{tabular}} \\
 & & Agarwal et al.~\cite{agarwal2014focused} \textbf{(D.)} & \\
 & & Giannakopoulos et al.~\cite{giannakopoulos2010multimodal} \textbf{(D.)} & \\
 & & Aggarwal et al.~\cite{aggarwal2014mining} \textbf{(D.)} & \\
 & & Mariconti et al.~\cite{mariconti2019you} \textbf{(D.)} & \\
 & & Jiang et al.~\cite{jiang2019bias} \textbf{(A.)} & \\
 & & Ottoni et al.~\cite{ottoni2018analyzing} \textbf{(A.)} & \\
 & & Moor et al.~\cite{moor2010flaming} \textbf{(A.)} & \\
 & & Alshamrani et al.~\cite{alshamrani2020investigating} \textbf{(A.)} & \\
 & & Yusha'u~\cite{yusha2015extremism} \textbf{(A.)} & \\
\midrule
BitChute & - & Trujillo et al.~\cite{trujillo2020bitchute} \textbf{(A.)} & - \\
\midrule
\multirow{3}{*}{\begin{tabular}[c]{@{}c@{}}Reddit\end{tabular}} & 
Massanari~\cite{massanari2017gamergate} \textbf{(A.)} & 
\multirow{3}{*}{\begin{tabular}[c]{@{}r@{}}Chandrasekharan~\cite{chandrasekharan2017you} \textbf{(A.)}\\Ribeiro et al.~\cite{manoel_bans} \textbf{(A.)}\end{tabular}} & \multirow{3}{*}{\begin{tabular}[c]{@{}r@{}}-\end{tabular}}\\
 & Farell et al.~\cite{farrell2019exploring} \textbf{(A.)} & & \\
 & Ribeiro et al.~\cite{ribeiro2020pick} \textbf{(A.)} & & \\
\midrule
\multirow{6}{*}{\begin{tabular}[c]{@{}c@{}}Twitter\end{tabular}} & 
\multirow{6}{*}{\begin{tabular}[c]{@{}r@{}}-\end{tabular}} & Silva et al.~\cite{silva2016analyzing} \textbf{(D.)} & 
\multirow{6}{*}{\begin{tabular}[c]{@{}r@{}}McCord et al.~\cite{mccord2011spam} \textbf{(D.)}\\Wang~\cite{wang2010don} \textbf{(D.)}\\Potthast~\cite{potthast2016clickbait} \textbf{(D.)}\end{tabular}} \\
 & & Ribeiro et al.~\cite{ribeiro2018characterizing} \textbf{(D.)} & \\
 & & Chatzakou et al.~\cite{chatzakou2017measuring} \textbf{(A.)} & \\
 & & Founta et al.~\cite{founta2018large} \textbf{(D.)} & \\
 & & Davidson et al.~\cite{davidson2017automated} \textbf{(D.)} & \\
 & & Davidson et al.~\cite{davidson2019racial} \textbf{(D.)} & \\
\midrule
\multirow{5}{*}{\begin{tabular}[c]{@{}c@{}}Other\end{tabular}} & 
\multirow{5}{*}{\begin{tabular}[c]{@{}r@{}}Baele et al.~\cite{baele2019incel} \textbf{(A.)}\\Lilly~\cite{lilly2016world} \textbf{(A.)}\end{tabular}} & 
\multirow{5}{*}{\begin{tabular}[c]{@{}r@{}}Zannettou et al.~\cite{zannettou2018gab} \textbf{(A.)}\\Hosseini et al.~\cite{hosseini2017deceiving} \textbf{(A.)}\\Gr{\"o}ndahl et al.~\cite{grondahl2018all} \textbf{(D.)}\end{tabular}} & 
Cao et al.~\cite{cao2012aiding} \textbf{(D.)}\\
 & & & Chen et al.~\cite{chen2015misleading} \textbf{(D.)}\\
 & & & Chakraborty et al.~\cite{chakraborty2016stop} \textbf{(D.)}\\
 & & & Biyani et al.~\cite{biyani20168} \textbf{(D.)} \\
 & & & Anand et al.~\cite{anand2017we} \textbf{(D.)} \\
\bottomrule
\end{tabular}
\caption{Studies that focus on misogyny and other harmful activity. The reported studies are separated based on their main methodology, as well as the considered OSNs. (D.) stands for the particular type of harmful content and activity, and (A.) stands for analysis and measurement studies.}
\label{tab:misogyny_harmful_activity_studies_overview}
\end{table}

\subsection{Misogyny and Other Harmful Activity on YouTube}
While YouTube has revolutionized the way people discover and consume video content online, it has also enabled the spread of hateful content.
It is also evident that misogyny, hate, and other types of harmful and malicious activity have ballooned out of proportion on the platform~\cite{alexander2019youtubehate}.
A large body of previous work focused on understanding and detecting such activity on the platform.
In this section, we review the most relevant studies on misogynistic and hateful content, as well as other harmful activity on YouTube.

\descr{Misogyny on YouTube.} %
Pratana~\cite{pratama2018identifying} focuses on identifying sexist language in the YouTube comments section.
She collects and analyzes 420 comments on videos that become popular popularity and she finds $13\%$ of them to be sexist.
In a similar context, Zahir et al.~\cite{zahir2020arabic} propose an approach for the automatic detection and assessment of attitudes towards violence against women and women's rights, by analyzing YouTube comments written in Arabic.
To do this, they collect a set of videos featuring positive and negative options on violence against women and women's rights and use them to train multiple classifiers that detect violent comments against women.

Additional studies point to the need for a better understanding of misogynistic content on YouTube. 
Wotanis et al.~\cite{wotanis2014performing} analyze hostile and sexist comments on the platform.
They claim that a large number of YouTubers are male and that more negative feedback is given to females than male YouTubers.
Moreover, they examine which kinds of negative (including hostile and sexist) and positive video comments were addressed at Jenna Mourey and Ryan Higa, the two most prominent US YouTubers within the comedy genre.
They find that $18\%$ of the comments that Jenna Mourey received under her videos are negative, while her male counterpart Ryan Higa received only $4\%$.
This finding is in line with other studies.
Tucker-McLaughlin~\cite{tucker2013youtube} analyzes popular videos on YouTube and finds that roughly $25\%$ of the most-viewed videos on the platform include misogynistic language, violence, or both, while the primary actors in those videos are male.
Furthermore, D{\"o}ring et al.~\cite{doring2019male} build on the study of Wotanis et al.~\cite{wotanis2014performing} by empirically investigating male dominance and sexism on YouTube.
They perform quantitative content analysis and they find that male YouTubers dominate the platform.
More precisely, they find that among the top 100 most subscribed YouTube channels in nine different countries, female video producers were strongly underrepresented. By also analyzing 2.4K video comments, they find that female content producers are prone to receiving more negative and hostile comments. They conclude that female YouTubers possibly attract more negative comments only if they display their sexuality or address feminist topics, but not if they conform to gender role expectations.

\descr{Hateful and Other Harmful Activity on YouTube.} %
Machine learning techniques have been extensively used by the research community to detect hateful, toxic, extremist, and violent content on YouTube.
Sureka et al.~\cite{sureka2010mining} focus on the detection of extremism on YouTube,
They propose a semi-assisted system that uses data mining and social network analysis techniques to discover hateful YouTube videos.
The proposed system can be used by law enforcement and intelligence agencies to assist the detection of hate videos, users, and hidden communities in the ecosystem where hate speech flourishes.
Agarwal et al.~\cite{agarwal2014focused} formulate the problem of identification of malicious videos disseminating hatred against a particular religion, country, or person on YouTube as a search problem.
They present a focused-crawler approach which consists of three phases: a training profile collection, statistical model building, and a crawler biased towards retrieving hateful videos. The proposed approach can be used for identifying YouTube videos that promote hate and extremism.
Giannakopoulos et al.~\cite{giannakopoulos2010multimodal} use video, audio, and textual features of videos uploaded on YouTube to train a k-nearest classifier for detecting violent videos on the platform.
Aggarwal et al.~\cite{aggarwal2014mining} use video features for detecting videos violating privacy or promoting harassment.
Mariconti et al.~\cite{mariconti2019you} propose an ensemble of machine learning classifiers to predict, during upload time, whether or not a YouTube video will be targeted with hateful comments or dislikes by coordinated attacks.

Prior work has also characterized and measured hateful, toxic, and extreme content on YouTube.
Jiang et al.~\cite{jiang2019bias} investigate how channel partisanship and video misinformation affect comment moderation on YouTube, finding that comments are more likely to be moderated if the video channel is ideologically extreme.
Ottoni et al.~\cite{ottoni2018analyzing} perform an in-depth analysis of 7K videos and 17M video comments retrieved from alt-right channels on YouTube to understand possible issues related to hate, violence, and discriminatory bias. They investigate similarities and differences between user comments and video content and compare it to a baseline set using a three-layered approach, in which they analyze the use of language, the topics discussed, and implicit biases present in the texts. They conclude that the comments of a video are a better indicator for detecting alt-right videos when compared to the video’s title. Moor et al.~\cite{moor2010flaming} study the use of offensive language on YouTube using surveys of users of the platform.
Although many of the participants claimed that they do not use insulting, swearing, or otherwise offensive language, the study points that offensive language is common on YouTube. Interestingly, they also find that users are reluctant to upload videos on the platform because of the offensive comments that they may receive. They conclude that although some users use offensive language for entertainment, this type of language is more often intended to express disagreement or as a response to a perceived offense by others.

Alshamrani et al.~\cite{alshamrani2020investigating} investigate the correlation of different toxic behaviors, such as identity hate and obscenity, in users' interactions on popular videos on YouTube. They collect 7.3M comments and more than 10K news videos transcript, annotate some of them and use them to train an ensemble of classifiers that can identify toxic comments on YouTube with high accuracy. They then use the trained classifier to detect toxic comments under news videos posted on YouTube and they find that religion and crime-related news have the highest rate of toxic comments, while economy-related news has the lowest rate. In another study, Yusha'u~\cite{yusha2015extremism} study user comments under YouTube videos posted by popular news media organizations to report the Norwegian attacks by Anders Breivik. Focusing on Islamophobia, they use critical discourse analysis to understand the Islamophobic themes that emerged in the reporting of the story, and how the user-generated comments can contribute to the understanding of the rise of Islamophobia in the West. They conclude that there is a need to address the way stories about Muslims are reported in the West.

\descr{Spam and Other Malicious Activity on YouTube.} %
Machine learning techniques have been extensively used by the research community to detect and mitigate spam and other malicious activity on YouTube.
Chowdury et al.~\cite{chowdury2013data} explore various video attributes that may enable the detection of spam videos and their uploaders on YouTube. 
They collect and manually classify videos from YouTube into spam and legitimate videos, and use them to train multiple machine learning classifiers using a variety of video features to identify the most indicative ones. They find that features like the comment count and the average rating of a video are among the most indicative features for effective spam detection.
Sureka et al.~\cite{sureka2011mining} propose a method for automatically detecting comment spammers on YouTube based on mining the comment activity log of a user and extracting patterns that indicate spam behavior like the time interval between subsequent comments posted by a user who is likely to be a spammer. Using similar features, Bulakh et al.~\cite{bulakh2014identifying} characterize and identify fraudulently promoted YouTube videos.
They collect 3.3K fraudulently promoted videos and 500 bot profiles that promote them and they train supervised machine learning classifiers that can successfully detect fraudulently promoted videos and the malicious user profiles that promote them.

In the same context, Benevenuto et al.~\cite{benevenuto2011practical} propose two supervised classification algorithms to detect spammers, promoters, and legitimate YouTube users.
Also, to improve the performance of spam filtering on the platform,~\cite{benevenuto2011practical} test numerous approaches and propose a tool, based on Naive Bayes, that filters spam comments on YouTube. Chaudhary et al.~\cite{chaudhary2013contextual} use only video features, and propose a one-class classifier approach for detecting spam videos. They divide the problem into three sub-problems: 1) promotional video recognition; 2) pornographic or dirty video recognition; and 3) botnet uploader recognition.
For each sub-problem, they collect videos and perform a characterization, which reveals that discourse, temporal features, popularity-based features,
and time-based features can be used to predict the type of the video. Ezpeleta et al.~\cite{ezpeleta2018mood} perform mood analysis of comments posted on YouTube videos and use it as a feature in social spam filtering classifiers to show that mood information can improve detection accuracy.

Finally, Zannettou et al.~\cite{zannettou2018good} leverage deep learning techniques to detect videos on YouTube that use manipulative techniques to increase their views (i.e., clickbait). Specifically, they propose a semi-supervised model based on variational auto-encoders.  Their evaluation indicates that they can detect clickbait videos with acceptable performance and that YouTube’s recommendation engine does not suppress clickbait videos.

\subsection{Misogyny and Hateful Activity on the Rest of the Web} 
In one of our lines of work in this thesis, we focus on characterizing hateful and misogynistic content through the lens of the Incel Community.
Hence, in this section, we also review prior work that focuses on the Incel community, as well as other misogynistic and hateful communities on the Web.

\descr{Misogyny on the Web.} %
Massanari~\cite{massanari2017gamergate} performs a qualitative study of how Reddit's algorithms, policies, and general community structure enables, and even supports, toxic culture.
She focuses on the \#GamerGate and Fappening incidents, both of which had primarily female victims, and argues that specific design decisions make it even worse for victims. 
For instance, the default ordering of posts on Reddit favors mobs of users promoting content over a smaller set of victims attempting to have it removed.
She notes that these issues are exacerbated in the context of online misogyny because many of the perpetrators are extraordinarily techno-literate and thus able to exploit more advanced features of social media platforms.

Baele et al.~\cite{baele2019incel} study content shared by members of the Incel community in the online forum Incels.me, focusing on how support and motivation for violence resulting from their worldview. They find a link between the Incels' worldview and violence, and that violence within the Incel ideology is not only seen as acceptable but also as the only possible way to solve the crisis endpoint in which society is supposedly stuck. They conclude that the Incel community is an extremist ideology both in terms of its logic categorization and explanation. Farell et al.~\cite{farrell2019exploring} perform a large-scale quantitative study of the misogynistic language across the Manosphere on Reddit. 
They create nine lexicons of misogynistic terms to investigate how misogynistic language is used in 6M posts from Manosphere-related subreddits.
Jaki et al.~\cite{jaki2019online} study misogyny on the Incels.me forum, analyzing users’ language and detecting misogyny instances, homophobia, and racism using a deep learning classifier that achieves up to $95\%$ accuracy.

\revision{Furthermore, Ribeiro et al.~\cite{ribeiro2020pick} focus on the evolution of the broader Manosphere and perform a large-scale characterization of multiple Manosphere communities mainly on Reddit and six other Web forums associated with these communities.
They find that older Manosphere communities, such as Men's Rights Activists and Pick Up Artists, are becoming less popular and active. 
In comparison, newer communities like Incels and MGTOWs attract more attention. 
They also find a substantial migration of users from old communities to new ones, and that newer communities harbor more toxic and extreme ideologies.
In another study, Ribeiro et al.~\cite{manoel_bans} investigate whether platform migration of toxic online communities compromises content moderation.
To do this, they focus on two communities on Reddit, namely, /r/Incels and /r/The\_Donald, and use them to assess whether community-level moderation measures were effective in reducing the negative impact of toxic communities.
They conclude that a given platforms’ moderation measures may create even more radical communities on other platforms.}

Finally, Lilly~\cite{lilly2016world} focuses on the Manosphere, and using mixed-methods critical discourse analysis they analyze the discourse of the two primary subcultures of the Manosphere.
She presents a taxonomy of the communities that together form the Manosphere, while also highlighting the key ideas behind the Manosphere.
Her taxonomy considers the four most popular communities of the Manosphere, Men’s Rights Activists (MRA), Men Going Their Own Way (MGTOW), Pick Up Artists (PUA), and Involuntary Celibates (Incels).

\descr{Hateful and Other Harmful Activity on the Web.} %
A large body of prior work studied how to detect and measure hate speech on the Web.
Silva et al.~\cite{silva2016analyzing} perform a large-scale measurement study of the main targets of hate speech in Whisper and Twitter.
They do this they develop a methodology to identify hate speech on both these systems based on the sentences' structure by matching expressions towards something.
Trujillo et al.~\cite{trujillo2020bitchute} focus on BitChute, a video-sharing platform alternative to YouTube.
BitChute is part of an ecosystem of alternative, low content moderation platforms, including Gab, Voat, and 4chan.
Trujillo et al. perform a characterization of BitChute finding that the platform facilitates conspiracy theories and a high rate of hate speech, mostly anti-Semitic.
Interestingly, they also find that while some BitChute content promoters have been banned from other platforms, they continue to maintain profiles on YouTube and other mainstream social media platforms.
Zannettou et al.~\cite{zannettou2018gab} provide a large-scale characterization of Gab, an alternative social network of Twitter with low content moderation.
They collect and analyze 22M posts produced by 336K users between August 2016 and January 2018 finding that attracts alt-right users, conspiracy theorists, and other trolls.
They also find that hate speech is much more prevalent in Gab than in Twitter but lower than 4chan’s Politically Incorrect board.
Chandrasekharan~\cite{chandrasekharan2017you} focuses on the efficacy of banning as a moderation approach in OSNs.
In particular, they focus on Reddit and they analyze 100M posts and comments from two banned subreddits, /r/fatpeoplehate and /r/CoonTown.
They first create lexicons of hate speech and use causal inference methods to examine the usage of hate speech in other subreddits after the ban.
They find that ban worked for Reddit as that hate speech usage decreased on the platform, while other related subreddits saw no significant changes in hate speech usage.

The research community also focused on understanding, detecting, and measuring hateful content and activity on Twitter.
Ribeiro et al.~\cite{ribeiro2018characterizing} perform a characterization of hateful users on Twitter.
They collect and analyze a sample of 100K users on the platform finding that hateful/suspended users differ from normal/active ones in terms of their activity patterns, word usage, and network structure.
By exploiting Twitter’s users' connections network, they also find that a node embedding algorithm outperforms content-based approaches for detecting both hateful and suspended users.
Chatzakou et al.~\cite{chatzakou2017measuring} study the \#Gamergate controversy focusing on Twitter.
They perform a measurement analysis of a dataset of 340K unique users and 1.6M tweets to study the properties of these users, the content they post,
and how they differ from other users on the platform.
They find that users involved in the \#Gamergate controversy tend to be more engaged on Twitter having more friends and followers, and they tend to post tweets with negative sentiment and more hate attitudes compared to what they post to other users on the platform. 
Founta et al.~\cite{founta2018large} perform a large-scale characterization of various forms of abusive behavior on Twitter, including offensive, abusive, and hateful language.
They propose an incremental and iterative methodology that leverages the power of crowdsourcing to annotate a large collection of tweets with a set of abuse-related labels.

Furthermore, other studies highlight that detecting and measuring hateful content and harmful activity is hard, as hate speech is contextual and subjective in nature~\cite{schmidt2017survey}.
Davidson et al.~\cite{davidson2017automated} show that automatic hate speech detection in social media is not a trivial task.
They train a multi-class classifier to distinguish between tweets that contain hate speech, only offensive language, and those with neither.
By analyzing the predictions of their classifier they find that racist and homophobic tweets are more likely to be classified as hate speech but that sexist tweets are generally classified as offensive. 
They also find that machine learning algorithms struggle to identify tweets that express hate but do not include explicit hate keywords.
Hosseini et al.~\cite{hosseini2017deceiving} point to possible ways to deceive well-established APIs that are used to detect toxic language, like Google's Perspective API.
Specifically, they show how an adversary can subtly modify a highly toxic phrase in a way that the system assigns a significantly lower toxicity score to it.
Davidson et al.~\cite{davidson2019racial} point to racial bias in datasets used for hate speech and abusive language detection.
They train machine learning classifiers using five different sets of Twitter data annotated for hate speech and abusive
language and they find that such hate speech detection classifiers tend to predict tweets written in African-American English as abusive at a substantially higher rate due to racial bias in the datasets used to train them.
Gr{\"o}ndahl et al.~\cite{grondahl2018all} focus on the difficulty in automatically detecting hate speech in social networks.
They reproduce seven state-of-the-art hate speech machine learning classifiers proposed in prior work~\cite{davidson2017automated,wulczyn2017ex,zhang2018detecting,badjatiya2017deep} and they highlight their inability to generalize in unseen data as all of them perform well only when tested on the datasets used to train them.

Finally, Ribeiro et al.~\cite{manoel_bans} investigate whether platform migration of toxic online communities compromises content moderation focusing on /r/The\_Donald and /r/Incels communities on Reddit. 
They conclude that a given platforms’ moderation measures may create even more radical communities on other platforms.

\descr{Spam and Other Malicious Activity.}
Several studies focus on the detection of spam and other types of malicious activity on the Web using machine learning and other techniques.
Cao et al.~\cite{cao2012aiding} investigate the detection of fake accounts in large-scale social online services.
They propose SybilRank, a tool that relies on social graph properties to rank users according to their perceived likelihood of being fake.
They evaluate SybilRank on Tuenti, Spain's largest social network, and they find that almost $90\%$ of the 200K accounts that SybilRank flagged as most likely to be fake were indeed fake.
Others focus on spam detection in Twitter using machine learning.
McCord et al.~\cite{mccord2011spam} analyze user- and content-based features that are different between spammers and legitimate users and use them to
train four convention machine learning classifiers for spam detection on Twitter.
They find that the Random Forest classifier produces the best results from all the four evaluated classifiers and using the suggested features can achieve $95.7\%$ precision.
Wang~\cite{wang2010don} collects around 25K users, 500K tweets, and 49M follower/friend relationships from publicly available data on Twitter and uses them to train a Bayesian classification algorithm that can detect malicious distinguish the suspicious behaviors from normal ones on Twitter with $89\%$ precision.

Furthermore, several studies focus on the detection of clickbait on the Web using machine learning techniques.
Chen et al.~\cite{chen2015misleading} consider clickbait as a deception problem and propose the use of SVM and Naive Bayes classifiers for the automatic detection of clickbait on the Web.
Similarly, Chakraborty et al.~\cite{chakraborty2016stop} propose the use of SVM for the automatic detection of clickbait news articles.
They then build a browser extension that leverages the developed classifier to detect clickbait and warn the readers about the possibility of being baited while accessing different media sites.
The evaluation of the proposed clickbait detection approach shows that it can achieve $93\%$ accuracy in detecting and $89\%$ accuracy in blocking clickbait news articles.
In another study, Potthast~\cite{potthast2016clickbait} proposes the use of Random Forest for detecting clickbait tweets. 
The proposed model is trained based on 215 features and can detect clickbait tweets with $76\%$ precision and $79\%$ ROC-AUC.
Moreover, Biyani et al.~\cite{biyani20168} propose the use of Gradient Boosted Decision Trees for clickbait detection in news articles.
They also highlight that the degree of the informality of a webpage increases the likelihood of the specific webpage being clickbait. 
Last, Anand et al.~\cite{anand2017we} is among the first studies that leverage deep learning and propose a neural network architecture based on
Recurrent Neural Networks (RNNs) in conjunction with word2vec embeddings~\cite{mikolov2013distributed} for detecting clickbait news articles on the Web.

\subsection{Misogyny and Other Harmful Activity - Remarks}
\revision{
The main insights from the review of the existing research on misogyny and harmful activity on YouTube and the Web in general are:
1) machine learning techniques can be used to effectively detect misogyny, hate, and other types of malicious and harmful activity expressed in textual format. However, for some types of harmful content, like hate speech, the proposed classifiers fail to generalize or can be easily tricked by malicious actors; and 
2) Extreme misogynistic ideologies with their own lingo, such as the Incel ideology, have not been satisfactorily studied on video-sharing platforms like YouTube, despite evidence that the members of this community are extremely toxic against women and are prone to radicalization \textbf{(RQ1 and RQ3)}.
}

\section{Pseudoscientific Misinformation and Conspiracy Theories}
\label{sec:literature_misinformation_pseudoscience}

Online Social Networks and video-generated platforms are often fertile ground for the dissemination of misleading and potentially harmful information like conspiracy theories and pseudoscientific misinformation.  
YouTube and other social media platforms have struggled with mitigating the harm from this type of content. 
The difficulty is partly due to the sheer scale and also because of the deployment of recommendation algorithms~\cite{fastcompeny2019conspiracies}.
In this section, we review the most relevant studies on conspiracy theories, misinformation, and auditing of recommendation algorithms on YouTube and other social networks.
Table~\ref{tab:pseudoscientific_misinformation_conspiracy_studies_overview} reports the reviewed work for each type of misleading information as well as the considered platform of each study.

\begin{table}[t!]
\footnotesize
\centering
\begin{tabular}{crr}
\toprule
\textbf{Platform} & \textbf{Data Analysis/Measurements} & \textbf{Detection and Containment}  \\
\midrule
\multirow{12}{*}{\begin{tabular}[c]{@{}c@{}}YouTube\end{tabular}} 
& Tuters~\cite{tuters20207} & 
\multirow{12}{*}{\begin{tabular}[c]{@{}r@{}}
    Heydari et al.~\cite{heydari2019youtube}\\Chen et al.~\cite{chen2021visual}\\Faddoul et al.~\cite{faddoul2020longitudinal}\\Hou~\cite{Hou2019TowardsAD}\\Serrano et al.~\cite{serrano2020nlp}\\Hussain et al.~\cite{hussain2018analyzing}
\end{tabular}}\\
 & Allgaier~\cite{allgaier2016science} & \\
 & Hussein et al.~\cite{hussein2020measuring} & \\
 & Paolillo~\cite{paolillo2018flat} & \\
 & Landrum et al.~\cite{landrum2019differential} & \\
 & Landrum et al.~\cite{landrum2020third} & \\
 & Mohammed~\cite{mohammed2019conspiracy} & \\
 & Li et al.~\cite{li2020youtube} & \\
 & Knuutila et al.~\cite{knuutila2020covid} & \\
 & Donzelli et al.~\cite{donzelli2018misinformation} & \\
 & Machado et al.~\cite{machado2020natural} & \\
 & Kim et al.~\cite{kim2020effects} & \\
\midrule
\multirow{2}{*}{\begin{tabular}[c]{@{}c@{}}Twitter\end{tabular}} 
& Havey~\cite{havey2020partisan} & Rajdev et al.~\cite{rajdev2015fake}\\
 & Singh et al.~\cite{singh2020understanding} & M{\o}nsted et al.~\cite{monsted2019algorithmic}\\
\midrule
\multirow{2}{*}{\begin{tabular}[c]{@{}c@{}}Facebook\end{tabular}} 
& Frigerri et al.~\cite{friggeri2014rumor} & \multirow{2}{*}{\begin{tabular}[c]{@{}r@{}}Conti et al.~\cite{conti2017s}\end{tabular}} \\
 & Johnson et al.~\cite{johnson2020online} & \\
\midrule
\multirow{6}{*}{\begin{tabular}[c]{@{}c@{}}Other\end{tabular}} 
& Samory et al.~\cite{samory2018conspiracies} & \multirow{6}{*}{\begin{tabular}[c]{@{}r@{}}Shahsavari et al.~\cite{shahsavari2020conspiracy}\\Chakraborty et al.~\cite{chakraborty2020around}\end{tabular}}\\
 & Zannettou et al.~\cite{zannettou2017web} & \\
 & Albright~\cite{albright2016data} & \\
 & Ferrara et al.~\cite{ferrara2020misinformation} & \\
 & Zannettou et al.~\cite{zannettou2019towards} & \\
 & Jennings et al.~\cite{jennings2021lack} & \\
\bottomrule
\end{tabular}
\caption{Studies that focus on pseudoscientific misinformation and conspiracy theories. The reported studies are separated based on their approach, as well as the considered OSNs.}
\label{tab:pseudoscientific_misinformation_conspiracy_studies_overview}
\end{table}

\subsection{Pseudoscientific Misinformation and Conspiracy Theories on YouTube} 
The emergence of social media and video-sharing platforms enabled the rapid circulation of conspiracy theories and pseudoscientific misinformation.
Such theories have detrimental effects on the cognitive foundations of society, especially during crises like the COVID-19 pandemic.
Specifically, YouTube got openly accused for the circulation of pseudoscientific misinformation around vaccination and several other pseudoscientific topics, as well as for the circulation of several conspiracy theories since the beginning of the pandemic~\cite{springyoutubefalse,lynas2020covid5g}.

\descr{Data Analysis/Measurements.}
Several studies focus on the dissemination of misinformation on YouTube on the political stage.
Tuters~\cite{tuters20207} studies fake news on the political stage by focusing on how fake news on YouTube frames the Dutch political debate. 
He collects and analyzes a list of YouTube channels related to Dutch political parties and their relevant channels as returned by YouTube's recommendation algorithm.
The analysis of how YouTube connects channels to the Dutch political parties' channels shows that the algorithm associates parties to a collection of large Dutch commercial and public media channels. 
In addition, he finds a network of several alternative YouTube channels, which tend to post misinformative videos, to be associated with different clusters of political parties' channels.

The scientific community also focuses on analyzing and measuring pseudoscientific misinformation on YouTube.
Allgaier~\cite{allgaier2016science} studies misinformation around climate change and climate manipulation on YouTube.
To do this, he first performs search queries using a list of search terms related to climate science, climate change, and climate manipulation, and he collects a set of 140 videos related to climate. 
He then inspects the collected videos and finds that YouTube mostly returns legitimate science videos when searching using generic climate-related terms (e.g., climate science). 
However, when searching using more controversial search terms like "climate hacking" or "chemtrails", YouTube returns videos that challenge the scientific consensus on climate change or promote conspiracy theories about science and technology. 
Hussein et al.~\cite{hussein2020measuring} measure misinformation on YouTube focusing on five popular topics namely, 9/11, chemtrails, flat earth, the moon landing, and vaccine controversies, while also investigating whether personalization contributes to amplifying misinformation. 
Among other things, they find that watching videos on YouTube that promote misinformation related to all the above topics, except vaccines, leads to more misinformative recommendations.

Furthermore, other scholars focus on misinformation and conspiracy theories on YouTube related to specific topics.
Paolillo~\cite{paolillo2018flat} studies the flat earth phenomenon on YouTube.
He collects videos uploaded on YouTube related to Flat Earth and performs a characterization of the phenomenon over time.
He also analyzes the discourse dynamics of the Flat Earth ``movement''. 
Landrum et al.~\cite{landrum2019differential} investigate how users with varying science comprehension and attitude towards conspiracies are susceptible to Flat Earth arguments on YouTube. They find that users with lower science intelligence and higher conspiracy mentality are more likely to be recommended Flat Earth-related videos.
In a subsequent study, Landrum et al.~\cite{landrum2020third} assess the degree to which people find Flat Earth arguments presented in YouTube videos convincing, and they compare this to these people's expectations for how convincing others might find the arguments. 
They find that most people believe that subscribers of a specific religion and political orientation are more influenced by Flat Earth arguments on YouTube. 
Mohammed~\cite{mohammed2019conspiracy} studies YouTube videos related to Flat Earth and finds that these videos significantly
outnumbered debunking videos; they were almost twice as long on average and were more likely to include conspiracy concepts and science denial.

Since the COVID-19 outbreak, the research interest on health-related misinformation on YouTube increased.
Li et al.~\cite{li2020youtube} study misinformation related to the COVID-19 pandemic on YouTube; they search YouTube using the terms "coronavirus" and "COVID-19," and analyze the top 75 viewed videos from each search term, finding $27.5\%$ of them to be misinformation. 
Knuutila et al.~\cite{knuutila2020covid} study the spread of COVID-related misinformation on YouTube and the effectiveness of the platform's policies.
They collect 8K COVID-related videos shared on other social networks but have been removed from YouTube due to violating the platform's community guidelines. 
They find that misinformative videos were uploaded on YouTube numerous times after their deletion, and that misinformative videos attract a worryingly large number of engagement on YouTube and other social in terms of views, reactions, and comments. 
More importantly, they find that Facebook is the most significant channel through which the misinformation videos spread.
Donzelli et al.~\cite{donzelli2018misinformation} focus on misinformation surrounding vaccines by performing a quantitative analysis of 560 YouTube videos related to the link between vaccines and autism. 
The analysis revealed that most of the videos were negative in tone and that the number of uploaded misinformative videos has increased since early 2017.

Machado et al.~\cite{machado2020natural} analyze alternative health services and distrust spread about vaccines on YouTube. 
They conclude that alternative health channels tend to spread doubt about traditional institutions; those alternative health channels promote themselves as trusted sources for their audience and thereby profit from alternative health services. 
Kim et al.~\cite{kim2020effects} collect and analyze videos related to vaccination that have been demonetized by YouTube or the platform has added a Wikipedia link and page preview to the page about "Vaccine Hesitancy" (in early 2019) as a result of an informational intervention implemented by YouTube to tackle vaccine misinformation. 
They find that these two measures reduced traffic to the affected vaccination videos.

\descr{Detection and Containment.}
The research community investigated the use of machine learning techniques and other methodologies for detecting and containing misinformative content on YouTube around specific topics.
Heydari et al.~\cite{heydari2019youtube} perform a discourse analysis of comments posted under misinformative and politically-related videos on YouTube.
Interestingly, they find that conspiracy and pseudoscientific channels, as well as questionable sources, receive much more comments and replies per view on the videos they posted compared to apolitical and pro-science YouTube channels. 
They also analyze the thread length, comments lengths, and profanity rates across the videos posted by their channels of interest and use them to train machine learning classifiers, which can predict the bias category of a video.
Chen et al.~\cite{chen2021visual} analyze the visual framing in science conspiracy videos by analyzing millions of frames from conspiracy and counter-conspiracy YouTube videos.
Their analysis shows that conspiracy videos tend to use lower color variance and brightness, especially in thumbnails and earlier parts of the videos.
They also demonstrate how textual and visual features can be used for the identification of conspiracies on social media.

Faddoul et al.~\cite{faddoul2020longitudinal} develop a classifier to detect conspiratorial videos on YouTube and use it to perform a longitudinal analysis of conspiracy videos simulating YouTube’s autoplay feature, without user personalization.
Hou~\cite{Hou2019TowardsAD} automatic detection of misinformation in online medical videos related to cancer.
They collect 250 misinformative videos related to prostate cancer and propose an SVM-based classifier trained with a combination of linguistic, acoustic, and user engagement features, which can detect misinformative videos on YouTube with $74\%$ accuracy.
Serrano et al.~\cite{serrano2020nlp} propose a Natural Language Processing (NLP) based methodology for detecting COVID-19 related misinformation by analyzing comments posted under videos on YouTube.
Specifically, they use transfer learning and they propose a multi-class classifier that can categorize conspiratorial content.
The proposed classifier uses the percentage of misinformation comments on each video as an additional feature. 
By also adding the first hundred comments on a video as tf-idf features they achieve $89.4\%$ accuracy.
However, the fact that this classifier is mainly based on the comments of a video is problematic; a malicious actor can manipulate the comments of a video while maintaining its content intact.
Hussain et al.~\cite{hussain2018analyzing} propose the use of video metadata for detecting and containing disinformation and radicalizing content on YouTube.
They collect and analyzed the metadata of 4K videos and 16.4K comments posted on a YouTube channel that posts content promoting conspiracy theories regarding World War III. 
Specifically, they analyze user engagement and apply social network analysis techniques to identity inorganic and deviant behavior.

\subsection{Pseudoscientific Misinformation and Conspiracy Theories on the Web}

\descr{Data Analysis/Measurements.}
Understanding how false information proliferates on social networks and why users are deceived by such information and the way it is presented to them is central to the development of efficient algorithms and tools that can detect and contain such information.
The scientific community has extensively studied the phenomenon of misinformation and the credibility issues of online content~\cite{zannettou2019web,zannettou2019towards}.
Kumar et al.~\cite{kumar2018false} provide a survey of existing research on understanding and detecting false information on the Web, as well as the actors involved in spreading false information.

Several researchers analyzed misinformation and disinformation spread on social media platforms.
Samory et al.~\cite{samory2018conspiracies} study r/conspiracy, a popular Reddit community dedicated to discussing conspiracy theories, and they analyze user comments by focusing on four tragic real-world events.
Using casual inference they analyze the comments around the time of the events and they find that discussions happening after the event exhibit signs of emotional shock, increased language complexity, and simultaneous expressions of certainty and doubtfulness.
They conclude that the use of network analysis to analyze online conspiracist communities may help detect new conspiracy theories that emerge ensuing dramatic events, and stop them before they spread.
Frigerri et al.~\cite{friggeri2014rumor} track the propagation of false information on Facebook and find that false information reshare cascades spread much deeper compared to that of the true reference cascades.
In other words, they are more likely to be reshared at greater depth and thus reach more people.

While the above studies focus on analyzing misinformation in specific platforms, recent studies mapped the false information ecosystem across social media platforms. Specifically, Zannettou et al.~\cite{zannettou2017web} study the temporal and causal relations between posts of the same false information appearing on Reddit, Twitter, and 4chan.
They found that false information is more likely to spread across platforms compared to true information, and it is also more likely to spread faster than true information.
They also find that false information "flow" from one platform to another, with Reddit to Twitter to 4chan being the most common route.
Albright~\cite{albright2016data} collects all the links referenced in 117 false information websites and creates a network from the referenced domains.
He finds that right-wing news websites tend to refer to each other and that YouTube is the most linked website overall, suggesting high use of videos in conveying false information messages.

As with other types of events with high impact on society, since the emerge of the COVID-19 pandemic several conspiracy theories and health-related misinformation started spreading in OSNs and the Web in general~\cite{ferrara2020misinformation}.
Havey~\cite{havey2020partisan} investigates how political learning influences the participation in the discussions of several COVID-19 misinformation narratives on Twitter.
He finds that conservative users dominated most of these discussions and pushed diverse conspiracy theories.
Johnson et al.~\cite{johnson2020online} focus on the opposition between pro- and anti-vaccination views on the Web.
By clustering users based on the content they share on Facebook, they find that anti-vaccination clusters manage to become highly entangled with clusters of undecided users in the main online network, whereas pro-vaccination clusters are more peripheral.
They conclude that anti-vaccination views will probably dominate social networks in a decade.

Singh et al.~\cite{singh2020understanding} investigate URL sharing patterns during the COVID-19 pandemic on Twitter.
They categorize URLs as either related to traditional news sources, authoritative health sources, or misinformation news sources.
Then, they build a network of shared URLs and they find that authoritative health sources and misinformation ones are shared much less than traditional news sources.
They also find that COVID-19 misinformation is shared at a higher rate than other sources and that the COVID-19 misinformation network appears to be tightly connected and disassortative.
Jennings et al.~\cite{jennings2021lack} investigate how the lack of trust and social media echo chambers affect vaccine hesitancy during the COVID-19 pandemic. 
They perform a survey of 1.4K adults in the UK and they find that vaccine hesitancy is driven by a misunderstanding of herd immunity as
providing protection, fear of rapid vaccine development and its possible side effects, and a belief that the virus is manmade and related to population control.
Interestingly, they find that those who obtain information from social media sources, such as YouTube, are less likely to be willing to become vaccinated.

\descr{Detection and Containment.}
Various types of techniques have been proposed to detect pseudoscientific misinformation and conspiracy theories in OSNs and the Web in general.
Conti et al.~\cite{conti2017s} focus on identifying conspiracy theories in OSNs by taking into account the structural features of the information cascade. 
They do this based on the idea that the structural features cannot be tampered with by malicious users who try to evade detection from classification algorithms. 
They use a Facebook dataset that includes scientific articles and conspiracy theories and they train conventional machine learning models finding that is quite hard to distinguish a conspiracy theory from a scientific article by only considering structural dynamics.
Rajdev et al.~\cite{rajdev2015fake} focus on understanding and detecting fake and spam content promoted by malicious users on Twitter during natural disasters. 
Using NLP techniques they perform feature engineering and use them to develop flat and hierarchical classifiers that can detect fake and spam tweets with $91.7\%$ accuracy.

Other researchers focus on detecting conspiracy theories and pseudoscientific misinformation related to the COVID-19 pandemic.
Shahsavari et al.~\cite{shahsavari2020conspiracy} focus on conspiracy theories around 5G and Bill gates focusing on Reddit and 4chan.
Specifically, they devise a pipeline that uses machine learning techniques and use it to determine the narrative frameworks supporting the generation of these conspiracy theories.
Chakraborty et al.~\cite{chakraborty2020around} investigate the relationship between news sentiment and real-world events for a large set of worldwide news articles published during the COVID-19 pandemic.
Using transfer-learning and other unsupervised sentiment analysis techniques, they explore correlations between news sentiment scores and the global and local numbers of infected people and deaths.
M{\o}nsted et al.~\cite{monsted2019algorithmic} analyze discourse around vaccination denialism on Twitter and they propose a deep neural network that predicts tweet vaccine sentiment with $90.4\%$ accuracy.

\subsection{Pseudoscientific Misinformation and Conspiracy Theories - Remarks}
\revision{
The main findings from the literature review of pseudoscientific misinformation and conspiracy theories on YouTube and the Web include:
1) Machine learning techniques can assist in the detection of false information. However, most of the available classification models mostly rely on handcrafted features, they are effective for specific types of information, topics, and social networks, and it is unclear whether they can generalize on other datasets and the information ecosystem in general; and
2) YouTube is plagued by pseudoscientific misinformation and conspiratorial content and the platform is regularly linked from various false information websites (\textbf{(RQ1 and RQ4)}).
}

\section{Recommendation Algorithms and Audits}
\label{sec:literature_youtube_recommendation_audit}
In the last decade, OSNs and recommendation algorithms have been repeatedly accused of user radicalization and for promoting offensive and inappropriate content.
Understanding user personalization employed by the recommendation algorithms, and the potential role of the recommendation algorithms in user radicalization is of paramount importance.
Therefore, in this section, we review the most relevant studies on understanding user personalization on YouTube and the Web in general, as well as studies on user radicalization on YouTube.
Table~\ref{tab:recommendation_algorithms_audits_studies_overview} summarizes the reviewed work on understanding user personalization and user radicalization, as well as the considered platform.

\begin{table}[t!]
\footnotesize
\centering
\begin{tabular}{crr}
\toprule
\textbf{Platform/Recommendation Algorithm} & \textbf{User Personalization} & \textbf{User radicalization}  \\
\midrule
\multirow{6}{*}{\begin{tabular}[c]{@{}c@{}}YouTube\end{tabular}} & \multirow{6}{*}{\begin{tabular}[c]{@{}r@{}}Zhao et al.~\cite{zhao2012personalized}\\Zhao et al.~\cite{zhao2019recommending}\\Hussein et al.~\cite{hussein2020measuring} \end{tabular}}
& Kaiser et al.~\cite{kaiser2018unite}\\
 & & Ribeiro et al.~\cite{ribeiro2020auditing}\\
 & & O'Callaghan et al.~\cite{o2015down}\\
 & & Schmitt et al.~\cite{schmitt2018counter}\\
 & & Bessi et al.~\cite{bessi2016users}\\
 & & St{\"o}cker et al.~\cite{stocker2020riding}\\
\midrule
\multirow{3}{*}{\begin{tabular}[c]{@{}c@{}}Google Web Search\end{tabular}} & Hannak et al.~\cite{hannak2013measuring} & \multirow{3}{*}{\begin{tabular}[c]{@{}r@{}}-\end{tabular}}\\
 & Kliman-Silver et al.~\cite{kliman2015location} & \\
 & Robertson et al.~\cite{robertson2018auditing} & \\
\midrule
\multirow{3}{*}{\begin{tabular}[c]{@{}c@{}}Google News\end{tabular}} & Nguyen et al.~\cite{nguyen2014exploring} & \multirow{3}{*}{\begin{tabular}[c]{@{}r@{}}-\end{tabular}}\\
 & Haim et al.~\cite{haim2018burst} & \\
 & Le et al.~\cite{le2019measuring} & \\
\midrule
\multirow{3}{*}{\begin{tabular}[c]{@{}c@{}}Other\end{tabular}} & Pariser~\cite{pariser2011filter} & \multirow{3}{*}{\begin{tabular}[c]{@{}r@{}}-\end{tabular}}\\
 & Bruns~\cite{bruns2019filter} & \\
 & Holone~\cite{holone2016filter} & \\
\bottomrule
\end{tabular}
\caption{Studies that focus on auditing recommendation algorithms on the Web and on auditing YouTube's recommendation algorithm. The reported studies are separated based on those focusing on understanding user personalization and those investigating user radicalization, as well as the considered recommendation algorithm.}
\label{tab:recommendation_algorithms_audits_studies_overview}
\end{table}

\subsection{Understanding User Personalization on the Web}

Most of the work on user personalization focuses on recommendation algorithms and Web search engines and is mainly motivated by the concerns around the Filter Bubble effect~\cite{pariser2011filter,bruns2019filter}.
According to Pariser~\cite{pariser2011filter}, a filter bubble is in a state of intellectual isolation where a user is surrounded by viewpoints
that is a result of personalized information provided by a recommendation algorithm that selectively guesses what information a user would like to see. 
As a result, users get less exposure to conflicting viewpoints and are isolated in their own ideological bubble.

Nguyen et al.~\cite{nguyen2014exploring} explore the longitudinal impacts of a recommendation system that is based on collaborative filtering.
Specifically, they measure the filter bubble effect by measuring the diversity of the content suggested to the users.
They introduce a methodology to study the effects of recommendation systems on users and use it to analyze a movie recommendation system.
They find that recommendation systems expose users to slightly narrower content over time and that users who actually consume the information recommended to them receive a more positive experience than those who do not.
However, they find a general bias in that Google News recommends certain news outlets more frequently than other news outlets.
Holone~\cite{holone2016filter} discusses how personalization and the filter bubble effect affect the health-related information that a user can find on the Web, and he proposes some possible measures that can be taken by search engines to mitigate these effects and by the users to consume more credible and unbiased health-related information online.

A large body of work on user personalization and the Filter Bubble effect focus on Google search engine.
Haim et al.~\cite{haim2018burst} analyze the effects of user personalization on the diversity of Google News.
They analyze the effects of both implicit and explicit personalization on the content and source diversity of Google News, and they do not find strong evidence of the existence of the filter bubble effect.
Le et al.~\cite{le2019measuring} investigate whether politically oriented Google News search results are personalized based on the user’s browsing history; using a "sock puppet" audit system, they find significant personalization, which tends to reinforce the presumed partisanship of a user.
Hannak et al.~\cite{hannak2013measuring} develop a methodology to measure personalization in Google Web search that takes into account a range of user attributes that are subject to personalization.
They apply the developed methodology considering 200 users on Google Web search, and they find that $11.7\%$ of the search results that users receive show differences due to personalization.
In addition, by further investigating the causes of personalization, the only find measurable personalization as a result of searching Google using a logged-in profile and the IP address of the user.
Kliman-Silver et al.~\cite{kliman2015location} study the impact of location-based personalization on Google search results.
They propose a methodology for assessing the relationship between location and personalization and use it to collect 30 days of Google search results as a result of 240 search queries gathered from various locations around the US.
They find that differences in search results due to location-based personalization are highly dependent on what a user searches for.
Robertson et al.~\cite{robertson2018auditing} focus on the personalization and composition of politically-related search engine results, and they propose a methodology for auditing Google Search.
Specifically, they use a Web browser extension to collect Google search results and autocomplete suggestions as a result of searching Google using a predefined set of political queries.
By analyzing the collected data, they found significant differences in the composition and personalization of politically-related search results by query type, subjects’ characteristics, and date.
\subsection{Auditing YouTube's Recommendation Algorithm}
In the last decade, YouTube has been repeatedly accused of user radicalization and for promoting offensive and inappropriate content.
In this section, we review the most relevant studies on understanding the recommendation algorithm and user personalization on YouTube, as well as studies investigating user radicalization on the platform.
\descr{User Personalization on YouTube.} %
Zhao et al.~\cite{zhao2012personalized} propose an approach on how to consider a user's viewing history for personalized video recommendations
The proposed approach calculates, for each user, a recommendation score for each candidate video which considers the interest degree of this video by the user’s friends, and the taste similarities between the user and his friends.
Zhao et al.~\cite{zhao2019recommending} introduce a large-scale ranking system for YouTube recommendations, which ranks the candidate recommendations of a given video, taking into account user engagement and satisfaction metrics (e.g., video likes).
Hussein et al.~\cite{hussein2020measuring} focus on measuring misinformation on YouTube considering five popular misinformative topics to investigate whether personalization contributes to amplifying misinformation. 
They find that, once a user develops a watch history, the demographic attributes affect the extend of misinformation recommended to the users.
More importantly, they find a filter bubble effect in the video recommendations section for almost all the topics they analyze.
\revision{We note that part of this thesis is party inspired and complementary to Hussein et al.~\cite{hussein2020measuring}.}

\descr{User Radicalization.}
The research community has repeatedly focused on analyzing user radicalization on YouTube, as well as on understanding the role of recommendation algorithms in amplifying user radicalization focusing on specific topics.
Kaiser et al.~\cite{kaiser2018unite} investigate far-right activity on YouTube and find that YouTube's recommendation algorithm actively contributes to the rise and unification of the far-right.
They also find that the recommendation algorithm suggests videos promoting bizarre conspiracy theories from channels with little or no political content.
Ribeiro et al.~\cite{ribeiro2020auditing} perform a large-scale audit of user radicalization on YouTube: they analyze videos from Intellectual Dark Web, Alt-lite, and Alt-right channels, showing that they increasingly share the same user base. 
They also analyze YouTube's recommendation algorithm finding that Alt-right channels can be reached from both Intellectual Dark Web and Alt-lite channels.
O'Callaghan et al.~\cite{o2015down} perform categorization of far-right content on YouTube and use it to demonstrate the political articulations of YouTube's recommendation algorithm.
They also collect and analyze data from English and German language far-right YouTube channels and they find that users are likely to be recommended further far-right content when they are watching a far-right video, hence leading them in an ideological bubble in just a few short clicks.
Schmitt et al.~\cite{schmitt2018counter} focus on counter- and extremist content on YouTube and investigate the extent to which YouTube's recommendation algorithm impacts the inter-relatedness of these types of content.
By creating and analyzing snapshots of YouTube's recommendation graph constructed mainly with videos of two counter-extremism campaigns on YouTube, they find that counter-extremism videos are closely or even directly connected to extremist content.

Furthermore, Bessi et al.~\cite{bessi2016users} attempt to "reverse" engineer the recommendation algorithm of YouTube and Facebook by examining user polarization on both platforms.
Specifically, they perform a comparison study on how similar videos are consumed on Facebook and YouTube from a sample of 12M users.
They find that content drives the emergence of echo chambers on both platforms and that the users' commenting patterns can be used to accurately predict the formation of echo chambers.
St{\"o}cker et al.~\cite{stocker2020riding} analyze the effect of extreme recommendations on YouTube. 
They develop a simulation system to analyze YouTube recommendations finding that YouTube’s auto-play feature is problematic.
They also state that completely preventing inappropriate recommendations is technically hard due to the nature of recommendation algorithms to suggest content similar to users' interests.

\subsection{Recommendation Algorithms and Audits - Remarks}
\revision{
In this section, we provided an overview of the existing work that focuses on understanding user personalization and recommendation algorithms, as well as studies that investigate the role of YouTube's recommendation algorithm in amplifying user radicalization.
Some of the main findings of this literature review are:
1) Recommendation algorithms tend to suggest narrower content over time, and several studies find significant personalization, which tends to reinforce the presumed partisanship of a user;
2) User attributes such as demographics, location, and browsing do affect the content being recommended to users on Google News and Google Web search;
3) YouTube's recommendation algorithm regrettably suggests extreme content to users. Several studies find evidence of the filter bubble effect in the video recommendation sections of YouTube for certain topics; and
4) Several approaches exist for effectively measuring personalization in Google Web search while taking into account a range of user attributes that are subject to personalization. However, a methodology for assessing the effect that various user attributes and personalization factors have on extreme and inappropriate recommendations on YouTube is not available, yet \textbf{(RQ2 and RQ4)}.
}

\section{Other Related Work}
\label{sec:literature_other}
In this section, we present other related work on YouTube that does not directly fit in the aforementioned lines of work,
Specifically, we group these studies in the following categories: 
1) Understanding YouTube characteristics; and 2) YouTube in science and education.

\subsection{Understanding YouTube Characteristics}
Several studies attempt to understand the characteristics of YouTube and other video-sharing platforms.
Gill et al.~\cite{gill2007youtube} perform a characterization of traffic on YouTube during the early stages of the platform.
They characterize users' activity on YouTube, usage patterns, and the most popular videos of a period of three months, and perform a comparison between other traditional Web and media streaming platforms.
Cheng et al.~\cite{cheng2008statistics} perform a systematic and in-depth measurement analysis of the statistics of 3M YouTube videos.
They find that YouTube videos have noticeable differences compared to traditional streaming videos in terms of duration and view patterns, growth trend, and active life span. 
They also investigate the social networking aspect of YouTube finding, at the time of the study, that the links to related videos generated by uploaders' choices have clear small-world characteristics.
Che et al.~\cite{che2015survey} analyze the characteristics of YouTube videos and provide insights on how YouTube videos have developed through the years.
They find that YouTube exhibits unique characteristics compared to other video-sharing platforms in terms of average video duration, type of videos, and video resolution, and file size.

In the early stages of YouTube, considering a large number of videos available on the platform and the absence of an efficient recommendation algorithm at that time, the research community also proposed several methodologies for personalized video recommendations.
Baluja et al.~\cite{baluja2008video} propose Adsorption, a methodology for personalized video suggestions to users, and video search on YouTube based on the analysis of the entire user-video graph.
Qin et al.~\cite{qin2010recommender} propose a methodology for video recommendations on YouTube by extracting video relationships using a network of videos formed from reviews left as comments in the videos.
Specifically, they create a network of videos called YouTube Recommender Network (YRN) and use complex network analysis on this network as the basis of a recommender system.

\subsection{YouTube in Science and Education}
Since its inception scholars from a wide range of disciplines and critical perspectives have found YouTube useful as a source of examples and case studies.
While a substantial body of research focuses on describing and critically understanding YouTube’s practices and politics, others have experimented directly with the scholarly and educational potential of the platform itself.
The research community has repeatedly studied how YouTube can be used as a tool in education and science communication, as well as the impact of educational videos.
Duffy~\cite{duffy2008using} proposes possible strategies for educators to use Web 2.0 technologies and social networks, such as YouTube, as tools within education. 
Other researchers also show that YouTube videos can be used to enhance math and history learning~\cite{niess2009rock,white2009coffy}.
Jones et al.~\cite{jones2011youtube} investigate the potential and pitfalls of using YouTube in social studies instruction and as a teaching resource in the elementary classroom.
They also discuss the potential challenges of using YouTube in the classroom and propose ways for educators for overcoming those challenges.

Furthermore, Thelwall et al.~\cite{thelwall2012assessing} go beyond these studies and assess the impact and characteristics of online academic videos.
They find that the most popular videos produced by well-known academics are those aimed at a general audience and have been well-produced, but more importantly, they find that the audience for such videos is small and cannot justify the production cost of such videos.
Others also focus on the use of YouTube in research and science communication.
Kousha et al.~\cite{kousha2012role} investigate the role of online videos in research communication by analyzing YouTube videos regularly cited in various academic publications.
Specifically, they analyze 551 YouTube videos cited by research articles and they observe an increasing trend in citing YouTube videos in scientific publications.
They also find that in science, and medicine, and health sciences, cited videos contain direct scientific or scientific-related content, while in other domains videos mostly contain other themes (e.g., history, news, politics, etc.).
Welbourne et al.~\cite{welbourne2016science} focus on science communication videos on YouTube. 
They collect and characterize 390 YouTube videos and examine the factors that affect the popularity of science communication videos on the platform.
Allgaier~\cite{allgaier2013shoulders} analyzes music videos related to science on YouTube, while in a more recent study he summarizes existing research on scholarly communication, science, and medicine on the platform~\cite{allgaier2020science}.

\chapter{Characterizing and Detecting Inappropriate Videos Targeting Young Children on YouTube}
\label{chapter:disturbed_youtube}

\section{Motivation}
\begin{figure}[t!]
\centering
\includegraphics[width=\columnwidth]{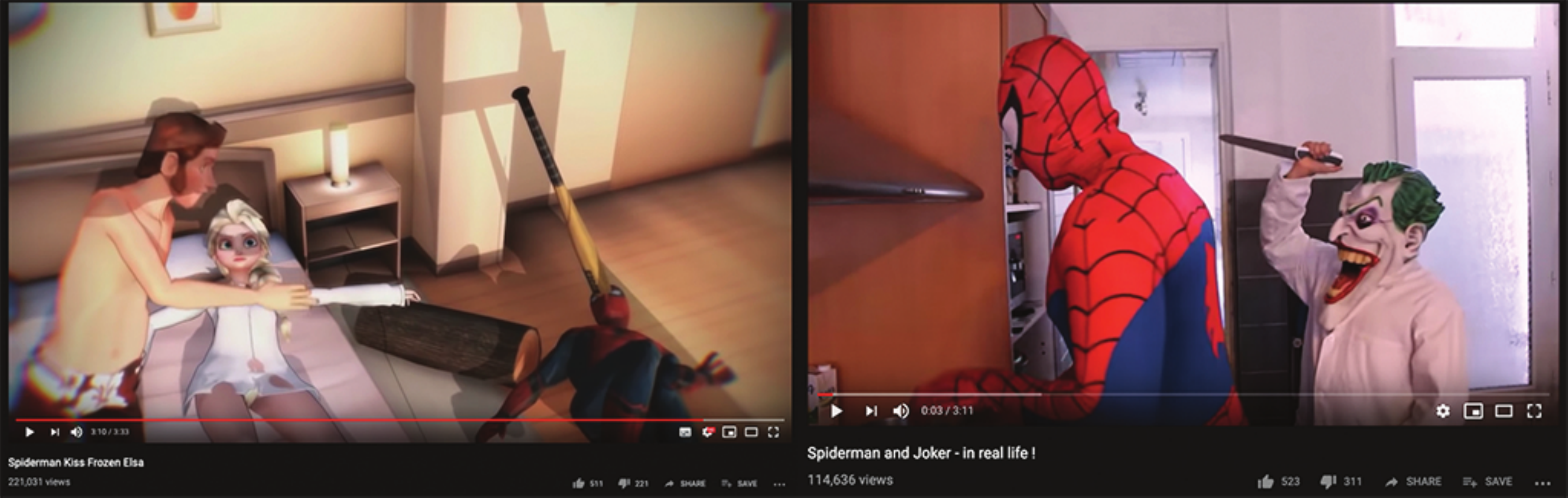}
\caption{Examples of disturbing videos, i.e. inappropriate videos that target toddlers.}
\label{fig:disturbing_video_example} 
\end{figure}

YouTube has emerged as an alternative to traditional children's TV, and a plethora of popular children’s videos can be found on the platform.
For example, consider the millions of subscribers that the most popular toddler-oriented YouTube channels have: ChuChu TV is the most-subscribed "child-themed" channel, with 44.8M subscribers as of January 2021~\cite{statista2021youtubekids}.
While most toddler-oriented content is inoffensive and is actually entertaining or educational, recent reports have highlighted the trend of inappropriate content targeting this demographic~\cite{bbdisturbingyoutube,nytiyoutubekids}.
Borrowing the terminology from the early press articles on the topic, we refer to this new class of content as \textit{disturbing}.
A prominent example of this trend is the Elsagate controversy~\cite{reddit2017elsagate,russell2017verge}, where malicious users uploaded videos featuring popular cartoon characters like Spiderman, Disney's Frozen, Mickey Mouse, etc., combined with disturbing content containing, for example, mild violence and sexual connotations.
These disturbing videos usually include an innocent thumbnail aiming at tricking the toddlers and their custodians.
Figure~\ref{fig:disturbing_video_example} shows examples of such videos. 
The issue at hand is that these videos have hundreds of thousands of views, more likes than dislikes, and have been available on the platform since 2016.

In an attempt to offer a safer online experience for its young audience, YouTube launched the YouTube Kids application~\cite{youtubekids}, which equips parents with several controls enabling them to decide what their children are allowed to watch on YouTube. 
Unfortunately, despite YouTube's attempts to curb the phenomenon of inappropriate videos for toddlers, disturbing videos still appear, even in YouTube Kids~\cite{weston2018youtubekids}, due to the difficulty of identifying them. 
An explanation for this may be that YouTube relies heavily on users reporting videos they consider disturbing~\cite{youtubehelp2021report}, and then YouTube employees manually inspecting them. 
However, since the process involves manual labor, the whole mechanism does not easily scale to the number of videos that a platform like YouTube serves.
\revision{
With this motivation in mind, in this chapter we study inappropriate or disturbing videos targeted at toddlers on YouTube. 
Guided by two of our main research questions \textbf{(RQ1 and RQ4)}, we set out to answer the following two research sub-questions: 
\begin{itemize}
    \item \textbf{RQ1.a:} Can we build a classifier that automatically detects disturbing child-oriented videos? (see RQ1 in Chapter~\ref{chapter:introduction})
    \item \textbf{RQ4.a:} How likely is it for a toddler to come across inappropriate videos when starting from a benign one? (see RQ4 in Chapter~\ref{chapter:introduction})
\end{itemize}
}

\descr{Contributions.}
In this thesis, we provide the first study of toddler-oriented disturbing content on YouTube.
First, we collect, manually review, and characterize toddler-oriented videos (both Elsagate-related and other child-related videos), as well as random and popular videos.
Our characterization confirms that unscrupulous and potentially profit-driven uploaders create disturbing videos with similar characteristics as benign toddler-oriented videos in an attempt to make them show up as recommendations to toddlers browsing the platform.

Second, we develop a deep learning classifier to automatically detect disturbing videos.
Even though this classifier performs better than baseline models, it still has a lower than desired performance. 
In fact, this low performance reflects the high degree of similarity between disturbing and suitable videos or restricted videos that do not target toddlers. 
It also reflects the subjectivity in deciding how to label these controversial videos, as confirmed by our trained annotators' experience.
For the sake of our analysis in the next steps, we collapse the initially defined labels into two categories and develop a more accurate classifier that can discern inappropriate from appropriate videos.
Our experimental evaluation shows that the developed binary classifier outperforms several baselines with an accuracy of $0.84$.

In the last phase, we leverage the developed classifier to understand how prominent the problem at hand is.
From our analysis of different subsets of the collected videos, we find that $1.1\%$ of the 233,337 Elsagate-related, and $0.5\%$ of the 154,957 other children-related collected videos are inappropriate for toddlers, which indicates that the problem is not negligible.
To further assess how safe YouTube is for toddlers, we run a live simulation in which we mimic a toddler randomly clicking on YouTube’s suggested videos. 
We find that there is a $3.5\%$ chance that a toddler following YouTube's recommendations will encounter an inappropriate video within ten hops if she starts from a video that appears among the top ten results of a toddler-appropriate keyword search (e.g., Peppa Pig).

\section{Datasets}
\label{sec:disturbed_datasets}

\subsection{Data Collection}
\label{subsec:disturbed_data_collection}
For our data collection, we use the YouTube Data API~\cite{youtubedataapi}, which provides metadata of videos uploaded on YouTube.
Unfortunately, YouTube does not provide an API for retrieving videos from YouTube Kids. 
We collect a set of seed videos using four different approaches. 
First, we use information from /r/ElsaGate, a subreddit dedicated to raising awareness about the disturbing videos problem~\cite{reddit2017elsagate}.
Second, we use information from /r/fullcartoonsonyoutube, a subreddit dedicated to listing cartoon videos available on YouTube. 
The other two approaches focus on obtaining a set of random and popular videos.

Specifically:
1) we create a list of 64 keywords\footnote{\url{https://tinyurl.com/yxpf73j4}} by extracting n-grams from the title of videos posted on /r/ElsaGate.
Subsequently, for each keyword, we obtain the first 30 videos as returned by YouTube's Data API search functionality.
This approach resulted in the acquisition of 893 seed videos.
Additionally, we create a list of 33 channels\footnote{\url{https://tinyurl.com/y5zhy4vt}}, which are mentioned by users on /r/ElsaGate because of publishing disturbing videos~\cite{russell2017verge,reddit2017elsagate}.
Then, for each channel we collect all their videos, hence acquiring a set of 181 additional seed videos;
2) we create a list of 83 keywords\footnote{\url{https://tinyurl.com/y23xxl3c}} by extracting n-grams from the title of videos posted on /r/fullcartoonsonyoutube. 
Similar to the previous approach, for each keyword, we obtain the first 30 videos as returned by the YouTube's Data API search functionality, hence acquiring another 2,342 seed videos;
3) to obtain a random sample of videos, we a REST API\footnote{\url{https://randomyoutube.net/api}} that provides random YouTube video identifiers which we then download using the YouTube Data API. This approach resulted in the acquisition of 8,391 seed random videos; and
4) we also collect the most popular videos in the USA, the UK, Russia, India, and Canada, between November 18 and November 21, 2018, hence acquiring another 500 seed videos.

\begin{table}[t!]
\footnotesize
\centering
\begin{tabular}{lrr}
\toprule
\textbf{Crawling Strategy} & \textbf{\#Seed Videos} & \textbf{\#Recommended Videos}\\
\midrule
Elsagate-related & 1,074 & 232,263 \\
Other Child-related	& 2,342	& 152,615 \\
Random & 8,391	& 473,516 \\
Popular & 500 	& 10,474 \\
\midrule
\textbf{Total} & 12,097 & 844,702 \\
\toprule
\end{tabular}%
\caption{Overview of the collected data: number of seed videos and number of their recommended videos acquired using each crawling strategy.}
\label{tab:disturbed_dataset_details}
\end{table}

Using these approaches, we collect 12,097 unique seed videos.
However, this dataset is not big enough to study the idiosyncrasies of this problem.
Therefore, to expand our dataset, for each seed video we iteratively collect the top 10 recommended videos associated with it, as returned by the YouTube Data API, for up to three hops within YouTube’s recommendation graph.
We note that for each approach we use API keys generated from different accounts.

Table~\ref{tab:disturbed_dataset_details} summarizes the collected dataset.
In total, our dataset comprises 12K seed videos and 844K videos that are recommended from the seed videos. 
Note, that there is a small overlap between the videos collected across the approaches, hence the number of total videos is slightly smaller than the sum of all videos for the four approaches. 

For each video in our dataset, we collect the following data descriptors: 1) title and description; 2) thumbnail; 3) tags; and 4) video statistics like the number of views, likes, dislikes, etc.
We chose to use these four data collection approaches for three reasons:
1) to get more breadth into children's content on YouTube, instead of only collecting Elsagate-related videos; 2) to examine and analyze different types of videos while also assessing the degree of the disturbing videos problem in these types of videos; and 3) to train a classifier for detecting disturbing videos able to generalize to the different types of videos available on YouTube.

\descr{Ethics.}
For this study, we only collect publicly available data, while not attempt to de-anonymize users.
In addition, all the manual annotators are informed adults.

\begin{table}[t!]
\footnotesize
\centering
\begin{tabular}{lrrrr}
\toprule
 & \textbf{\#Suitable} & \textbf{\#Disturbing} & \textbf{\#Restricted} & \textbf{\#Irrelevant} \\
\midrule
Elsagate-related & 805 & 857 & 324 & 394 \\
Other Child-related & 650 & 47 & 21 & 243 \\
Random & 27 & 5 & 67 & 867 \\
Popular & 31 & 20 & 7 & 432 \\
\midrule
\textbf{Total} & 1,513 & 929 & 419 & 1,936 \\
\toprule
\end{tabular}%
\caption{Summary of our final ground truth dataset.}
\label{tab:disturbed_final_groundtruth_dataset_details}
\end{table}

\subsection{Manual Annotation Process}
\label{subsec:disturbed_annotation_process}
To get labeled data, we manually review a 5K videos subset of the collected dataset by inspecting their video content, title, thumbnail, and tags.
Each video is presented to three annotators that inspect its content and metadata to assign one of the following labels:

\begin{enumerate}
    \item \textbf{Suitable.} A video is {\em suitable} when its content is appropriate for toddlers (aged 1-5 years) and it is relevant to their typical interests. Some examples include normal cartoon videos, children's songs, and educational videos (e.g., learning colors). In other words, any video that can be classified as G by the MPAA\footnote{MPAA stands for Motion Picture Association of America (MPAA) \url{https://www.mpaa.org/film-ratings/}} and its target audience is toddlers.

    \item \textbf{Disturbing.} A video is {\em disturbing} when it targets toddlers but it contains sexual hints, sexually explicit or abusive/inappropriate language, graphic nudity, child abuse (e.g., children hitting each other), scream and horror sound effects, scary scenes or characters (e.g., injections, attacks by insects, etc.). In general, any video targeted at toddlers that should be classified as PG, PG-13, NC-17, or R by MPAA is considered disturbing.

    \item \textbf{Restricted.} We consider a video {\em restricted} when it does not target toddlers and it contains content that is inappropriate for individuals under the age of 17 (rated as R or NC-17 according to MPAA). Such videos usually contain sexually explicit language, graphic nudity, pornography, violence (e.g., gaming videos featuring violence, or life-like violence, etc.), abusive/inappropriate language, online gambling, drug use, alcohol, or upsetting situations and activities.

    \item \textbf{Irrelevant.} We consider a video {\em irrelevant} when it contains appropriate content that is not relevant to a toddler's interests. That is, videos that are not disturbing or restricted but are only suitable for school-aged children (aged 6-11 years), adolescents (aged 12-17 years), and adults, like gaming videos (e.g., Minecraft) or music videos (e.g., a video clip of John Legend's song) reside in this category. In general, G, PG, and PG-13 videos that do not target toddlers are considered irrelevant.
\end{enumerate}

We elect to use these labels for our annotation process instead of adopting the five MPAA ratings for two reasons.
First, our scope is videos that would normally be rated as PG, PG-13, R, and NC-17 but target very young audiences.
We consider such targeting a malevolent activity that needs to be treated separately.
At the same time, we have observed that a significant portion of videos that would normally be rated as R or NC-17 are already classified by YouTube as ``age-restricted'' and target either adolescents or adults.
Second, YouTube does not use MPAA ratings to flag videos, thus, a ground truth dataset with such labels is not available.

\begin{table}[t!]
\footnotesize
\centering
\begin{tabular}{llrrrr}
\toprule
& \textbf{Category} & \multicolumn{1}{c}{\textbf{Suitable}} & \multicolumn{1}{l}{\textbf{Disturbing}} & \multicolumn{1}{l}{\textbf{Restricted}} & \multicolumn{1}{l}{\textbf{Irrelevant}} \\ \midrule
Elsagate-related & EN & 353 (44\%) & 208 (24\%) & 99 (31\%) & 98 (25\%) \\
 & F\&A & 130 (16\%) & 190 (22\%) & 39 (12\%) & 33 (8\%) \\
 & ED & 128 (16\%) & 21 (3\%) & 17 (5\%) & 16 (4\%) \\
 & P\&B & 109 (13\%) & 239 (28\%) & 71 (22\%) & 73 (19\%) \\
 & M & 21 (3\%) & 15 (2\%) & 8 (3\%) & 45 (11\%) \\ \midrule

Other & EN & 131 (20\%) & 7 (15\%) & 9 (43\%) & 51 (21\%) \\
Child-related & F\&A & 317 (49\%) & 27 (58\%) & 3 (14\%) & 26 (11\%) \\
 & ED & 27 (4\%) & 1 (2\%) & 2 (10\%) & 34 (14\%) \\
 & P\&B & 130 (20\%) & 4 (8\%) & 2 (9\%) & 35 (14\%) \\
 & M & 5 (1\%) & 2 (4\%) & 0 (0\%) & 26 (11\%) \\ \midrule

Random & EN & 4 (15\%) & 1 (20\%) & 3 (5\%) & 68 (8\%) \\
 & F\&A & 1 (4\%) & 1 (20\%) & 1 (2\%) & 18 (2\%) \\
 & ED & 1 (4\%) & 0 (0\%) & 0 (0\%) & 31 (4\%) \\
 & P\&B & 13 (48\%) & 3 (60\%) & 21 (31\%) & 354 (41\%) \\
 & M & 1 (3\%) & 0 (0\%) & 0 (0\%) & 79 (9\%) \\ \midrule

Popular & EN & 12 (39\%) & 9 (45\%) & 2 (29\%) & 168 (39\%) \\
 & F\&A & 9 (29\%) & 7 (35\%) & 0 (0\%) & 13 (3\%) \\
 & ED & 2 (7\%) & 0 (0\%) & 0 (0\%) & 11 (3\%) \\
 & P\&B & 2 (6\%) & 0 (0\%) & 0 (0\%) & 32 (7\%) \\
 & M & 0 (0\%) & 1 (5\%) & 0 (0\%) & 63 (15\%) \\ \midrule
\end{tabular}%
\caption{Number of videos in each category per class for each subset of videos in our ground truth dataset. EN: Entertainment, F\&A: Film \& Animation, ED: Education, P\&B: People \& Blogs, M: Music.}
\label{tab:disturbed_groundtruth_video_categories_stats}
\end{table}

\subsubsubsection{Sampling Process}
We aim to create a ground truth dataset that enables us to: 
1) understand the main characteristics of disturbing toddler-oriented videos compared to suitable children videos on YouTube; and 
2) train a deep learning model that will detect disturbing videos with an acceptable performance while being able to generalize to the various types of videos available on the platform.
To this end, we use the following videos for the annotation process.
1) We randomly select 1,000 of the 2,342 seed child-related videos aiming to get suitable videos.
2) Since the Elsagate-related collected videos are likely to include disturbing videos, we select all the seed Elsagate-related videos (1,074), as well as a small, randomly selected set (1,171) of their recommended videos.
3) To get a sample of restricted videos, we randomly select 500 of the 2,597 age-restricted videos in our dataset.
4) To ensure that we include irrelevant videos, we select all the seed popular videos (500) as well as a small set (1,000) of the 8,391 randomly collected videos.

\subsubsubsection{Manual Annotation}
The annotation process is carried out by two of the authors of this study and 76 undergraduate students (aged 20-24 years), both male and female.
Each video is annotated by the two authors and one of the undergraduate students.
The students come from different backgrounds and receive no specific training concerning our study.
To ease the annotation process, we develop a platform that includes a clear description of the annotation task, our labels, as well as all the video information that an annotator needs to inspect and correctly annotate a video.

After obtaining all the annotations, we compute the Fleiss agreement score ($\kappa$)~\cite{fleiss1971measuring} across all annotators: we find $\kappa=0.60$, which is considered ``moderate'' agreement.
We also assess the level of agreement between the two authors, as we consider them experienced annotators, finding $\kappa=0.69$, which is considered ``substantial'' agreement. 
Finally, for each video, we assign one of the labels according to the majority agreement of all the annotators, except a small percentage ($4\%$) where all annotators disagreed, which we also exclude from our ground-truth dataset.
Table~\ref{tab:disturbed_final_groundtruth_dataset_details} summarizes our ground truth dataset, which includes 1,513 suitable, 929 disturbing, 419 restricted, and 1,936 irrelevant videos.

\section{General Characterization}

\descr{Video Category.}
First, we look at the categories of the videos in our ground truth dataset. 
Table~\ref{tab:disturbed_groundtruth_video_categories_stats} reports the top five categories, for each subset of videos. 
Most of the disturbing and restricted in the Elsagate-related videos are in Entertainment ($24\%$ and $31\%$), People \& Blogs ($28\%$ and $22\%$), and Film \& Animation ($22\%$ and $12\%$). 
These results are similar to previous work~\cite{chaudhary2013contextual}. 
A similar trend is also observed in all the other sets of videos.
In addition, in the Elsagate-related videos, we find a non-negligible percentage of disturbing videos in seemingly innocent categories like Education ($3\%$) and Music ($2\%$). 
This is alarming since it indicates that disturbing videos ``infiltrate'' categories of videos that are likely to be selected by the toddler's parents. 
Unsurprisingly, after manually inspecting all the disturbing videos in the Education and Music categories, we find that the majority of them are nursery rhymes, ``wrong heads'', and ``peppa pig'' videos with disturbing content.

\begin{figure*}[t!]
\centering
\subfigure[]{\includegraphics[width=0.49\textwidth, height=2.45in]{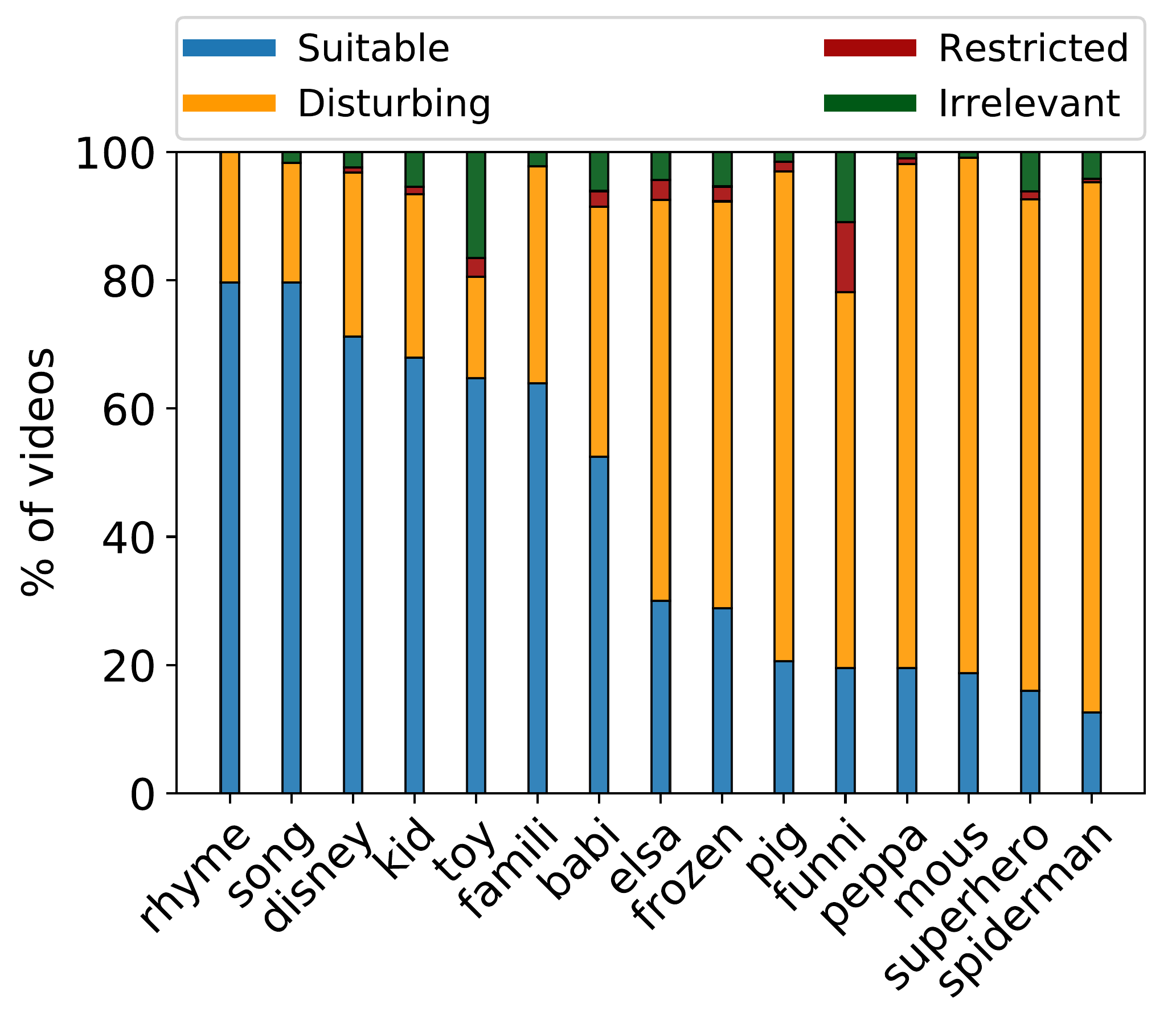}\label{fig:disturbed_headline_normalized_mean_scores_elsagate}}
\subfigure[]{\includegraphics[width=0.49\textwidth, height=2.45in]{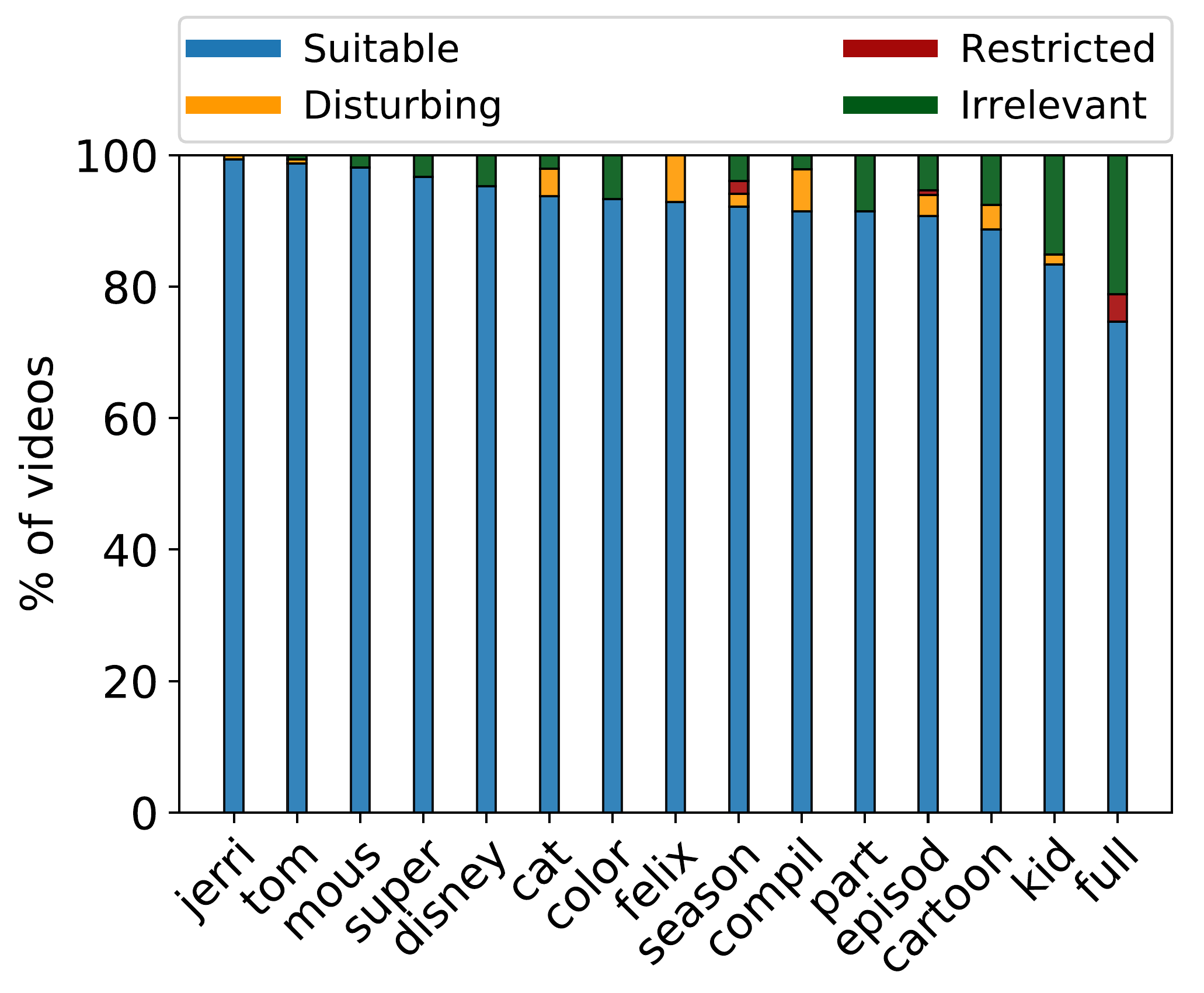}\label{fig:disturbed_headline_normalized_mean_scores_childrelated}}
\subfigure[]{\includegraphics[width=0.49\textwidth, height=2.45in]{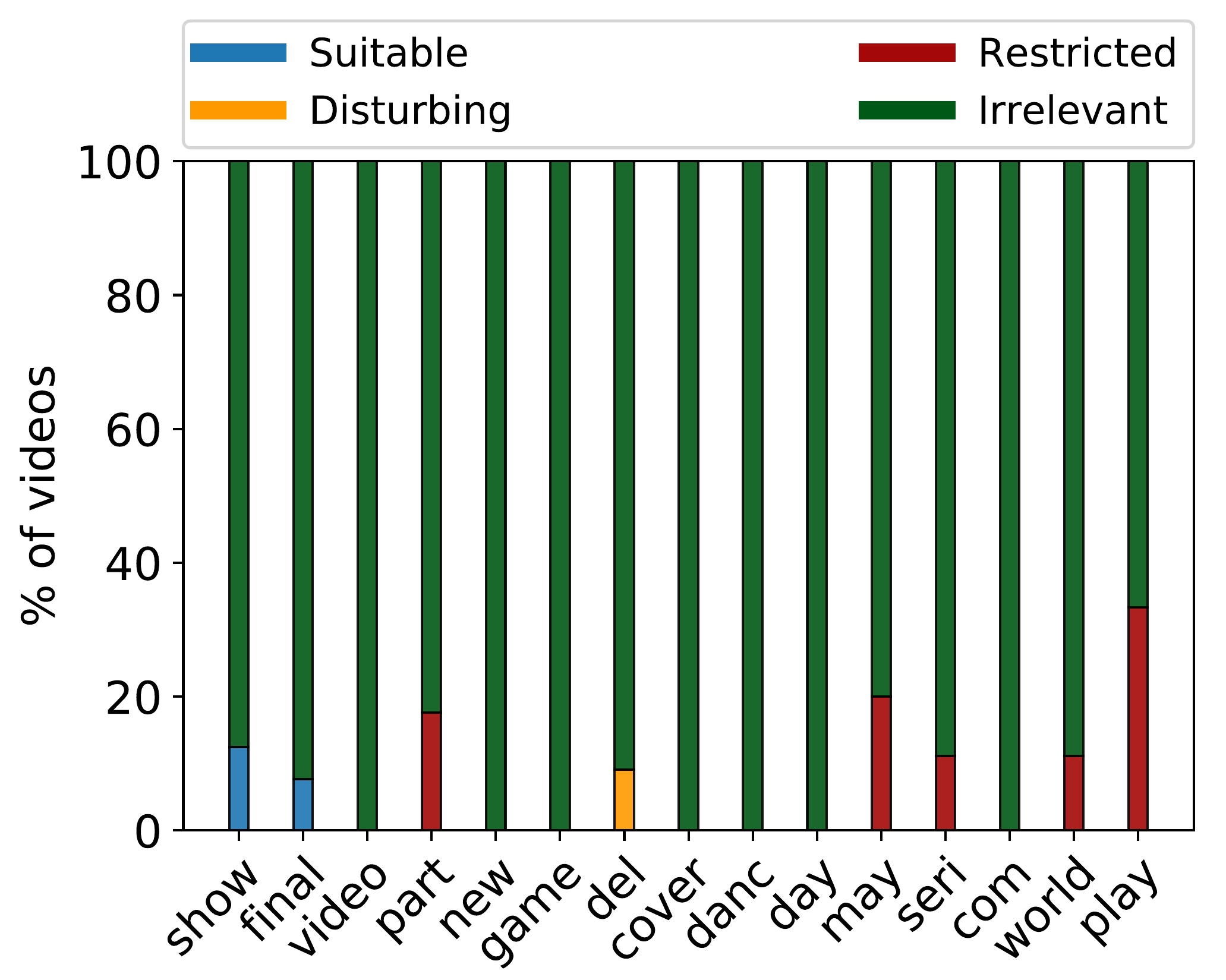}\label{fig:disturbed_headline_normalized_mean_scores_random}}
\subfigure[]{\includegraphics[width=0.49\textwidth, height=2.45in]{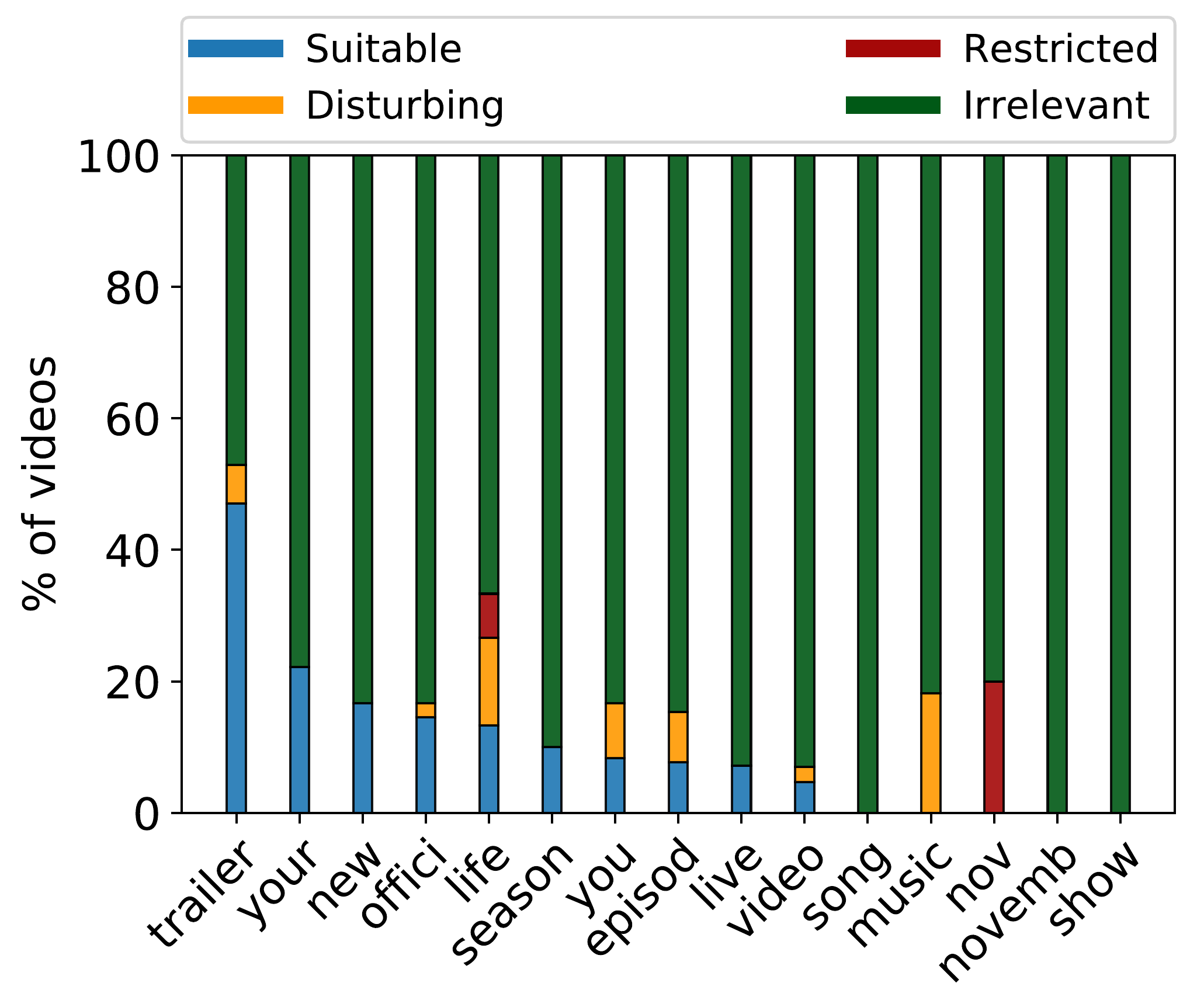}\label{fig:disturbed_headline_normalized_mean_scores_popular}}
\caption{Per class proportion of videos for top 15 stems found in titles of (a) Elsagate-related; (b) other child-related; (c) random; and (d) popular videos.}
\label{fig:disturbed_video_groundtruth_analysis_plots_headlines}
\end{figure*}

\descr{Title.}
The title of a video is an important factor that affects whether a particular video will be recommended when viewing other toddler-oriented videos.
Consequently, we study the titles in our ground truth dataset to understand the tactics and terms that are usually used by uploaders of disturbing or restricted videos on YouTube.
First, we pre-process the title by tokenizing the text into words and then we perform stemming using the Porter Stemmer~\cite{porter1980algorithm} algorithm.
Figure~\ref{fig:disturbed_video_groundtruth_analysis_plots_headlines} shows the top 15 stems found in titles along with their proportion for each class of videos for all the different sets of videos in our ground truth dataset.
Unsurprisingly, the top 15 stems of the Elsagate-related videos refer to popular cartoons like Peppa Pig, Mickey and Minnie Mouse, Elsa, and Spiderman (see Figure~\ref{fig:disturbed_headline_normalized_mean_scores_elsagate}). 
When looking at the results, we observe that a substantial percentage of the videos that include these terms in their title are actually disturbing.
For example, from the videos that contain the terms ``spiderman'' and ``mous'', $82.6\%$ and $80.4\%$, respectively, are disturbing.
Similar trends are observed with other terms like ``peppa'' ($78.6\%$), ``superhero''($76.7\%$), ``pig'' ($76.4\%$), ``frozen'' ($63.5\%$), and ``elsa'' ($62.5\%$).
Also, we observe a small percentage of the other child-related videos that contain the terms ``felix'' ($7.1\%$), ``cat'' ($4.2\%$), and ``cartoon'' ($3.8\%$) are also disturbing (see Figure~\ref{fig:disturbed_headline_normalized_mean_scores_childrelated}).
These results reveal that disturbing videos on YouTube refer to seemingly ``innocent'' cartoons in their title, but in reality, the content of the video is likely to be either restricted or disturbing. Note that we find these terms in suitable videos too.
This demonstrates that it is quite hard to distinguish suitable from disturbing videos by only inspecting their titles.

\begin{figure*}[t!]
\centering
\subfigure[]{\includegraphics[width=0.49\textwidth, height=2.45in]{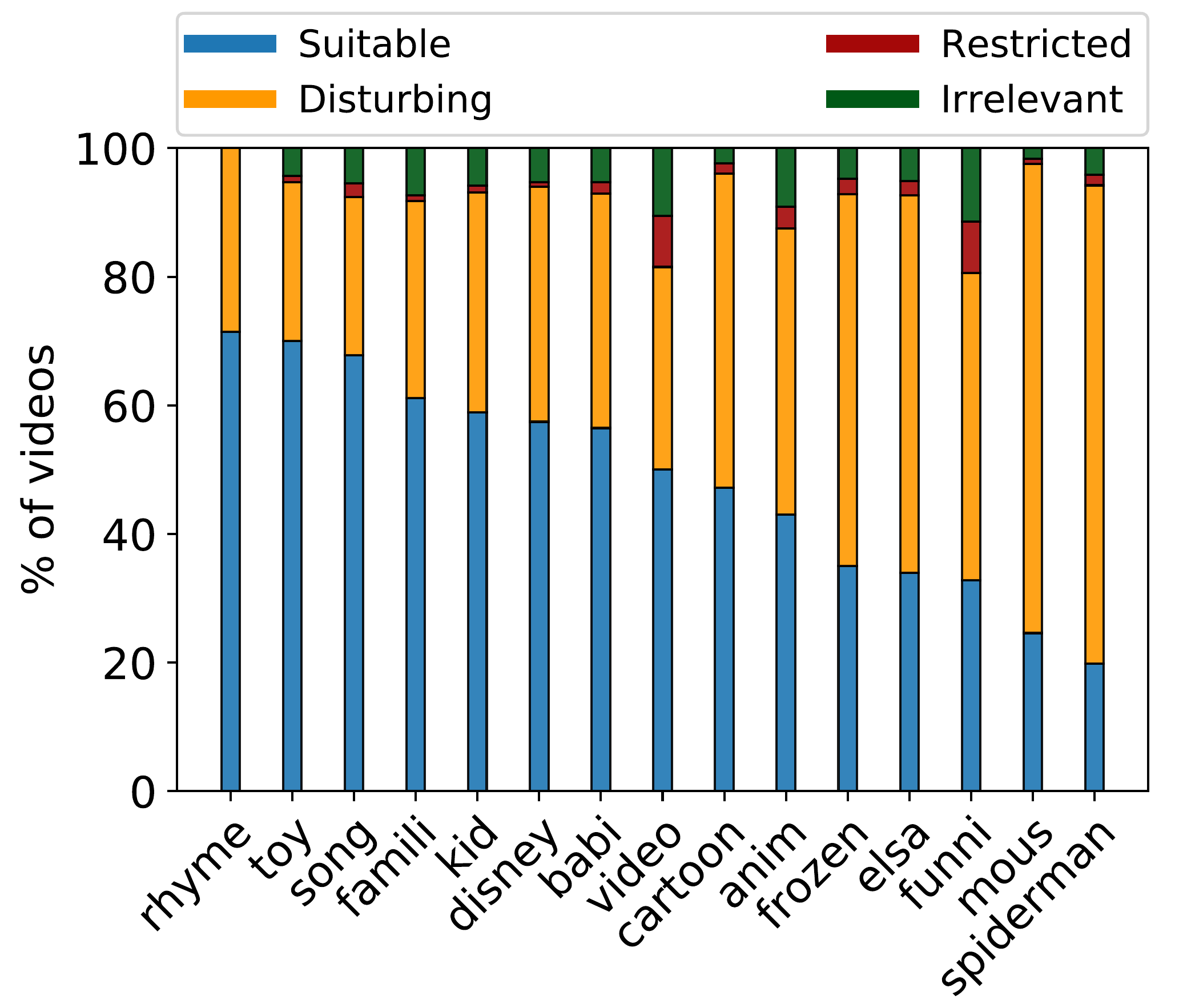}\label{fig:disturbed_video_tags_stems_normalized_mean_scores_elsagate}}
\subfigure[]{\includegraphics[width=0.49\textwidth, height=2.45in]{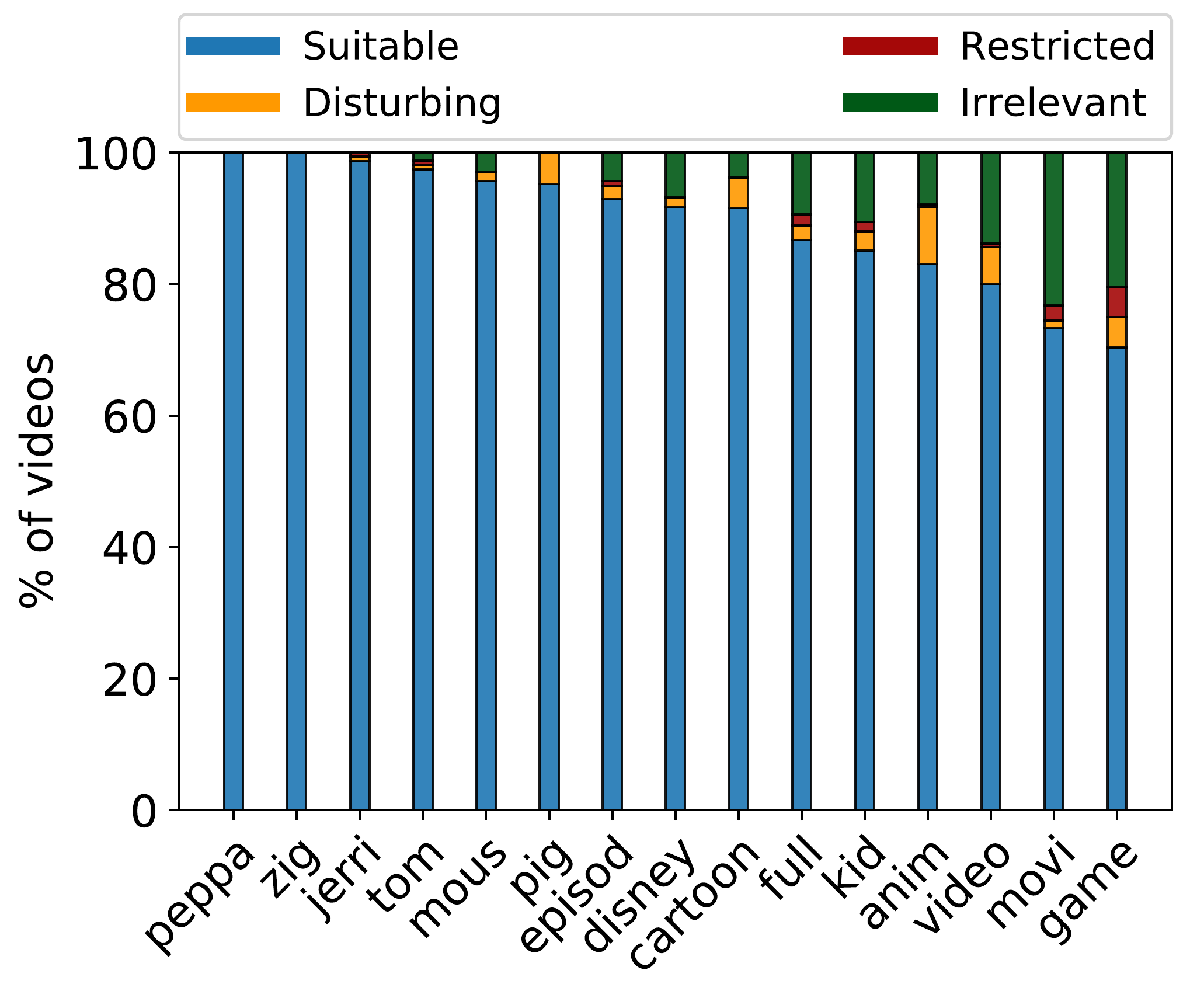}\label{fig:disturbed_video_tags_stems_normalized_mean_scores_childrelated}}
\subfigure[]{\includegraphics[width=0.49\textwidth, height=2.45in]{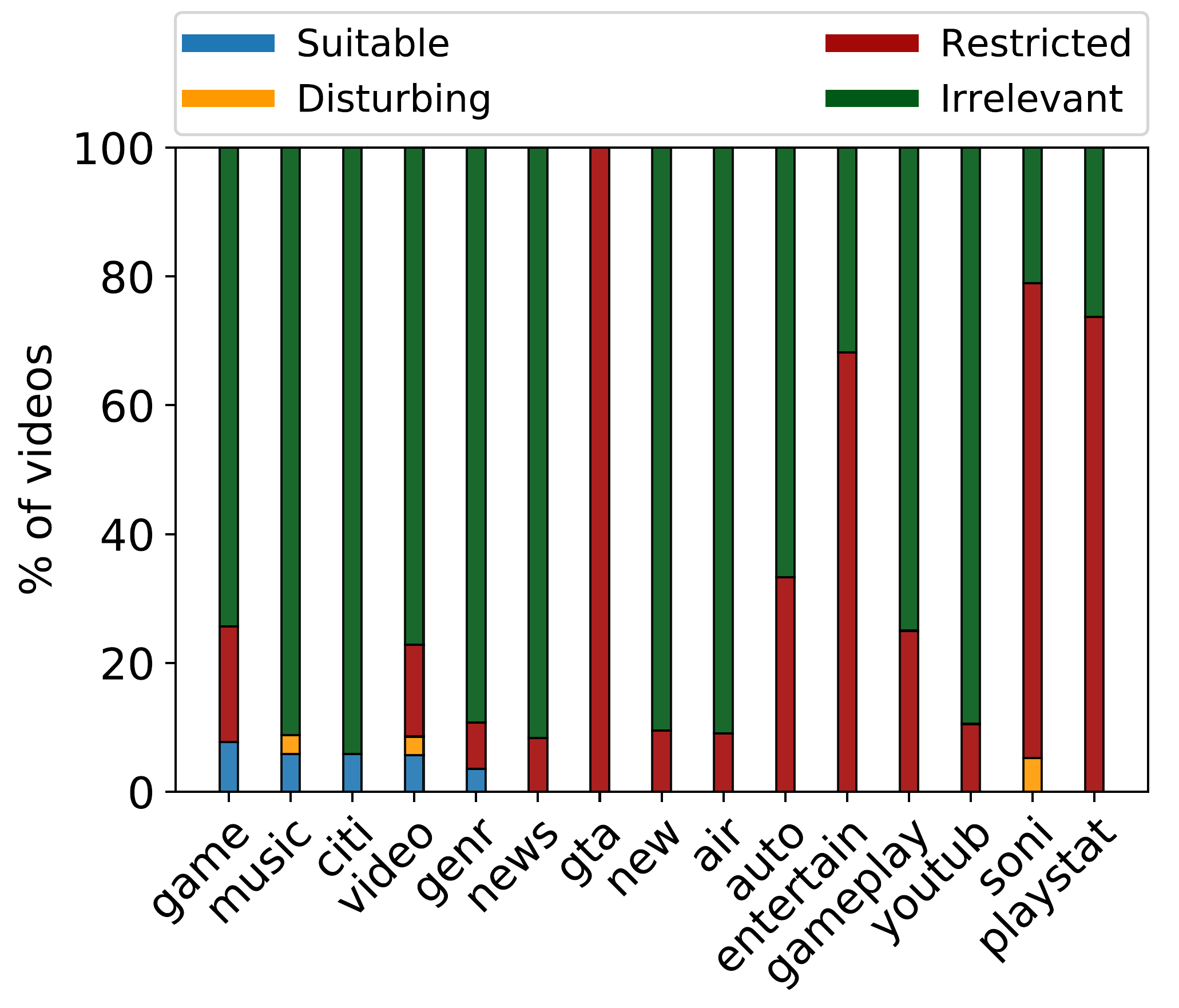}\label{fig:disturbed_video_tags_stems_normalized_mean_scores_random}}
\subfigure[]{\includegraphics[width=0.49\textwidth, height=2.45in]{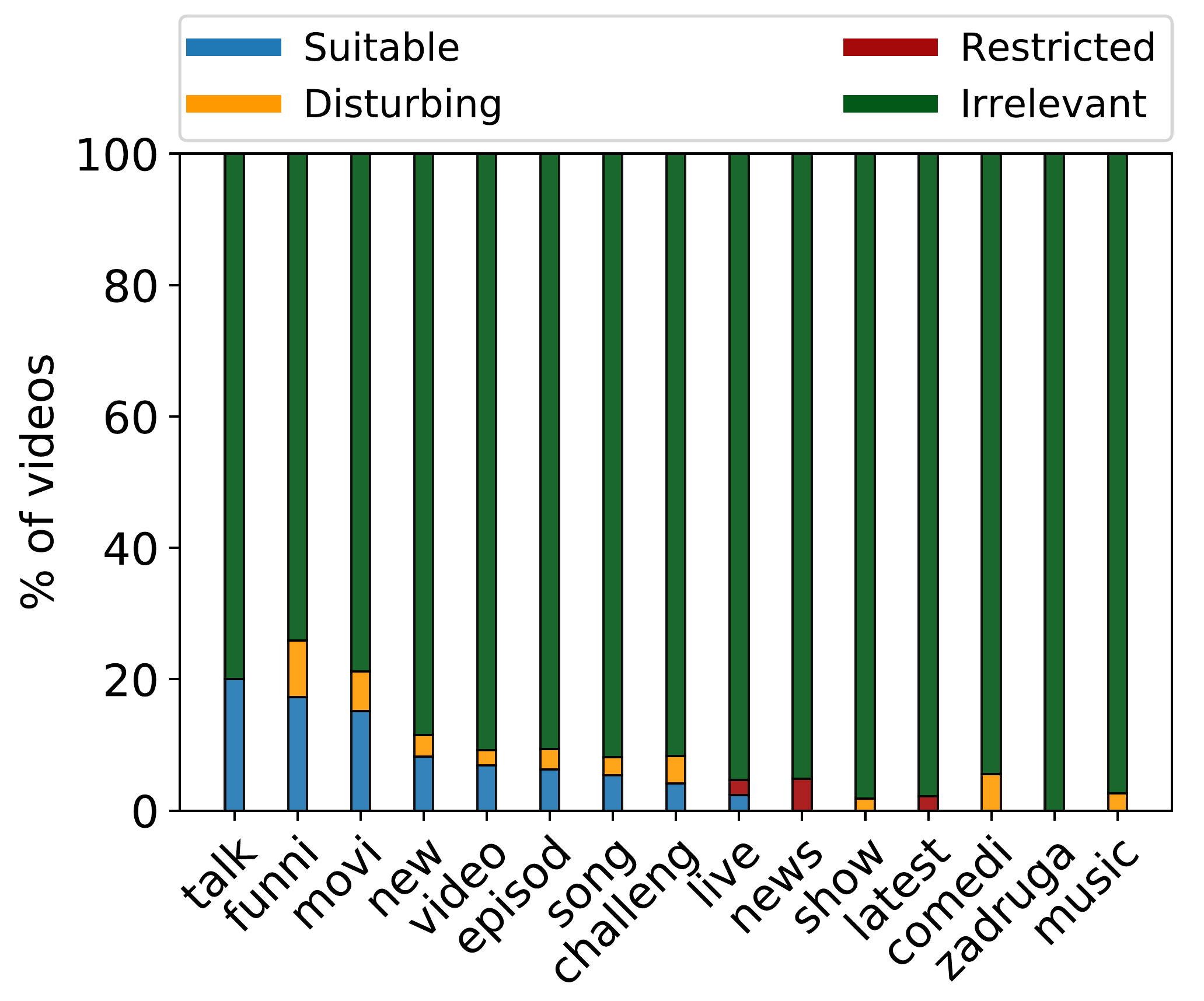}\label{fig:disturbed_video_tags_stems_normalized_mean_scores_popular}}
\caption{Per class proportion of videos for the top 15 stems found in video tags of (a) Elsagate-related; (b) other child-related; (c) random; and (d) popular videos.}
\label{fig:disturbed_video_groundtruth_analysis_plots_video_tags}
\end{figure*}

\descr{Video Tags.}
Tags are words that uploaders define when posting a video on YouTube. 
To study the effect of tags in this problem, we plot in Figure~\ref{fig:disturbed_video_groundtruth_analysis_plots_video_tags} the top 15 stems from tags found in each subset of videos in our ground truth dataset.
We make several observations: first, in the Elsagate-related and other child-related videos, there is a substantial overlap between the stems found in the tags and title (cf. Figure~\ref{fig:disturbed_video_groundtruth_analysis_plots_headlines} and Figure~\ref{fig:disturbed_video_groundtruth_analysis_plots_video_tags}).
Second, in the Elsagate-related videos, we find that suitable and disturbing classes have a considerable percentage for each tag, hence highlighting that Elsagate-related disturbing videos use the same tags as suitable videos. Inspecting these results, we find that the tags ``funni'' ($47.8\%$), ``elsa'' ($58.7\%$), ``frozen'' ($57.8\%$), ``cartoon'' ($48.8\%$), and ``anim'' ($44.5\%$) appear mostly in disturbing videos.
Also, ``spiderman'' ($74.4\%$) and ``mous'' ($73.0\%$) appear to have a higher portion of disturbing videos than the other tags (see Figure~\ref{fig:disturbed_video_tags_stems_normalized_mean_scores_elsagate}).
Third, we observe that the tags ``mous'' ($73.0\%$), ``anim'' ($44.5\%$), ``cartoon'' ($48.8\%$), ``video'' ($31.5\%$), ``disney'' ($36.5\%$), and ``kid'' ($34.2\%$) that appear in a considerable number of disturbing Elsagate-related videos also appear in a high portion of suitable other child-related videos (cf. Figure~\ref{fig:disturbed_video_tags_stems_normalized_mean_scores_elsagate} and Figure~\ref{fig:disturbed_video_tags_stems_normalized_mean_scores_childrelated}).
The main take-away from this analysis is that it is hard to detect disturbing content just by looking at the tags, and that popular tags are shared among disturbing and suitable videos.

\begin{figure*}[t!]
\centering
\subfigure[]{\includegraphics[width=0.49\textwidth, height=2.6in]{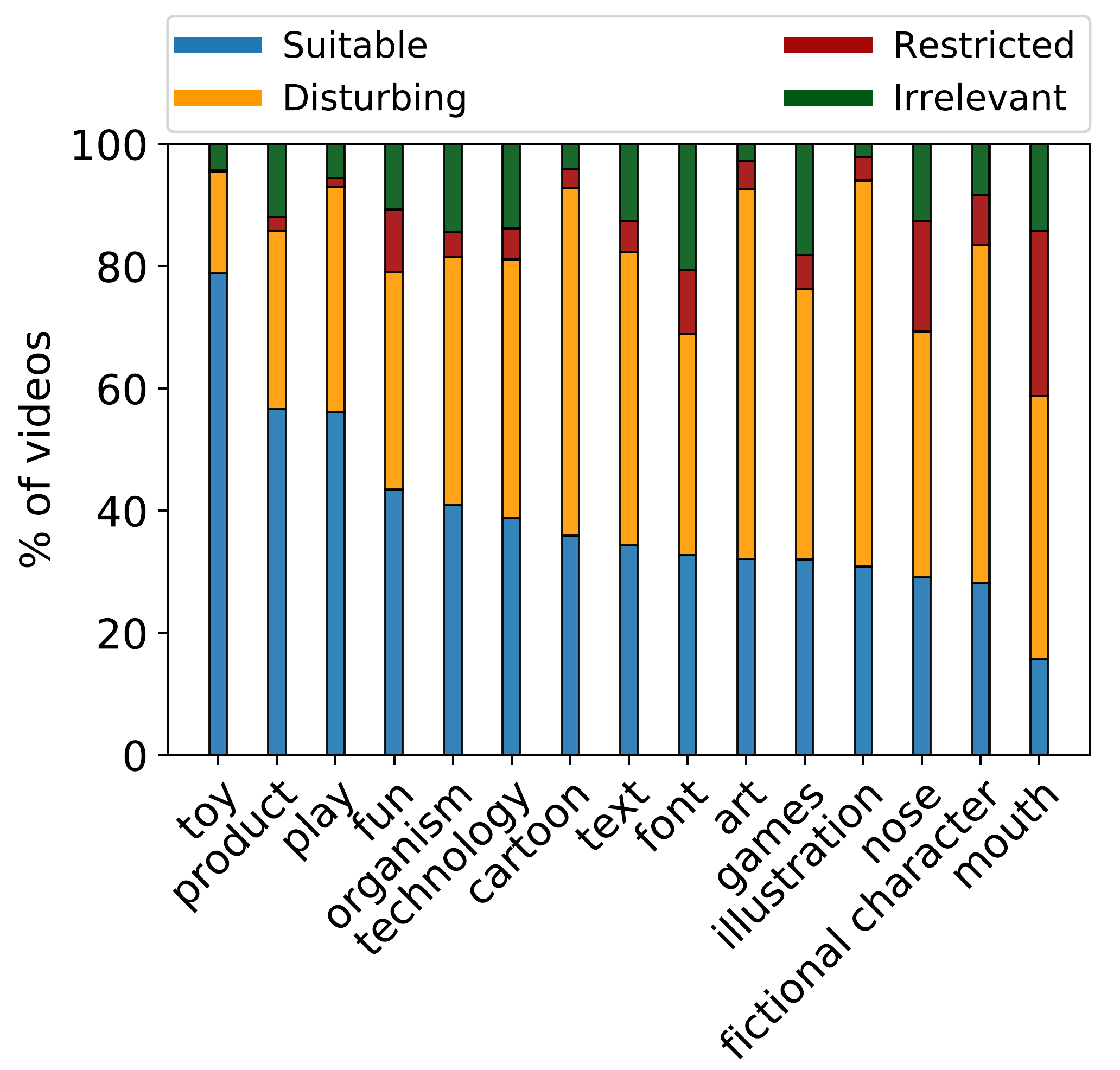}\label{fig:disturbed_thumbnails_labels_normalized_mean_scores_elsagate}}
\subfigure[]{\includegraphics[width=0.49\textwidth, height=2.6in]{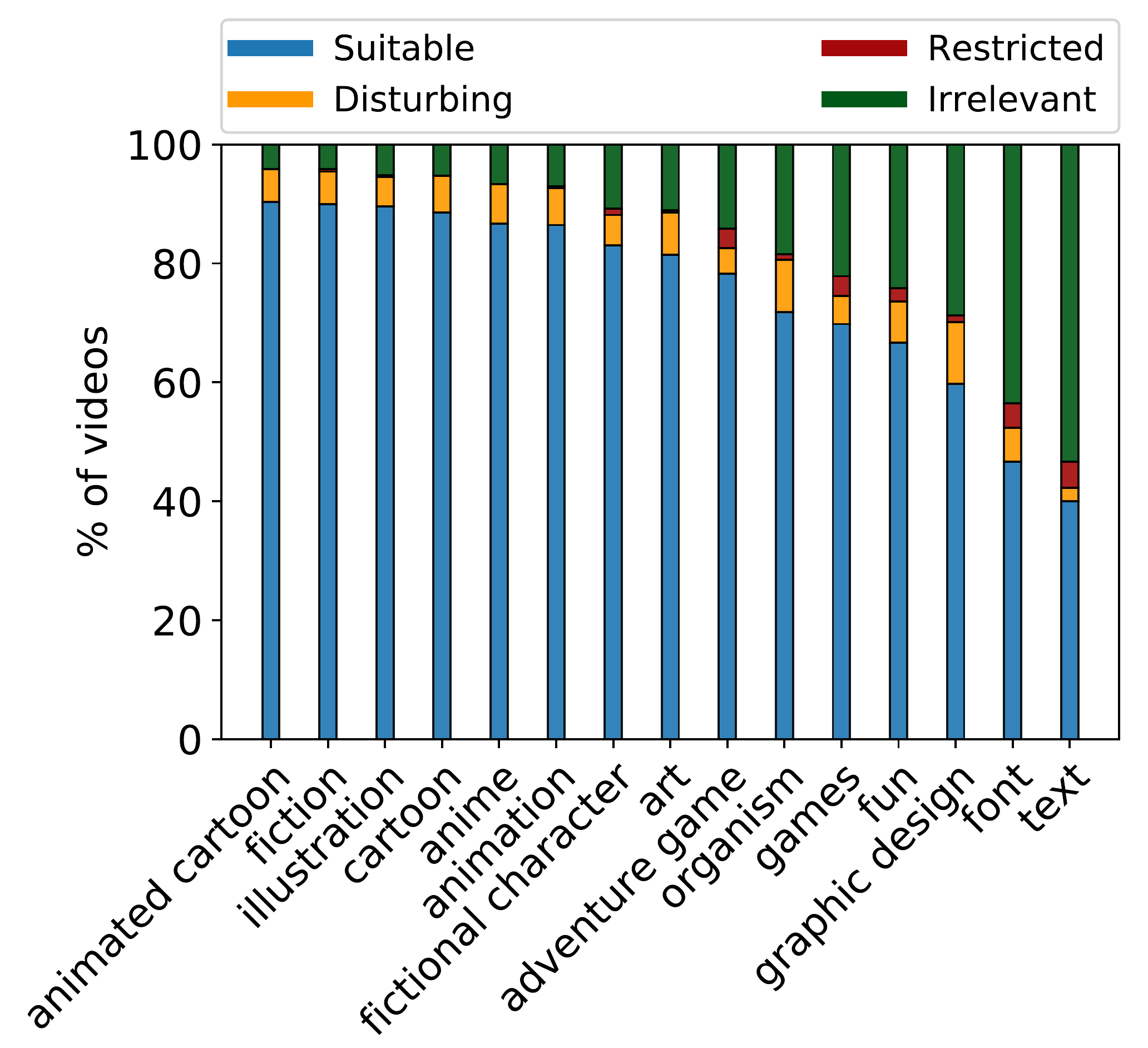}\label{fig:disturbed_thumbnails_labels_normalized_mean_scores_childrelated}}
\subfigure[]{\includegraphics[width=0.49\textwidth, height=2.6in]{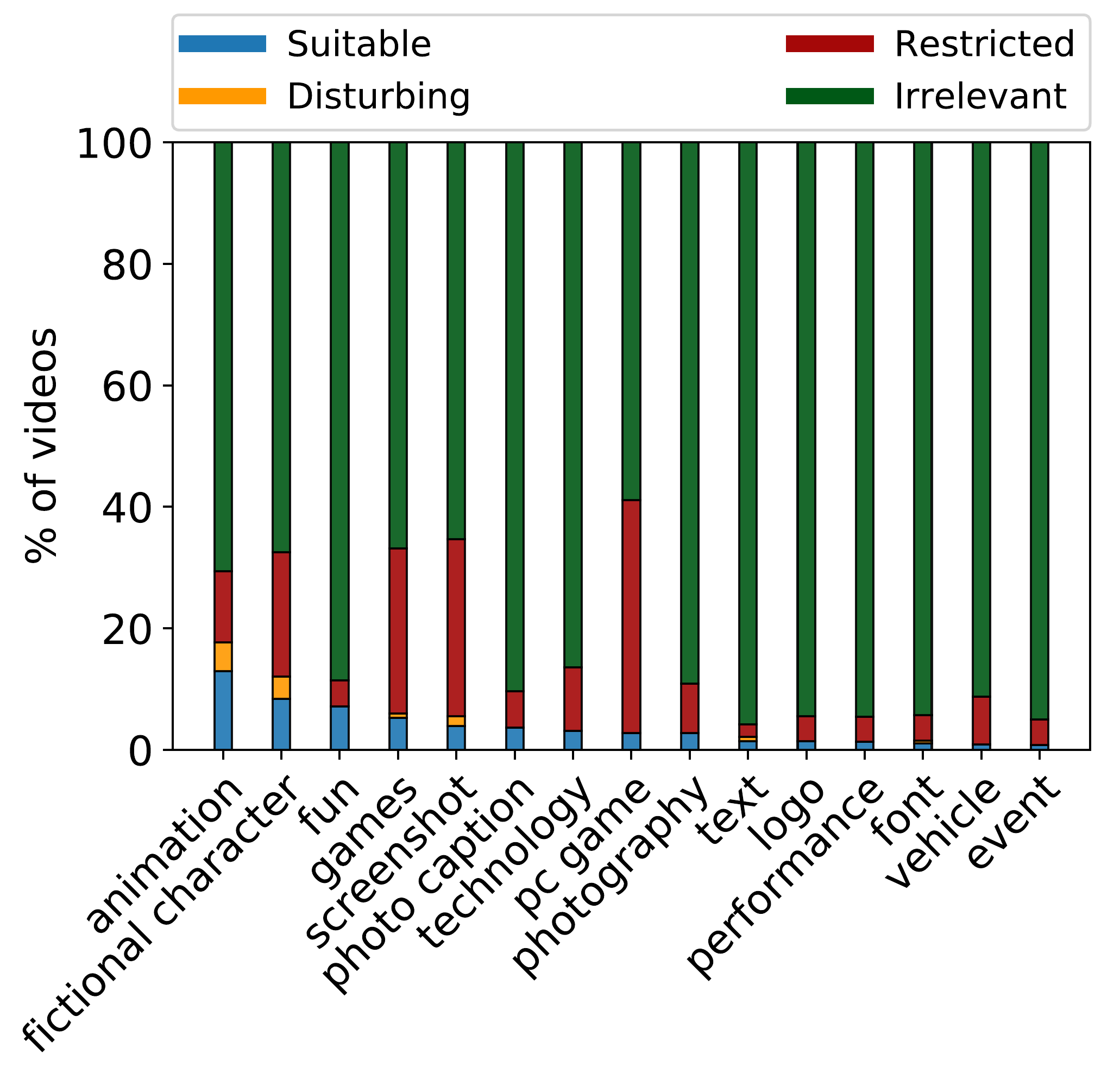}\label{fig:disturbed_thumbnails_labels_normalized_mean_scores_random}}
\subfigure[]{\includegraphics[width=0.49\textwidth, height=2.6in]{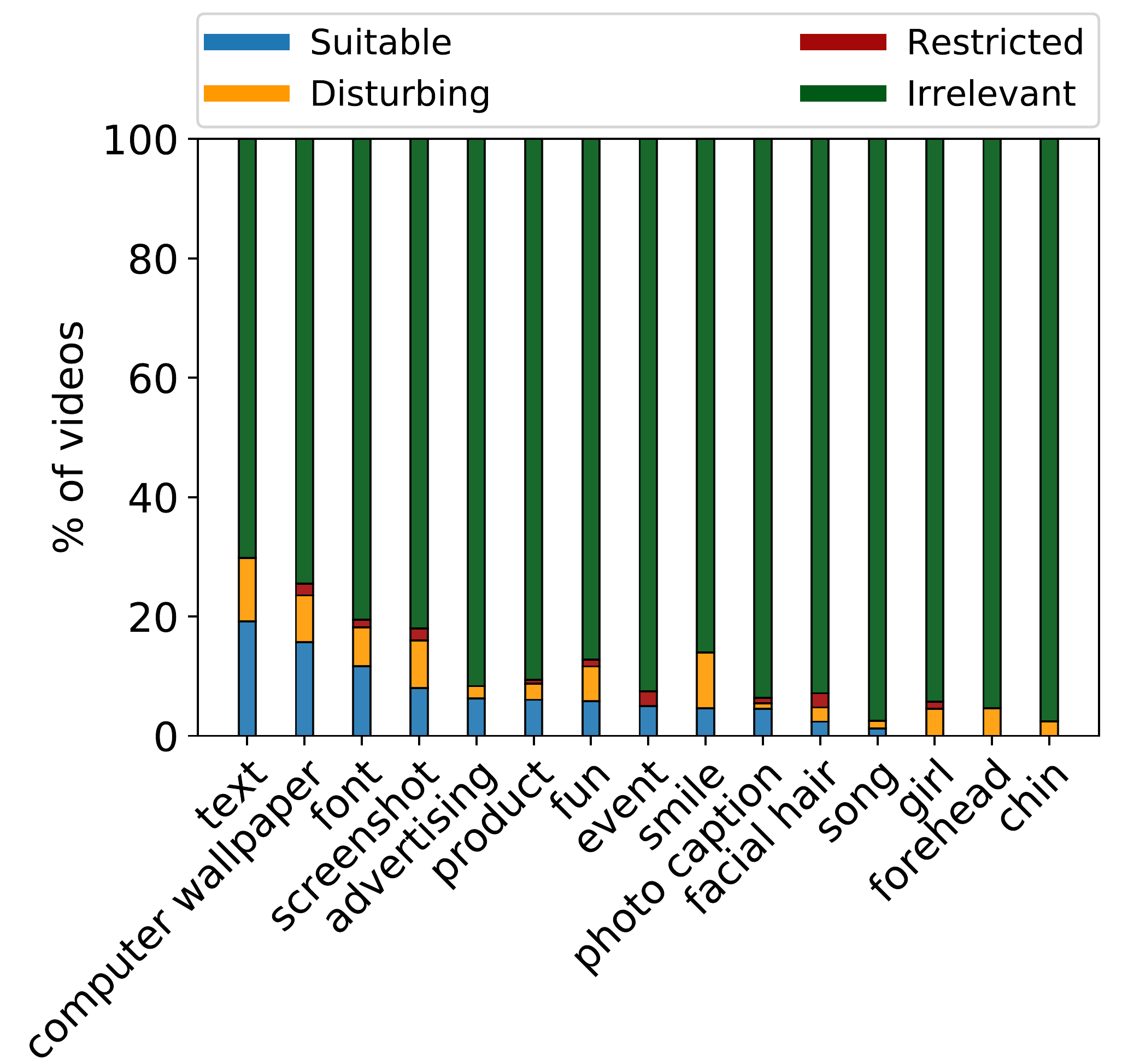}\label{fig:disturbed_thumbnails_labels_normalized_mean_scores_popular}}
\caption{Per class proportion of videos for the top 15 labels found in thumbnails of (a) Elsagate-related; (b) other child-related; (c) random; and (d) popular videos.}
\label{fig:disturbed_groundtruth_analysis_plots_thumbnails_stems}
\end{figure*}

\begin{figure*}[t!]
\centering
\subfigure[]{\includegraphics[width=0.49\textwidth]{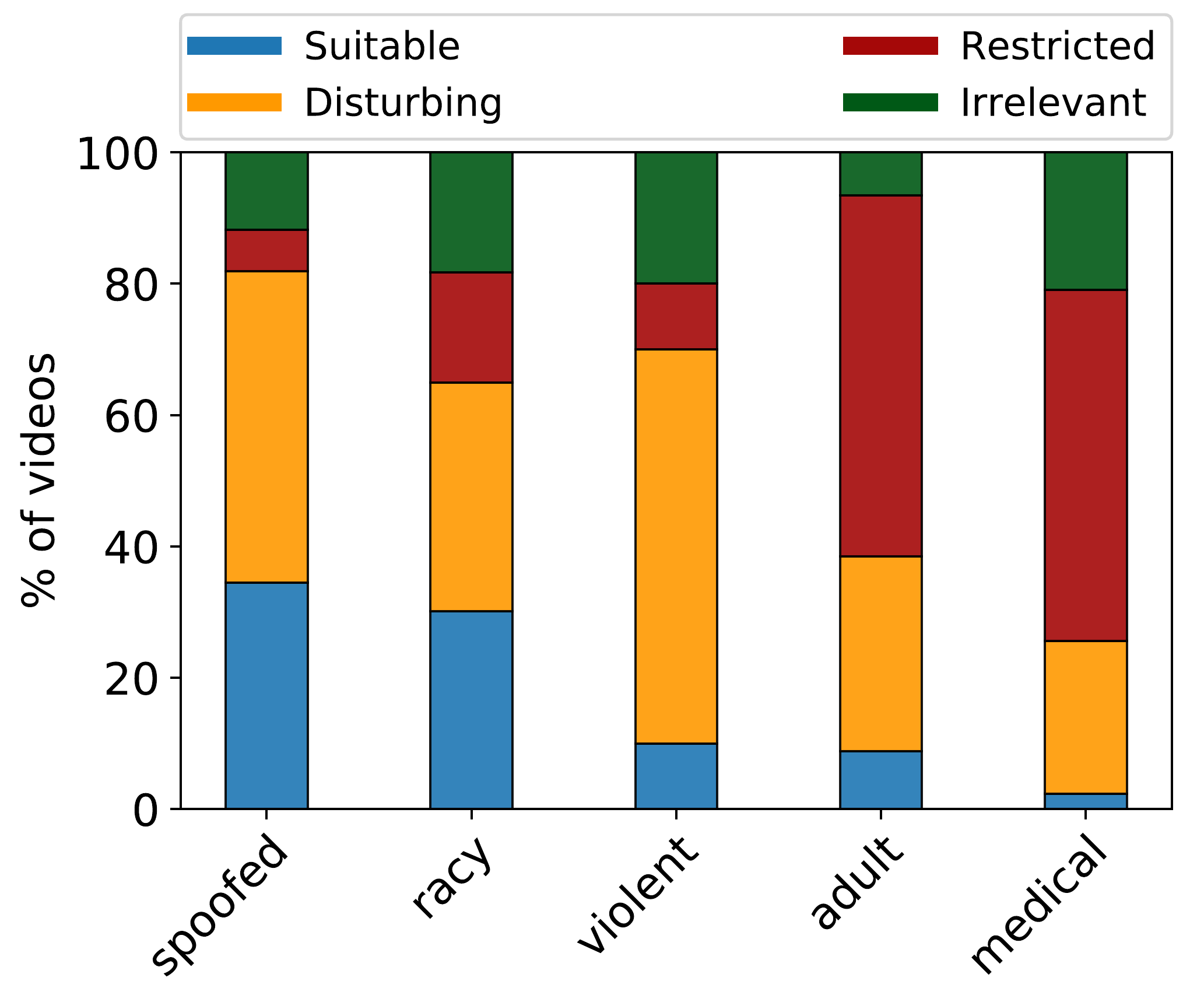}\label{fig:disturbed_thumbnails_safe_search_normalized_mean_scores_elsagate}}
\subfigure[]{\includegraphics[width=0.49\textwidth]{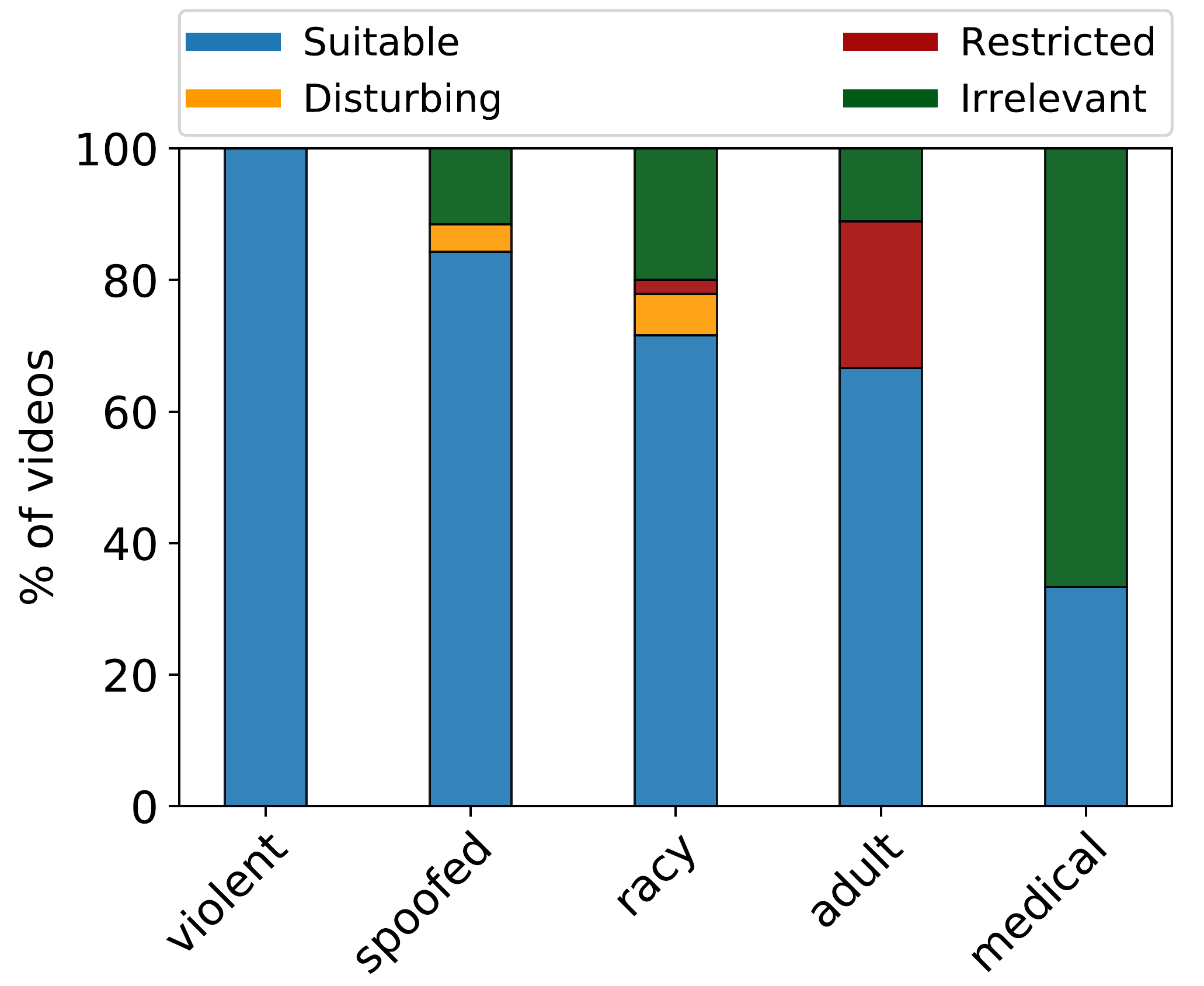}\label{fig:disturbed_thumbnails_safe_search_normalized_mean_scores_childrelated}}
\subfigure[]{\includegraphics[width=0.49\textwidth]{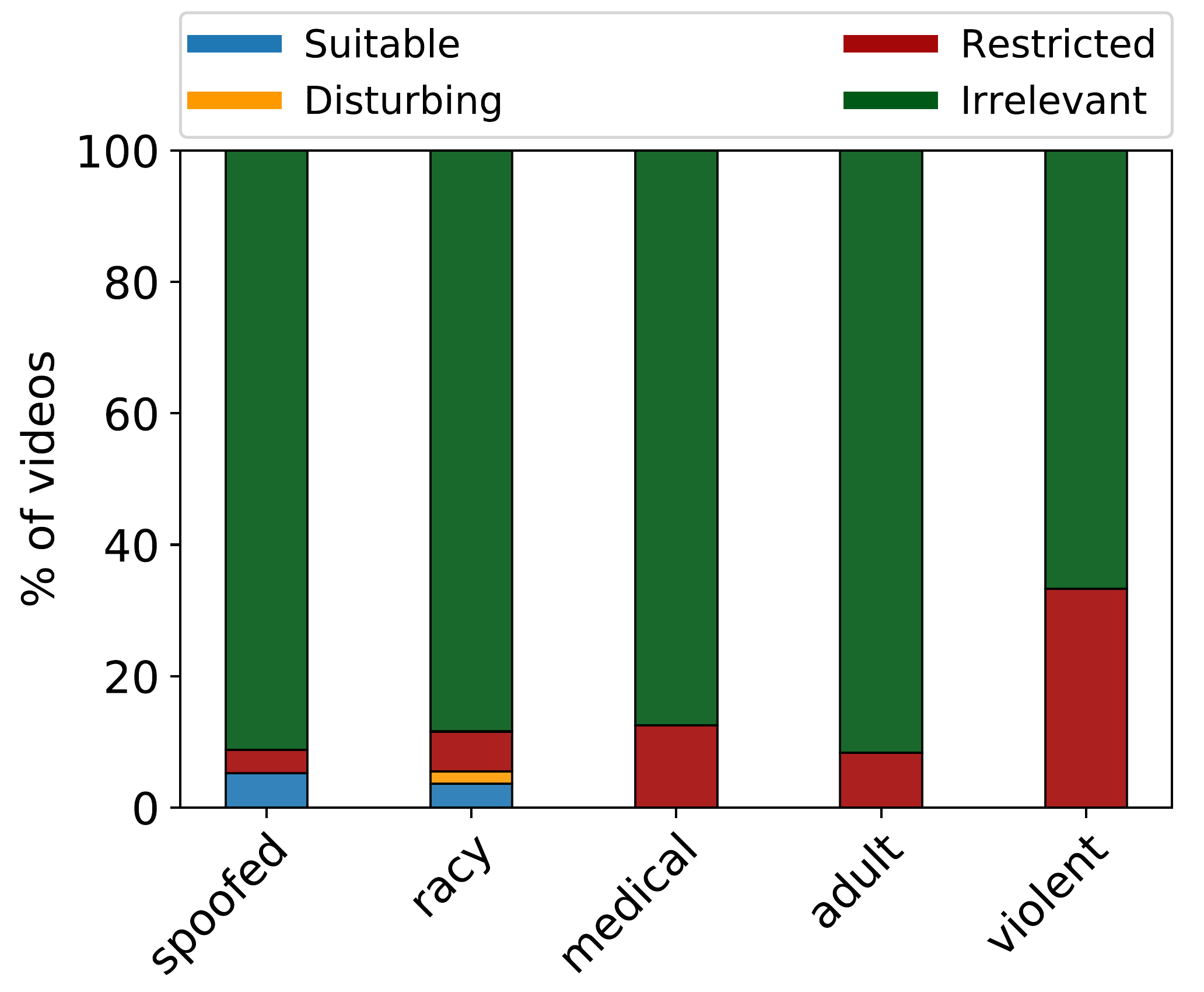}\label{fig:disturbed_thumbnails_safe_search_normalized_mean_scores_random}}
\subfigure[]{\includegraphics[width=0.49\textwidth]{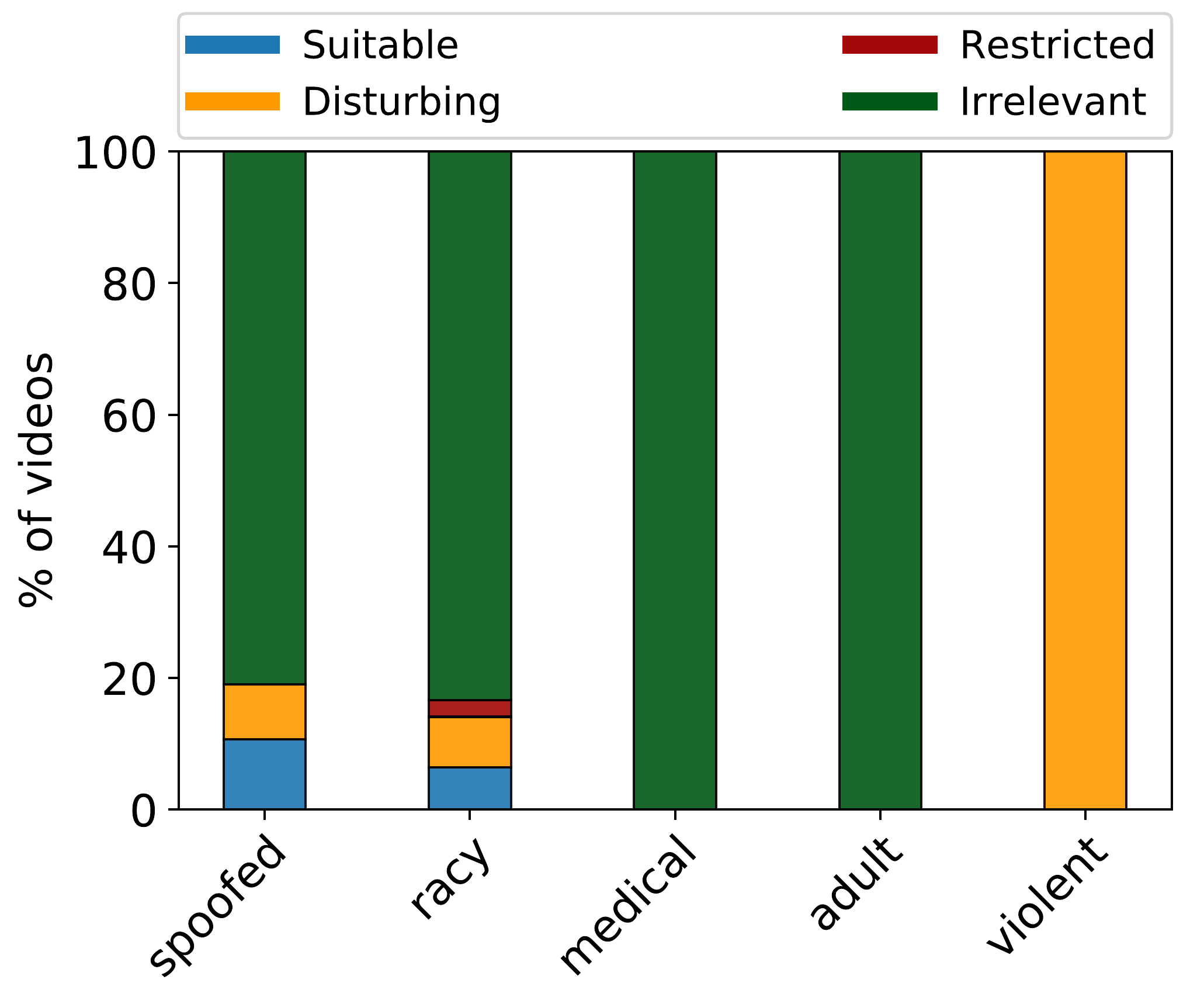}\label{fig:disturbed_thumbnails_safe_search_normalized_mean_scores_popular}}
\caption{Per class proportion of videos that their thumbnail contains spoofed, adult, medical, violent, and/or racy content for (a) Elsagate-related; (b) other child-related; (c) random; and (d) popular videos.}
\label{fig:disturbed_groundtruth_analysis_plots_thumbnails_safesearch_categories}
\end{figure*}

\descr{Thumbnails.}
To study the thumbnails of the videos in our ground truth dataset, we make use of the Google Cloud Vision API\footnote{\url{https://cloud.google.com/vision/}}, which is a RESTful API that derives useful insights from images using pre-trained machine learning models. 
Using this API we can: 1) extract descriptive labels for all the thumbnails in our ground truth; and 2) check whether a modification was made to a thumbnail and whether a thumbnail contains adult, medical-related, violent, and/or racy content. 
Figure~\ref{fig:disturbed_groundtruth_analysis_plots_thumbnails_stems} depicts the top 15 labels derived from the thumbnails of videos in our ground truth. 
In the Elsagate-related case, we observe that the thumbnails of disturbing videos contain similar entities as the thumbnails of both the Elsagate-related and other child-related suitable videos (cartoons, fictional characters, etc.).

Figure~\ref{fig:disturbed_groundtruth_analysis_plots_thumbnails_safesearch_categories} shows the proportion of each class for videos that contain spoofed, adult, medical-related, violent, and/or racy content.
As expected, most of the Elsagate-related videos whose thumbnails contain adult ($54.9\%$) and medical content ($53.5\%$) are restricted.
However, this is not the case for videos whose thumbnails contain spoofed ($47.4\%$), violent ($60.0\%$), or racy ($34.8\%$) content, where we observe a high number of disturbing videos (cf. Figure~\ref{fig:disturbed_thumbnails_safe_search_normalized_mean_scores_elsagate}).
Surprisingly, we notice that $100.0\%$ of the other child-related videos whose thumbnail contains violent content are suitable.
Nonetheless, after manually inspecting some of those thumbnails we notice that they depict mild cartoon violence (i.e., tom hitting jerry), which we consider as suitable.
In general, we observe that Elsagate-related videos whose thumbnail is modified with violent, racy, medical, and/or adult content are more likely to be restricted or disturbing, while this is not the case for the other child-related videos.

\begin{figure*}[t!]
\centering
\subfigure[]{\includegraphics[width=0.49\textwidth]{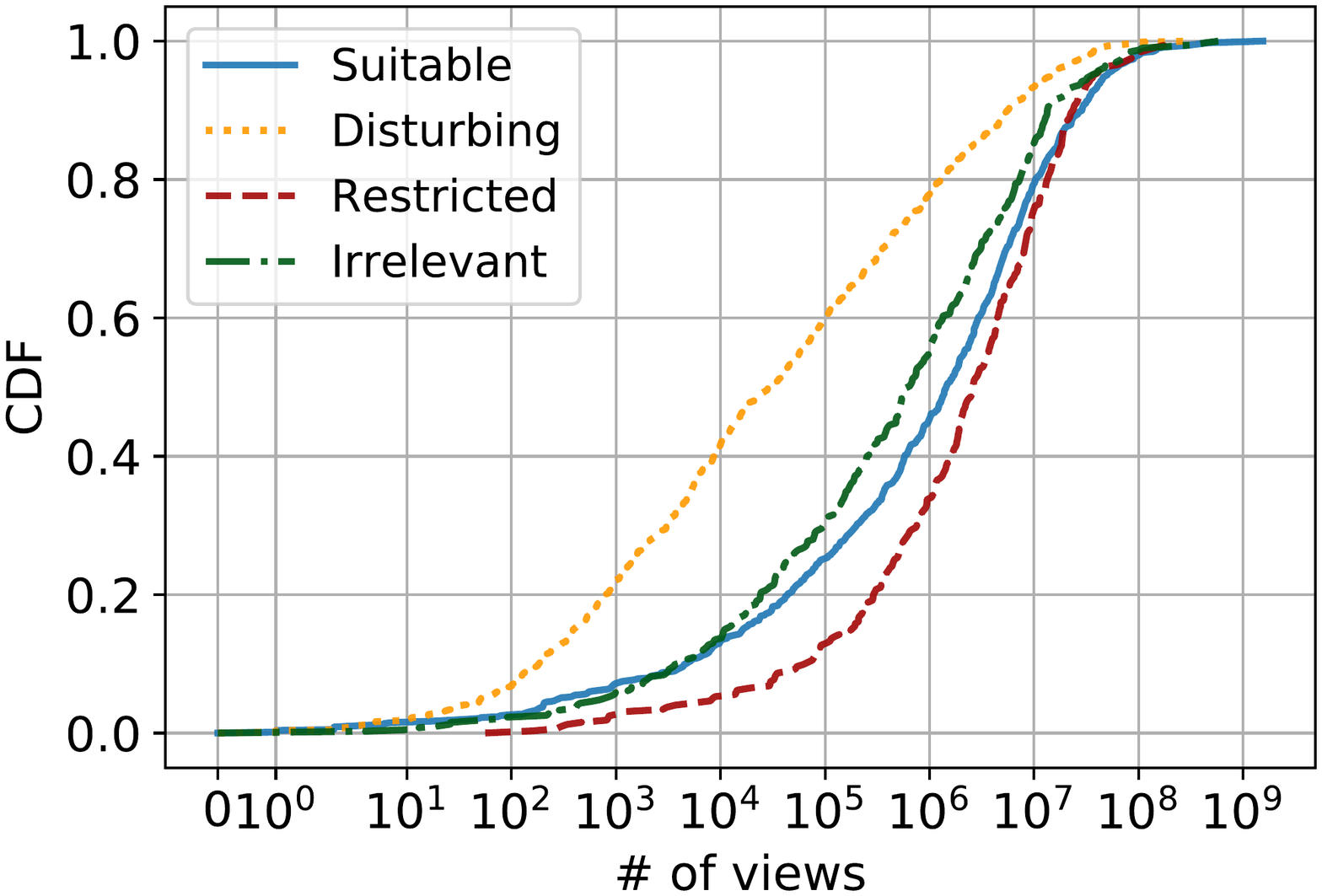}}
\subfigure[]{\includegraphics[width=0.49\textwidth]{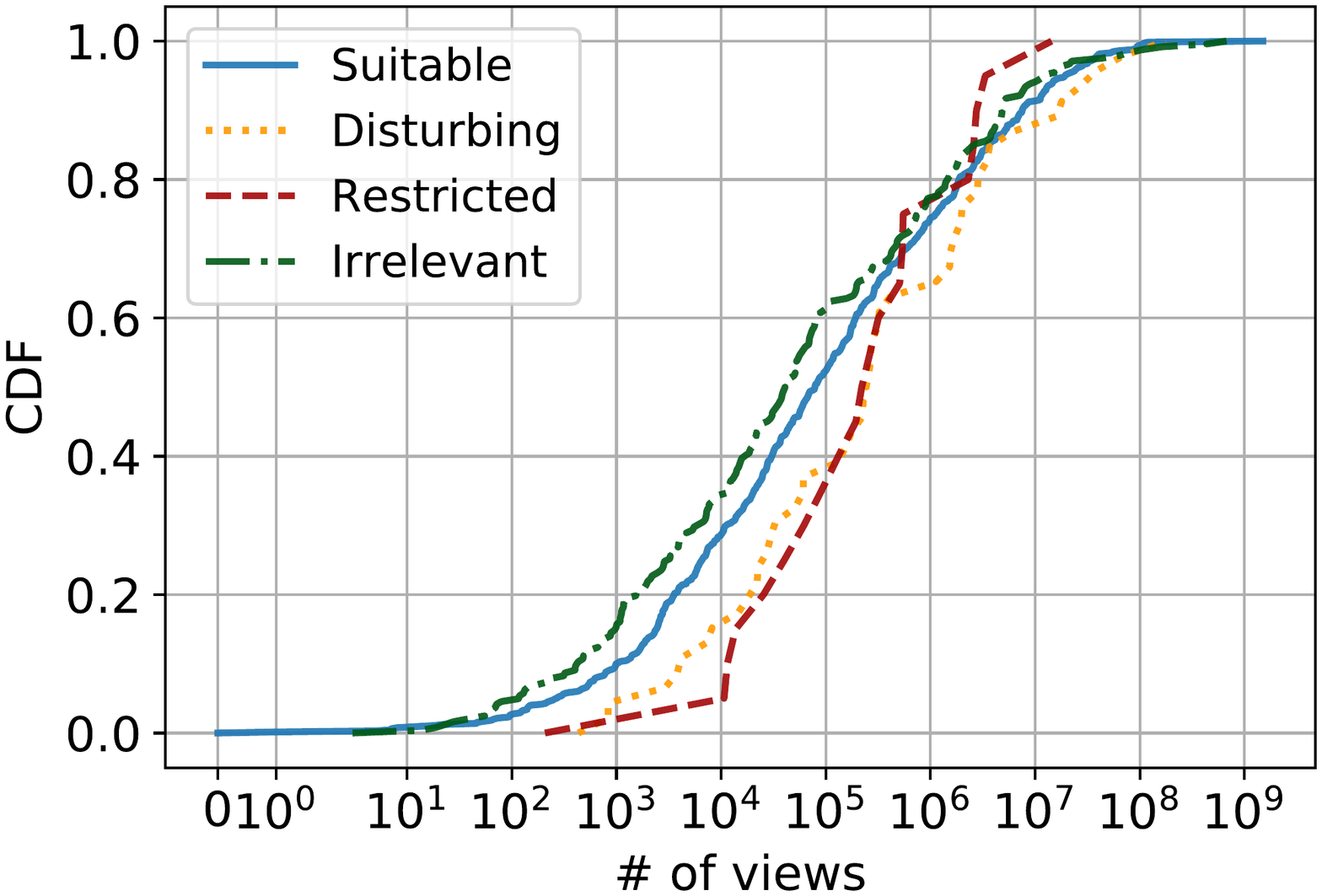}}
\subfigure[]{\includegraphics[width=0.49\textwidth]{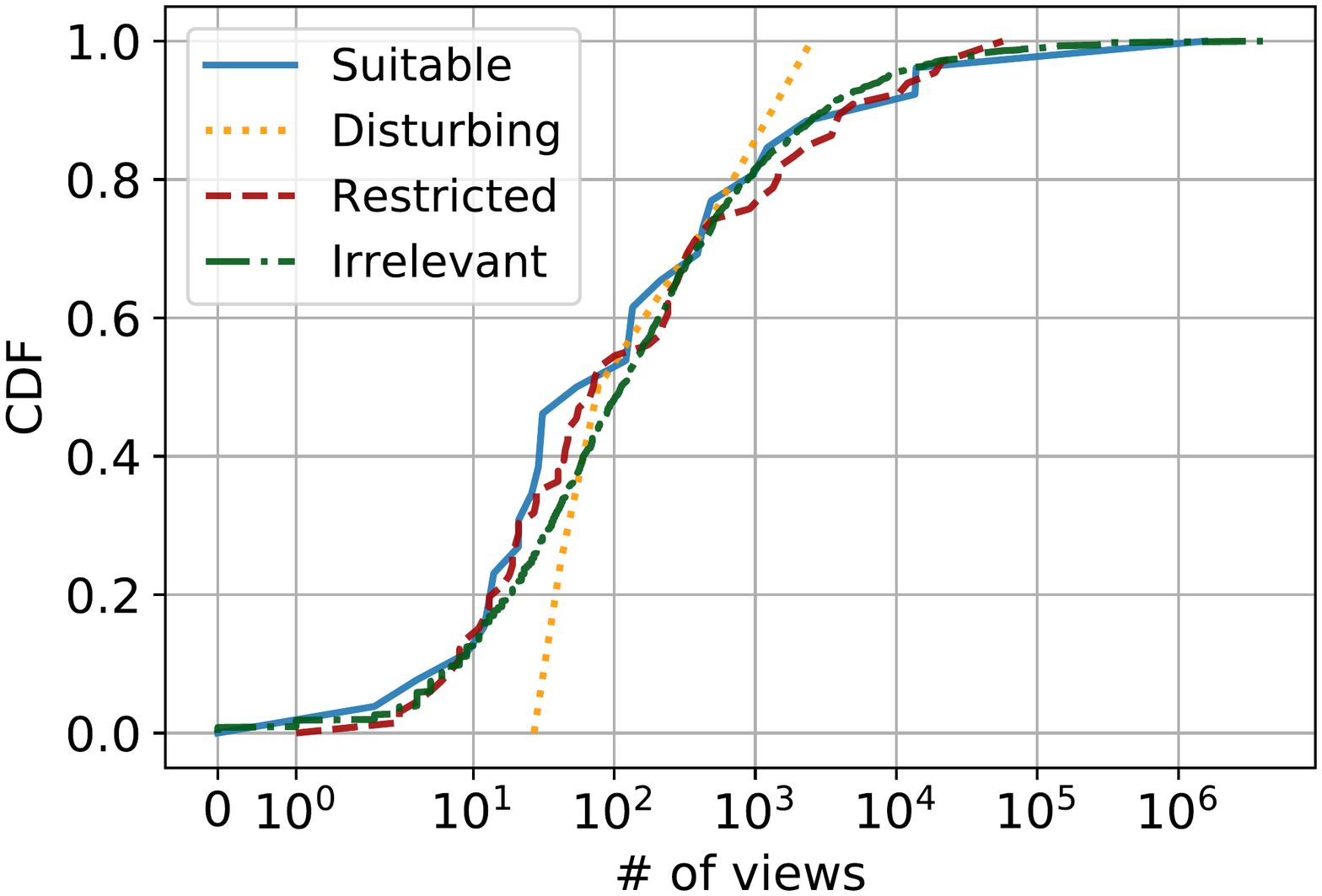}}
\subfigure[]{\includegraphics[width=0.49\textwidth]{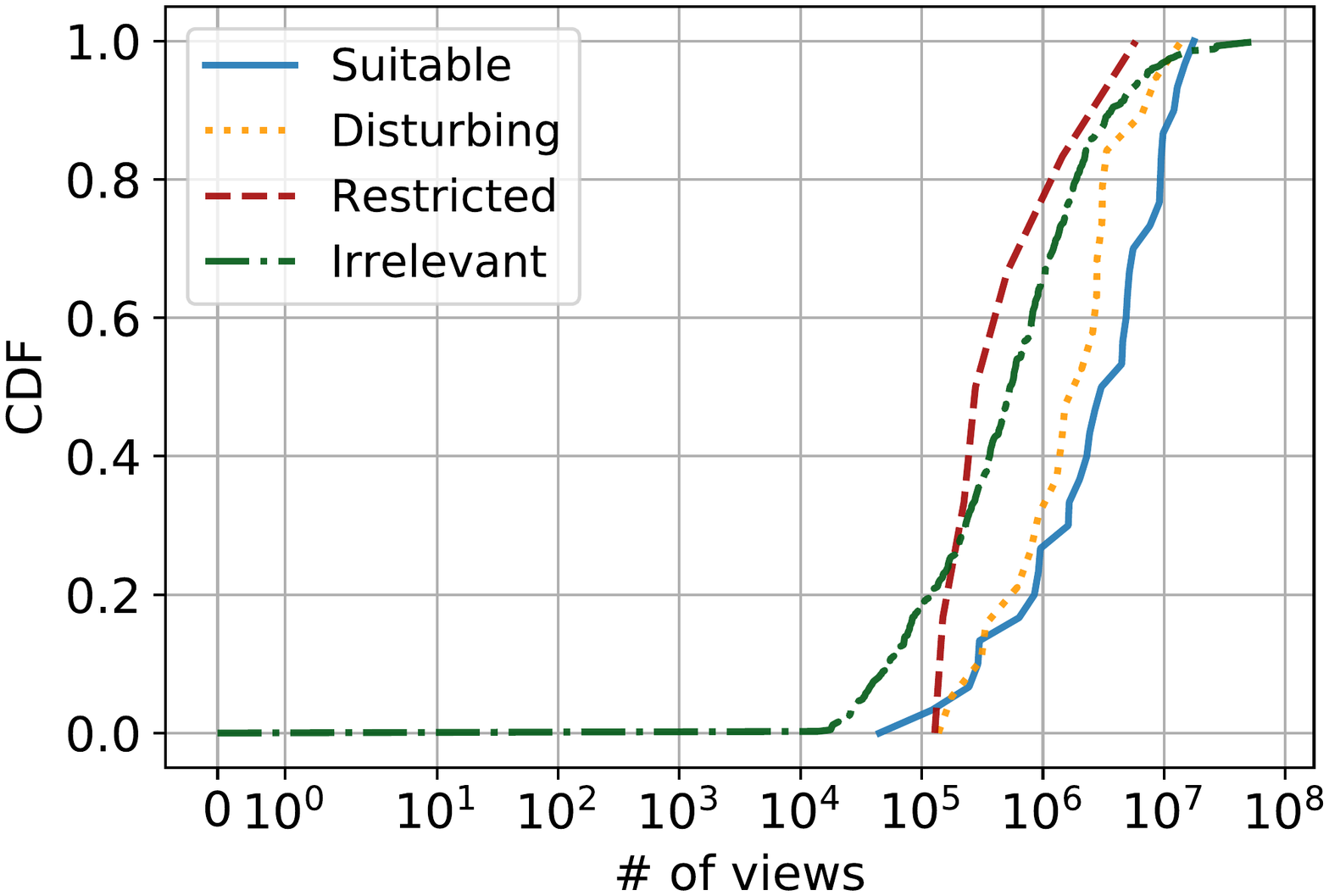}}
\caption{CDF of the number of views per class for (a) Elsagate-related (b) other child-related, (c) random, and (d) popular videos.}
\label{fig:disturbed_cdf_plots_views}
\end{figure*}

\begin{figure*}[t!]
\centering
\subfigure[]{\includegraphics[width=0.49\textwidth]{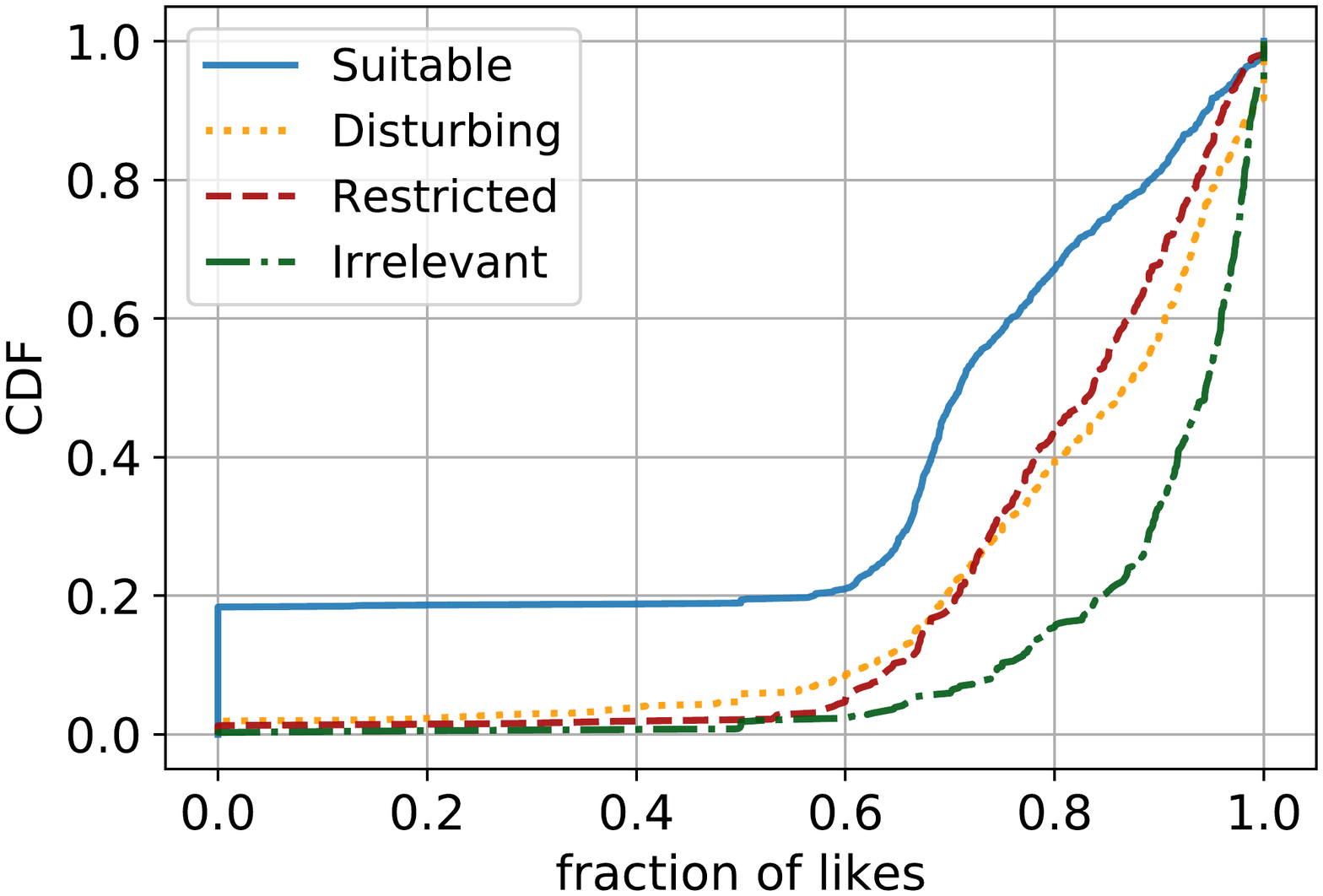}}
\subfigure[]{\includegraphics[width=0.49\textwidth]{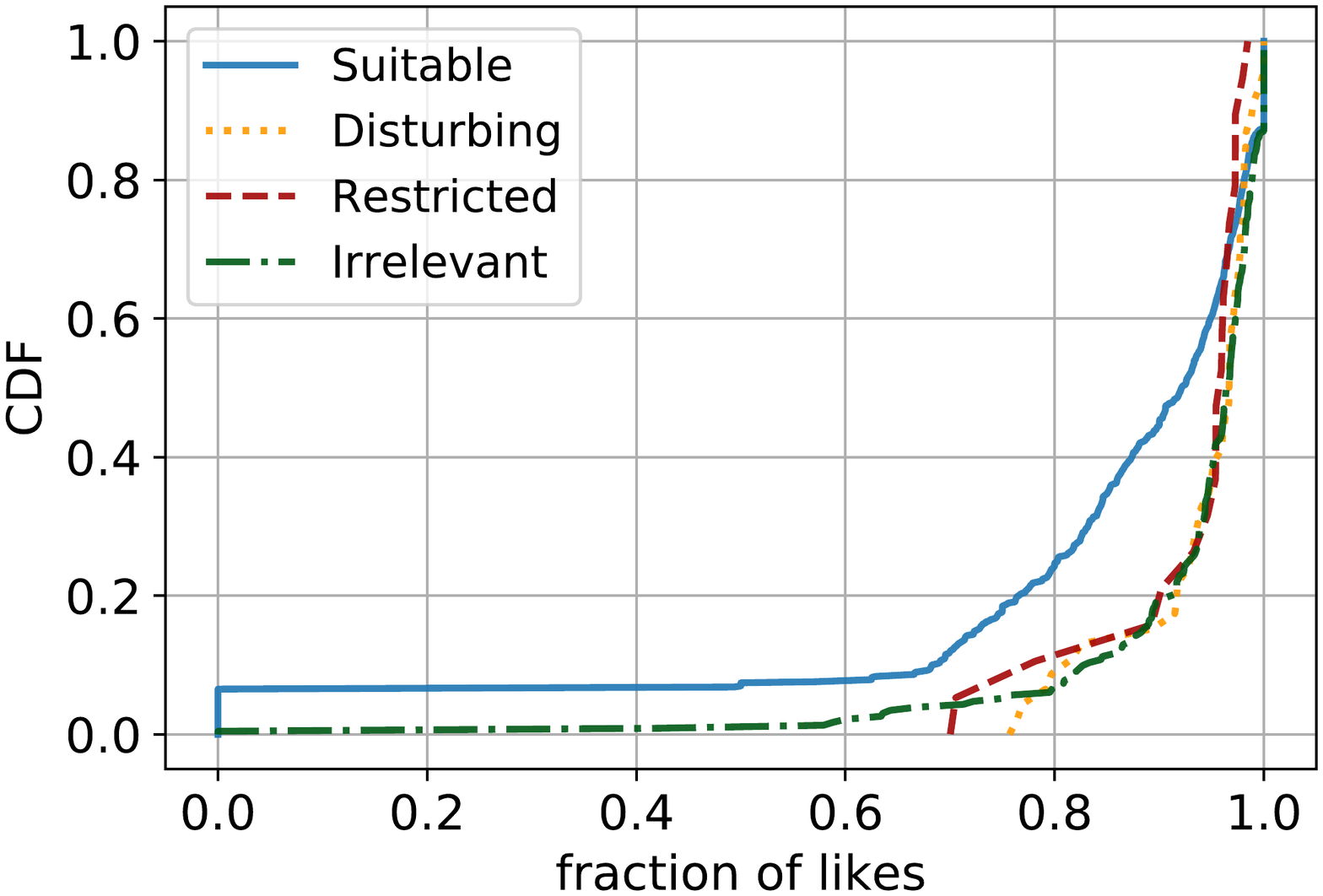}}
\subfigure[]{\includegraphics[width=0.49\textwidth]{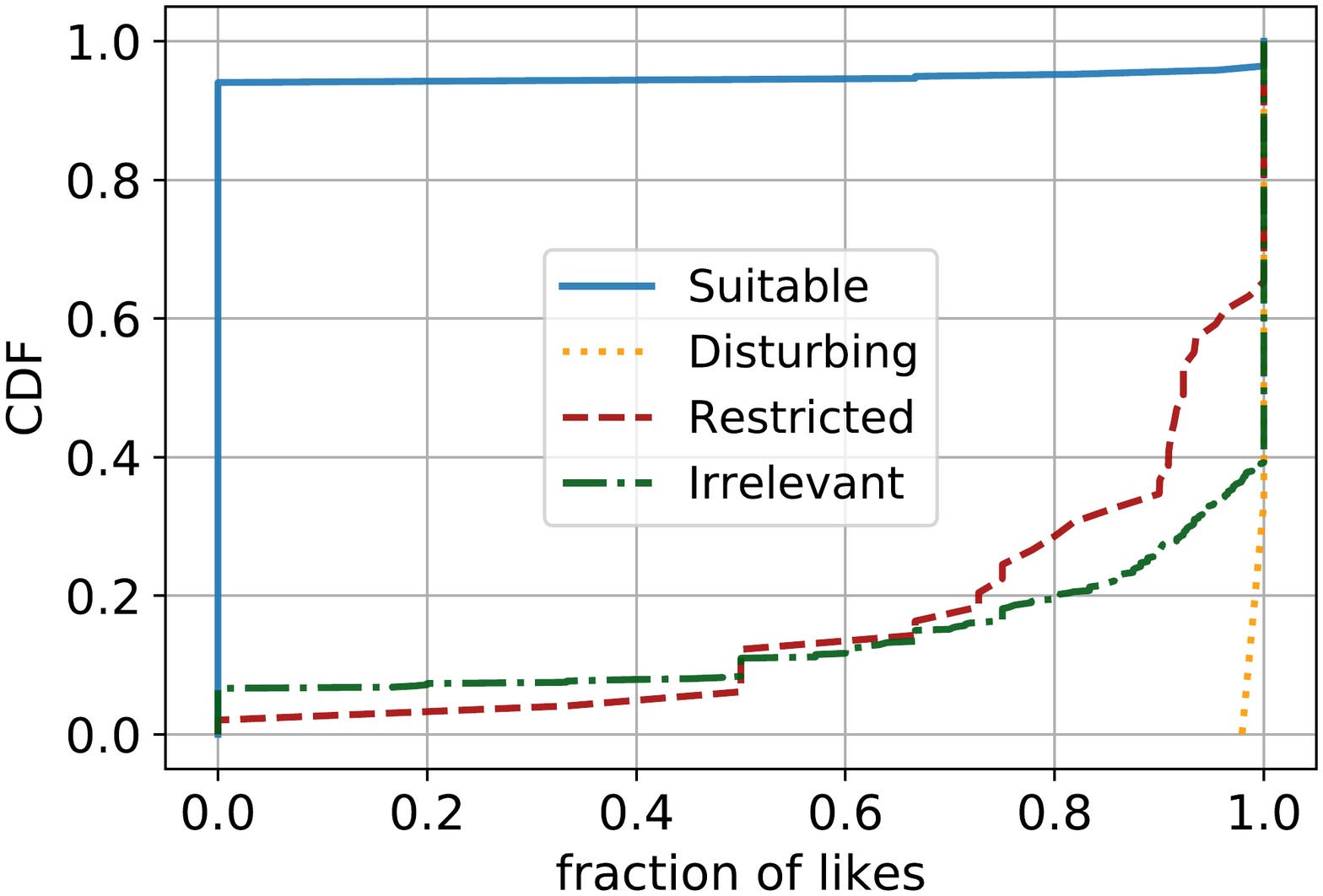}}
\subfigure[]{\includegraphics[width=0.49\textwidth]{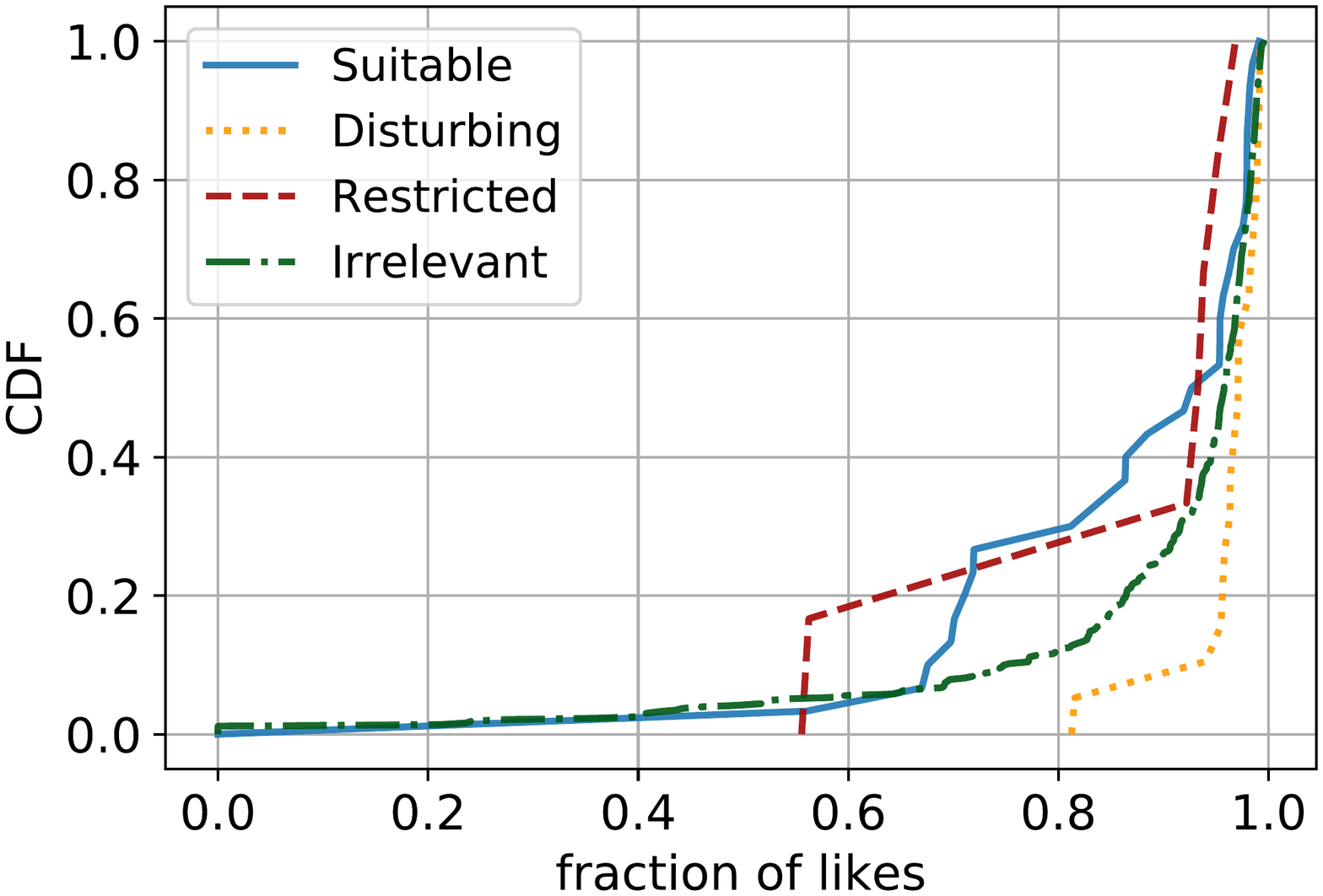}}
\caption{CDF of the fraction of likes to dislikes per class for (a) Elsagate-related (b) other child-related, (c) random, and (d) popular videos.}
\label{fig:disturbed_cdf_plots_likesdislikes_fraction}
\end{figure*}

\begin{figure*}[t!]
\centering
\subfigure[]{\includegraphics[width=0.49\textwidth]{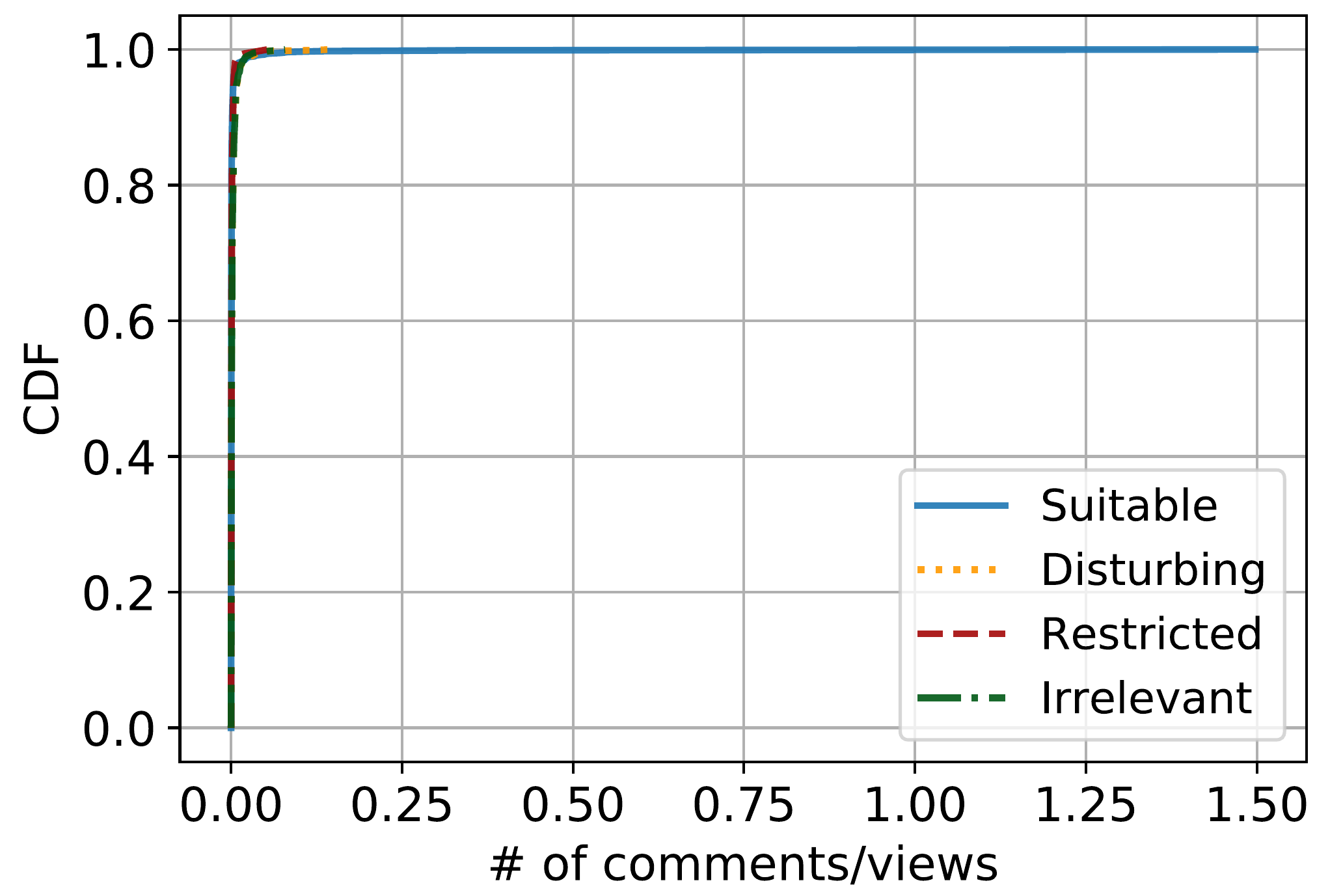}}
\subfigure[]{\includegraphics[width=0.49\textwidth]{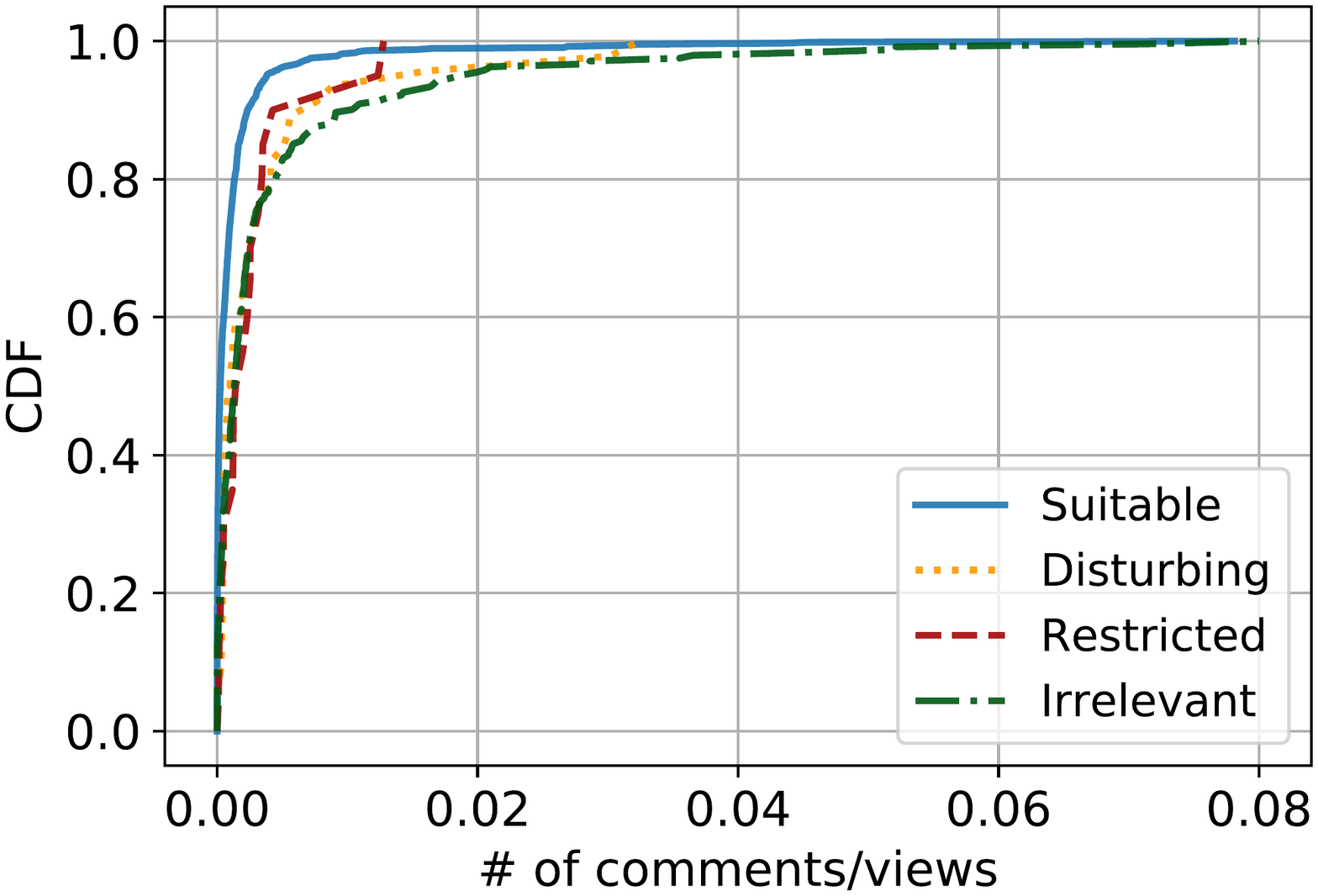}}
\subfigure[]{\includegraphics[width=0.49\textwidth]{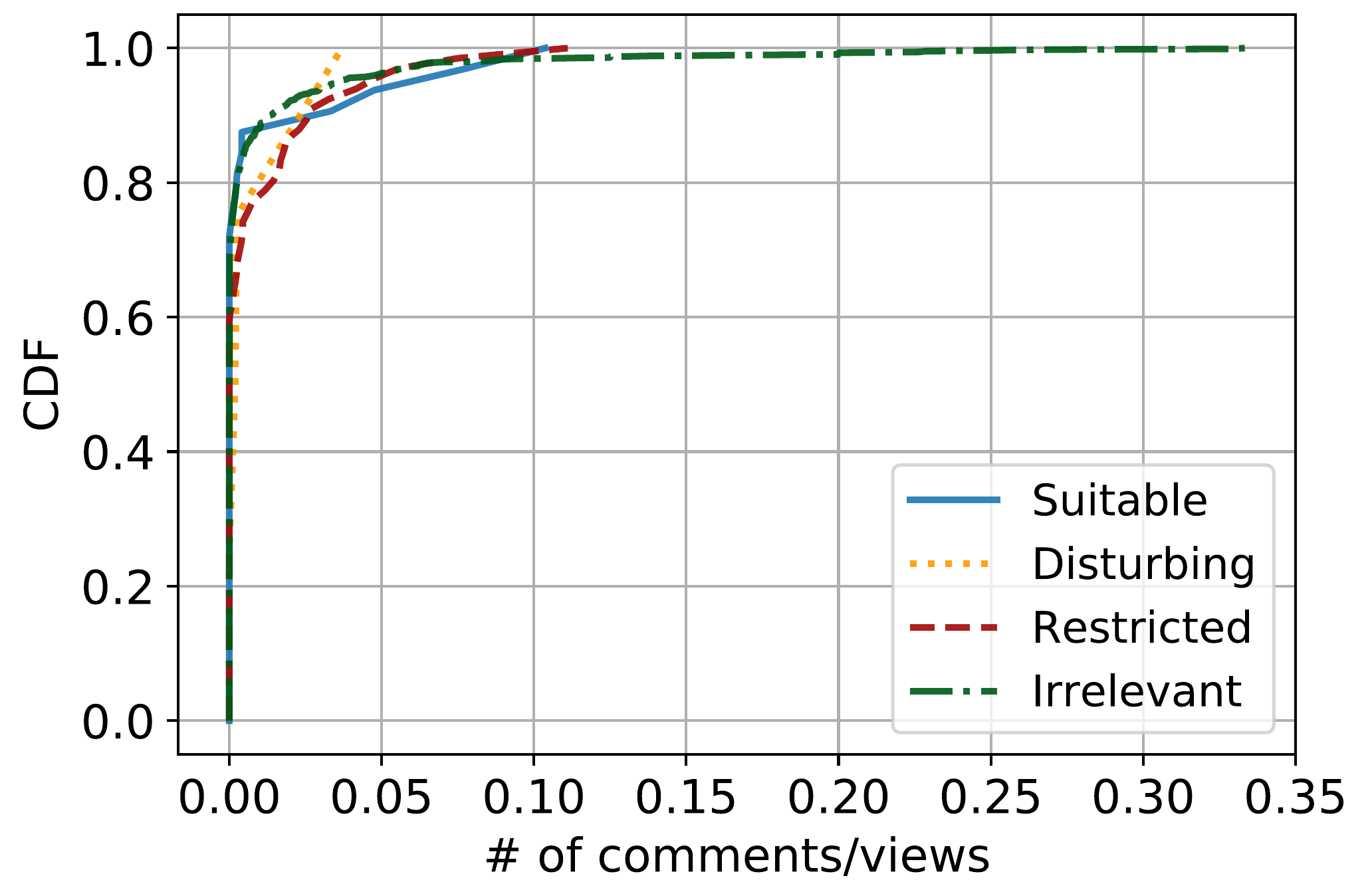}}
\subfigure[]{\includegraphics[width=0.49\textwidth]{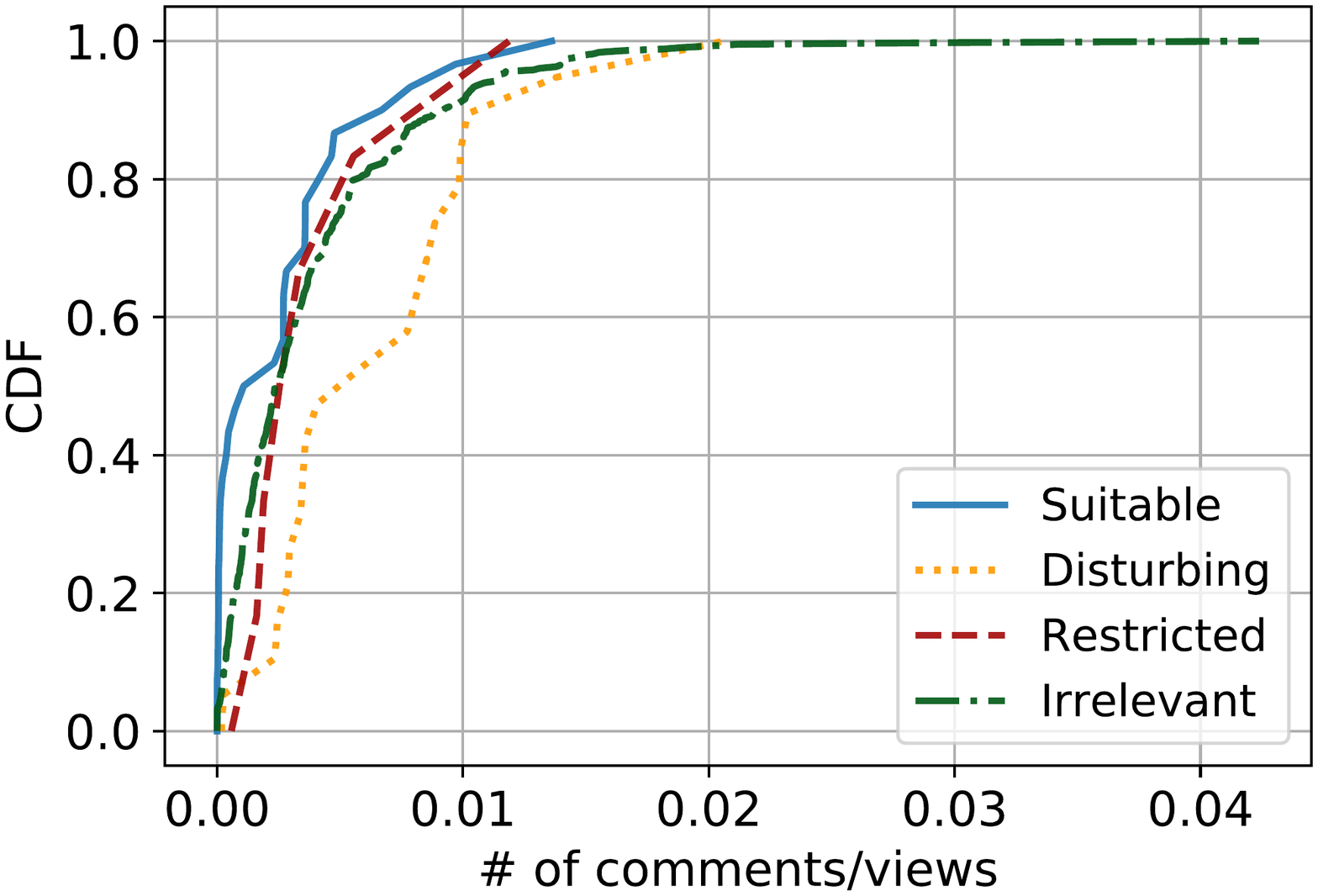}}
\caption{CDF of the number of comments/views per class for (a) Elsagate-related (b) other child-related, (c) random, and (d) popular videos.}
\label{fig:disturbed_cdf_plots_commentsviews_fraction}
\end{figure*}

\descr{Statistics.}
Next, we examine statistics that pertain to the videos in our ground truth dataset. 
Figure~\ref{fig:disturbed_cdf_plots_views} shows the CDF of the number of views of all the videos in each distinct subset of videos in our ground truth.
We observe that Elsagate-related suitable videos have more views than disturbing videos while this is not the case for all the other types of videos.
Figure~\ref{fig:disturbed_cdf_plots_likesdislikes_fraction} shows the CDF of the fraction of likes of all the videos in each subset. 
Interestingly, we observe that in all cases disturbing and restricted videos have a higher fraction of likes compared to suitable videos, which, particularly in the case of disturbing videos, indicates manipulation to boost their ranking. 
Lastly, Figure~\ref{fig:disturbed_cdf_plots_commentsviews_fraction} shows the CDF of the fraction of comments to views.
Although for the Elsagate-related videos the suitable and disturbing videos have a similar ratio of comments, the situation shifts when it comes to all the other types of videos where we observe a higher ratio of comments for disturbing and restricted videos compared to suitable videos.

A general take away from this ground truth analysis is that none of the videos' metadata can clearly indicate that a video is disturbing or not, thus, in most cases, one (e.g., a guardian) has to carefully inspect all the available video metadata, and potentially the actual video, to accurately determine if it is safe for a toddler to watch.

\descr{Assessing YouTube's Counter-measures.}
To assess how fast YouTube detects and removes inappropriate videos, we leverage the YouTube Data API to count the number of off-line videos (either removed by YouTube due to a Terms of Service violation or deleted by the uploader) in our manually reviewed ground-truth dataset.
We note that we do not consider the videos that were already marked as age-restricted since YouTube took the appropriate measures.

As of May 10, 2019, only $9.65\%$ of the suitable, $20.5\%$ of the disturbing, $2.5\%$ of the restricted, and $2.4\%$ of the irrelevant videos were removed, while from those that were still available, $0.0\%$, $6.9\%$, $1.3\%$, and $0.1\%$, respectively, were marked as age-restricted.
Alarmingly, the amount of deleted disturbing, and restricted videos is considerably low.
The same stands for the amount of disturbing and restricted videos marked as age-restricted.
A potential issue here is that the videos on our dataset were recently uploaded and YouTube simply did not have time to detect them.
To test this hypothesis, we calculate the mean number of days from publication up to May 2019: we find this hypothesis does not hold.
The mean number of days since being uploaded for the suitable, disturbing, restricted, and irrelevant videos is 866, 942, 1091, and 991, respectively, with a mean of 947 days across the entire manually reviewed ground-truth dataset.
This indicates that YouTube's deployed counter-measures eliminated some of the disturbing videos, but they are unable to tackle the problem promptly.

\begin{table}[t!]
\footnotesize
\centering
\begin{tabular}{ll}
\toprule
\textbf{Type} & \textbf{Style Features Description} \\
\midrule
\textbf{Video-related} & video category, video duration \\
\textbf{Statistics-related} & ratio of \# of likes to dislikes \\
\textbf{Title-related \& Description-related} & length of title, length of description,\\
 & ratio of description to title,\\
 & jaccard sim. of title \& description, \\
 & \# '!' and '?' in title \& description, \\
 & \# emoticons in title \& description, \\
 & \# bad words in title \& description, \\
 & \# child-related words in title \& description \\
\textbf{Tags-related} & \# tags, \# bad words in tags, \\ 
 & \# child-related words in tags, \\
 & jaccard sim. of tags \& title \\
\toprule
\end{tabular}
\caption{List of the style features extracted from the available metadata of a video.}
\label{tab:disturbed_style_features_description}
\end{table}

\section{Detection of Disturbing Videos}
\label{sec:disturbed_detectionofdisturbingvideos}
\revision{
In this section, we provide details on our deep learning model for detecting disturbing videos on YouTube.
Our ground truth analysis shows that none of the examined video's metadata can clearly indicate whether a video is disturbing or not.
Hence, we devise a deep learning model that considers all the available metadata of a given video and we perform experiments and an ablation study to quantify the importance and contribution of each input type in the classification task.
}

\subsection{Dataset and Feature Description}
To train and test our proposed deep learning model we use our ground truth dataset of 4,797 videos, summarized in Table~\ref{tab:disturbed_final_groundtruth_dataset_details}.
For each video in our ground truth dataset, our model processes the following: 

\descr{Title.}
Our model considers the text of the title by training an embedding layer, which encodes each word in the text in an N-dimensional vector space. 
The maximum number of words found in the title of videos in our ground truth is 21, while the size of the vocabulary is 12,023.

\descr{Tags.} 
Similarly to the title, we encode the video tags into an N-dimensional vector space by training a separate embedding layer. 
The maximum number of tags found in a video is 78, while the size of the word vocabulary is 40,096.

\descr{Thumbnail.} 
We scale down the thumbnail images to 299x299 while preserving all three color channels.

\descr{Statistics.} 
We consider all available statistical metadata for videos (number of views, likes, dislikes, and comments).

\descr{Style Features.} 
We consider some style features from the actual video (e.g., duration), the title (e.g., number of bad words), the video description (e.g., description length), and the tags (e.g., number of tags).
For this, we use features proposed in~\cite{kaushal2016kidstube} that help the model to better differentiate between the videos of each class. 
Table~\ref{tab:disturbed_style_features_description} summarizes the style features that we use. 

\revision{
We note that in the context of this line of work we do not consider the comments of a given video.
Instead, we focus on detecting disturbing videos only on their actual content as defined by the uploader of the video. 
Considering the comments of a video may bias the decision of our classifier, as comments with inappropriate content may also exist in suitable videos for toddlers.
}

\subsection{Model Architecture}
\begin{figure}[t!]
\centering
\includegraphics[width=\columnwidth]{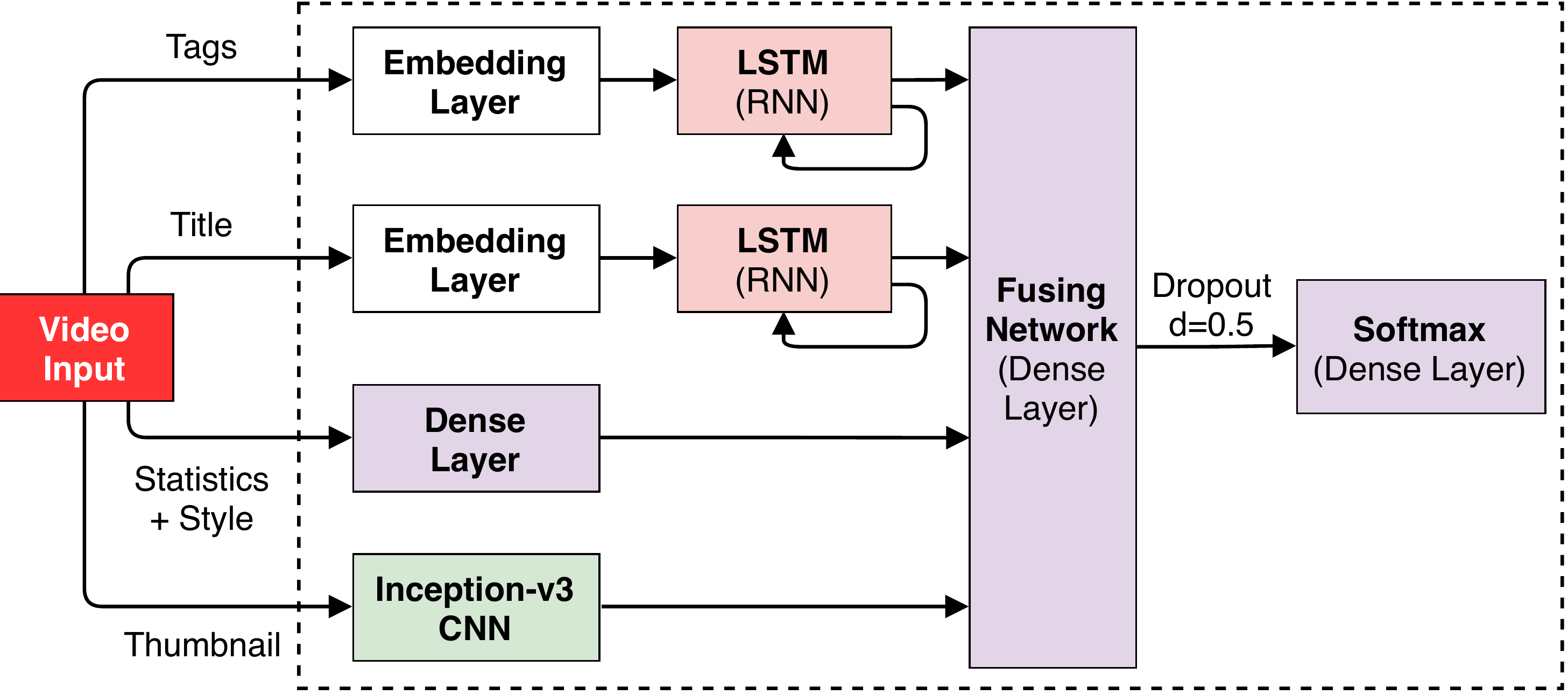}
\caption{Architecture of our deep learning model for detecting disturbing videos. The model processes almost all the video features: (a) tags; (b) title; (c) statistics and style; and (d) thumbnail.}
\label{fig:disturbed_model_architecture} 
\end{figure}

Fig.~\ref{fig:disturbed_model_architecture} depicts the architecture of our classifier, which combines the above-mentioned features.
Initially, the classifier consists of four different branches, where each branch processes a distinct feature type: title, tags, thumbnail, and statistics and style features.
Then the outputs of all the branches are concatenated to form a two-layer, fully connected neural network that merges their output and drives the final classification.

The title feature is fed to a trainable embedding layer that outputs a 32-dimensional vector for each word in the title text.
Then, the output of the embedding layer is fed to a Long Short-Term Memory (LSTM)~\cite{hochreiter1997long} Recurrent Neural Network (RNN) that captures the relationships between the words in the title.
For the tags, we use an architecturally identical branch trained separately from the title branch.

For thumbnails, due to the limited number of training examples in our dataset, we use transfer learning~\cite{oquab2014learning} and the pre-trained Inception-v3 Convolutional Neural Network (CNN)~\cite{Szegedy_2015_CVPR}, which is built from the large-scale ImageNet dataset.\footnote{\url{http://image-net.org/}}
We use the pre-trained CNN to extract a meaningful feature representation (2,048-dimensional vector) of each thumbnail.
Last, the statistics together with the style features are fed to a fully-connected dense neural network comprising 25 units.

The second part of our classifier is essentially a two-layer, fully-connected dense neural network. 
At the first layer, (dubbed Fusing Network), we merge the outputs of the four branches, creating a 2,137-dimensional vector.
This vector is subsequently processed by the 512 units of the Fusing Network.
Next, to avoid possible over-fitting issues we regularize via the prominent Dropout technique~\cite{srivastava2014dropout}.
We apply a Dropout level of $d=0.5$, which means that during each iteration of training, half of the units in this layer do not update their parameters.
Finally, the output of the Fusing Network is fed to the last dense-layer neural network of four units with softmax activation, which is essentially the probabilities that a particular video is suitable, disturbing, restricted, or irrelevant.

\begin{table}[t!]
\footnotesize
\centering
\begin{tabular}{lcccc}
\toprule
\textbf{Model} & \textbf{Accuracy} & \textbf{Precision} & \textbf{Recall} & \begin{tabular}[c]{@{}c@{}}\textbf{F1 Score}\end{tabular}  \\
\midrule
Naive Bayes & 0.34 & 0.34 & 0.32 & 0.30 \\
K-Nearest & 0.35 & 0.30 & 0.30 & 0.30 \\
Decision Tree & 0.38 & 0.32 & 0.32 & 0.32 \\
SVM & 0.41 & 0.39 & 0.26 & 0.17 \\
Random Forest & 0.57 & 0.47 & 0.42 & 0.35 \\
DDNN & 0.47 & 0.37 & 0.37 & 0.37 \\
CNN-DDNN & 0.54 & 0.48 & 0.48 & 0.47 \\
\textbf{Proposed Model} & \textbf{0.64} & \textbf{0.50} & \textbf{0.51} & \textbf{0.48} \\
\toprule
\end{tabular}%
\caption{Performance metrics for the evaluated baselines and for the proposed deep learning model.}
\label{tab:disturbed_performance_metrics_multiclass}
\end{table}

\subsection{Experimental Evaluation} 
We implement our model using Keras~\cite{keras2015application} with TensorFlow as the backend~\cite{abadi2016tensorflow}. 
To train our model we use five-fold stratified cross-validation~\cite{arlot2010survey} and we train and test our model using all the aforementioned features. 
\revision{For hyper-parameter tuning of our model we use the random search strategy~\cite{bergstra2012random}.}
To deal with the data imbalance problem we use the Synthetic Minority Over-sampling technique (SMOTE)~\cite{chawla2002smote} to over-sample the train set at each fold.

\begin{figure}[t!]
\centering
\includegraphics[width=0.5\linewidth]{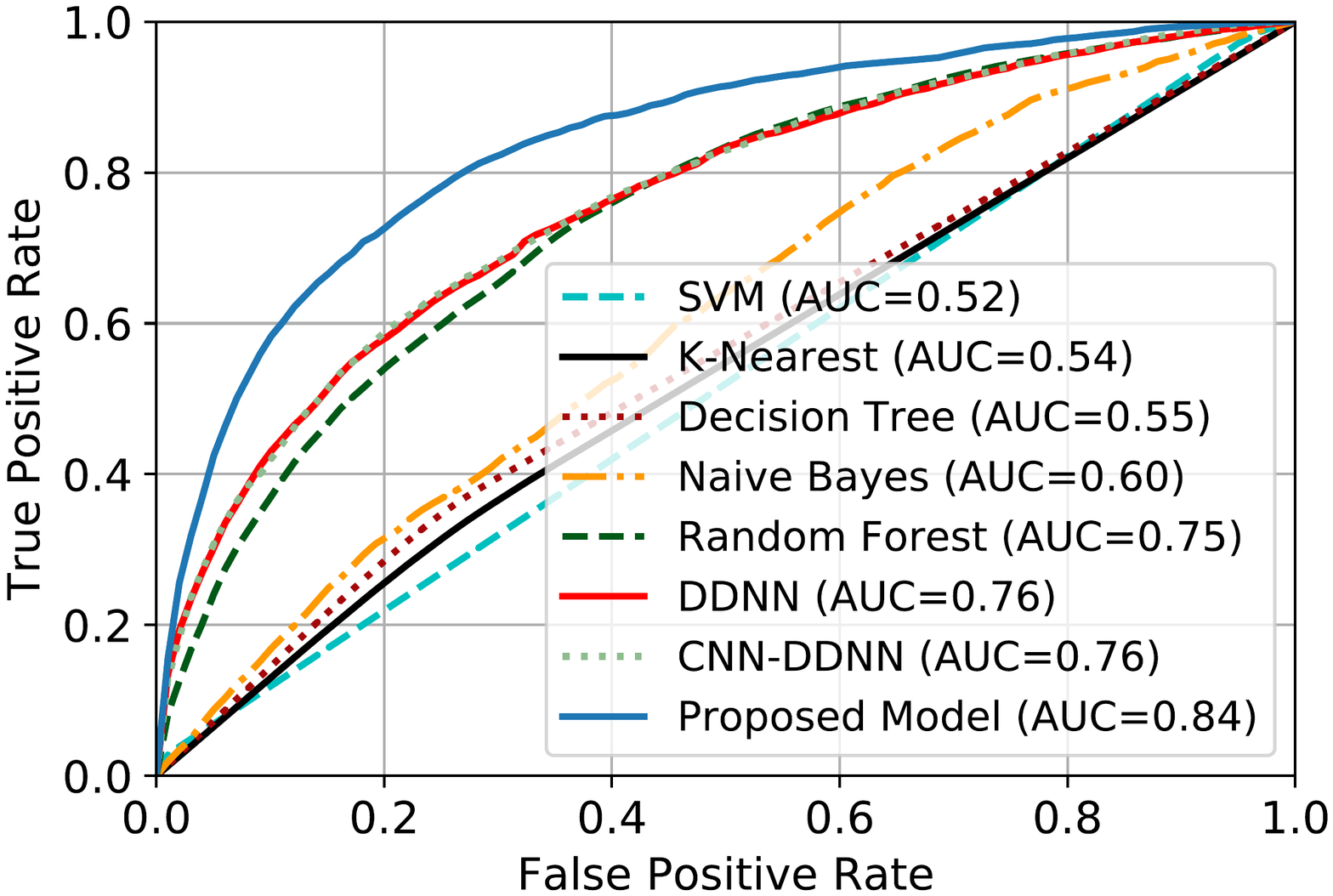}
\caption{ROC curves (and AUC) of all the baselines and of the proposed model trained for multi-class classification.}
\label{fig:disturbed_all_models_multiclass_roc}
\end{figure}

\begin{table}[t!]
\footnotesize
\centering
\begin{tabular}{lccccccc}
\toprule
\textbf{Input Features} & \textbf{Accuracy} & \textbf{Precision} & \textbf{Recall} & \begin{tabular}[c]{@{}c@{}}\textbf{F1 Score}\end{tabular}  \\
\midrule
Thumbnail & 0.64 & 0.48 & 0.50 & 0.45 \\
Title & 0.46 & 0.35 & 0.33 & 0.30 \\
Tags & 0.40 & 0.29 & 0.30 & 0.27 \\
Style \& Statistics & 0.43 & 0.35 & 0.35 & 0.29 \\
\midrule
Thumbnail, Title & 0.63 & 0.45 & 0.50 & 0.45 \\
Thumbnail, Tags & 0.63 & 0.47 & 0.49 & 0.45 \\ 
Thumbnail, Style\&Stats & 0.63 & 0.48 & 0.50 & 0.47 \\
Title, Tags & 0.49 & 0.40 & 0.38 & 0.36 \\ 
Title, Style \& Statistics & 0.44 & 0.39 & 0.37 & 0.34 \\
Tags, Style \& Statistics & 0.41 & 0.36 & 0.34 & 0.28 \\ 
\midrule
Title, Tags, Style \& Statistics & 0.46 & 0.39 & 0.48 & 0.36 \\
Thumbnail, Tags, Style \& Statistics & 0.64 & 0.48 & 0.51 & 0.48 \\ 
Thumbnail, Title, Style \& Statistics & 0.63 & 0.46 & 0.50 & 0.46 \\
Thumbnail, Title, Tags & 0.64 & 0.48 & 0.51 & 0.47 \\ 
\textbf{All Input Features} & \textbf{0.64} & \textbf{0.50} & \textbf{0.51} & \textbf{0.48} \\ 
\bottomrule
\end{tabular}%
\caption{Performance of the proposed model trained with all the possible combinations of the four input feature types.}
\label{tab:disturbed_ablation_study_details}
\end{table}

For the stochastic optimization of our model, we use the Adam algorithm with an initial learning rate of $1\mathrm{e}{-5}$, and $\epsilon=1\mathrm{e}{-8}$. 
To evaluate our model, we compare it in terms of accuracy, precision, recall, F1 score, and area under the ROC curve (AUC) against the following five baselines: 
1) a Support Vector Machine (SVM) with parameters $\gamma=auto$ and $C=10.0$;
2) a K-Nearest Neighbors classifier with $n=8$ neighbors and leaf size equal to $10$; 
3) a Bernoulli Naive Bayes classifier with $a=1.0$; 
4) a Decision Tree classifier with an entropy criterion; and 
5) a Random Forest classifier with an entropy criterion and number of trees equal to $100$.
To further evaluate the performance of our model, we also compare it with two deep neural networks: 
1) a simple double dense layer network (DDNN); and
2) a CNN combined with a double dense layer network (CNN-DDNN).
For hyper-parameter tuning of all the baselines, we use the grid search strategy, while for the deep neural networks we use the same hyper-parameters as with the proposed model.
For a fair comparison, we note that all the evaluated models use all the available input features.
Table~\ref{tab:disturbed_performance_metrics_multiclass} reports the performance of the proposed model as well as the $7$ baselines, while Figure~\ref{fig:disturbed_all_models_multiclass_roc} shows their ROC curves.
Although the proposed model outperforms all the baselines in all performance metrics, it still has poor performance. 

\revision{
In an attempt to achieve better accuracy, we consider the video itself as an additional input feature to our model.
We train our model considering 45 frames per video along with all the other types of input.
We then evaluate our model and we find that adding the video frames in the set of features of our model yields worse performance in terms of accuracy.
At the same time, the training of the model with the video frames is more than 50 times more computationally expensive compared to the model without the video frames. Hence, we decided to keep the previous model formulation ignoring the video frames.
}

\descr{Ablation Study.}
In an attempt to understand which of the input feature types contribute the most to the classification of disturbing videos we perform an ablation study. 
That is, we systematically remove each of the four input feature types (as well as their associated branch in the proposed model's architecture), while also training models with all the possible combinations of the four input feature types.
Again, to train and test these models we use five-fold cross-validation and the oversampling technique to deal with data imbalance.
Table~\ref{tab:disturbed_ablation_study_details} shows the performance metrics of all the models for each possible combination of inputs.
We observe that the thumbnail is more important than the other input feature types for good classification performance.

\descr{Binary Classification.}
To perform a more representative analysis of the inappropriate videos on YouTube, we need a more accurate classifier.
Thus, for the sake of our analysis in the next steps, we collapse our four labels into two general categories, by combining the suitable with the irrelevant videos into one ``appropriate'' category (3,499 videos) and the disturbing with the restricted videos into a second ``inappropriate'' category (1,348 videos).
We call the first category ``appropriate'' despite including PG and PG-13 videos because those videos are not aimed at toddlers (irrelevant). 
On the other hand, videos rated as PG or PG-13 that target toddlers (aged 1 to 5) are disturbing and fall under the inappropriate category. 
When such videos appear on the video recommendation list of toddlers, it is a strong indication that they are disturbing and our binary classifier is very likely to detect them as inappropriate.

\begin{figure}[t!]
\centering
\includegraphics[width=0.5\linewidth]{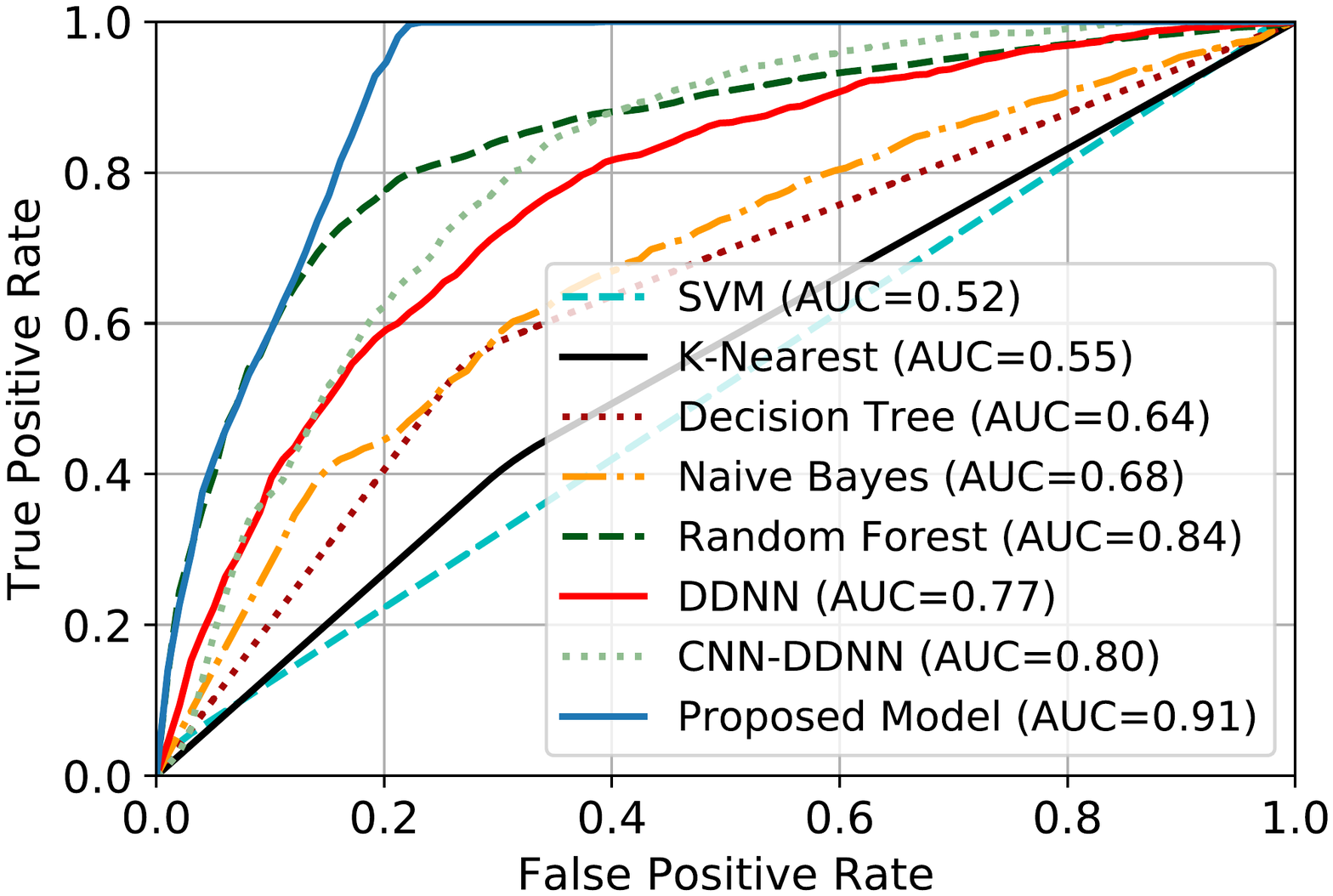}
\caption{ROC Curves of all the baselines and of the proposed model trained for binary classification.}
\label{fig:disturbed_all_models_binary_roc}
\end{figure}

We train and evaluate the proposed model for binary classification on our reshaped ground truth dataset following the same approach as the one presented above.
Table~\ref{tab:disturbed_performance_metrics_binary} reports the performance of our model as well as the baselines, while Figure~\ref{fig:disturbed_all_models_binary_roc} shows their ROC curves. 
We observe that our deep learning model outperforms all baseline models across all performance metrics. 
Specifically, our model substantially outperforms the CNN-DDNN model, which has the best overall performance from all the evaluated baselines, on accuracy, precision, recall, F1 score, and AUC by $0.12$, $0.13$, $0.17$, $0.14$, $0.11$ respectively.

\begin{table}[t!]
\footnotesize
\centering
\begin{tabular}{lrrrr}
\toprule
\textbf{Model} & \textbf{Accuracy} & \textbf{Precision} & \textbf{Recall} & \begin{tabular}[c]{@{}c@{}}\textbf{F1 Score}\end{tabular}  \\
\midrule
K-Nearest & 0.61 & 0.34 & 0.42 & 0.38 \\
Decision Tree & 0.68 & 0.44 & 0.56 & 0.50 \\
SVM & 0.72 & 0.47 & 0.03 & 0.05 \\
Naive Bayes & 0.73 & 0.52 & 0.36 & 0.43 \\
Random Forest & 0.80 & 0.74 & 0.46 & 0.57 \\
DDNN & 0.73 & 0.66 & 0.63 & 0.64 \\
CNN-DDNN & 0.72 & 0.69 & 0.72 & 0.69 \\
\textbf{Proposed Model} & \textbf{0.84} & \textbf{0.82} & \textbf{0.89} & \textbf{0.83} \\
\toprule
\end{tabular}%
\caption{Performance of the evaluated baselines trained for binary classification and of our proposed binary classifier.}
\label{tab:disturbed_performance_metrics_binary}
\end{table}

\section{Analysis}
\label{sec:disturbed_analysis}
In this section, we study the interplay of appropriate and inappropriate videos on YouTube using our binary classifier.
First, we assess the prevalence of inappropriate videos in each subset of videos in our dataset and investigate how likely it is for YouTube to recommend an inappropriate video.
Second, we perform live random walks on YouTube's recommendation graph to simulate the behavior of a toddler that selects videos based on the recommendations.

\begin{table}[t!]
\footnotesize
\centering
\begin{tabular}{lrr}
\toprule
\textbf{Videos subset} & \textbf{Appropriate (\%)} & \textbf{Inappropriate (\%)} \\ 
\midrule
Elsagate-related & 230,890 (98.95\%) & 2,447 (1.05\%) \\
Other Child-related & 154,262 (99.55\%) & 695 (0.45\%) \\
Random & 478,420 (99.28\%) & 3,487 (0.72\%) \\
Popular & 10,947 (99.75\%) & 27 (0.25\%) \\
\toprule
\end{tabular}%
\caption{Number of appropriate and inappropriate videos found in each subset of videos in our dataset.}
\label{tab:disturbed_appropriate_inappropriate_videos_in_dataset}
\end{table}

\subsection{Recommendation Graph Analysis}
First, we investigate the prevalence of inappropriate videos in each subset of videos in our dataset by running our binary classifier on the whole dataset, which allows us to find which videos are inappropriate or appropriate.
Table~\ref{tab:disturbed_appropriate_inappropriate_videos_in_dataset} shows the number of appropriate and inappropriate videos found in each subset.
For the Elsagate-related videos, we find 231K ($98.9\%$) appropriate videos and 2.5K ($1.1\%$) inappropriate videos, while the proportion of inappropriate videos is a bit lower in the set of other child-related videos ($0.4\%$ inappropriate and $99.5\%$ appropriate).
These findings highlight the gravity of the problem: a parent searching on YouTube with simple toddler-oriented keywords and casually selecting from the recommended videos, is likely to expose their child to inappropriate videos.

But what is the interplay between the inappropriate and appropriate videos in each subset?
To shed light on this question, we create a directed graph for each subset of videos, where nodes are videos, and edges are recommended videos (up to 10 videos due to our data collection methodology).
For instance, if $video_2$ is recommended via $video_1$ then we add an edge from $video_1$ to $video_2$.
Then, for each video in each graph, we calculate the out-degree in terms of appropriate and inappropriate labeled nodes.
From here, we can count the number of \emph{transitions} the graph makes between differently labeled nodes.
Table~\ref{tab:disturbed_graph_transitions_in_dataset_videos} summarizes the percentages of transitions between the two classes of videos in each subset.
Unsurprisingly, we find that most of the transitions in each subset $(98\%$-$99\%$), are between appropriate videos, which is mainly because of the large number of appropriate videos in each set.
We also find that when a toddler watches an Elsagate-related benign video, if she randomly follows one of the top ten recommended videos, there is a $0.6\%$ probability that she will end up at a disturbing or restricted video.
Taken altogether, these findings show that the problem of toddler-oriented inappropriate videos on YouTube is notable, especially when considering YouTube's massive scale and the large number of views of toddler-oriented videos.
That is, there is a non-negligible chance that a toddler will be recommended an inappropriate video when watching an appropriate video.

\begin{table*}[t!]
\footnotesize
\centering
\resizebox{\textwidth}{!}{%
\begin{tabular}{llrrrr}
\toprule
\textbf{Source} & \textbf{Destination} & \textbf{Elsagate-related} & \textbf{Other Child-related} & \textbf{Random} & \textbf{Popular} \\ 
\midrule
Appropriate 	& Appropriate 	& 917,319 (97.80\%) & 648,406 (99.49\%) & 1,319,518 (98.82\%) & 34,764 (99.12\%) \\
Appropriate 	& Inappropriate 	& 5,951 (0.64\%) 	& 1,681 (0.26\%) 	& 7,014 (0.53\%)  & 64 (0.18\%) \\
Inappropriate 	& Appropriate 	& 14,202 (1.51\%) 	& 1,542 (0.24\%) 	& 7,946 (0.59\%) 	& 246 (0.70\%) \\
Inappropriate 	& Inappropriate 	& 478 (0.05\%) 		& 72 (0.01\%) 	& 831 (0.06\%) 	& 0 (0.00\%) \\ 
\toprule
\end{tabular}%
}
\caption{Number of transitions between appropriate and inappropriate videos for each subset of videos in our dataset.}
\label{tab:disturbed_graph_transitions_in_dataset_videos}
\end{table*}

\subsection{How likely is it for a toddler to come across inappropriate videos?}
\label{subsec:disturbed_random_walks_analysis}
In the previous section, we showed that the problem of toddler-oriented videos is prevalent enough to be cause for concern. 
However, it is unclear whether the previous results generalize to YouTube at large since our dataset is based on a snowball sampling of up to three hops from a set of seed videos. In reality, though, YouTube comprises billions of videos, which are recommended over many hops within YouTube's recommendation graph.
Therefore, to assess how prominent the problem is on a larger scale, we perform live random walks on YouTube's recommendation graph.
This allows us to simulate the behavior of a ``random toddler'' who searches the platform for a video and then he watches several videos according to the recommendations.
To do this, we use the lists of Elsagate-related and other child-related seed keywords used for constructing our dataset, as well as a list of sanitized Elsagate-related seed keywords which we construct by stripping all the inappropriate words from all the Elsagate-related keywords using a dictionary of inappropriate words\footnote{\url{https://tinyurl.com/yxb4kmxg}}.
We do this to assess the degree of the problem around Elsagate-related videos while ensuring that we are not biasing the search results with any sensitive words.

\begin{figure}[t!]
\centering
\subfigure[]{\includegraphics[width=.49\linewidth]{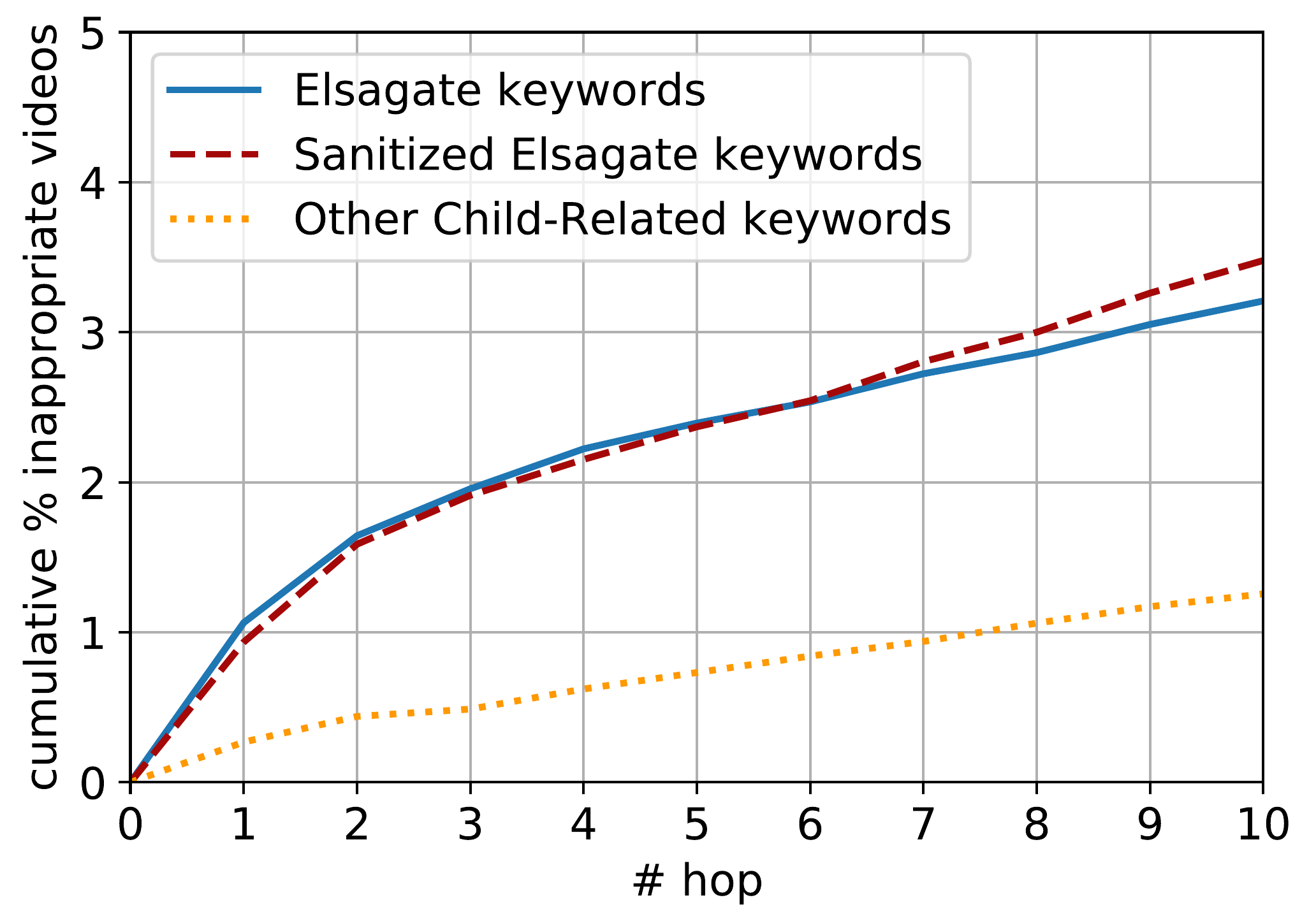}\label{fig:disturbed_cumulative_percentage_disturbing_in_hops_classes}}
\subfigure[]{\includegraphics[width=.49\linewidth]{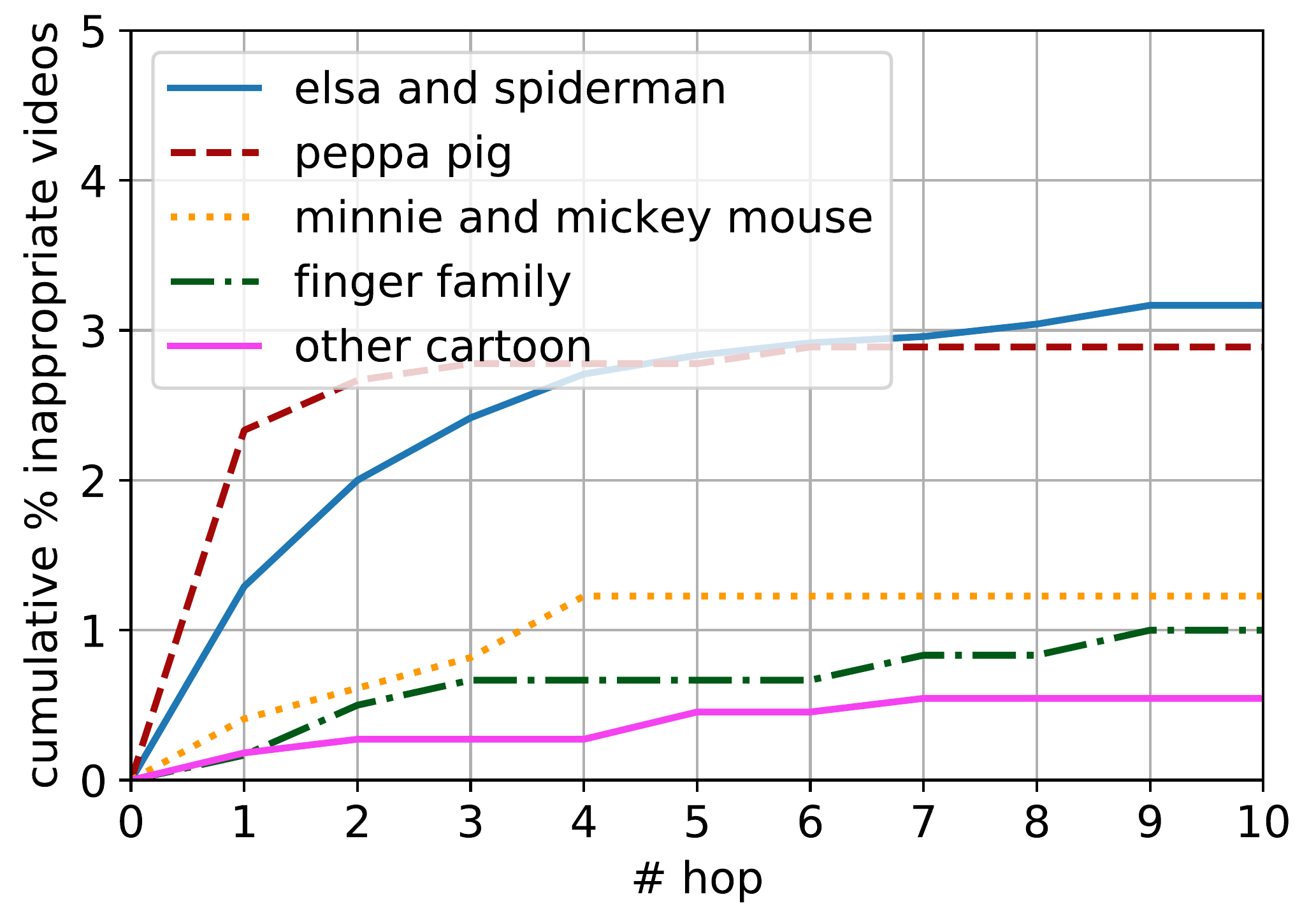}\label{fig:disturbed_cumulative_percentage_disturbing_in_hops_clusters}}
\caption{Cumulative percentage of inappropriate videos encountered at each hop for: (a) Elsagate-related, sanitized Elsagate-related, and other child-related seed keywords; and (b) clusters of seed keywords.}
\label{fig:disturbed_all_random_walks_plots}
\end{figure}

For each seed keyword, we initially perform a search query on YouTube and randomly select one video from the top ten results.
Then, we obtain the recommendations of the video and select one randomly. 
We iterate with the same process until we reach ten hops, which constitutes the end of a single random walk.
We repeat this operation for 100 random walks for each seed keyword, while at the same time classifying each video we visit, using our binary classifier.

First, we group the random walks based on the keywords used to seed them.
Figure~\ref{fig:disturbed_cumulative_percentage_disturbing_in_hops_classes} shows the cumulative percentage of inappropriate videos encountered at each hop of the random walks for Elsagate-related, sanitized Elsagate-related, and other child-related search keywords.
We observe that, when using sanitized Elsagate-related keywords, we find at least one inappropriate video in $3.5\%$ of the walks, while for the other child-related keywords we find at least one inappropriate video in $1.3\%$ of the walks.
We also observe that most of the inappropriate videos are found early in our random walks (i.e., at the first hop) and this number decreases as the number of hops increases.
These findings highlight that the problem of inappropriate videos on YouTube emerges quite early when users are browsing the platform starting from benign toddler-oriented search terms.

Next, to assess whether our results change according to the content of the videos we use the k-means clustering algorithm~\cite{hartigan1979algorithm} to create clusters from all the seed keywords.
Then, we manually inspect the clusters and associate a label to each cluster.
We create five clusters: 1) ``Elsa and Spiderman'' (24 keywords); 2) ``Peppa Pig'' (9 keywords); 3) ``Finger Family'' (6 keywords); 4) ``Minnie and Mickey mouse'' (5 keywords); and 5) ``Other Cartoon'' (11 keywords).
Then, based on the clusters we group the random walks.
Fig.~\ref{fig:disturbed_cumulative_percentage_disturbing_in_hops_clusters} shows the cumulative percentage of inappropriate videos encountered at each hop for each cluster.
We observe interesting differences across the clusters: specifically, we observe the higher percentages in the ``Elsa and Spiderman'' ($3.2\%$), and ``Peppa pig'' ($2.9\%$) cluster, whereas for the clusters ``finger family'' ($1.0\%$) and ``other cartoon'' ($0.5\%$) we observe a lower percentage of walks with inappropriate videos.
Also, we find that most of the inappropriate videos are found at the beginning of the random walks in particular for the clusters ``Peppa pig'' ($2.3\%$) and ``Elsa and Spiderman'' ($1.3\%$) (see the first hop in Fig.~\ref{fig:disturbed_cumulative_percentage_disturbing_in_hops_clusters}).

Note that, by merging the two classes in the binary classifier while seeking out disturbing videos with short random walks from suitable videos, we correct for the misclassification of disturbing videos as restricted. 
That is, an NC-17 video in the proximity of benign toddler-oriented videos could be erroneously classified as restricted by the multi-class classifier (because of similarities with the NC-17 videos that do not target toddlers).
However, due to this proximity, this is probably an NC-17 video that targets toddlers and should have therefore been classified as disturbing. 
Thus, the vast majority of inappropriate videos detected during the random walks are expected to be disturbing. 
In fact, $84.6\%$ of the detected inappropriate videos are disturbing (obtained by inspecting all the 338 detected inappropriate videos).
On the other hand, videos that would be classified as irrelevant by the multi-class classifier, fall under the appropriate category of the binary classifier. 
However, the training set for the appropriate category includes irrelevant videos, which include PG and PG-13 videos that do not target toddlers. 
Therefore, the binary classifier may classify such a video that is in the proximity of suitable videos as appropriate.
However, a PG and PG-13 video in the proximity of suitable videos is likely to actually be disturbing, thus inappropriate. 
This negatively affects the accuracy of the binary classifier. 
Yet, only $1.6\%$ of the videos encountered during the random walks and classified as appropriate were in fact disturbing (obtained by sampling 300 of the 176,619 detected appropriate videos).

\section{Challenges and Limitations}
\revision{In this section, we discuss the technical challenges we faced, how we addressed them, and we highlight the limitations of this line of work.}

\subsection{Challenges}
\revision{
Our data collection and video annotation methodology faced many challenges.
First, there was no available dataset related to the Elsagate phenomenon nor there is a dataset with appropriate and inappropriate videos targeting toddlers on YouTube.
Following different approaches, we compile lists of keywords from videos posted on Reddit and use them to search YouTube and collect videos.
Second, devising a classifier that will detect disturbing videos targeting young children is not trivial.
To do this, we initially devised a multi-class deep learning classifier that takes into account the various metadata of a given video. 
However, due to the low performance of the multi-class classifier, we decide to collapse our labels into two and train a binary classifier that can discern appropriate from inappropriate videos with rather acceptable accuracy.
}

\subsection{Limitations}
\revision{
Unfortunately, this line of work is not without limitations.
First, we collect and analyze videos only from YouTube and not YouTube Kids. 
This is because YouTube does not provide an open API for collecting videos that appear on YouTube Kids. 
However, according to YouTube, only videos marked as age-restricted are excluded from YouTube Kids unless specific settings are set by the parent\footnote{\url{https://support.google.com/youtubekids/answer/6130562?hl=en&ref_topic=7556083}}.
Second, we acknowledge that the performance of our classifier is highly affected by the small training size: we were unable to provide a larger annotated dataset mainly due to the lack of resources for the annotation process.
}

\revision{
Moreover, the use of the YouTube Data API lets us analyze YouTube's recommendation algorithm's output based on content relevance and the user base’s engagement in aggregate. Hence, in this line of work, we do not consider per-user personalization; the video recommendations we collect represent only some of the recommendation system’s facets. However, we believe that our approach still allows us to understand how YouTube’s recommendation system is behaving in our scenario. 
}

\section{Remarks}
In this work, we presented the first characterization of inappropriate or disturbing videos targeted at toddlers on YouTube.
We collected thousands of Elsagate-related, other child-related, as well as random and popular videos available on YouTube.
We manually reviewed 5K of them, and we use them to develop a deep learning classifier that achieves an accuracy of $0.84$.
Our analysis on the manually reviewed videos showed that none of the videos’ metadata can clearly indicate that a video is disturbing or not, thus, in most cases one (e.g., a guardian) has to carefully inspect all the available video metadata, and potentially the actual video, to accurately determine if it is safe for a toddler to watch.

We then leveraged this classifier to perform a large-scale study of toddler-oriented content on YouTube. 
We found $1.05\%$ of the 233,337 Elsagate-related videos in our dataset to be inappropriate, which indicates that the problem of toddler-oriented videos is prevalent enough to be cause for concern.
We also used the developed classifier to assess how prominent the problem is on a larger scale.
Specifically, we performed a live simulation of a toddler who searches the platform for a video and then he watches several videos according to the recommendations, finding a $3.5\%$ chance of a toddler who starts by watching appropriate videos to be recommended inappropriate ones within ten recommendations.

To the best of our knowledge, our analysis is the first large-scale attempt to characterize disturbing videos targeted at toddlers on YouTube and to assess the chance of a toddler who casually browses YouTube to come across such videos.
Overall, our classifier and the insights gained from our analysis can be used as a starting point to gain a deeper understanding and begin mitigating this issue.

\chapter{Characterizing Hateful and Misogynistic Content on YouTube Through the Lens of the Incel Community}
\label{chapter:incels_youtube}

\revision{
In the previous chapter, we show that inappropriate content exists on YouTube and we find a non-negligible chance that its recommendation system will recommend such problematic content to the wrong audience. At the same time, the platform and in particular its recommendation algorithm has been repeatedly accused of promoting hateful and dangerous content, and even for helping radicalize users~\cite{rooseyoutuberadical,mozillaregrets,ribeiro2020auditing}.
Hence, in this chapter, we focus on characterizing hateful and misogynistic content on YouTube to investigate these claims and to examine the communities that the recommendation algorithm might create by promoting such problematic content.
}

\section{Motivation}
One fringe community active on YouTube that has often been linked with sharing and publishing hateful and misogynistic content are the so-called Involuntary Celibates or \emph{Incels}~\cite{martineau2020youtubeincels}.
While not particularly structured, Incel's ideology revolves around the idea of the ``blackpill'' -- a bitter and painful truth about society -- which roughly postulates that life trajectories are determined by how attractive one is. 
For example, Incels often deride the alleged rise of {\em lookism}, whereby things that are largely out of personal control, like facial structure, are more ``valuable'' than those under our control, like the fitness level.
Incels are one of the most extreme communities of the Manosphere~\cite{bbc2018incel}, a larger collection of movements discussing men's issues~\cite{ging2019alphas} (see Section~\ref{sec:incels_background}).
When taken to the extreme, these beliefs can lead to a dystopian outlook on society, where the only solution is a radical, potentially violent shift towards traditionalism, especially in terms of women's role in society~\cite{cook2018looksmaxing}.

Overall, Incel ideology is often associated with misogyny and anti-feminist viewpoints, and it has also been linked to multiple mass murders and violent offenses~\cite{splc2019malesupremacy,fifthestate2020incels}.
In May 2014, Elliot Rodger killed six people and himself in Isla Vista, CA.
This incident was a harbinger of things to come.
Rodger uploaded a video on YouTube with his ``manifesto,'' as he planned to commit mass murder due to his belief in what is now generally understood to be Incel ideology~\cite{wiki2019rodgers}.
He served as an apparent ``mentor'' to another mass murderer who shot nine people at Umpqua Community College in Oregon the following year~\cite{harpermercer_2015}.
In 2018, another mass murderer drove his van into a crowd in Toronto, killing nine people, and after his interrogation, the police claimed he had been radicalized online by Incel ideology~\cite{cecco2019toronto}. 
Thus, while the concepts underpinning Incels' principles may seem absurd, they also have grievous real-world consequences~\cite{hoffman2020assessing,insideincels2018washington,incel2018vox}.

Online platforms like Reddit became aware of the problem and banned several Incel-related communities on the platform~\cite{incelssubredditbanned_2017}.
However, prior work suggests that banning subreddits and their users for hate speech does not solve the problem, but instead makes these users someone else's problem~\cite{chandrasekharan2017you}, as banned communities migrate to other platforms~\cite{newell2016user}.
Indeed, the Incel community comprising several banned subreddits ended up migrating to various other online communities such as new subreddits, blogs, stand-alone forums, and YouTube channels~\cite{ribeiro2020pick}.

The research community has mostly studied the Incel community and the broader Manosphere on Reddit, 4chan, and online discussion forums like Incels.me or Incels.co~\cite{farrell2019exploring,nagle2015investigation,jaki2019online,ribeiro2020pick,maxwell2020short,manoel_bans}.
However, the fact that YouTube has been repeatedly accused of user radicalization and promoting offensive and inappropriate content~\cite{rooseyoutuberadical,ribeiro2020auditing,rabbithole2019donovan,kaiser2018unite} prompts the need to study the extent to which Incels are exploiting the YouTube platform to spread their views.
\revision{
With this motivation in mind, and guided by two of our main research questions \textbf{(RQ1 and RQ3)}, in this chapter we explore the footprint of the Incel community on YouTube. 
More precisely, we formulate the following two sub-questions:
\begin{itemize}
    \item \textbf{RQ1.b:} How has the Incel community evolved on YouTube over the last decade? (see RQ1 in Chapter~\ref{chapter:introduction})
    \item \textbf{RQ3.b:} Does YouTube’s recommendation algorithm contribute to steering users towards Incel communities? Does the frequency with which Incel-related videos are recommended increase for users who choose to see the content? (see RQ3 in Chapter~\ref{chapter:introduction})
\end{itemize}
}

\descr{Methods.}
In this thesis, we provide, to the best of our knowledge, the first study of the Incel community on YouTube.
We collect a set of 6.5K YouTube videos shared on Incel-related subreddits (e.g., /r/incels, /r/braincels, etc.), as well as a set of 5.7K random videos as a baseline.
We then use a lexicon of Incel-related terms to label the collected videos as “Incel-related,” based on the appearance of terms in the transcript, which describes the video’s content, and comments on the videos. 
Next, we use several tools, including temporal and graph analysis, to investigate the evolution of the Incel community on YouTube and whether YouTube’s recommendation algorithm contributes to steering users towards Incel content. 
To build our graphs, we use the YouTube Data API, which lets us analyze YouTube’s recommendation algorithm’s output based on video item-to-item similarities, as well as general user engagement and satisfaction metrics~\cite{zhao2019recommending}.

\descr{Findings.}
Overall, the main findings of our analysis are:
\begin{enumerate}
    \item We find an increase in Incel-related activity on YouTube over the past few years and in particular concerning Incel-related videos, as well as comments that include pertinent terms. This indicates that Incels are increasingly exploiting the YouTube platform to broadcast and discuss their views.
    
    \item Random walks on the YouTube's recommendation graph using the Data API and without personalization reveal that with a $6.3\%$ probability a user will encounter an Incel-related video within five hops if they start from a random non-Incel-related video posted on Reddit. Simultaneously, Incel-related videos are more likely to be recommended within the first two to four hops than in the subsequent hops. 

    \item We also find a $9.4\%$ chance that a user will encounter an Incel-related video within three hops if they have visited Incel-related videos in the previous two hops. This means that a user who purposefully and consecutively watches two or more Incel-related videos is likely to continue being recommended such content and with higher frequency.
\end{enumerate}

\revision{
Overall, our findings indicate that Incels are increasingly exploiting YouTube to spread their ideology and express their misogynistic views. 
They also indicate that the threat of recommendation algorithms nudging users towards extreme content is real and that platforms and researchers need to address and mitigate these issues. 
}

\section{Background}
\label{sec:incels_background}
Incels are a part of the broader "Manosphere," a loose collection of groups revolving around a common shared interest in “men’s rights” in society~\cite{ging2019alphas}. 
While we focus on Incels, understanding the overall Manosphere movement provides relevant context. 
In this section, we provide background information about Incels and the Manosphere.
We also review related work focusing on understanding Incels on the Web.

\descr{The Manosphere.}
The emergence of the so-called Web 2.0 and popular social media platforms have been crucial in enabling the Manosphere~\cite{marwick2017media}.
Although the Manosphere had roots in anti-feminism~\cite{farrell1996myth, messner1998limits}, it is ultimately a reactionary community, with its ideology evolving and spreading mostly on the Web~\cite{ging2019alphas}.
Coston et al.~\cite{coston2012white} argue about the growth of feminism: "If women were imprisoned in the home [...] then men were exiled from the home, turned into soulless robotic workers, in harness to a masculine mystique, so that their only capacity for nurturing was through their wallets."
Further, Blais et al.~\cite{blais2012masculinism} analyze the beliefs concerning the Manosphere from a sociological perspective and refer to it as masculinism.
They conclude that masculinism is: "a trend within the anti-feminist counter-movement mobilized not only against the feminist movement but also for the defense of a non-egalitarian social and political system, that is, patriarchy." 
Subgroups within the Manosphere actually differ significantly.
For instance, Men Going Their Own Way (MGTOWs) are hyper-focused on a particular set of men’s rights, often in the context of a bad relationship with a woman. 
These subgroups should not be seen as distinct units.
Instead, they are interconnected nodes in a network of misogynistic discourses and beliefs~\cite{bratich2019pick}.
According to Marwick and Lewis~\cite{marwick2017media}, what binds the manosphere subgroups is “the idea that men and boys are victimized; that feminists, in particular, are the perpetrators of such attacks.”

Overall, research studying the Manosphere has been mostly theoretical and qualitative in nature~\cite{ging2019alphas,gotell2016sexual,hunte2019female,lin2017antifeminism}. 
These qualitative studies are important because they guide our study in terms of framework and conceptualization while motivating large-scale data-driven work like ours.

\descr{Incels.}
Incels are arguably the most extreme subgroup of the Manosphere~\cite{bbc2018killer}.
Incels appear disarmingly honest about what is causing their grievances compared to other radical ideologies. 
They openly put their sexual deprivation, which is supposedly caused by their unattractive appearance, at the forefront, thus rendering their radical movement potentially more persuasive and insidious~\cite{maxwell2020short}.
Incel ideology differs from the other Manosphere subgroups in the significance of the "involuntary" aspect of their celibacy. 
They believe that society is rigged against them in terms of sexual activity, and there is no solution at a personal level for the systemic dating problems of men~\cite{incelswikiblackpill,menzie2020stacys,rationalwiki2019incel}.
Further, Incel ideology differs from, for example, MGTOW, in the idea of voluntary vs. involuntary celibacy.
MGTOWs are choosing to not partake in sexual activities, while Incels believe that society adversarially deprives them of sexual activity. 
This difference is crucial, as it gives rise to some of their more violent tendencies~\cite{ging2019alphas}.

Incels believe to be doomed from birth to suffer in a modern society where women are not only able but encouraged to focus on superficial aspects of potential mates, e.g., facial structure or racial attributes.
Some of the earliest studies of "involuntary celibacy" note that celibates tend to be more introverted and that, unlike women, celibate men in their 30s tend to be poorer or even unemployed~\cite{kiernan1988remains}.
In this distorted view of reality, men with these desirable attributes (colloquially nicknamed Chads by Incels) are placed at the top of society's hierarchy. 
While a perusal of influential people in the world would perhaps lend credence to the idea that "handsome" white men are indeed at the top, the Incel ideology takes it to the extreme.
Incels rarely hesitate to call for violence~\cite{baele2019incel}.
For example, seeking advice by other Incels based on their physical appearance using the phrase "How over is it?," they may be encouraged to “rope” (to hang oneself), or commit suicide~\cite{conti2018decode}. 
Occasionally they also approach calls for outright gendercide. 
Zimmerman et al.~\cite{zimmerman2018recognizing} associate Incel ideology with white supremacy, highlighting how it should be taken as seriously as other forms of violent extremism.

\section{Datasets}
\label{sec:incels_datasets}
We now present our data collection and annotation process to identify Incel-related videos.

\subsection{Data Collection}
\label{subsec:incels_data_collection}
\begin{figure}[t!]
\centering
\includegraphics[width=\columnwidth]{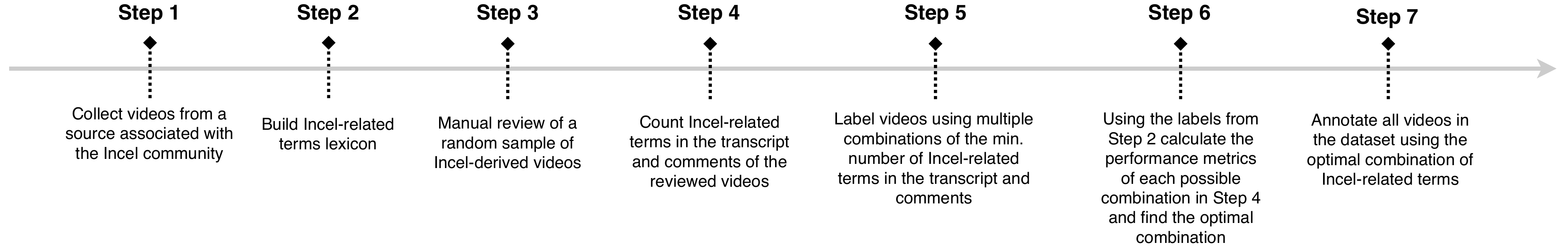}
\vspace{-0.25cm}
\caption{\revision{Overview of our data collection and video annotation methodology.}}
\label{fig:video_annotation_methodology}
\end{figure}

\revision{
To collect Incel-related videos on YouTube, we look for YouTube links on Reddit (Step 1 in Figure~\ref{fig:video_annotation_methodology}), since recent work highlighted that Incels are particularly active on Reddit~\cite{ribeiro2020pick}.
We start by creating a list of subreddits that are relevant to Incels.}
To do so, we inspect around 15 posts on the Incel Wiki~\cite{incelswiki_2019} looking for references to subreddits, and compile a list of 19 Incel-related subreddits.
This list also includes a set of communities broadly relevant to Incel ideology (even possibly anti-incel like /r/Inceltears) to capture a broader collection of relevant videos. 
The list of subreddits and where on the Incel Wiki we find them is available from~\cite{incelsubredditslist_2019}.

We collect all submissions and comments made between June 1, 2005, and April 30, 2019, on the 19 Incel-related subreddits using the Reddit monthly dumps from Pushshift~\cite{baumgartner2020pushshift}. 
We parse them to gather links to YouTube videos, extracting 5M posts, including 6.5K unique links to YouTube videos that are still online and have a transcript available by YouTube to download. Next, we collect the metadata of each YouTube video using the YouTube Data API~\cite{youtubedataapi}.
Specifically, we collect: 1) transcript; 2) title and description; 3) a set of tags defined by the uploader; 4) video statistics such as the number of views, likes, etc.; and 5) the top 1K comments, defined by YouTube's relevance metric, and their replies. Throughout the rest of this paper, we refer to this set of videos, which is derived from Incel-related subreddits, as {\bf\em ``Incel-derived''} videos.

Table~\ref{tab:incels_reddit_dataset_overview} reports the total number of users, posts, linked YouTube videos, and the period of available information for each subreddit.
Although recently created, /r/Braincels has the largest number of posts and YouTube videos. Also, even though it was banned in November 2017 for inciting violence against women~\cite{incelssubredditbanned_2017}, /r/Incels is fourth in terms of YouTube videos shared. Lastly, note that most of the subreddits in our sample were created between 2015 and 2018, which already suggests a trend of increasing popularity for the Incel community.

\begin{table}[t!]
\resizebox{\columnwidth}{!}{
\begin{tabular}{lrrrrrrr}
\toprule
\textbf{Subreddit} & \textbf{\#Videos} & \textbf{\#Users} & \textbf{\#Posts} & \textbf{Min. Date} & \textbf{Max. Date} & \textbf{\#Incel-related} & \textbf{\#Other}\\
 & & & & & & \textbf{Videos} & \textbf{Videos} \\
\midrule
Braincels       & 2,744 & 2,830,522 & 51,443  & 2017-10 & 2019-05 & 175 & 2,569\\
ForeverAlone    & 1,539 & 1,921,363 & 86,670    & 2010-09 & 2019-05 & 45 & 1,494\\
IncelTears      & 1,285 & 1,477,204 & 93,684  & 2017-05 & 2019-05 & 56 & 1,229\\
Incels        & 976   & 1,191,797 & 39,130  & 2014-01 & 2017-11 & 48 & 928\\
IncelsWithoutHate   & 223   & 163,820 & 7,141   & 2017-04 & 2019-05 & 16 & 207\\
ForeverAloneDating  & 92  & 153,039 & 27,460  & 2011-03 & 2019-05 & 0 & 92\\
askanincel      & 25  & 39,799  & 1,700   & 2018-11 & 2019-05 & 2 & 23\\
BlackPillScience  & 25  & 9,048   & 1,363   & 2018-03 & 2019-05 & 5 & 20\\
ForeverUnwanted   & 23  & 24,855  & 1,136   & 2016-02 & 2018-04 & 4 & 19\\
Incelselfies    & 17  & 60,988  & 7,057   & 2018-07 & 2019-05 & 1 & 16\\
Truecels      & 15  & 6,121   & 714   & 2015-12 & 2016-06 & 1 & 14\\
gymcels       & 5   & 1,430   & 296   & 2018-03 & 2019-04 & 2 & 3\\
MaleForeverAlone  & 3   & 6,306   & 831     & 2017-12 & 2018-06 & 0 & 3\\
foreveraloneteens   & 2   & 2,077   & 450   & 2011-11 & 2019-04 & 0 & 2\\
gaycel        & 1   & 117     & 43    & 2014-02 & 2018-10 & 0 & 1\\
SupportCel      & 1   & 6,095   & 474     & 2017-10 & 2019-01 & 0 & 1\\
Truefemcels     & 1   & 311     & 95    & 2018-09 & 2019-04 & 0 & 1\\
Foreveralonelondon  & 0   & 57    & 19    & 2013-01 & 2019-01 & 0 & 0\\
IncelDense      & 0   & 2,058   & 388     & 2018-06 & 2019-04 & 0 & 0\\
\midrule
\textbf{Total (Unique)}    & \textbf{6,977} & \textbf{7,897,007} & \textbf{320,094} & \textbf{-} & \textbf{-} & \textbf{290} & \textbf{6,162}\\ 
\bottomrule
\end{tabular}}
\caption{\revision{Overview of our Reddit dataset. We also include, for each subreddit, the number of videos from our Incel-derived labeled dataset.
The total number of videos reported in the individual subreddits differs from the unique videos collected since multiple videos have been shared in more than one subreddit.}}
\label{tab:incels_reddit_dataset_overview}
\end{table}

\descr{Control Set.} 
\label{subsec:control_videos}
We also collect a dataset of random videos and use it as a control to capture more general trends on YouTube videos shared on Reddit as the Incel-derived set includes only videos posted on Incel communities on Reddit. 
To collect Control videos, we parse all submissions and comments made on Reddit between June 1, 2005, and April 30, 2019, using the Reddit monthly dumps from Pushshift, and we gather all links to YouTube videos.  
From them, we randomly select 5,793 links shared in 2,154 subreddits for which we collect their metadata using the YouTube Data API. 
We provide a list of these subreddits and the number of control videos shared in each subreddit at~\cite{controlsubreddits}.

\revision{
We choose to use a randomly selected set of videos shared on Reddit as our Control set for a more fair comparison since our Incel-derived set also includes videos shared on this platform.
We collect random videos instead of videos relevant to another sensitive topic because this allows us to study the amount of Incel-related content that can generally be found on YouTube.
At the same time, videos relevant to another sensitive topic or community (e.g., MGTOW) may have strong similarities with Incel-related videos, hence they may not be able to capture more general trends on YouTube.
}

\begin{table}[t!]
\footnotesize
\centering
\begin{tabular}{rr|rrrr}
\toprule
\multicolumn{2}{c|}{\bf \#Incel-related Terms} & & & &\\
  {\bf\em in Transcript} & {\bf\em in Comments} &
{\textbf{Accuracy}} &
{\textbf{Precision}} &
{\textbf{Recall}} &
{\textbf{F1 Score}} \\ 
\midrule
$\geq$0          & $\geq$7          & 0.81          & 0.77          & 0.80          & 0.78          \\
$\geq$1          & $\geq$1          & 0.82          & 0.78          & 0.82          & 0.79          \\
$\geq$0          & $\geq$3          & 0.79          & 0.79          & 0.79          & 0.79          \\
$\geq$1          & $\geq$2          & 0.83          & 0.78          & 0.83          & 0.79          \\
\textbf{$\geq$1} & \textbf{$\geq$3} & \textbf{0.83} & \textbf{0.79} & \textbf{0.83} & \textbf{0.79} \\
\bottomrule
\end{tabular}%
\caption{Performance metrics of the top combinations of the number of Incel-related terms in a video's transcript and comments.}
\label{tab:incels_thresholds_performance_metrics}
\end{table}

\subsection{Video Annotation}
\label{subsec:incels_video_annotation}
The analysis of Incel-related content on YouTube differs from analyzing other types of inappropriate content on the platform.
So far, there is no prior study exploring the main themes involved in videos that Incels find of interest. 
This renders the task of annotating the actual video rather cumbersome. 
Besides, annotating the video footage does not by itself allow us to study the footprint of the Incel community on YouTube effectively. 
When it comes to this community, it is not only the video's content that may be relevant. 
Rather, the language that the community members use in their videos or comments for or against their views is also of interest. For example, there are videos featuring women talking about feminism, which are heavily commented on by Incels.

\descr{Building a Lexicon.} 
To capture the variety of aspects of the problem, we devise an annotation methodology based on a lexicon of terms that are routinely used by members of the Incel community and use it to annotate the videos in our dataset.
\revision{
Figure~\ref{fig:video_annotation_methodology} depicts the individual steps that we follow in the devised video annotation methodology.
}

\revision{
To create the lexicon (Step 2 in Figure~\ref{fig:video_annotation_methodology}), we first crawl the ``glossary'' available on the Incels Wiki page~\cite{incelglossary_2019}, gathering 395 terms. 
Since the glossary includes several words that can also be regarded as general-purpose (e.g., fuel, hole, legit, etc.), we employ three human annotators to determine whether each term is specific to the Incel community.
}

\revision{
We note that all annotators label all the 395 terms of the glossary. 
The three annotators are authors of this paper and they are familiar with scholarly articles on the Incel community and the Manosphere in general.
Before the annotation task, a discussion took place to frame the task and the annotators were told to consider a term relevant only if it expresses hate, misogyny, or is directly associated with Incel ideology.
For example, the phrase ``Beta male'' or any Incel-related incident (e.g., ``supreme gentleman,''  an indirect reference to the Isla Vista killer Elliot Rodgers~\cite{wiki2019rodgers}).
We note that, during the labeling, the annotators had no discussion or communication whatsoever about the task at hand.
}

We then create our lexicon by only considering the terms annotated as relevant, based on all the annotators' majority agreement, which yields a 200 Incel-related term dictionary. 
We also compute the Fleiss' Kappa Score~\cite{fleiss1971measuring} to assess the agreement between the annotators, finding it to be $0.69$, which is considered ``substantial'' agreement~\cite{landis1977measurement}. 
The final lexicon with all the relevant terms is available from~\cite{inceltermslexicon_2019}.

\descr{Labeling.} 
Next, we use the lexicon to label the videos in our dataset. 
We look for these terms in the transcript, title, tags, and comments of our dataset videos. 
Most matches are from the transcript and the videos' comments; thus, we decide to use these to determine whether a video is Incel-related.
To select the minimum number of Incel-related terms that transcripts and comments should contain to be labeled as ``Incel-related,'' we devise the following methodology:
\begin{enumerate}
	\item \revision{We randomly select 1K videos from the Incel-derived set, which the first author of this paper manually annotates as ``Incel-related'' or ``Other'' by watching them and looking at the metadata. Note that Incel-related videos are a subset of Incel-derived ones (Step 3 in Figure~\ref{fig:video_annotation_methodology}).}

	\item \revision{We count the number of Incel-related terms in the transcript and the annotated videos' comments (Step 4 in Figure~\ref{fig:video_annotation_methodology}).}

	\item \revision{For each possible combination of the minimum number of Incel-related terms in the transcript and the comments, we label each video as Incel-related or not, and calculate the accuracy, precision, recall, and F1 score based on the labels assigned to the videos during the manual annotation (Steps 5 and 6 in Figure~\ref{fig:video_annotation_methodology}).}
\end{enumerate}
Table~\ref{tab:incels_thresholds_performance_metrics} shows the performance metrics for the top five combinations of the number of Incel-related terms in the transcript and the comments. 
\revision{
We pick the one yielding the best F1 score (to balance between false positives and false negatives), which is reached if we label a video as Incel-related when there is at least one Incel-related term in the transcript and at least three in the comments. 
Using this rule, we annotate all the videos in our dataset (Steps 6 and 7 in Figure~\ref{fig:video_annotation_methodology}).
}

\revision{Table~\ref{tab:incels_reddit_dataset_overview} reports the label statistics of the Incel-derived videos per subreddit.}
Our final labeled dataset includes 290 Incel-related and $6,162$ Other videos in the Incel-derived set and 66 Incel-related and $5,727$ Other videos in the Control set.

\subsection{Ethics} 
\revision{
Overall, we follow standard ethical guidelines~\cite{dittrich2012menlo,rivers2014ethical} regarding information research and the use of shared measurement data.
In this work, we only collect and process publicly available data, make no attempt to de-anonymize users, and our data collection does not violate the terms of use of the APIs we employ. 
More precisely, we ensure compliance with GDPR's ``Right of Access''~\cite{gdpr2018rightaccess} and ``Right to be Forgotten''~\cite{gdpr2018rightforgotten} principles. 
For the former, we give users the right to obtain a copy of any data that we maintain about them for this research, while for the former we ensure that we delete and not share with any unauthorized party any information that has been deleted from the public repositories from which we obtain our data.
We also note that we do not share with anyone any sensitive personal data, such as the actual content of the comments that we analyze or the usernames of the commenting users.
Instead, we make publicly available for reproducibility and research purposes\footnote{\revision{We make publicly available all the metadata of the collected and annotated videos and the identifiers of the comments that we analyze~\cite{incelsdataset} while ensuring that we abide by GDPR's ``Right to be Forgotten''~\cite{gdpr2018rightforgotten} principle}} the identifiers of the comments we analyze along with all the metadata of the videos we analyze that do not include any personal data.
}

\revision{
Furthermore, our video annotation methodology abides by the ethical guidelines defined by the Association of Internet Researchers (AoIR) for the protection of researchers~\cite{aoir2019ethicalguidelines}.
Note that in our video annotation methodology we do not engage any human subjects other than the three authors of this paper.
Since the annotators are authors of this paper, we do not take into consideration harmful effects on random human annotators due to inappropriate content. However, we still consider the effect of the content that we study on the authors and especially the student authors. 
We address this with continuous monitoring and open discussions with members of the research team, as
well as by properly applying best practices from the psychological and social scientific literature on the topic, e.g.,~\cite{beale2004impact,bashir2018doing,blagden2010challenge}. 
One of the primary goals is to minimize the risk of the researchers becoming desensitized to such content.
Finally, we believe that studying misogynistic and hateful communities in depth is bound to be beneficial for society at large, as well as for victims of such abuse.
}

\section{Analysis}

\subsection{Evolution of Incel Community on YouTube}
This section explores how the Incel communities on YouTube and Reddit have evolved in terms of videos and comments posted.

\begin{figure}[t!]
\centering
\subfigure[]{\includegraphics[width=0.85\columnwidth]{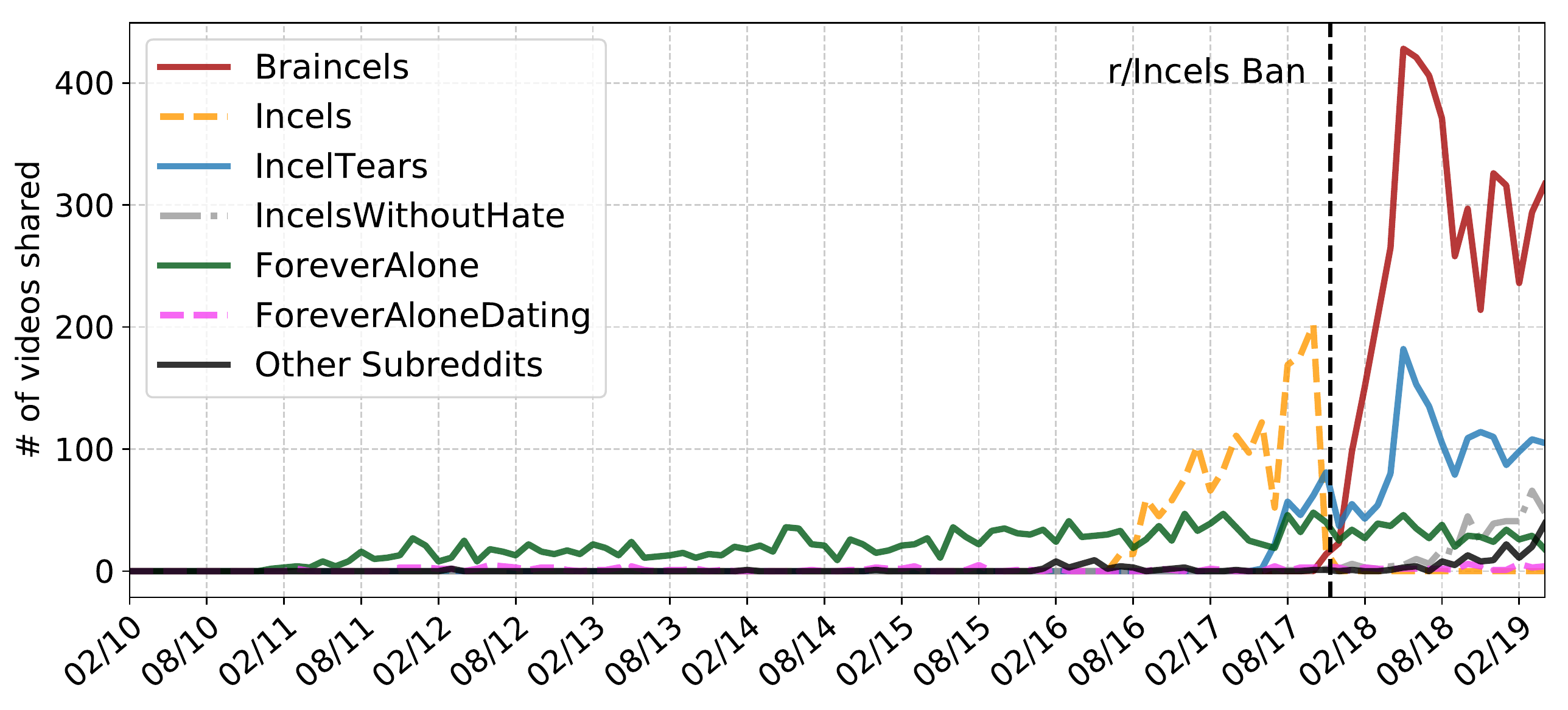}\label{fig:incels_subreddits_youtube_links_shared_per_year}}
\subfigure[]{\includegraphics[width=0.85\columnwidth]{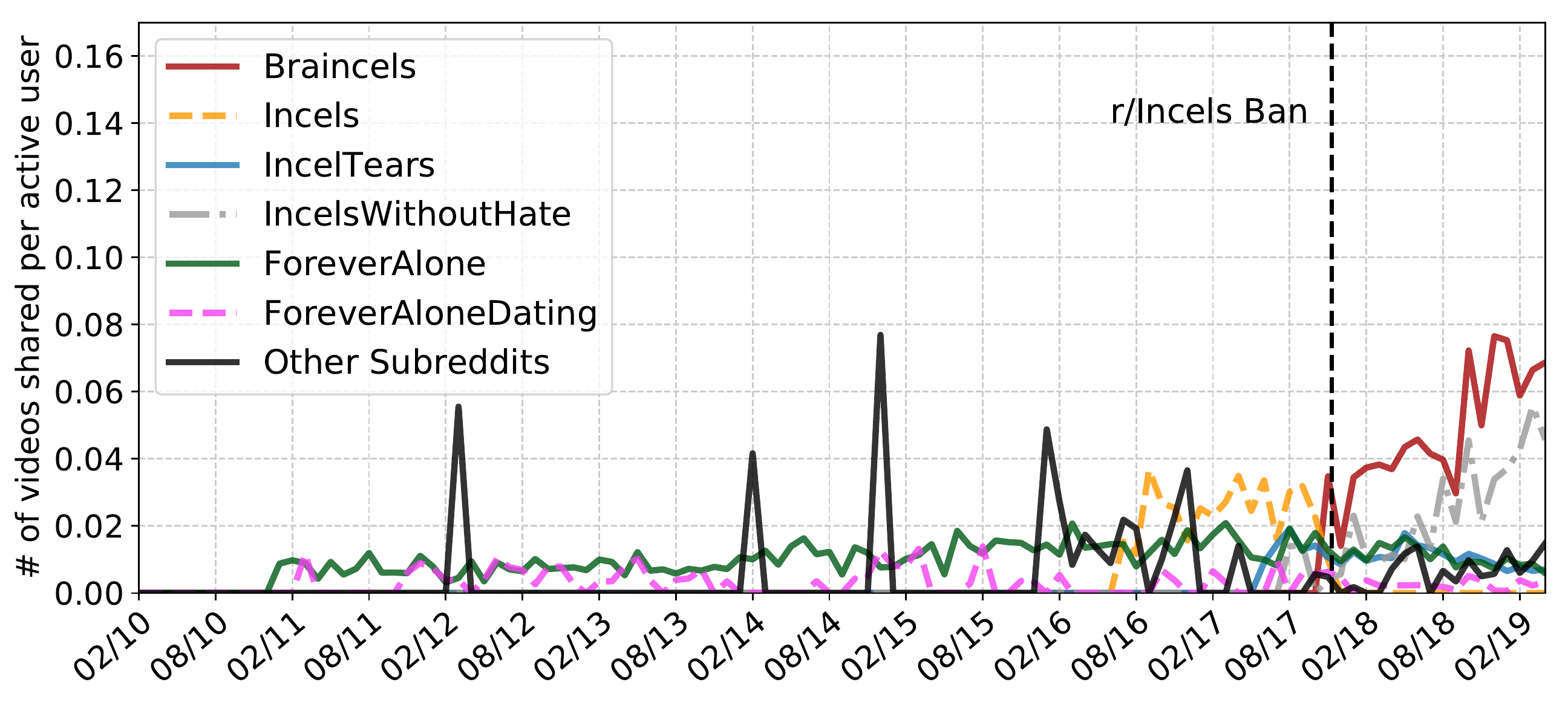}\label{fig:incels_normalized_subreddits_youtube_links_shared_per_year}}
\vspace{-0.25cm}
\caption{\revision{Temporal evolution of the number of YouTube videos shared on each subreddit per month. (a) depicts the absolute number of videos shared in each subreddit and (b) depicts the normalized number of videos shared per active user of each subreddit.
We annotate both figures with the date when Reddit decided to ban the /r/Incels subreddit.
}}
\label{fig:incels_subreddits_youtube_links_shared_plot}
\end{figure}

\descr{Videos.}
We start by studying the ``evolution'' of the Incel communities concerning the number of videos they share.
First, we look at the frequency with which YouTube videos are shared on various Incel-related subreddits per month; see Figure~\ref{fig:incels_subreddits_youtube_links_shared_plot}.
\revision{Figure~\ref{fig:incels_subreddits_youtube_links_shared_plot} shows the absolute number of videos shared in each Incel-related subreddit per month, while Figure~\ref{fig:incels_normalized_subreddits_youtube_links_shared_per_year} shows the number of videos shared in each subreddit per active user of each community.}
After June 2016, we observe that Incel-related subreddits users start linking to YouTube videos more frequently and more in 2018. 
\revision{
This trend is more pronounced on /r/Braincels in both the absolute number of videos shared and the number of videos shared per active user; see /r/Braincels in Figure~\ref{fig:incels_subreddits_youtube_links_shared_per_year} and Figure~\ref{fig:incels_normalized_subreddits_youtube_links_shared_per_year}.}
This indicates that the use of YouTube to spread Incel ideology is increasing.

\revision{
Note that the sharp drop of /r/Incels activity is due to Reddit's decision to ban this subreddit for inciting violence against women in November 2017~\cite{incelssubredditbanned_2017} (see annotation in Figure~\ref{fig:incels_subreddits_youtube_links_shared_per_year} and Figure~\ref{fig:incels_normalized_subreddits_youtube_links_shared_per_year}).
However, the sharp increase of /r/Braincels activity after this period questions the efficacy of Reddit's decision to ban /r/Incels, and it can be considered as evidence that the ban was ineffective in terms of suppressing the activity of the Incel community on the platform.
}
It is also worth noting that Reddit decided to ban /r/Braincels in September 2019~\cite{theverge2019raincelsban}.

\begin{figure}[t!]
\centering
\includegraphics[width=0.85\columnwidth]{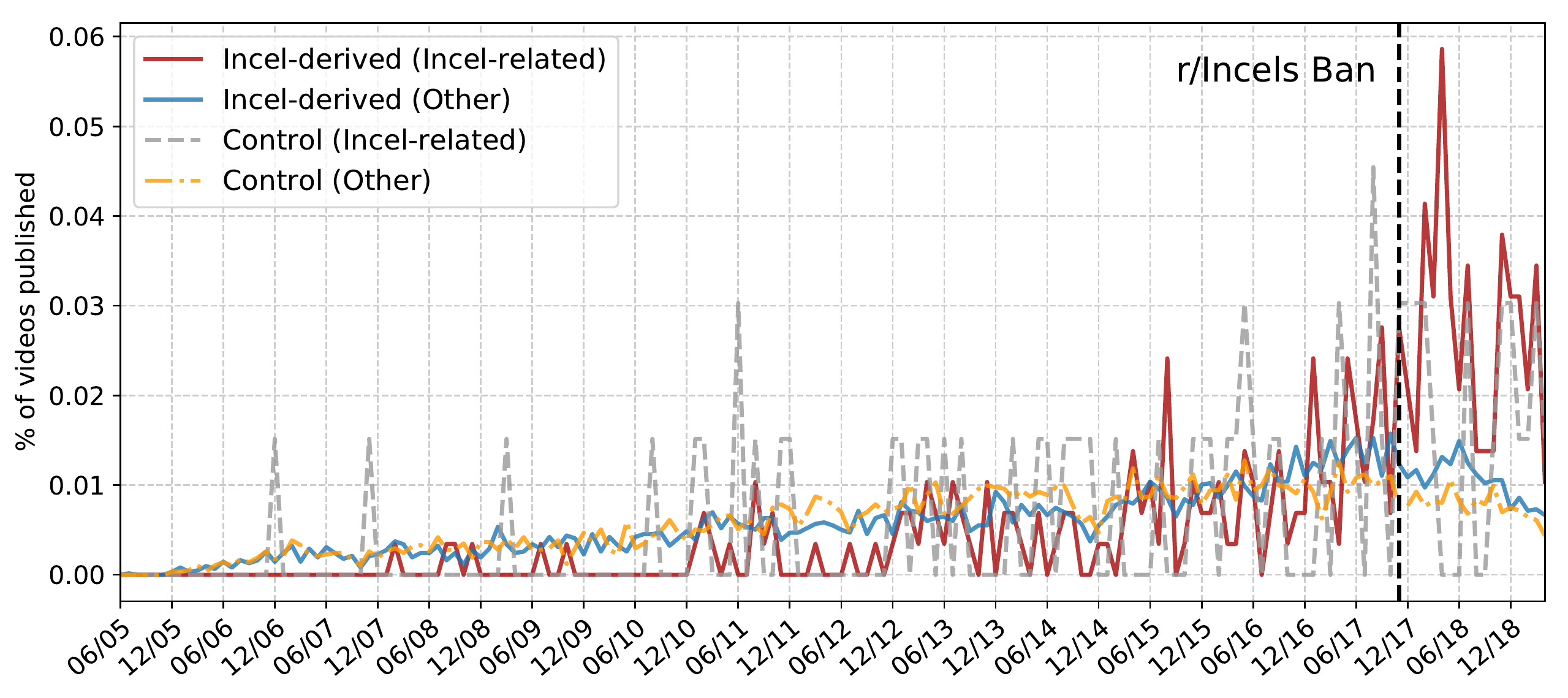}
\caption{\revision{Percentage of videos published per month for both Incel-derived and Control videos. We also depict the date when Reddit decided to ban the /r/Incels subreddit.}}
\label{fig:incels_videos_published_per_date}
\end{figure}

\revision{
In Figure~\ref{fig:incels_videos_published_per_date}, we plot the percentage of videos published per month for both Incel-derived and Control videos, while we also depict the date when Reddit decided to ban the /r/Incels subreddit.
}
While the increase in the number of Other videos remains relatively constant over the years for both sets of videos, this is not the case for Incel-related ones, as $81\%$ and $64\%$ of them in the Incel-derived and Control sets, respectively, were published after December 2014.
Overall, there is a steady increase in Incel activity, especially after 2016, which is particularly worrisome as we have several examples of users who were radicalized online and have gone to undertake deadly attacks~\cite{cecco2019toronto}.
\revision{
An even higher increase in Incel-activity is also observed after the ban of the /r/Incels subreddit.
}

\begin{figure*}[t!]
\centering
\includegraphics[width=0.85\columnwidth]{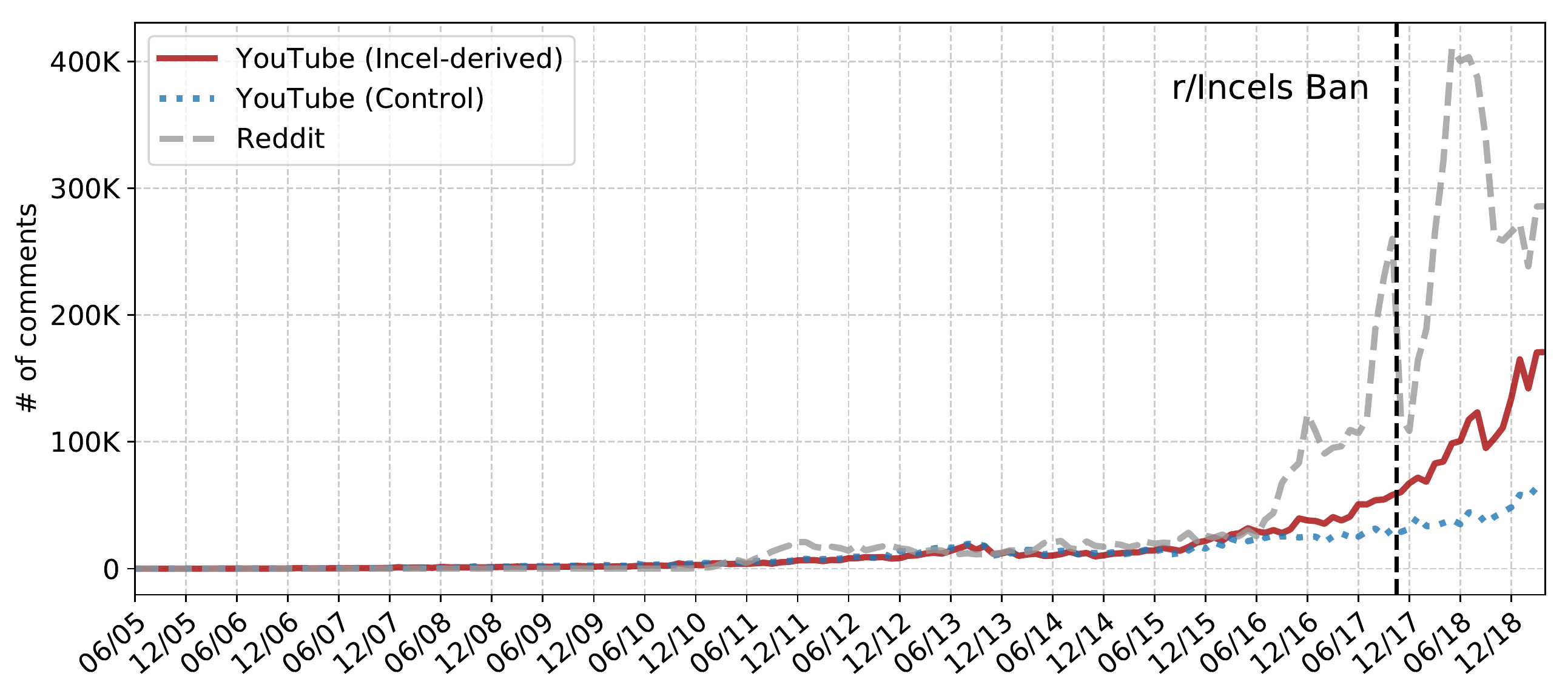}
\caption{\revision{Temporal evolution of the number of comments per month. We also depict the date when Reddit decided to ban the /r/Incels subreddit.}}
\label{fig:incels_number_of_comments_per_date_plot}
\end{figure*}

\descr{Comments.}
Next, we study the commenting activity on both Reddit and YouTube.
Figure~\ref{fig:incels_number_of_comments_per_date_plot} shows the number of comments posted per month for both YouTube Incel-derived and Control videos, as well as Reddit.
\revision{
Activity on both platforms starts to markedly increase after 2016, and more after the ban of /r/Incels in November 2017, with Reddit and YouTube Incel-derived videos having substantially more comments than the Control videos.
}
Once again, the sharp increase in the commenting activity over the last few years signals an increase in the Incel user base's size.

\begin{figure*}[t!]
\centering
\includegraphics[width=0.85\columnwidth]{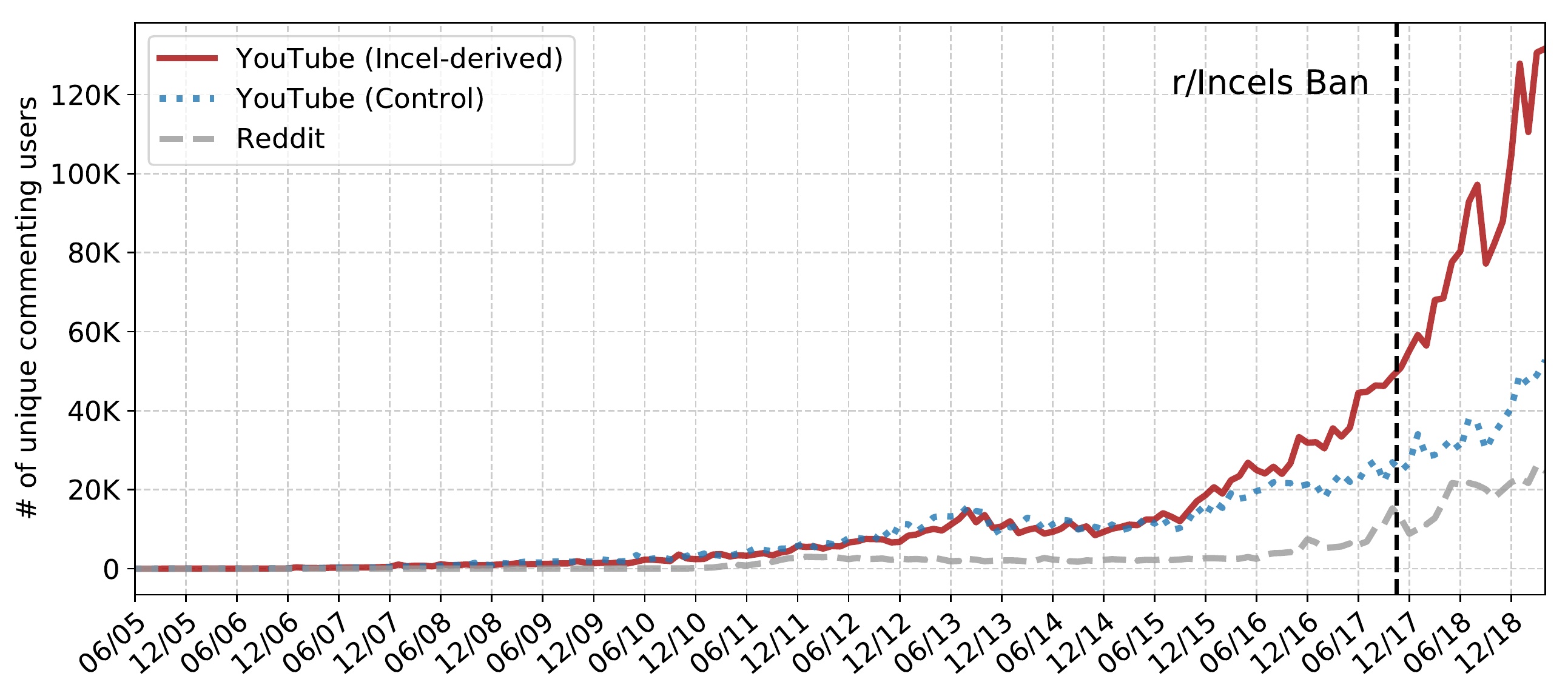}
\caption{\revision{Temporal evolution of the number of unique commenting users per month. We also depict the date when Reddit decided to ban the /r/Incels subreddit.}}
\label{fig:incels_unique_commenting_users_per_date_plot}
\end{figure*}

To further analyze this trend, we look at the number of {\em unique} commenting users per month on both platforms; see Figure~\ref{fig:incels_unique_commenting_users_per_date_plot}.
On Reddit, we observe that the number of unique users remains steady over the years, increasing from 10K in August 2017 to 25K in April 2019. 
This is mainly because most of the subreddits in our dataset ($58\%$) were created after 2016.
On the other hand, for the Incel-derived videos on YouTube, there is a substantial increase from 30K in February 2017 to 132K in April 2019. 
We also observe an increase of the Control videos' unique commenting users (from 18K in February 2017 to 53K in April 2019).
However, the increase is not as sharp as that of the Incel-derived videos; $483\%$ vs. $1,040\%$ increase in the average unique commenting users per month after January 2017 in Control and Incel-derived videos, respectively.

\begin{figure}[t!]
\centering
\includegraphics[width=.85\columnwidth]{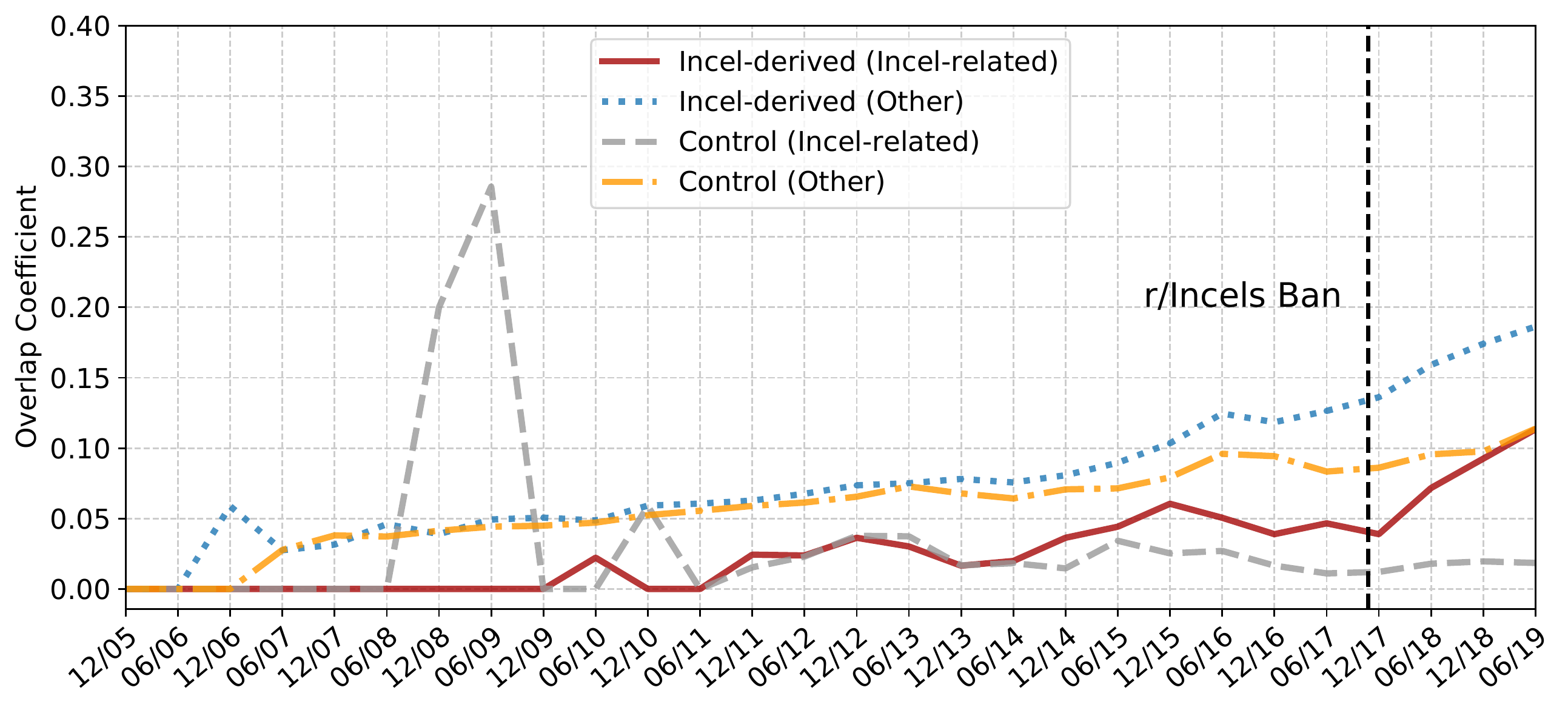}
\caption{\revision{Self-similarity of commenting users in adjacent months for both Incel-derived and Control videos. We also depict the date when Reddit decided to ban the /r/Incels subreddit.}}
\label{fig:incels_overlap_videos_users_similarity_per_year_plot}
\end{figure}

\revision{
To assess whether the sharp increase in unique commenting users of the Incel-derived and Control videos after 2017 is due to the increased interest by random users or to an increased interest in those videos and their discussions by the same users over the years, we use the Overlap Coefficient similarity metric~\cite{vijaymeena2016survey}; it measures user retention over time for the videos in our dataset. 
Specifically, we calculate, for each month, the similarity of commenting users with those doing so {\em the month before}, for both Incel-related and Other videos in the Incel-derived and Control sets. 
Note that if the set of commenting users of a specific month is a subset of the previous month’s commenting users or the converse, the overlap coefficient is equal to 1.
The results of this calculation are shown in Figure~\ref{fig:incels_overlap_videos_users_similarity_per_year_plot}, in which we again depict the date when Redid decided to ban /r/Incels.
Interestingly, for the Incel-related videos of the Incel-derived set, we find a sharp growth in user retention right after the ban of the /r/Incels subreddit in November 2017, while this is not the case for the Incel-related videos of the Control set.
For the Incel-related videos of the Control set, we observe a more steady increase in user retention over time.
Once again, this might be related to the increased popularity of the Incel communities and might indicate that the ban of /r/Incels energized the community and made participants more persistent.
Also, the higher user retention of Other videos in both sets is likely due to the much higher proportion of Other videos in each set.
}

\subsection{Does YouTube's Recommendation Algorithm steer users towards Incel-related videos?}
\label{subsec:incels_recommendations_analysis}
Next, we present an analysis of how YouTube's recommendation algorithm behaves with respect to Incel-related videos.
More specifically, 1) we investigate how likely it is for YouTube to recommend an Incel-related video; and
2) We simulate the behavior of a user who views videos based on the recommendations by performing random walks on YouTube's recommendation graph to measure the probability of such a user discovering Incel-related content.

\begin{table}[t!]
\centering
\small
\begin{tabular}{lrr}
\toprule
\textbf{Recommendation Graph} & \textbf{Incel-related} & \textbf{Other} \\
\midrule
Incel-derived & 1,074 ($2.9\%$) & 36,673 ($97.1\%$) \\
Control & 428 ($1.5\%$) & 28,866 ($98.5\%$) \\
\bottomrule
\end{tabular}%
\caption{Number of Incel-related and Other videos in each recommendation graph.}
\label{tab:incels_videos_in_recommendations_rounds}
\end{table}

\subsubsection{Recommendation Graph Analysis}
\revision{
To build the recommendation graphs used for our analysis, we use functionality provided by the YouTube Data API. 
For each video in the Incel-derived and Control sets, we collect the top 10 recommended videos associated with it.
Note that the use of the YouTube Data API is associated with a specific account only for authentication to the API and that the API does not maintain a watch history nor any cookies.
Thus, our data collection does not capture how specific account features or the viewing history affect personalized recommendations.  
Instead, the YouTube Data API allows us to collect recommendations provided by YouTube's recommendation algorithm based on video item-to-item similarity, as well as general user engagement and satisfaction metrics~\cite{zhao2019recommending}. 
The collected recommendations are similar to the recommendations presented to a non-logged-in user who watches videos on YouTube.
We collect the recommendations for the Incel-derived videos between September 20 and October 4, 2019, and the Control between October 15 and November 1, 2019. 
To annotate the collected videos, we follow the same approach described in Section~\ref{subsec:incels_video_annotation}. 
}
Since our video annotation is based on the videos' transcripts, we only consider the videos that have one when building our recommendations graphs.

Next, we build a directed graph for each set of recommendations, where nodes are videos (either our dataset videos or their recommendations), and edges between nodes indicate the recommendations between all videos (up to ten).
For instance, if \textit{video2} is recommended via \textit{video1}, then we add an edge from \textit{video1} to \textit{video2}. 
Throughout the rest of this paper, we refer to each set of videos' collected recommendations as separate \emph{recommendation graphs}.

First, we investigate the prevalence of Incel-related videos in each recommendation graph.
Table~\ref{tab:incels_videos_in_recommendations_rounds} reports the number of Incel-related and Other videos in each graph.
For the Incel-derived graph, we find 36,7K ($97.1\%$) Other and 1K ($2.9\%$) Incel-related videos, while in the Control graph, we find 28,9K ($98.5\%$) Other and 428 ($1.5\%$) Incel-related videos. 
These findings highlight that despite the proportion of Incel-related video recommendations in the Control graph being smaller, there is still a non-negligible amount recommended to users. 
Also, note that we reject the null hypothesis that the differences between the two graphs are due to chance via the Fisher's exact test ($p<0.001$)~\cite{fisher1922interpretation}.

\begin{table}[t!]
\footnotesize
\centering
\begin{tabular}{llrr}
\toprule
\textbf{Source} & \textbf{Destination} & \textbf{Incel-derived} & \textbf{Control} \\ 
\midrule
Incel-related   & Incel-related & 889 ($0.8\%$)   		& 89 ($0.2\%$) \\
Incel-related   & Other      	& 3632 ($3.2\%$)   	& 773 ($1.4\%$) \\
Other      		& Other    		& 104,706 ($93.2\%$)  	& 54,787 ($97.0\%$) \\
Other      		& Incel-related & 3,160 ($2.8\%$)   	& 842 ($1.5\%$) \\ 
\bottomrule
\end{tabular}%
\caption{Number of transitions between Incel-related and Other videos in each recommendation graph.}
\label{tab:incels_graph_transitions_in_recommendation_rounds}
\end{table}

\descr{How likely is it for YouTube to recommend an Incel-related Video?}
Next, to understand how frequently YouTube recommends an Incel-related video, we study the interplay between the Incel-related and Other videos in each recommendation graph. For each video, we calculate the out-degree in terms of Incel-related and Other labeled nodes.
We can then count the number of \textit{transitions} the graph makes between differently labeled nodes.
Table~\ref{tab:incels_graph_transitions_in_recommendation_rounds} reports the percentage of each transition between the different types of videos for both graphs.
Perhaps unsurprisingly, most of the transitions, $93.2\%$ and $97.0\%$, respectively, in the Incel-derived and Control graphs are between Other videos, but this is mainly because of the large number of Other videos in each graph. We also find a high percentage of transitions between Other and Incel-related videos.
When a user watches an Other video, if they randomly follow one of the top ten recommended videos, there is a $2.8\%$ and $1.5\%$ probability in the Incel-derived and Control graphs, respectively, that they will end up at an Incel-related video.
\revision{
Interestingly, in both graphs, Incel-related videos are more often recommended by Other videos than by Incel-related videos.
On the one hand, this might be due to the larger number of Other videos compared to Incel-related videos in both recommendation graphs.
On the other hand, this may indicate that YouTube's recommendation algorithm cannot discern Incel-related videos, which are likely misogynistic.
}

\begin{figure}[t!]
\centering
\subfigure[]{\includegraphics[width=.49\linewidth]{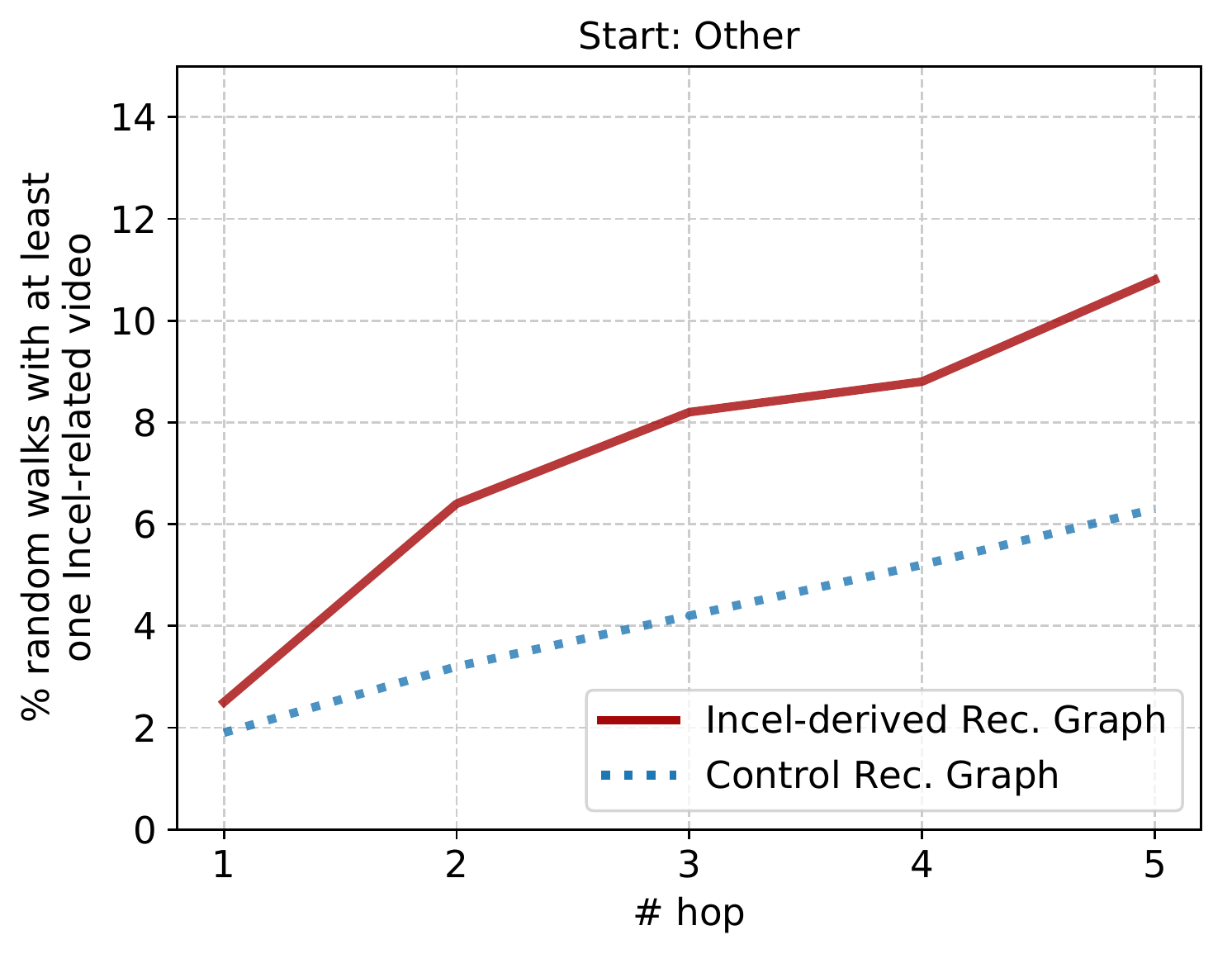}
\label{fig:incels_at_least_one_random_walks_plot_started_from_irrelevant}}
\subfigure[]{\includegraphics[width=.49\linewidth]{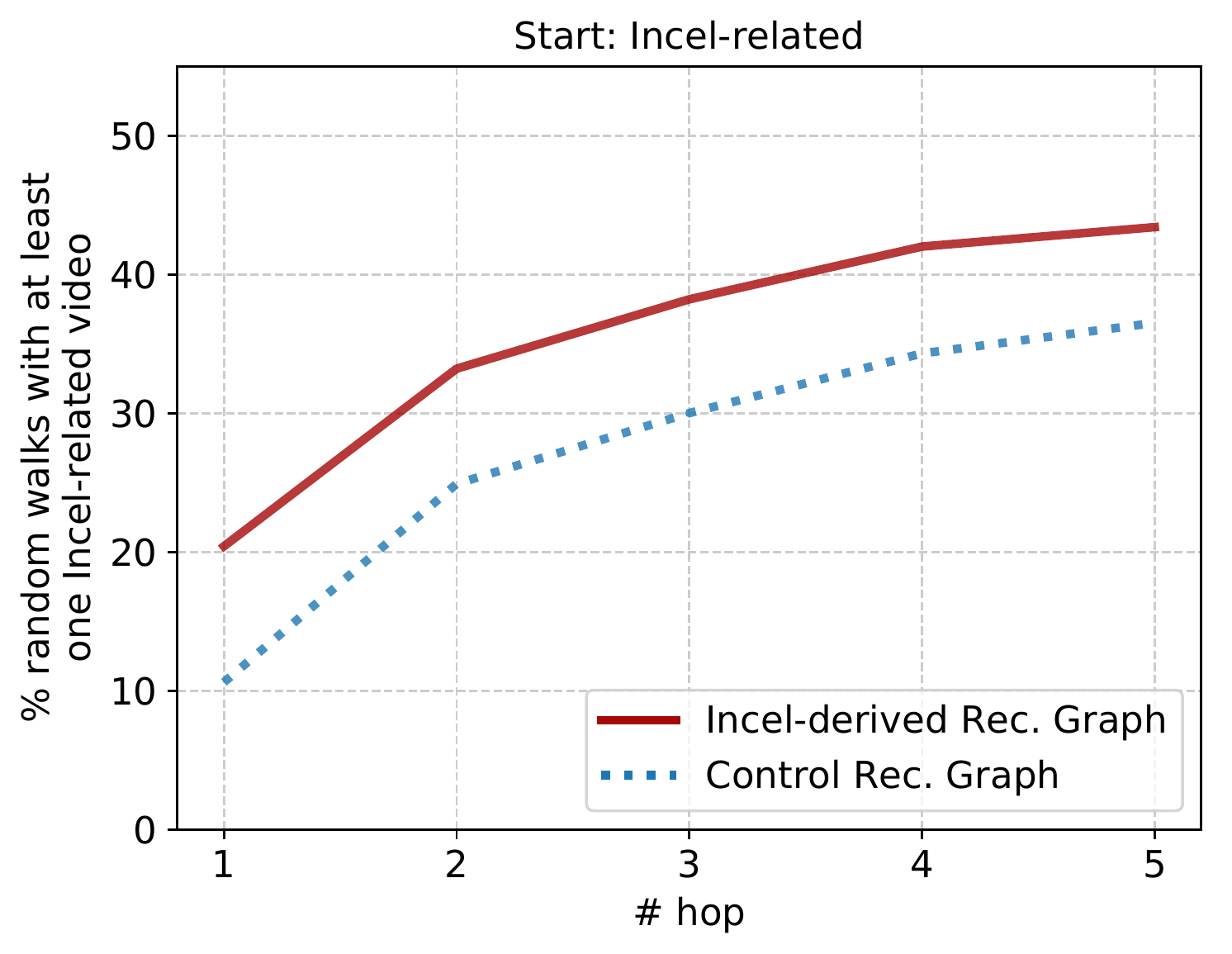}
\label{fig:incels_at_least_one_random_walks_plot_started_from_incels_related}}
\caption{\revision{Percentage of random walks where the random walker encounters at least one Incel-related video for both starting scenarios. Note that the random walker selects, at each hop, the next video to watch at random.}}
\label{fig:incels_random_walks_plots_at_least_one}
\end{figure}

\subsubsection{Does YouTube’s recommendation algorithm contribute to steering users towards Incel communities?}
\label{subsubsec:random_walks_recommendation_graph}
We then study how YouTube's recommendation algorithm behaves with respect to discovering Incel-related videos.
Through our graph analysis, we showed that the problem of Incel-related videos on YouTube is quite prevalent.
However, it is still unclear how often YouTube's recommendation algorithm leads users to this type of content.

To measure this, we perform experiments considering a ``random walker.''
This allows us to simulate a random user who starts from one video and then watches several videos according to the recommendations.
\revision{
More precisely, since we build our recommendation graphs using the YouTube Data API, the random walker simulates non-logged-in users who watch videos on YouTube.
}
The random walker begins from a randomly selected node and navigates the graph choosing edges at random for five hops. %
We repeat this process for $1,000$ random walks considering two starting scenarios. In the first scenario, the starting node is restricted to Incel-related videos. In the second, it is restricted to Other.
We perform the same experiment on both the Incel-derived and Control recommendations graphs.

\begin{figure}[t!]
\centering
\subfigure[]{\includegraphics[width=.49\columnwidth]{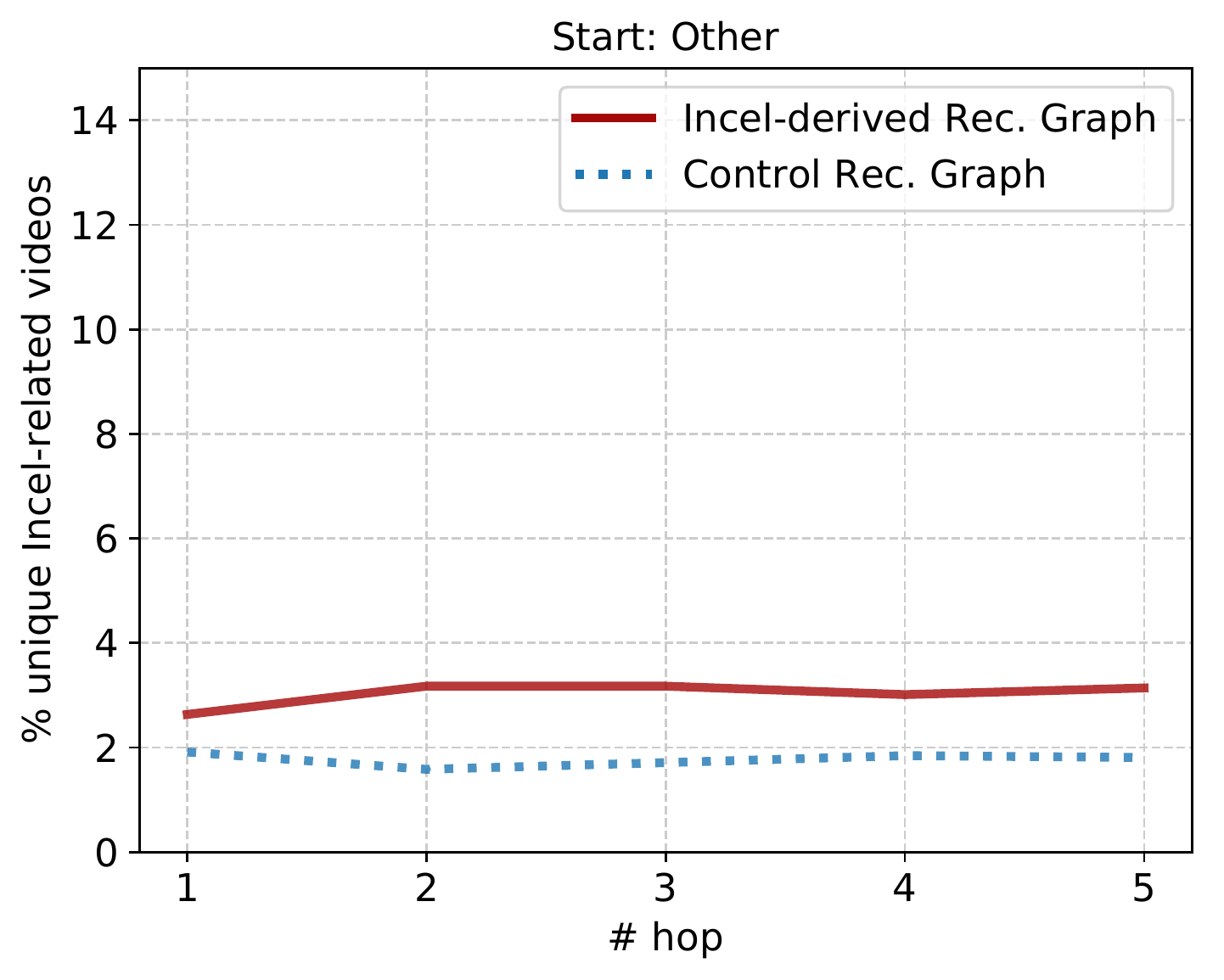}
\label{fig:incels_random_walks_plot_started_from_irrelevant}}
\subfigure[]{\includegraphics[width=.49\columnwidth]{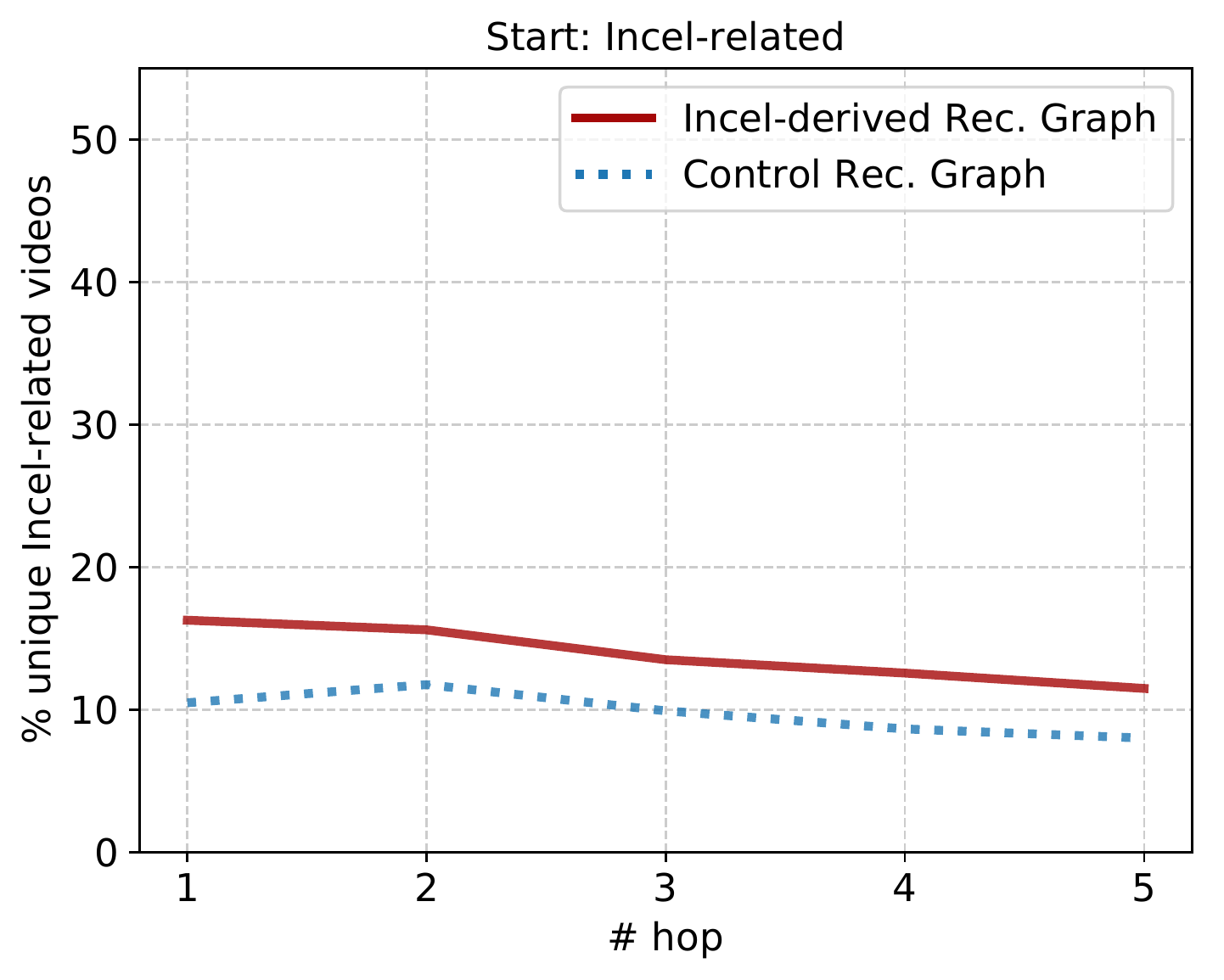}
\label{fig:incels_random_walks_plot_started_from_incels_related}}
\caption{\revision{Percentage of Incel-related videos across all unique videos that the random walk encounters at hop $k$ for both starting scenarios. Note that the random walker selects, at each hop, the next video to watch at random.}}
\label{fig:incels_random_walks_plots}
\end{figure}

Next, for the random walks of each recommendation graph, we calculate two metrics:
1) the percentage of random walks where the random walker finds at least one Incel-related video in the $k$-th hop; 
and 2) the percentage of Incel-related videos over all the unique videos that the random walker encounters up to the $k$-th hop for both starting scenarios.
The two metrics, at each hop, are shown in Figure~\ref{fig:incels_random_walks_plots_at_least_one} and~\ref{fig:incels_random_walks_plots} for both recommendation graphs.

When starting from an Other video, there is, respectively, a $10.8\%$ and $6.3\%$ probability to encounter at least one Incel-related video after five hops in the Incel-derived and Control recommendation graphs (see Figure~\ref{fig:incels_at_least_one_random_walks_plot_started_from_irrelevant}). 
When starting from an Incel-related video, we find at least one Incel-related in $43.4\%$ and $36.5\%$ of the random walks performed on the Incel-derived and Control recommendation graphs, respectively (see Figure~\ref{fig:incels_at_least_one_random_walks_plot_started_from_incels_related}).
Also, when starting from Other videos, most of the Incel-related videos are found early in our random walks (i.e., at the first hop), and this number remains almost the same as the number of hops increases (see Figure~\ref{fig:incels_random_walks_plot_started_from_irrelevant}). 
The same stands when starting from Incel-related videos, but in this case, the percentage of Incel-related videos decreases as the number of hops increases for both recommendation graphs (see Figure~\ref{fig:incels_random_walks_plot_started_from_incels_related}).

As expected, in all cases, the probability of encountering Incel-related videos in random walks performed on the Incel-derived recommendation graph is higher than in the random walks performed on the Control recommendation graph. 
We also verify that the difference between the distribution of Incel-related videos encountered in the random walks of the two recommendation graphs is statistically significant via the Kolmogorov-Smirnov test~\cite{massey1951kolmogorov} ($p<0.05$).
Overall, we find that Incel-related videos are usually recommended within the two first hops. 
However, in subsequent hops, the number of encountered Incel-related videos decreases. 
This indicates that in the absence of personalization (e.g., for a non-logged-in user), a user casually browsing YouTube videos is unlikely to end up in a region dominated by Incel-related videos.

\begin{table*}[t!]
\footnotesize
\centering
\begin{tabular}{>{\raggedleft\arraybackslash}p{1cm}
>{\raggedleft\arraybackslash}p{2.8cm}
>{\raggedleft\arraybackslash}p{2.8cm}
>{\raggedleft\arraybackslash}p{2.8cm}
>{\raggedleft\arraybackslash}p{2.8cm}}
\toprule
 & \multicolumn{2}{c}{\textbf{Incel-derived Recommendation Graph}} & \multicolumn{2}{c}{\textbf{Control Recommendation Graph}} \\
\midrule
\textbf{\#hop (M)} & \textbf{In next 5-M hops, $\geq$1 Incel-related} & \textbf{In next hop, 1 Incel-related} & \textbf{In next 5-M hops, $\geq$1 Incel-related} & \textbf{In next hop, 1 Incel-related} \\
\midrule
1 & $43.4\%$ 	& $4.1\%$ 	& $36.5\%$ & $2.1\%$ \\
2 & $46.5\%$ 	& $9.4\%$ 	& $38.9\%$ & $5.4\%$ \\
3 & $49.3\%$ 	& $11.4\%$ 	& $41.6\%$ & $5.0\%$ \\
4 & $49.7\%$ 	& $18.9\%$ 	& $42.0\%$ & $11.2\%$ \\
5 & $47.9\%$ 	& $30.1\%$ 	& $39.7\%$ & $17.7\%$ \\
\bottomrule
\end{tabular}%
\caption{\revision{Probability of finding (a) at least one Incel-related video in the next $5-M$ hops having already watched M consecutive Incel-related videos; and (b) an Incel-related video at hop $M+1$ assuming the user already watched $M$ consecutive Incel-related videos for both the Incel-derived and Control recommendation graphs. Note that in this scenario the random walker chooses to watch Incel-related videos.}}
\label{tab:incels_rabbit_hole_consecutive_hops_incel_related}
\end{table*}

\subsubsection{Does the frequency with which Incel-related videos are recommended increase for users who choose to see the content?}
So far, we have simulated the scenario where a user browses the recommendation graph randomly, i.e., they do {\em not} select Incel-related videos according to their interests or other cues nudging them to view certain content.
Next, we simulate the behavior of a user who chooses to watch a few Incel-related videos and investigate whether or not they will get recommended Incel-related videos with a higher probability within the next few hops.

Table~\ref{tab:incels_rabbit_hole_consecutive_hops_incel_related} reports how likely it is for a user to encounter Incel-related videos assuming he has already watched a few. 
To do so, we use the random walks performed on the Incel-derived and Control recommendation graphs in section~\ref{subsubsec:random_walks_recommendation_graph}. 
We consider only the random walks started from an Incel-related video, and we zero in on those where the user watches consecutive Incel-related videos.
Specifically, we report two metrics: 1)~the probability that a user encounters at least one Incel-related video in $5-M$ hops, having already seen M consecutive Incel-related videos; 
and 2)~the probability that the user will encounter an Incel-related video on the $M+1$ hop, assuming they have already seen $M$ consecutive Incel-related videos. 
\revision{
Note that, at each hop $M$ of a random walk, we calculate both metrics by only considering the random walks for which all the videos encountered in the first M hops of the walk were Incel-related.
}
These metrics allow us to understand whether the recommendation algorithm keeps recommending Incel-related videos to a user that starts watching a few of them.

At every hop M, there is a $\geq43.4\%$ and $\geq36.5\%$ chance to encounter at least one Incel-related video within $5-M$ hops in the Incel-derived and Control recommendation graphs, respectively. 
Furthermore, by looking at the probability of encountering an Incel-related video at hop $M+1$, having already watched $M$ Incel-related videos (third and right-most column in Table~\ref{tab:incels_rabbit_hole_consecutive_hops_incel_related}), we find an increasingly higher chance as the number of consecutive Incel-related increases. 
Specifically, for the Incel-derived recommendation graph, the probability rises from $4.1\%$ at the first hop to $30.1\%$ for the last hop. For the Control recommendation graph, it rises from $2.1\%$ to $17.7\%$.

\begin{figure}
\centering
\includegraphics[width=.5\linewidth]{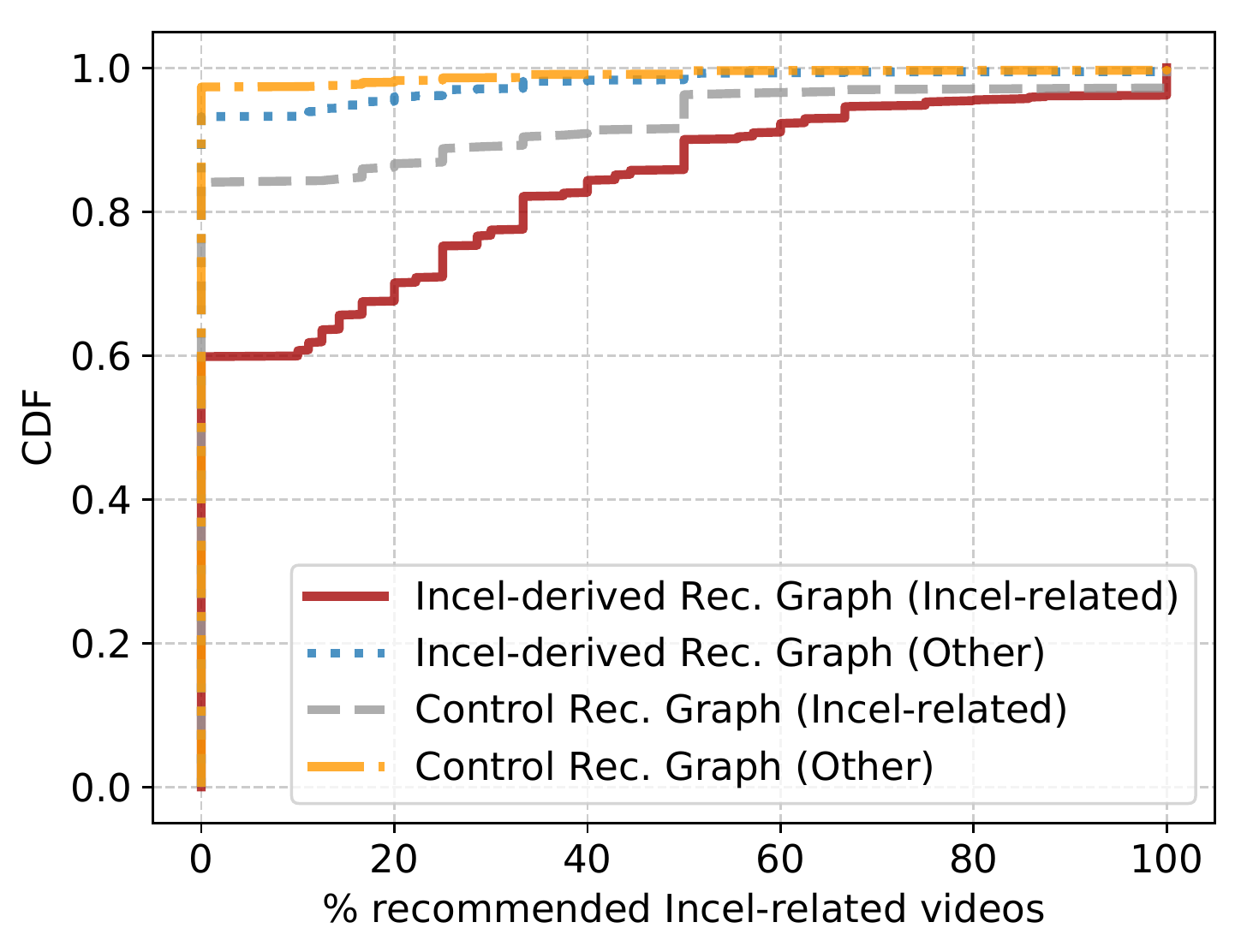}
\caption{CDF of the percentage of recommended Incel-related videos per video for both Incel-related and other videos in the Incels-derived and Control recommendation graphs.}
\label{fig:incels_graphs_perc_recommended_incel_related_per_video_cdf}
\end{figure}

These findings unveil that as users watch Incel-related videos, the algorithm recommends other Incel-related content with increasing frequency.
In Figure~\ref{fig:incels_graphs_perc_recommended_incel_related_per_video_cdf}, we plot the CDF of the percentage of Incel-related recommendations for each node in both recommendation graphs. 
In the Incel-derived recommendation graph, $4.6\%$ of the Incel-related videos have more than $80\%$ Incel-related recommendations, while $10\%$ of the Incel-related videos have more than $50\%$ Incel-related recommendations. 
The percentage of Other videos that have more than $50\%$ Incel-related recommendations is negligible. 
Although the percentage of Incel-related recommendations is lower, we see similar trends for the Control recommendation graph: $8.6\%$ of the Incel-related videos have more than $50\%$ Incel-related recommendations.

\revision{
Arguably, the effect we observe may be a contributor to the anecdotally reported echo chamber effect. 
This effect entails a viewer who begins to engage with this type of content and likely falls into an algorithmic rabbit hole, with recommendations becoming increasingly dominated by such harmful content and beliefs, which also becomes increasingly extreme~\cite{cinelli2021echo,mozillaregrets,rooseyoutuberadical,buzzeefrabbithole,ribeiro2020auditing}. 
However, the degree to which the above-inferred algorithm characteristics contribute to a possible echo chamber effect depends on: 1) personalization factors; and 2) the ability to measure whether recommendations become increasingly extreme.
}

\section{Challenges and Limitations}
\revision{In this section, we discuss the technical challenges we faced and how we addressed them and highlight the limitations of this line of work.}

\subsection{Challenges}
\revision{
Our data collection and annotation efforts faced several challenges. 
First, there was no available dataset of YouTube videos related to the Incel community or any other Manosphere groups. 
Guided by other studies using Reddit as a source for collecting and analyzing YouTube videos~\cite{papadamou2020disturbed}, and based on evidence suggesting that Incels are particularly active on Reddit~\cite{farrell2019exploring,ribeiro2020pick}, we build our dataset by collecting videos shared on Incel-related communities on Reddit.
Second, devising a methodology for the annotation of the collected videos is not trivial. 
Due to the nature of the problem, we hypothesize that using a classifier on the video footage will not capture the various aspects of Incel-related activity on YouTube. This is because the misogynistic views of Incels may force them to heavily comment on a seemingly benign video (e.g., a video featuring a group of women discussing gender issues)~\cite{doring2019male}. 
Hence, we devise a methodology to detect Incel-related videos based on a lexicon of Incel-related terms that considers both the video's transcript and its comments. 
}

\revision{
We believe that the scientific community can use our text-based approach to study other misogynistic ideologies on the platform, which tend to have their particular glossary; see Figure~\ref{fig:video_annotation_methodology}.
}

\subsection{Limitations}
\revision{
Unfortunately, this line of work is not without limitations.
Our video annotation methodology might flag some benign videos as Incel-related. 
This can be a false positive or due to Incels that heavily comment on (or even raid~\cite{mariconti2019you}) a benign video (e.g., a video featuring a group of women discussing gender issues). 
However, by considering the video's transcript in our video annotation methodology, we can achieve an acceptable detection accuracy that uncovers a substantial proportion of Incel-related videos (see Section~\ref{subsec:incels_video_annotation}). 
Despite this limitation, we believe that our video annotation methodology allows us to capture and analyze various aspects of Incel-related activity on the platform. 
Another limitation of this approach is that we may miss some Incel-related videos. 
Notwithstanding such limitation, our approach approaches the lower bound of the Incel-related videos available in our dataset, allowing us to conclude that the implications of YouTube's recommendation algorithm on disseminating misogynistic content are at least as profound as we observe. 
}

\revision{
Moreover, our work does not consider per-user personalization; the video recommendations we collect represent only some of the recommendation system's facets. 
More precisely, we analyze YouTube recommendations generated based on content relevance and the user base's engagement in aggregate.
However, we believe that the recommendation graphs we obtain do allow us to understand how YouTube's recommendation system is behaving {\em in our scenario.}
Also, note that a similar methodology for auditing YouTube's recommendation algorithm has been used in previous work~\cite{ribeiro2020auditing}.
}

\section{Remarks}
In this line of work, we presented a large-scale data-driven characterization of the Incel community on YouTube.
We collected 6.5K YouTube videos shared by users in Incel-related communities within Reddit. 
We used them to understand how Incel ideology spreads on YouTube and study the evolution of the community. 
We found a non-negligible growth in Incel-related activity on YouTube over the past few years, both in terms of Incel-related videos published and comments likely posted by Incels. 
This result suggests that users gravitating around the Incel community are increasingly using YouTube to disseminate their views.

Overall, our study is a first step towards understanding the Incel community and other misogynistic ideologies on YouTube. 
We argue that it is crucial to protect potential radicalization “victims” by developing methods and tools to detect Incel-related videos and other misogynistic activities on YouTube. 
Our analysis shows growth in Incel-related activities on Reddit and highlights how the Incel community operates on multiple platforms and Web communities. 
This also prompts the need to perform more multi-platform studies to understand Manosphere communities further.

We also analyzed how YouTube’s recommendation algorithm behaves with respect to Incel-related videos.
By performing random walks on YouTube’s recommendation graph, we estimated a $6.3\%$ chance for a user who starts by watching non-Incel-related videos to be recommended Incel-related ones within five recommendation hops.
At the same time, users who have seen two or three Incel-related videos at the start of their walk see recommendations that consist of $9.4\%$ and $11.4\%$ Incel-related videos, respectively. 
Moreover, the portion of Incel-related recommendations increases substantially as the user watches an increasing number of consecutive Incel-related videos.

Our results highlight the pressing need to further study and understand the role of YouTube's recommendation algorithm in users' radicalization and content consumption patterns. 
Ideally, a recommendation algorithm should avoid recommending potentially harmful or extreme videos. 
However, our analysis confirms prior work showing that this is not always the case on YouTube~\cite{ribeiro2020auditing}.

\chapter{Assessing the Effect of Watch History on YouTube’s Pseudoscientific Video Recommendations}
\label{chapter:pseudoscience_youtube}

\revision{
So far, we do not take user personalization into account in our analysis, hence it is hard to derive conclusions about YouTube at large.
Instead, we obtain and analyze recommendation graphs using the YouTube Data API, which allows us to analyze YouTube recommendations generated based on content relevance and general user engagement and satisfaction metrics on YouTube.
However, one of the main reasons behind YouTube's success is user personalization.
YouTube provides personalized recommendations to users based on their interests and various other engagement and satisfaction metrics like, for example, the watch history of the user.
Hence, to get a better insight into how the user’s watch history affects YouTube recommendations, this line of work has a special focus on how the user’s watch history affects YouTube’s video recommendations.
}

\section{Motivation}
Platforms like YouTube are often fertile ground for the spread of misleading and potentially harmful information like conspiracy theories and health-related disinformation~\cite{carneyoutubeconspricacies}.
YouTube and other social media platforms have struggled with mitigating the harm from this type of content. 
The difficulty is partly due to the sheer scale and also because of the deployment of recommendation algorithms~\cite{fastcompeny2019conspiracies}.
Purely automated moderation tools have thus far been insufficient to moderate content, and human moderators had to be brought back into the loop~\cite{vincent2020humanmoderators}.
Additionally, the machine learning algorithms that YouTube relies on to recommend content to users also recommend potentially harmful content~\cite{papadamou2020disturbed}, and their opaque nature makes them difficult to audit.

For certain types of content, e.g., health-related topics, harmful videos can have devastating effects on society, especially during crises like the COVID-19 pandemic~\cite{springyoutubefalse}.
For instance, conspiracy theories have suggested that COVID-19 is caused by 5G~\cite{covid5g2020conspiracy} or Bill Gates~\cite{covid2020billgates}, hindering social distancing, masking, and vaccination efforts~\cite{enserink2020fact}.
Conspiracy theories are usually built on tenuous connections between various events, with little to no actual evidence to support them; on user-generated content platforms like YouTube, these are often presented as facts, regardless of whether they are supported by facts and even though they have been widely debunked.
Motivated by the pressing need to mitigate the spread of pseudoscientific content, we focus on detecting and characterizing pseudoscientific and conspiratorial content on YouTube.
In particular, we aim to: 1) assess how likely it is for users with different watch histories to come across pseudoscientific content on YouTube, and 2) analyze how YouTube's recommendation algorithm contributes to promoting pseudoscience.
\revision{
To do this, guided by three of our main research questions \textbf{(RQ1, RQ2, and RQ4)}, we set out to answer the following three research sub-questions:
\begin{itemize}
    \item \textbf{RQ1.c:} Can we effectively detect and characterize pseudoscientific content on YouTube? (see RQ1 in Chapter~\ref{chapter:introduction})
    \item \textbf{RQ2.c:} Can we effectively quantify the influence of YouTube's recommendation algorithm in the dissemination of pseudoscientific content? What is the effect of a user's watch history on pseudoscientific video recommendations? (see RQ2 in Chapter~\ref{chapter:introduction})
    \item \textbf{RQ4.c:} What is the proportion of pseudoscientific content on the homepage of a YouTube user, in search results, and the video recommendations section of YouTube? (see RQ4 in Chapter~\ref{chapter:introduction})
\end{itemize}
}

\descr{Methodology.}
In this line of work, we focus on four pseudoscientific topics: 1)~COVID-19, 2)~Flat Earth theory, 3)~anti-vaccination, and 4)~anti-mask movement.
We collect 6.6K unique videos and use crowdsourcing to label them in three categories: science, pseudoscience, or irrelevant.
\revision{We then assign labels to each video based on the majority agreement of the annotators. 
We excluded videos where all the annotators disagreed resulting in a final dataset of 5.7K videos.}
We then train a deep learning classifier to detect pseudoscientific content across multiple topics on YouTube.

Next, we use three carefully crafted user profiles, each with a different watch history, while all other account information remains the same, to simulate logged-in users.
We also perform a set of experiments using a browser without a Google account to simulate non-logged-in users and another set using the YouTube Data API exclusively.
To populate the watch history of the three user profiles, we devise a methodology to identify the minimum amount of videos that must be watched by a user before YouTube's recommendation algorithm starts generating substantially personalized recommendations.
We build three distinct profiles:
1) a user interested in scientific content;
2) a user interested in pseudoscientific content; and
3) a user interested in both scientific and pseudoscientific content.
Using these profiles, we perform three experiments to quantify the user's exposure to pseudoscientific content on various parts of the platform and how this exposure changes based on a user's watch history. Note that we manually review all the videos classified as pseudoscientific in all experiments.

\descr{Findings.}
Overall, our study leads to the following findings:
\begin{enumerate}
    \item We can detect pseudoscientific content, as our deep learning classifier yields $0.79$ accuracy and outperforms SVM, Random Forest, and BERT-based classifiers.

    \item \revision{We find that the minimum amount of videos a user needs to watch before YouTube learns her interests and starts generating more personalized science and pseudoscience-related recommendations is 22.}

    \item The watch history of the user substantially affects search results and related video recommendations. At the same time, pseudoscientific videos are more likely to appear in search results than in the video recommendations section or the user's homepage.

    \item In traditional pseudoscience topics (e.g., Flat Earth), there is a higher rate of recommended pseudoscientific content than in more recent issues like COVID-19, anti-vaccination, and anti-mask. For COVID-19, we find an even smaller amount of pseudoscientific content being suggested. This indicates that YouTube took partly effective measures to mitigate pseudoscientific misinformation related to the COVID-19 pandemic.

    \item The YouTube Data API results are similar to those of the non-logged-in profile with no watch history (using a browser); this indicates that recommendations returned using the API are not subject to personalization.
\end{enumerate}

\descr{Contributions.}
\revision{To the best of our knowledge, we present the first study focusing on multiple health-related pseudoscientific topics on YouTube pertaining to the COVID-19 pandemic.
Inspired by the literature, we develop a complete framework that allows us to assess the prevalence of pseudoscientific content on various parts of the YouTube platform (i.e., homepage, search results, video recommendations) while accounting for the effect of a user's watch history.}
Our methodology can be re-used for other studies focusing on other topics of interest.
We are confident that this will help the research community shed additional light on YouTube's recommendation algorithm and its potential influence.

\section{Datasets}
In this section, we present our data collection and crowdsourced annotation methodology.
We collect a set of YouTube videos related to science and then use crowdsourcing to annotate videos as pseudoscientific or not.

\subsection{Data Collection}
\label{subsec:pseudoscience_data_collection}

Since we aim to detect pseudoscientific video content automatically, we collect a set of YouTube videos related to four, arguably relevant, topics:
1)~COVID-19~\cite{buzzfeedcoronavirus_2020};
2)~the anti-vaccination movement~\cite{antiVaxx2020}; 
3)~the anti-mask movement~\cite{antiMask2020}; and 
4)~the Flat Earth theory~\cite{guardianflatearth_2019}.
We focus on COVID-19 and the anti-mask movement because both are timely topics of great societal interest.
We also choose anti-vaccination because it is both an increasingly popular and traditional pseudoscientific topic. 
Last, we include the Flat Earth theory because it is a ``long-standing'' pseudoscientific subject.

\revision{
Then, for each topic of interest, we define search queries and use them to search YouTube and collect videos.
For COVID-19 we search using the terms ``COVID-19'' and ``coronavirus,'' and for the anti-vaccination movement we use the terms ``anti-vaccination'' and ``anti-vaxx''. 
On the other hand, for the anti-mask movement and the Flat Earth theory we only use the terms ``anti-mask'' and ``flat earth,'' respectively, since there are no other terms that point to the same definition as is the case for the other two pseudoscientific topics.
}

Next, we search YouTube using the YouTube Data API~\cite{youtubedataapi} and the search queries defined for each topic.
For each search query of each selected topic we obtain the first 200 videos as returned by YouTube's Data API search functionality. 
We refer to those videos as the ``seed'' videos of our data collection methodology.
Additionally, for each seed video, we collect the top 10 recommended videos associated with it, as returned by the YouTube Data API.
We perform our data collection on August 1-20, 2020, collecting 6.6K unique videos (1.1K seed videos and 5.5K videos recommended from the seed videos).
Table~\ref{tab:pseudoscience_dataset_overview} summarizes our dataset.
For each video in our dataset, we collect:
1) the transcript of the video;
2) the video title and description;
3) a set of tags defined by the uploader;
4) video statistics such as the number of views, likes, etc.; and
5) the 200 top comments, defined by YouTube's relevance metric, without their replies.

\begin{table}[t!]
\footnotesize
\centering
\begin{tabular}{lrrrr}
\toprule
\textbf{Pseudoscientific Topic} & \textbf{\#Seed} & \textbf{\#Recommended}  \\
\midrule
COVID-19 & 378 & 1,645 \\
Anti-vaccination & 346 & 1,759 \\
Anti-mask & 199 & 912 \\
Flat Earth & 200 & 1,211 \\
\midrule
\textbf{Total} & \textbf{1,123} & \textbf{5,527} \\
\bottomrule
\end{tabular}%
\caption{Overview of the collected data: number of seed videos and number of their recommended videos.}
\label{tab:pseudoscience_dataset_overview}
\end{table}

\subsection{Crowdsourcing Data Annotation}
\label{subsec:pseudoscience_data_annotation}
To create a ground-truth dataset of scientific and pseudoscientific videos, we use the Appen platform~\cite{appen_2020} to get crowdsourced annotations for all the collected videos. 
We present each video to three annotators who inspect its content and metadata to assign one of three labels:

\begin{enumerate}
    \item \textbf{Science.} The content is related to any scientific field that systematically studies the natural world's structure and the behavior of humanity's artifacts (e.g., Chemistry, Biology, Mathematics, Computer Science, etc.). Videos that debunk science-related conspiracy theories (e.g., explaining why 5G technology is not harmful) also fall in this category. For example, a COVID-19 video with an expert estimating the total number of cases or excess deaths falls in this category if the estimation rests on the scientific consensus and official data.

    \item \textbf{Pseudoscience.} The video meets at least one of the following criteria: a) holds a view of the world that goes against the scientific consensus (e.g., anti-vaccine movement); b) comprises statements or beliefs that are self-fulfilling or unfalsifiable (e.g., Meditation)~\cite{mediation2017}); c) develops hypotheses that are not evaluated following the scientific method (e.g., Astrology); or d) explains events as secret plots by powerful forces rather than overt activities or accidents (e.g., the 5G-coronavirus conspiracy theory).

    \item \textbf{Irrelevant.} The content is not relevant to any scientific field and does not fall in the Pseudoscience category. For example, music videos and cartoon videos are considered irrelevant. Conspiracy theory debunking videos that are not relevant to a scientific field are deemed irrelevant (e.g., a video debunking the Pizzagate conspiracy theory).
\end{enumerate}

\descr{Annotation.}
The annotation process is carried out by 992 annotators, both male and female, recruited through the Appen platform. 
We give annotators instructions on what constitutes scientific and pseudoscientific content using appropriate descriptions and several examples. They are offered $\$0.03$ per annotation. 
Three annotators label each video. To ease the annotation process, we provide a clear description of the task and our labels, and all video information that an annotator needs to inspect and correctly annotate a video.
Screenshots of the instructions are available from~\cite{pseudoscienceresources}.

Appen provides no demographic information about the annotators, other than an assurance that they are experienced and attained high accuracy in other tasks.
To assess the annotators' quality, before allowing them to submit annotations, we ask them to annotate 5 test videos randomly selected from a set of 54 test videos (20 science, 21 pseudoscience, and 13 irrelevant) annotated by the first author of this paper.
An annotator can submit annotations only when she labels at least 3 out of the 5 test videos correctly.
\revision{This initial test guarantees that our annotators are more likely to have a scientific rather than conspiratorial pseudoscientific outlook, which would probably pollute our results.}

\revision{
Furthermore, using the collected annotations, we calculate the Fleiss' Kappa Score ($k$)~\cite{fleiss1971measuring} to assess the annotators' agreement.
We get $k=0.14$, which is considered ``slight'' agreement.
To mitigate the effect of the low agreement score on our results, we first exclude from our dataset all the 915 videos ($13.8\%$) where all annotators disagreed with each other and we calculate again the agreement score.
We get $k=0.24$, which is considered ``fair'' agreement.
Next, we assign labels to each video in our ground-truth dataset based to the majority agreement of all the annotators resulting in a ground-truth dataset that includes 1,197 science, 1,325 pseudoscience, and 3,212 irrelevant videos (see Table~\ref{tab:pseudoscience_groundtruth_dataset_overview}).}

\revision{
Last, to further mitigate the effects of the low agreement score of our crowdsourced annotation, we collapse our three labels into two, combining the science with the irrelevant videos into an ``Other'' category.
This yields a final ground-truth dataset with 1,325 pseudoscience and 4,409 other videos (see Table~\ref{tab:final_groundtruth_dataset_overview}).
}

\begin{table}[t!]
\footnotesize
\centering
\begin{tabular}{lrrrr}
\toprule
\textbf{Topic} & \textbf{\#Science} & \textbf{\#Pseudoscience} & \textbf{\#Irrelevant}  \\
\midrule
COVID-19 & 607 & 368 & 721 \\
Anti-vaccination & 363 & 394 & 1,060 \\
Anti-mask & 65 & 188 & 724 \\
Flat Earth & 162 & 375 & 707 \\
\midrule
\textbf{Total} & \textbf{1,197} & \textbf{1,325} & \textbf{3,212} \\
\bottomrule
\end{tabular}%
\caption{Overview of our ground-truth dataset.}
\label{tab:pseudoscience_groundtruth_dataset_overview}
\end{table}

\begin{table}[t!]
\centering
\small
 \setlength{\tabcolsep}{3pt}
\begin{tabular}{lrrr}
\toprule
\textbf{Topic} & \textbf{\#Pseudoscience} & \textbf{\#Other}  \\
\midrule
COVID-19 & 368 & 1,328 \\
Anti-vaccination & 394 & 1,423 \\
Anti-mask & 188 & 789 \\
Flat Earth & 375 & 869 \\
\midrule
\textbf{Total} & \textbf{1,325} & \textbf{4,409} \\
\bottomrule
\end{tabular}%
\caption{Overview of our final ground-truth dataset.}
\label{tab:final_groundtruth_dataset_overview}
\end{table}

\descr{Performance Evaluation.}
\revision{
To evaluate our crowdsourced annotation performance, we randomly select 600 videos from our ground-truth dataset and manually annotate them.
Using the first author's annotations as ground-truth, we calculate the precision, recall, and F1 score of our crowdsourced annotation, yielding respectively $0.92$, $0.91$, and $0.92$.
We argue that this represents an acceptable performance given the subjective nature of scientific and pseudoscientific content.}

We only collect publicly available data, we do not attempt to de-anonymize users and overall follow standard ethical guidelines~\cite{dittrich2012menlo,rivers2014ethical}.
We also note that we obtained institutional ethics approval from the first author's national ethics committee to ensure that our crowdsourced annotation process does not pose risks to the annotators. 

\subsection{Ethics}
\revision{We only collect publicly available data and we make no attempt to de-anonymize users. 
Overall, we also follow standard ethical guidelines~\cite{dittrich2012menlo,rivers2014ethical} regarding information research and the use of shared measurement data.
More precisely, we ensure compliance with GDPR's~\cite{gdpr2018eu} ``Right to be Forgotten'' and ``Right of Access'' principles.
}

\revision{
Furthermore, we note that we obtained ethics approval from the first author's national ethics committee to ensure that our crowdsourced annotation process does not pose risks to the annotators. 
Nevertheless, we consider the detrimental effects of the controversial content we study.
For this reason we inform our annotators and enable them to stop the classification task and opt-out our annotation process at any time.
Finally, we acknowledge that the price offered per annotation is quite low and this is mainly because of the large number of videos we needed to annotate, and the number of annotations (3) required per video.
However, this price allows us to acquire all the required annotations within the budget allocated for this research.
}

\section{Detection of Pseudoscientific Videos}
\label{sec:pseudoscience_videos_detection}
In this section, we present our classifier geared to detect pseudoscientific videos.
To train and test it, we use our ground-truth dataset of 5,734 videos. 
Below we describe the input features and the architecture of our proposed classifier. 
We perform an experimental evaluation to assess the classifier's performance and an ablation study to understand which of the input features contribute the most to the classification task.

\subsection{Classifier Architecture}
\begin{figure}[t!]
\centering
\includegraphics[width=1.0\linewidth]{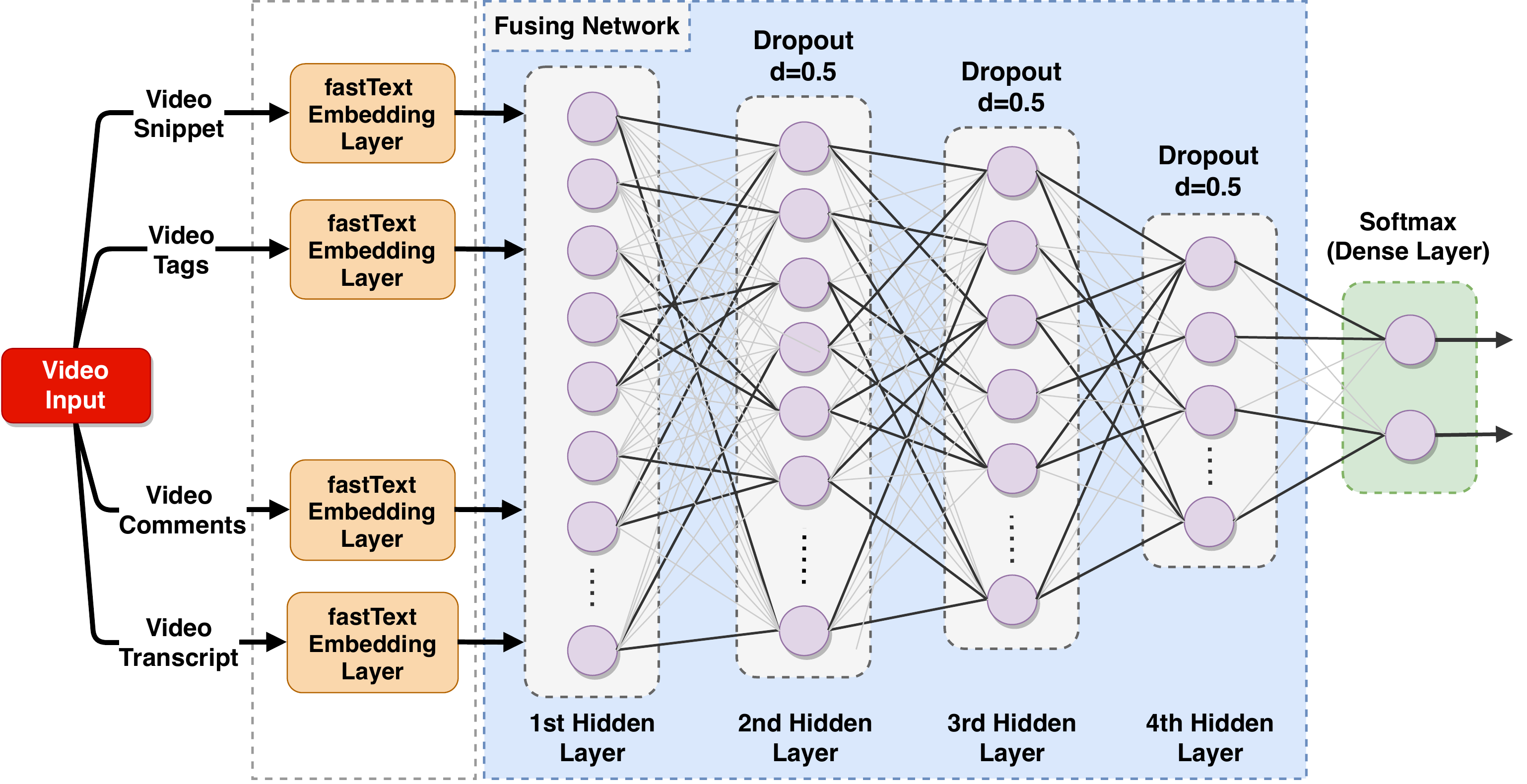}
\caption{Architecture of our deep learning classifier for the detection of pseudoscientific videos.}
\label{fig:pseudoscience_model_architecture}
\end{figure}

\revision{Figure~\ref{fig:pseudoscience_model_architecture} depicts the architecture of our classifier for detecting videos with pseudoscientific content.}
The classifier consists of four different branches, each processing a distinct input feature type: snippet, video tags, transcript, and the top 200 comments of a video. 
Then, all four branches' outputs are concatenated to form a five-layer, fully-connected neural network that merges their output and drives the final classification.
\revision{
This classifiers differs from the classifier developed in Section~\ref{sec:disturbed_detectionofdisturbingvideos} for the detection of disturbing videos targeting toddlers. For the detection of pseudoscientific content, we choose to build a classifier that analyzes the textual metadata (e.g., title, transcript, etc.) of a video and the discussions associated with it (i.e., comments) because we believe that they can provide a more meaningful signal about the pseudoscientific stance of a video than other types of input (e.g., thumbnail, video frames, etc.).
Also, the transcript of the video allows us to also consider the main themes discussed in the actual video by the creator/uploader of the video.
}

The classifier uses fastText~\cite{fasttext_2020}, a library %
for efficient learning of word/document-level vector representations and sentence classification, to generate vector representations (embeddings) for all the available video metadata in text.
\revision{
For each type of input feature, we use the pre-trained fastText embeddings released in~\cite{mikolov2018advances} and fine-tune them for our text classification task using each of our input features.
These fine-tuned models extract a 300-dimensional vector representation for each of the following input features of our dataset:
}
\begin{itemize}
    \item {\bf Snippet.} Concatenation of the title and the description of the video. 

    \item {\bf Tags.} Words defined by the uploader of a video to describe the content of the video.

    \item  {\bf Transcript.} Naturally, this is one of the most important features, as it describes the video's actual content. (It includes the subtitles uploaded by the creator of the video or auto-generated by YouTube.) The classifier uses the fine-tuned model to learn a vector representation of the concatenated text of the transcript.

    \item {\bf Comments.} We consider the top 200 comments of the video as returned by the YouTube Data API. We first concatenate each video's comments and use them to fine-tune the fastText model and extract vector representations.
\end{itemize}

The second part of the classifier (the ``Fusing Network'' in Figure~\ref{fig:pseudoscience_model_architecture}) is essentially a four-layer, fully-connected, dense neural network. 
We use a Flatten utility layer to merge the outputs of the four branches of the first part of the classifier, creating a 1200-dimensional vector.
This vector is processed by the four subsequent layers comprising 256, 128, 64, and 32 units, respectively, with ReLU activation.
To avoid overfitting, we regularize using the Dropout technique~\cite{srivastava2014dropout};
at each fully connected layer, we apply a Dropout level of $d=0.5$, i.e., during each iteration of training, half of each layer's units do not update their parameters. 
Finally, the Fusing Network output is fed to the last neural network of two units with softmax activation, which yields the probabilities that a particular video is pseudoscientific or not.
We implement our classifier using Keras~\cite{keras2015application} with Tensorflow as the back-end~\cite{abadi2016tensorflow}.

\begin{table}[t!]
\footnotesize
\centering
\begin{tabular}{lrrrr}
\toprule
\textbf{Classifier} & \textbf{Accuracy} & \textbf{Precision} & \textbf{Recall} & \textbf{F1 Score}  \\
\midrule
SVM & 0.68 & 0.72 & 0.68 & 0.70 \\
Random Forest & 0.72 & 0.70 & 0.72 & 0.71 \\
BERT-based Classifier & 0.73 & 0.64 & 0.73 & 0.67 \\
\midrule
\textbf{Proposed Classifier} & \textbf{0.76} & \textbf{0.74} & \textbf{0.76} & \textbf{0.74} \\
\midrule
\textbf{Proposed Classifier (threshold-moving)} & \textbf{0.79} & \textbf{0.77} & \textbf{0.79} & \textbf{0.74} \\
\bottomrule
\end{tabular}%
\caption{Performance of the evaluated baselines and of the proposed deep learning classifier.}
\label{tab:pseudoscience_performance_metrics}
\end{table}

\subsection{Experimental Evaluation}
We use ten-fold stratified cross-validation~\cite{arlot2010survey}, training and testing the classifier for binary classification using all the aforementioned input features.
To deal with data imbalance, we use the Synthetic Minority Oversampling Technique~\cite{chawla2002smote} and over-sample only the training set at each fold.
\revision{To select and tune the hyper-parameters of our model we use the random search strategy~\cite{bergstra2012random}.}
For stochastic optimization, we use Adam with an initial learning rate of $1\mathrm{e}{-3}$, and $\epsilon=1\mathrm{e}{-8}$.

We then compare the performance of the classifier, in terms of accuracy, precision, recall, and F1 score, using three baselines:
1) a Support Vector Machine (SVM) classifier with parameters $\gamma=0.1$ and $C=10$, 
2) a Random Forest classifier with an entropy criterion and number of minimum samples leaf equal to 2, and
3) a neural network with the same architecture as our classifier that uses a pre-trained BERT model~\cite{turc2019} to learn document-level representations from all the available input features (BERT-based).
For hyper-parameter tuning of baselines (1) and (2), we use the grid search strategy, while for (3), we use the same hyper-parameters as the proposed classifier.
Note that all evaluated models use all available input features.

Table~\ref{tab:pseudoscience_performance_metrics} reports the performance of all classifiers.
We observe that our classifier outperforms all baseline models across all performance metrics.
To further reduce false positives and improve the performance of our classifier, we apply a threshold-moving approach, which tunes the threshold used to map probabilities to class labels~\cite{provost2000machine}. 
We use the grid-search technique to find the optimal lower bound probability above which we consider a video pseudoscientific, and find it to be $0.7$.
Using this threshold, we train and re-evaluate the proposed classifier, which yields, respectively, $0.79$, $0.77$, $0.79$, and $0.74$ on the accuracy, precision, recall, and F1 score (see the last row in Table~\ref{tab:pseudoscience_performance_metrics}).

\descr{Ablation Study.}
To understand which of the input features contribute the most to the classification of pseudoscientific videos, we perform an ablation study. We systematically remove each of the four input feature types (and their branch in the classifier) and retrain the classifier.
Again, we use ten-fold cross-validation and oversampling to deal with data imbalance and use the classification threshold of $0.7$.
Table~\ref{tab:pseudoscience_ablation_study_details} reports the performance metrics for each combination of inputs.
Video tags and transcripts yield the best performance, indicating that they are the most informative input features.
However, using all the available input features yields better performance, which indicates that all four input features are ultimately crucial for the classification task.

\begin{table}[t!]
\footnotesize
\centering
\begin{tabular}{lrrrr}
\toprule
\textbf{Input Features} & \textbf{Accuracy} & \textbf{Precision} & \textbf{Recall} & \textbf{F1 Score}  \\
\midrule
Snippet 						& 0.78 & 0.76 & 0.78 & 0.71 \\
Tags 							& 0.78 & 0.77 & 0.78 & 0.72 \\
Transcript 						& 0.78 & 0.74 & 0.78 & 0.71 \\
Comments 						& 0.78 & 0.71 & 0.77 & 0.68 \\
\midrule
Snippet, Tags 					& 0.78 & 0.76 & 0.78 & 0.72 \\
Snippet, Transcript 			& 0.78 & 0.75 & 0.78 & 0.72 \\ 
Snippet, Comments 				& 0.78 & 0.77 & 0.78 & 0.71 \\
Tags, Transcript 				& 0.79 & 0.77 & 0.79 & 0.73 \\ 
Tags, Comments 					& 0.78 & 0.76 & 0.78 & 0.72 \\
Transcript, Comments 			& 0.78 & 0.76 & 0.78 & 0.73 \\ 
\midrule
Snippet, Tags, Transcript 		& 0.78 & 0.75 & 0.78 & 0.72 \\
Snippet, Tags, Comments 		& 0.78 & 0.75 & 0.78 & 0.72 \\ 
Snippet, Transcript, Comments\hspace*{-1cm} 	& 0.78 & 0.76 & 0.78 & 0.73 \\
Tags, Transcript, Comments 		& 0.78 & 0.76 & 0.78 & 0.73 \\ 
\midrule
\textbf{All Features} & \textbf{0.79} & \textbf{0.77} & \textbf{0.79} & \textbf{0.74} \\ 
\bottomrule
\end{tabular}%
\caption{Performance of the proposed classifier (considering the $0.7$ classification threshold) trained with all the possible combinations of the four input feature types.}
\label{tab:pseudoscience_ablation_study_details}
\end{table}

\descr{Remarks.} 
Although our classifier outperforms all the baselines, ultimately, its performance ($0.74$ F1-score) reflects the subjective nature of pseudoscientific vs. scientific content classification on YouTube. 
This relates to our crowdsourced annotation's relatively low agreement score, which highlights the difficulty in identifying whether a video is pseudoscientific. It is also evidence of the hurdles in devising models that automatically discover pseudoscientific content.
Nonetheless, we argue that our classifier is only the first step in this direction and can be further improved; overall, it does provide a meaningful signal on whether a video is pseudoscientific. 
\revision{
It can also be used to derive a lower bound of YouTube’s recommendation algorithm's tendency to recommend pseudoscience; we uncover a substantial portion of pseudoscientific videos while also eliminating all false positives with manual review of all the videos classified as pseudoscientific (see Section~\ref{subsec:pseudoscience_experiments_design}).
}

\begin{figure*}[t!]
\centering
\subfigure[Homepage]{\includegraphics[width=0.428\linewidth]{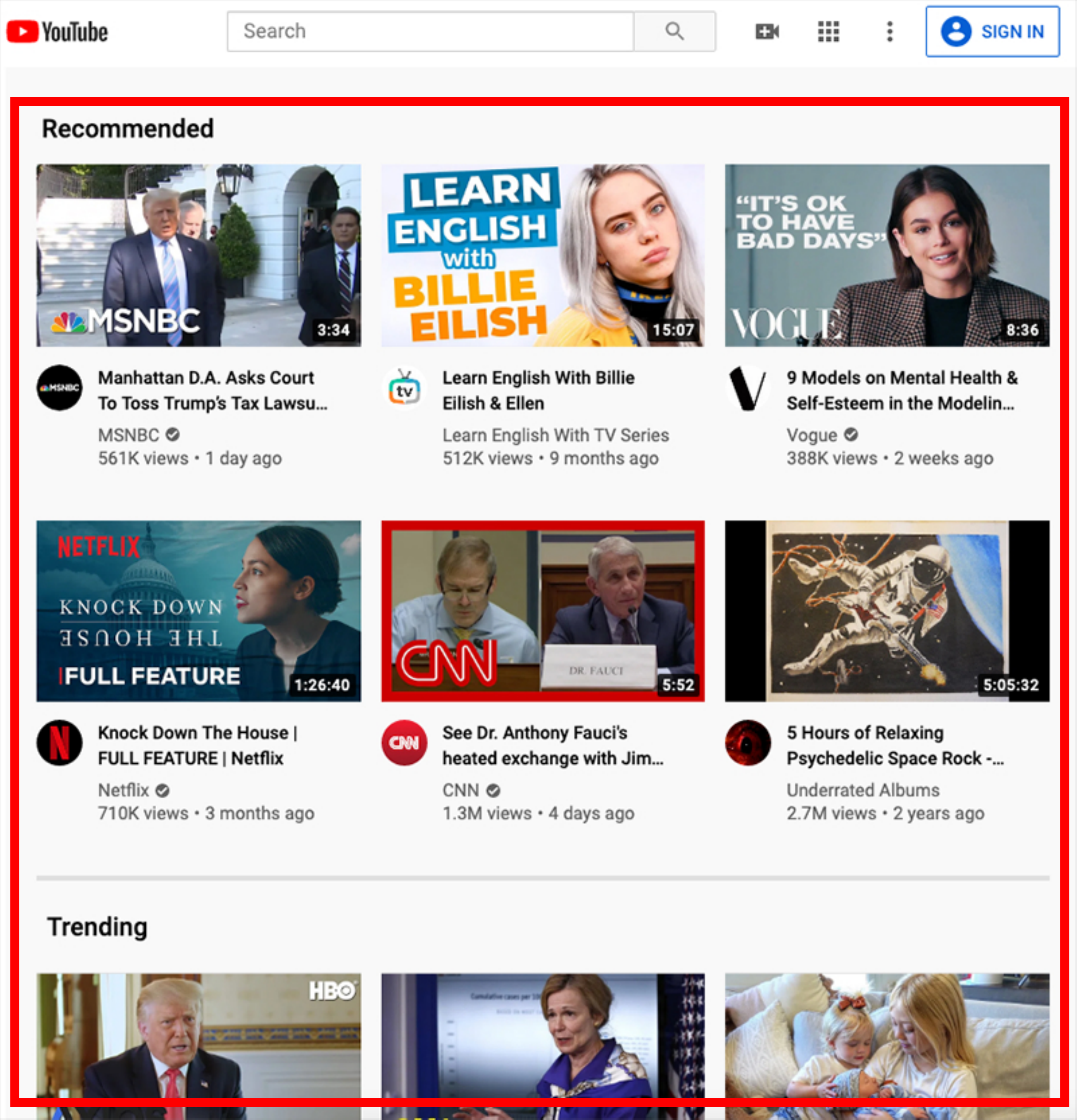}
\label{fig:pseudoscience_example_youtube_homepage}}
\subfigure[Search Results]{\includegraphics[width=0.49\linewidth]{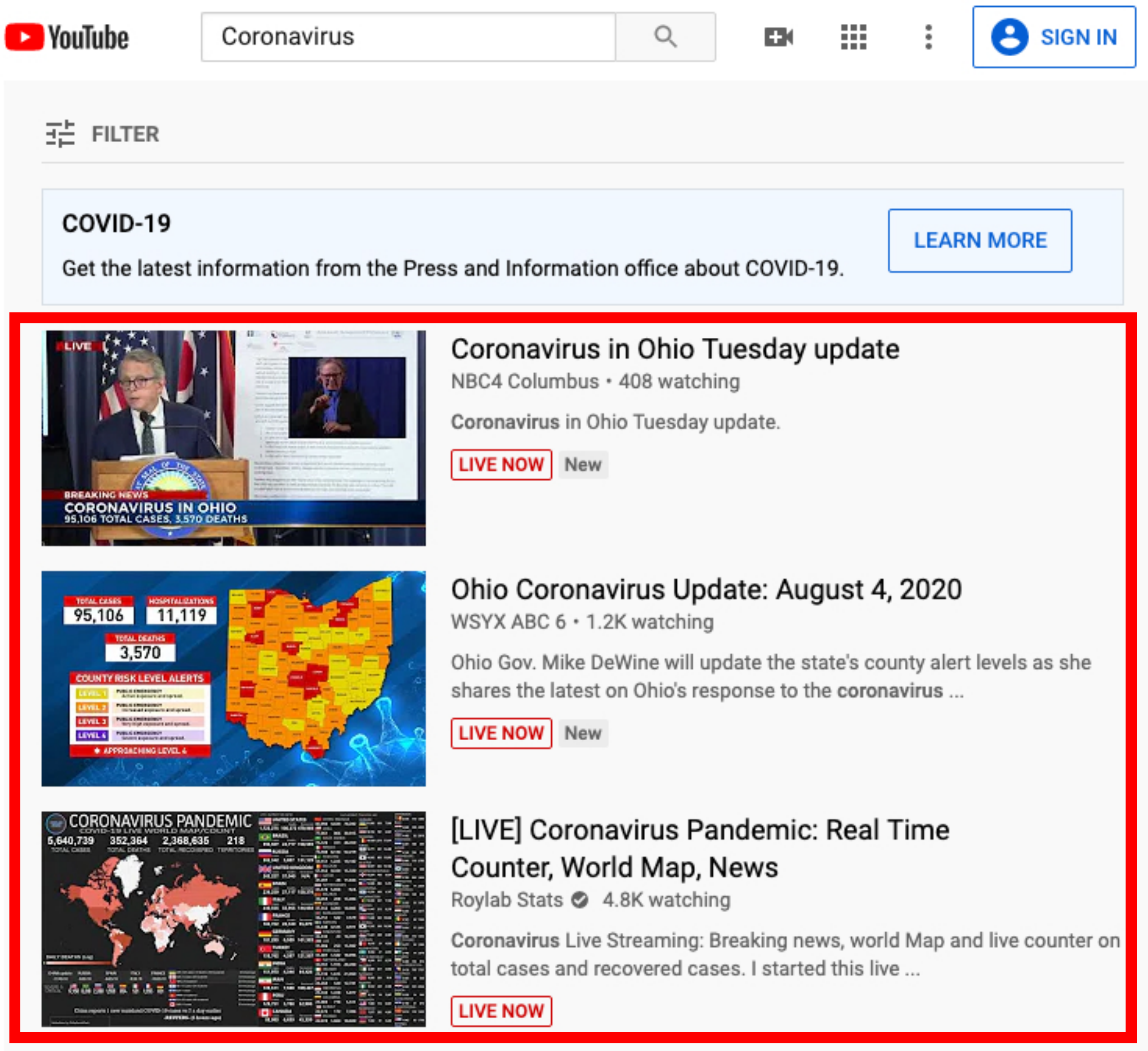}
\label{fig:pseudoscience_example_youtube_search}}
\subfigure[Video Recommendations]{\includegraphics[width=0.6\linewidth]{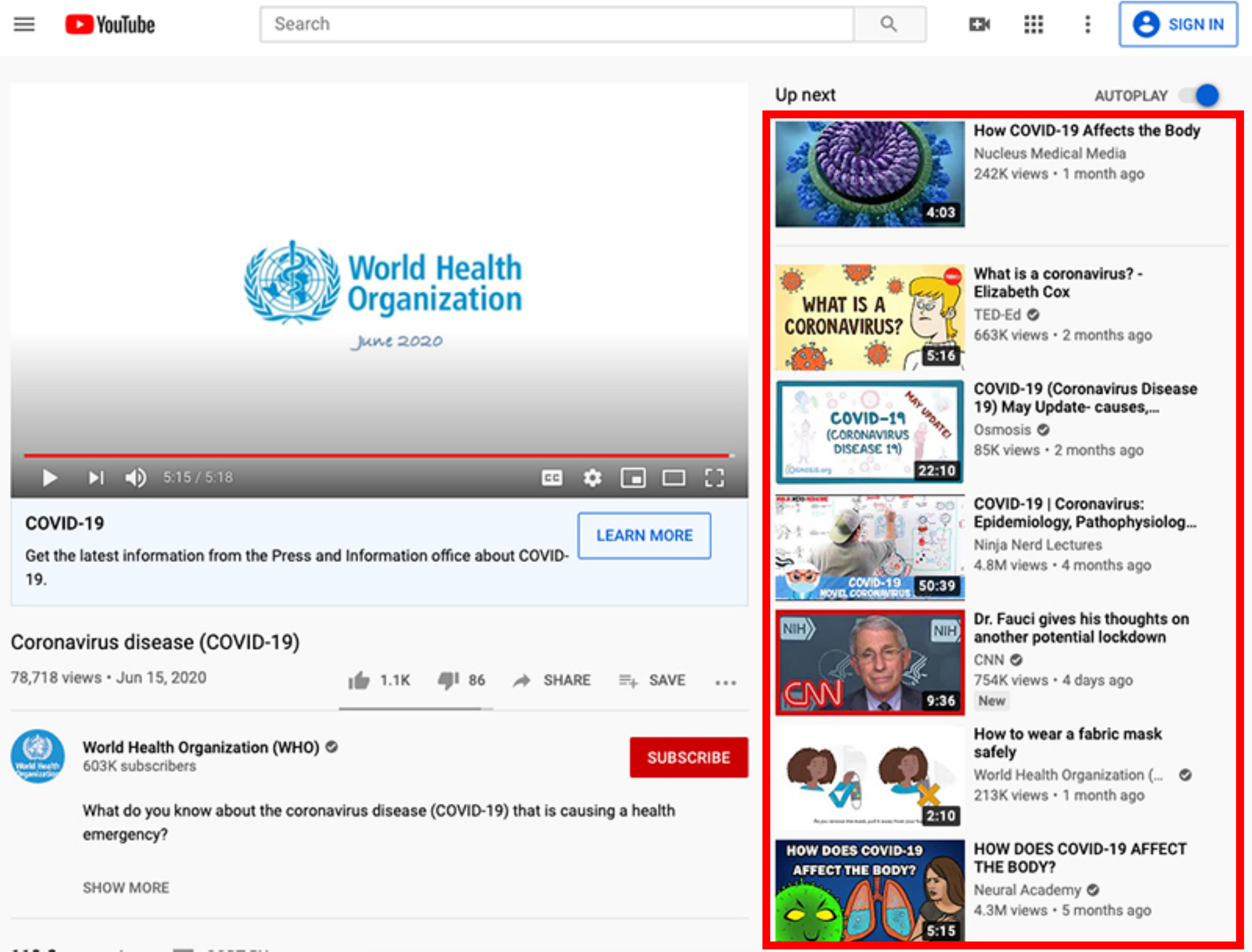}
\label{fig:pseudoscience_example_video_recommendations}}
\caption{The three main parts of the YouTube platform that we consider in our experiments: (a) homepage; (b) search results; and (c) video recommendations section.}
\label{fig:pseudoscience_youtube_components_examples}
\end{figure*}

\section{Analysis}
In this section, we analyze the prominence of pseudoscientific videos on various parts of the platform.

\subsection{Experimental Design}
\label{subsec:pseudoscience_experiments_design}
We focus on three parts of the platform: 1) the homepage; 2) the search results page; and 3) the video recommendations section (recommendations when watching videos). 
Figure~\ref{fig:pseudoscience_youtube_components_examples} shows an example of each part.
We aim to simulate the logged-in and non-logged-in users' behavior with varying interests and measure how the watch history affects pseudoscientific content recommendation.
To do so, we create three different Google accounts, each one with a different watch history, while all the other account information is the same to avoid confounding effects caused by profile differences.
Additionally, we perform experiments on a browser without a Google account to simulate not logged-in users. Moreover, we perform experiments using the YouTube Data API (when the API provides the required functionality) to investigate the differences between YouTube as an application and the API.

\begin{algorithm}
\centering
\small
 \setlength{\tabcolsep}{1pt}
\caption{Minimum number of videos needed to build the watch history of a user profile.}
\label{alg:findminimunwvideosalgorithm}
\begin{algorithmic}[1]
	\State Let $\mathbf{S}$ be a set of 100 randomly selected COVID\-19 pseudoscientific videos
	\State Let $\mathbf{V}_{ref}$ be a randomly selected COVID\-19 pseudoscientific video
	\State Let $\mathbf{V}_{refRec}$ be the top $10$ recommendations of $\mathbf{V}_{ref}$
	\State $\mathbf{RH}_{recs} \gets \{\mathbf{V}_{refRec}$\}
	\State $\mathbf{S}_{threshold} \gets 1.0$
	\State $\mathbf{W} \gets 0$ \Comment{Number of videos watched}
	\For{\textbf{each} video $\mathbf{V}$ in $\mathbf{S}$}
		\State Watch video $\mathbf{V}$
		\State $W\gets W + 1$
		\State Get the top 10 recommendations $\mathbf{R}$ of $\mathbf{V}_{ref}$
		\State Calculate the Overlap Coefficient $\mathbf{O}_{coef}$ between \\
		\hskip\algorithmicindent $\mathbf{R}$ and $\mathbf{RH}_{recs}$
		\If {$\mathbf{O}_{coef} \geq \mathbf{S}_{threshold}$}
			\State \textbf{return} $\mathbf{W}$
		\Else
			\State Add $\mathbf{R}$ to the set of recommendations $\mathbf{RH}_{recs}$ \\
			\hskip\algorithmicindent\hskip\algorithmicindent retrieved in the previous iterations
		\EndIf
	\EndFor
\end{algorithmic} %
\end{algorithm}

\descr{User Profile Creation.}
According to~\cite{hussein2020measuring}, once a user forms a watch history, user profile attributes (i.e., demographics) affect future video recommendations.
Hence, since we are only interested in the watch history, each of the three accounts has the same profile: 30 years old and female. 
To decrease the likelihood of Google automatically detecting our user profiles, we carefully crafted each one assigning them a unique name and surname and performed standard phone verification.
None of the created profiles were banned or flagged by Google during or after our experiments.

\descr{Watch History.} 
Next, we build the watch history of each profile, aiming to create the following three profiles: 
1)~a user interested in legitimate science videos (``Science Profile'');
2)~a user interested in pseudoscientific content (``Pseudoscience Profile''); and
3)~a user interested in both science and pseudoscience videos (``Science/Pseudoscience Profile'').

\begin{figure}[t!]
\centering
\includegraphics[width=0.85\columnwidth]{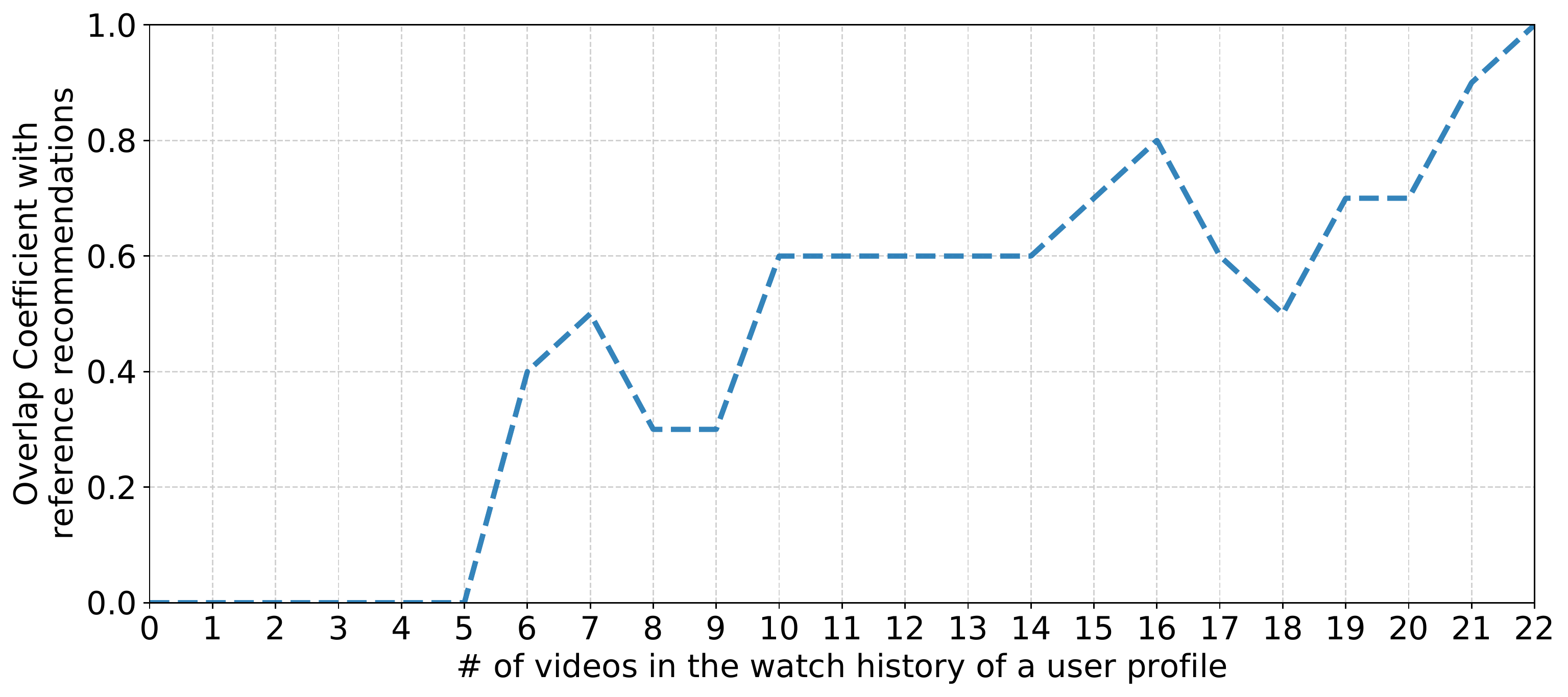}
\caption{Overlap Coefficient with the reference recommendations of previous iterations of the devised algorithm as a function of the number of videos in the user's watch history.}
\label{fig:similarity_build_watch_history}
\end{figure}

\revision{
To find the minimum number of videos a profile needs to watch before YouTube learns the user's interests and starts generating more personalized recommendations, we use a newly created Google account with no watch history, and we devise and execute the following algorithm (see Algorithm~\ref{alg:findminimunwvideosalgorithm}).
First, we randomly select a video, which we refer to as the ``reference'' one, from the COVID-19 pseudoscientific videos of our ground-truth dataset, and we collect its top 10 recommended videos.
Next, we create a list of 100 randomly selected COVID-19 pseudoscientific videos, excluding videos exceeding five minutes in duration, and we repeat the following process iteratively:}
\begin{enumerate}
    \item We start by watching a video from the list of the randomly selected pseudoscientific videos;

    \item \revision{We visit the reference video, and we collect the top 10 recommendations, store them, and compare them using the Overlap Coefficient with all the recommendations of the reference video collected in the previous iterations;}

    \item \revision{If all the recommended videos of the reference video at the current iteration have also been recommended in the previous iterations (Overlap Coefficient = 1.0), we stop our experiment. Otherwise, we increase the number of videos watched and proceed to the next iteration.}
\end{enumerate}
\revision{\noindent Using this algorithm, we find that the minimum amount of videos required to be watched by a user for YouTube to start generating more personalized recommendations is 22.
Figure~\ref{fig:similarity_build_watch_history} depicts the overlap coefficient between the recommendations at each iteration and the reference recommendations of previous iterations as a function of the number of videos in the user's watch history.}
However, to create more representative watch histories and get even more personalized recommendations, we increase this number to 100. %
Finally, we select the most popular science and pseudoscience videos from the ground-truth dataset, based on the number of views, likes, comments, etc., and use them to personalize the three Google accounts' profiles.
Since it is not clear how YouTube measures the satisfaction score on videos and how watch time affects this score, during profile training, we always watch the same proportion of the video ($50\%$ of the total duration).

\descr{Controlling for noise.}
Some differences in search results and recommendations are likely due to factors other than the user's watch history and personalization in general.
To reduce the possibility of this noise affecting our results, we take the following steps: 
1)~We execute, in parallel, experiments with identical search queries for all accounts to avoid updates to search results over time for specific search queries;
\revision{2)~All requests to YouTube are sent from the same geographic location (through the same US-based Proxy Server) to avoid location-based differentiation};
3)~We perform all experiments using the same browser user-agent and operating system;
4)~To avoid the carry-over effect (the previous search and watch activity affecting subsequent searches and recommendations), at each repetition of our experiments, we use the ``Delete Watch and Search History'' function to erase the activity of the user on YouTube from the date after we built the user profiles; and
5)~Similarly to the profiles' watch history creation, we always watch the same proportion of the video ($50\%$ of the total duration).

\descr{Implementation.} The experiments are written as custom scripts using Selenium in Python 3.7.
For each Google account, we create a separate Selenium instance for which we set a custom data directory, thus being able to perform manual actions on the browser before starting our experiments, e.g., performing Google authentication, installing AdBlock Plus to prevent advertisements within YouTube videos from interfering with our simulations, etc.
Finally, for all our experiments, we use Chromedriver 83.0.4 that runs in headless mode and stores all received cookies.

\descr{Video Annotation.} 
\revision{
Here, we describe how we use our classifier in our experiments.
In particular, we initially use our classifier to annotate all the videos encountered in our experiments and identify videos that are more likely to be pseudoscientific.
Then, the first author of this paper manually inspects all the videos classified as pseudoscientific to confirm that they are indeed pseudoscientific. 
Following this approach, we eliminate all the false positives.
}

\subsection{Pseudoscientific Content on Homepage, Search Results, and Video Recommendations}

\subsubsection{Homepage}
\label{subsubsec:pseudoscience_homepage_experiment}
We begin by assessing the magnitude of the pseudoscientific content problem on the YouTube homepage.
To do so, we use each one of the three user profiles (Science, Pseudoscience, and Science/Pseudoscience), as well as another user with no account (No Profile) that simulates the behavior of not logged-in users. We then visit each profile's homepage to collect and classify the top 30 videos as ranked by YouTube.
Note that we cannot perform this experiment using the YouTube Data API since it does not support this functionality.
We repeat the same experiment 50 times with a waiting time of 10 minutes between each repetition because YouTube shows different videos on the homepage each time a user visits YouTube.
\revision{We perform this experiment during December, 2020.}

\begin{figure}[t!]
\centering
\includegraphics[width=0.5\columnwidth]{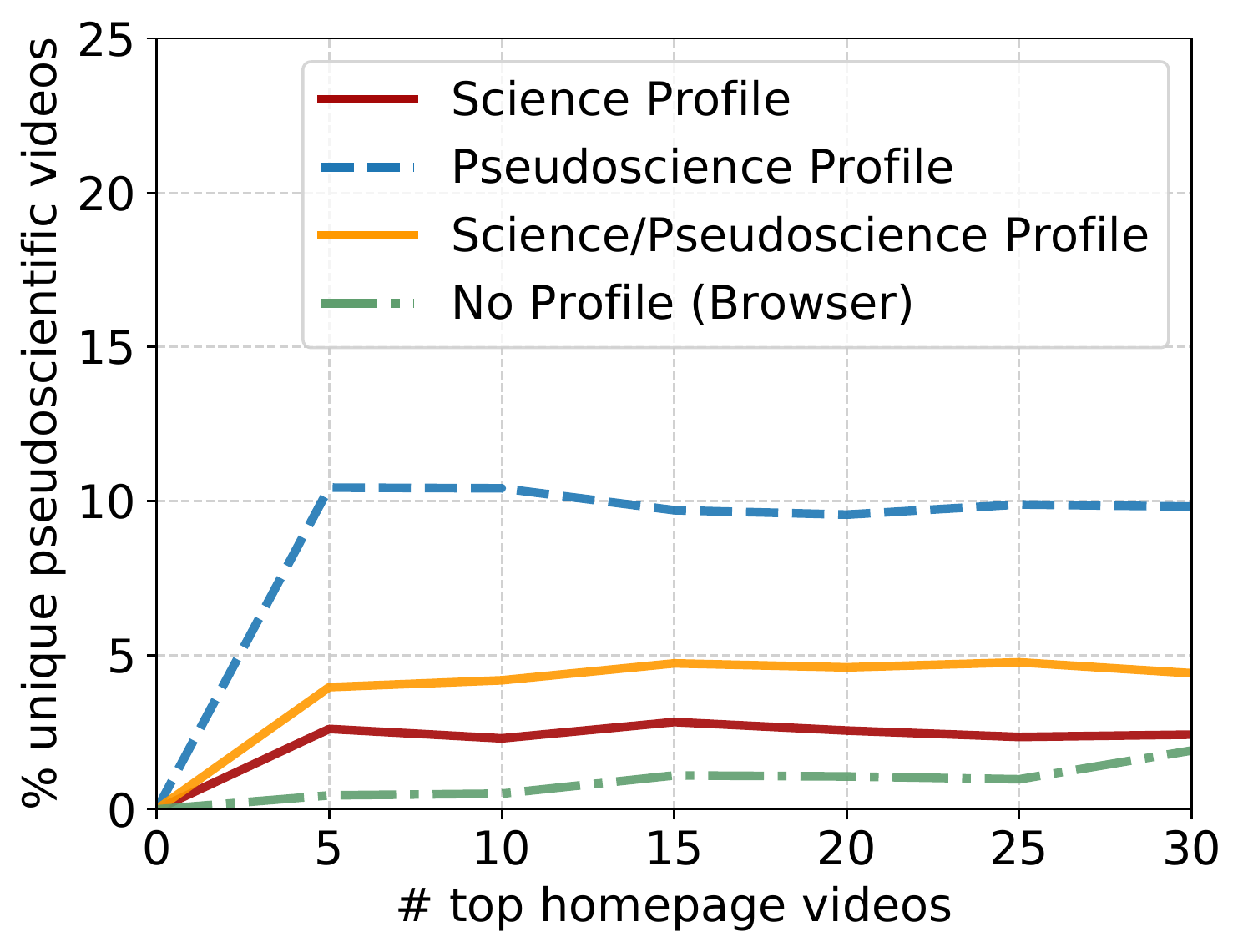}
\caption{Percentage of pseudoscience videos found in the homepage of each user profile.}
\label{fig:pseudoscience_homepage_experiment_plot}
\end{figure}

Figure~\ref{fig:pseudoscience_homepage_experiment_plot} shows the percentage of unique pseudoscientific videos on the homepage of each user profile.
We find that $2.4\%$, $9.8\%$, $4.4\%$, and $1.9\%$ of all the unique videos found in the top 30 videos of the homepage of the  Science, Pseudoscience, Science/Pseudoscience, and the No profile (browser) users, respectively, are pseudoscientific.
Overall, the Pseudoscience and the Science/Pseudoscience profile get shown a higher amount of pseudoscientific content on their homepage.
We also verify the significance of the difference in the amount of pseudoscientific content in the homepage of the Pseudoscience and the Science/Pseudoscience profiles compared to the one of the No profile (browser) using the Fisher's Exact test ($p<0.05$)~\cite{fisher1922interpretation}. 
We obtain similarly high significance ($p<0.05$) when we compare the Pseudoscience and Science/Pseudoscience profiles with the Science profile.
This indicates that the users' watch history substantially affects the number of pseudoscientific recommendations on their homepage. Nevertheless, users who are not interested in this type of content (i.e., science profile) still receive a non-negligible amount of pseudoscientific content.
We also observe that as the number of videos on the user's homepage increases (e.g., when a user scrolls down), the pseudoscientific videos' percentage remains approximately identical.

\begin{figure*}[t!]
\centering
\includegraphics[width=1.0\linewidth]{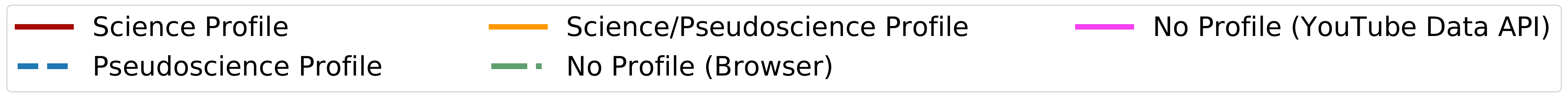}\\
\includegraphics[width=0.49\linewidth]{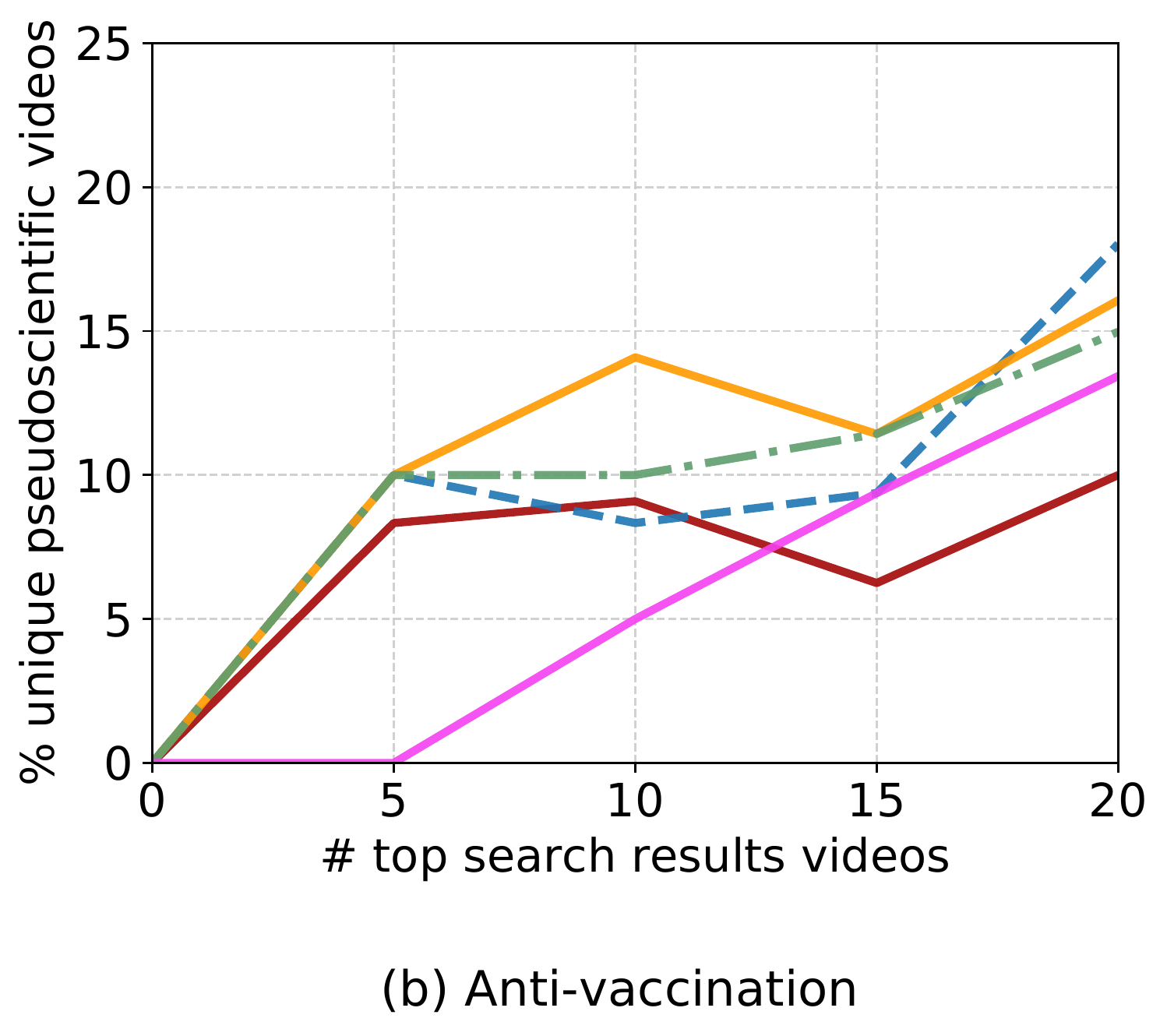}
\includegraphics[width=0.49\linewidth]{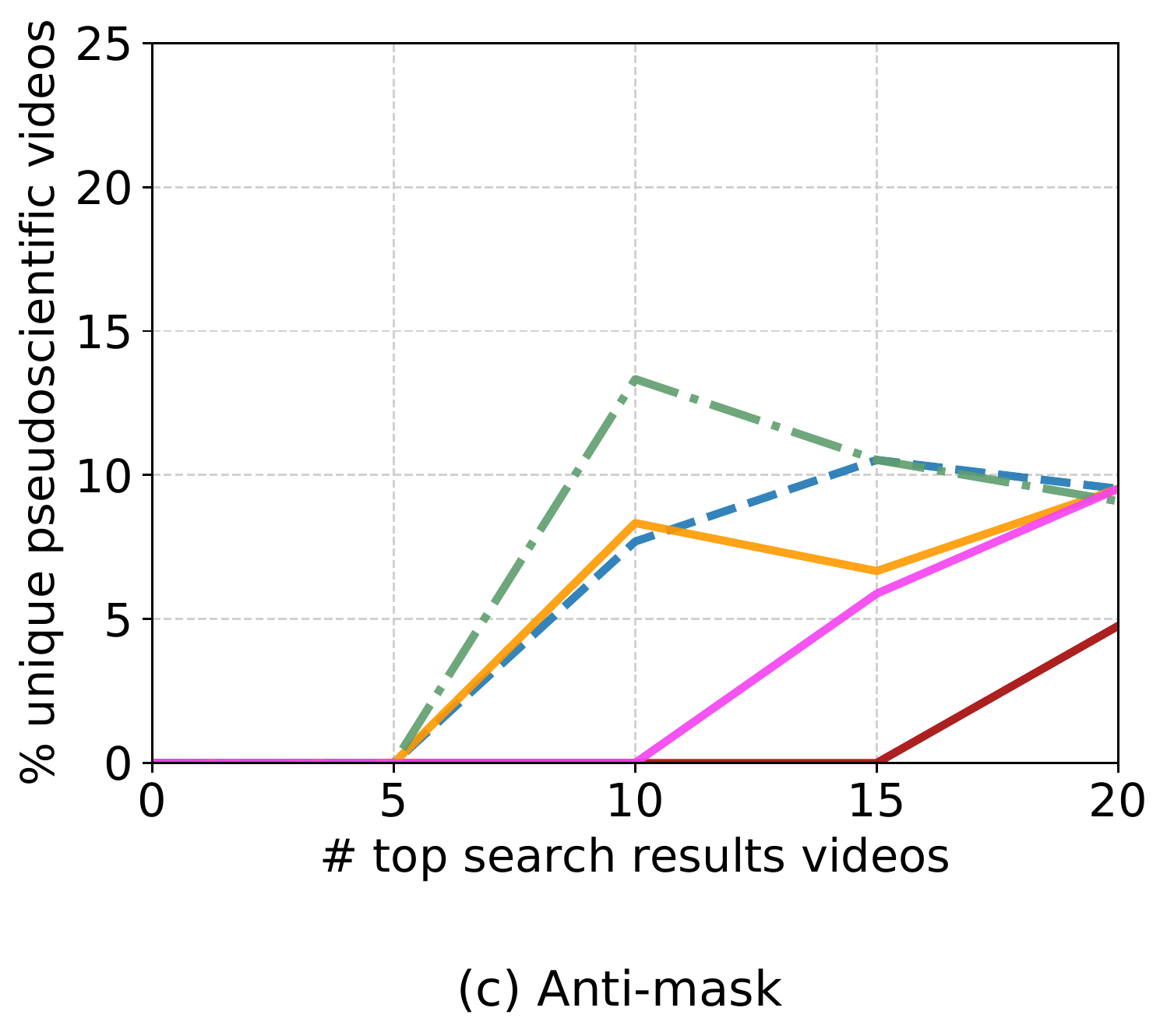}
\includegraphics[width=0.49\linewidth]{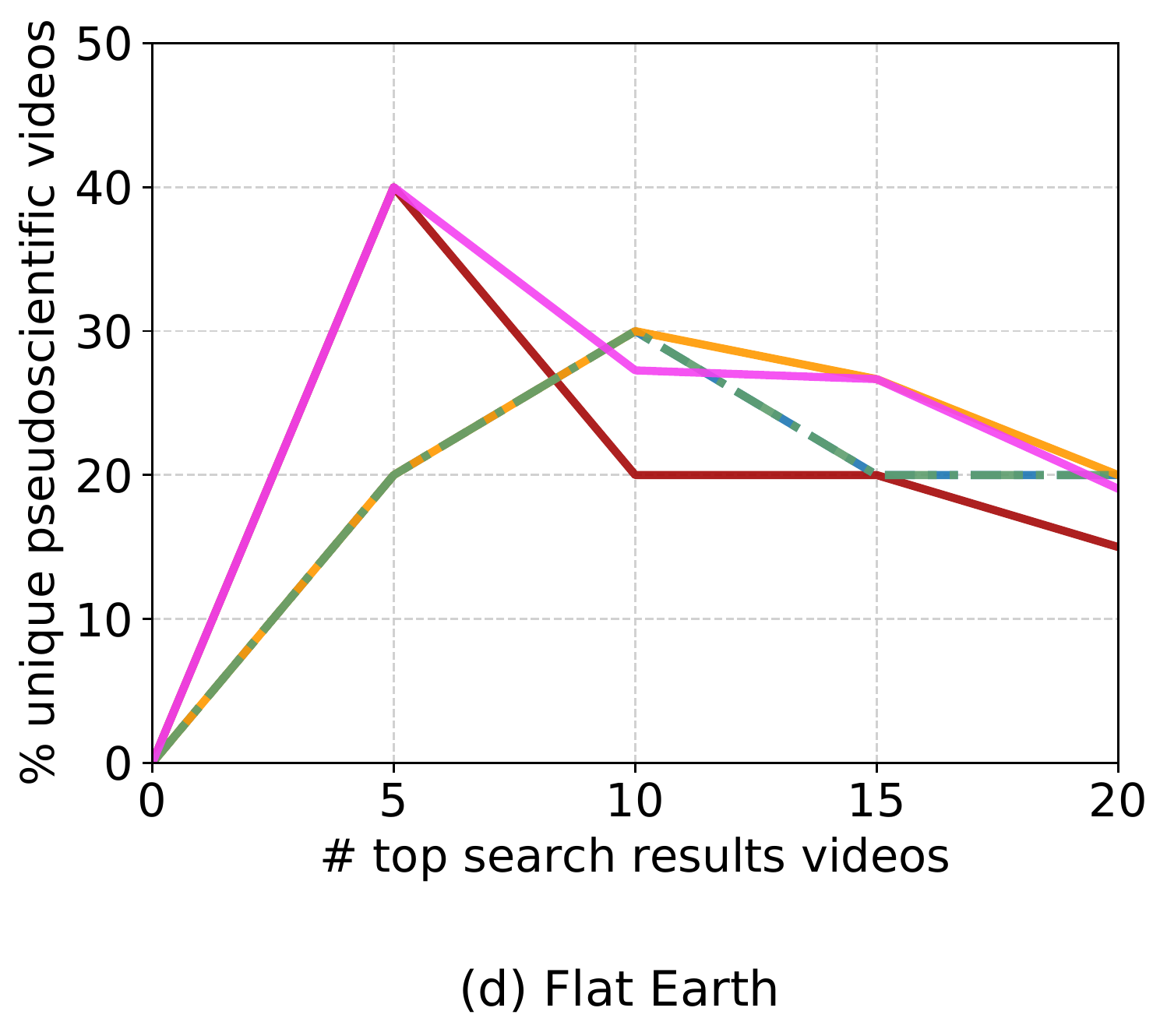}
\caption{Percentage of unique pseudoscience videos found in the search results of each user profile.}
\label{fig:pseudoscience_youtube_search_experiment_plot_all_terms}
\end{figure*}

\subsubsection{Search Results}
\label{subsubsec:pseudoscience_search_experiment}
Next, we focus on quantifying the prevalence of pseudoscientific content when users search for videos on YouTube.
For this experiment, we use the four pseudoscientific topics in our ground-truth dataset, and we perform search queries on YouTube for each topic.
\revision{For topics with two search queries (i.e., COVID-19), we perform the experiment twice and average their results.}
We retrieve the top 20 videos for each search query and use our classifier to classify each video in the result set.
We repeat this experiment 50 times for each pseudoscientific topic using all three user profiles and two non-logged-in users with no profile (one using a browser and another using YouTube's Data API).
Recall that we delete the user's watch history at each experiment repetition and between those performed with different search queries to ensure that future search results are not affected by previous activity other than our controlled watch history.
\revision{We perform this experiment in December, 2020.}

Overall, we find a large variation in the results across pseudoscientific topics (see Figure~\ref{fig:pseudoscience_youtube_search_experiment_plot_all_terms}).
For more traditional pseudoscientific topics like Flat Earth, YouTube search returns even more pseudoscientific content.
In particular, when searching for Flat earth, the Science profile, Pseudoscience profile, Science/Pseudoscience profile, no profile (browser), and the YouTube Data API encounter, respectively, $5.0\%$, $2.0\%$, $3.9\%$, $5.0\%$, and $5.6\%$ more unique pseudoscientific content than when searching for Anti-vaccination.
\revision{
In fact, Anti-vaccination is the topic with the second-highest amount of pseudoscientific content across all profiles.
For topics like COVID-19, all the recommended videos are \emph{not} pseudoscientific, suggesting that YouTube's recommendation algorithm does a better job in recommending less harmful videos---at least for COVID-19.}
This also signifies that YouTube has made substantial efforts to tackle COVID-related misinformation~\cite{youtubecovid2020tackle}, establishing an official, dedicated policy for that~\cite{youtubecovid2020policy}.
However, this is not the case for other controversial and timely pseudoscientific topics like Anti-vaccination or Anti-mask.
\revision{An explanation of the differences observed between COVID-19 and Anti-mask lies in that COVID-19 has a longer timeline than the masks-related problem.
The Anti-mask movement gained attraction after a few months from the emergence of the COVID-19 pandemic and YouTube might need some more time to develop effective moderation strategies to tackle misinformation surrounding the use of masks.}
Nevertheless, YouTube has recently announced that they will also attempt to target COVID-19 vaccine misinformation~\cite{youtubevaccine2020tackle}.

For Anti-vaccination, Anti-mask, and Flat earth searches, YouTube outputs more pseudoscientific content to the Pseudoscience and Science/Pseudoscience profiles than to the Science one.
Specifically, the amount of unique pseudoscientific videos in the top 20 search results of the Pseudoscience profile is, respectively, $18.0\%$, $9.5\%$, and $20.0\%$ for Anti-vaccination, Anti-mask, and Flat Earth.
For the Science/Pseudoscience profile, it is $16.1\%$, $9.5\%$, and $20.0\%$, while for the Science one is $10.0\%$, $4.8\%$, and $18.0\%$.

\revision{Furthermore, when taking into account the ranking of the search results, as the number of search results increases for Anti-vaccination and Anti-mask so does the percentage of unique pseudoscientific videos, which might indicate that YouTube does a good job in ranking content with higher quality on top for this topics.
On the other hand, for Flat Earth more of the pseudoscientific content is observed in the top five search results.}

\begin{figure*}[t!]
\centering
\includegraphics[width=1.0\linewidth]{figures/pseudoscience_random_walks_legend.pdf}\\
\includegraphics[width=0.49\linewidth]{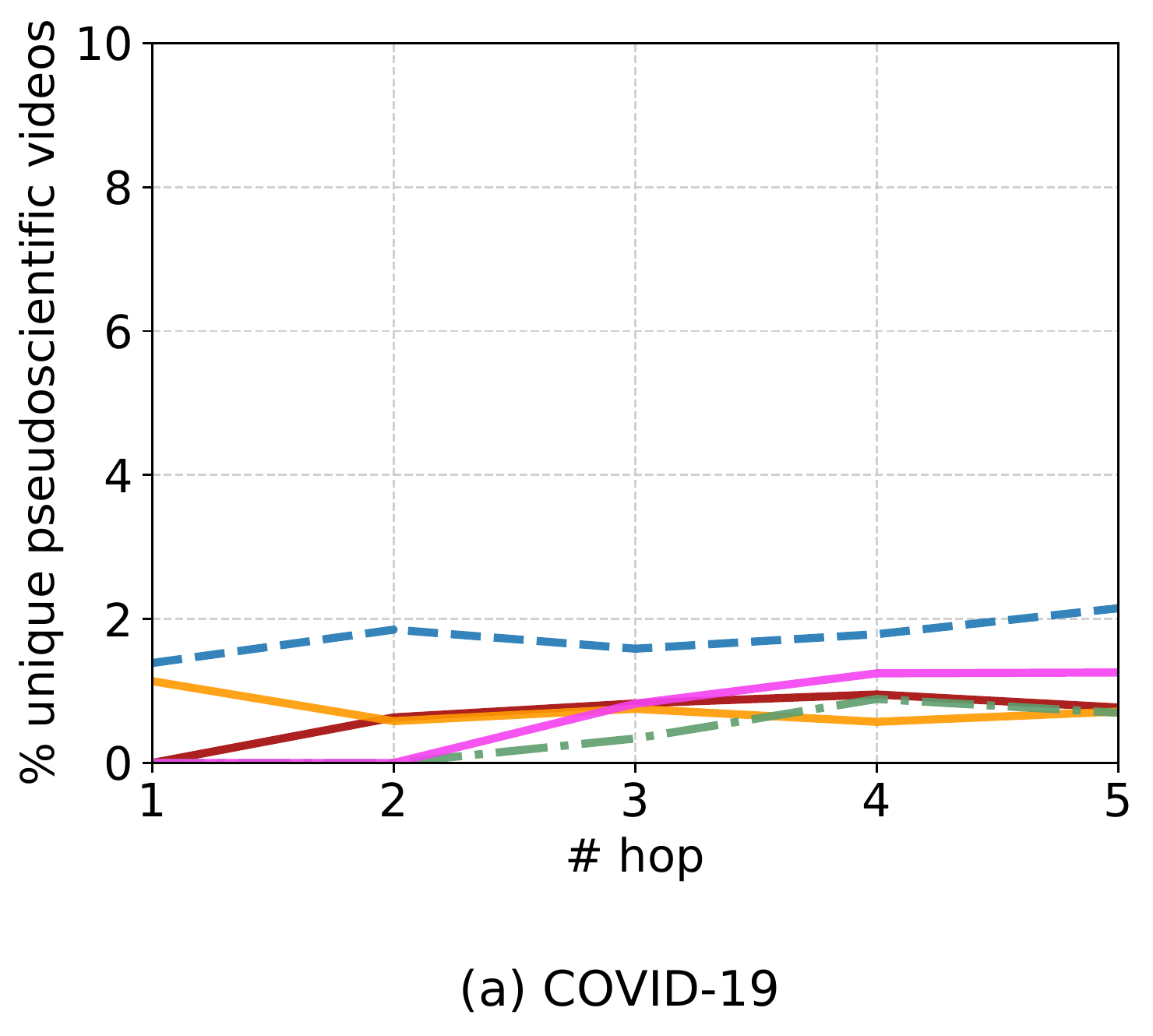}
\label{fig:pseudoscience_random_walks_covid}
\includegraphics[width=0.49\linewidth]{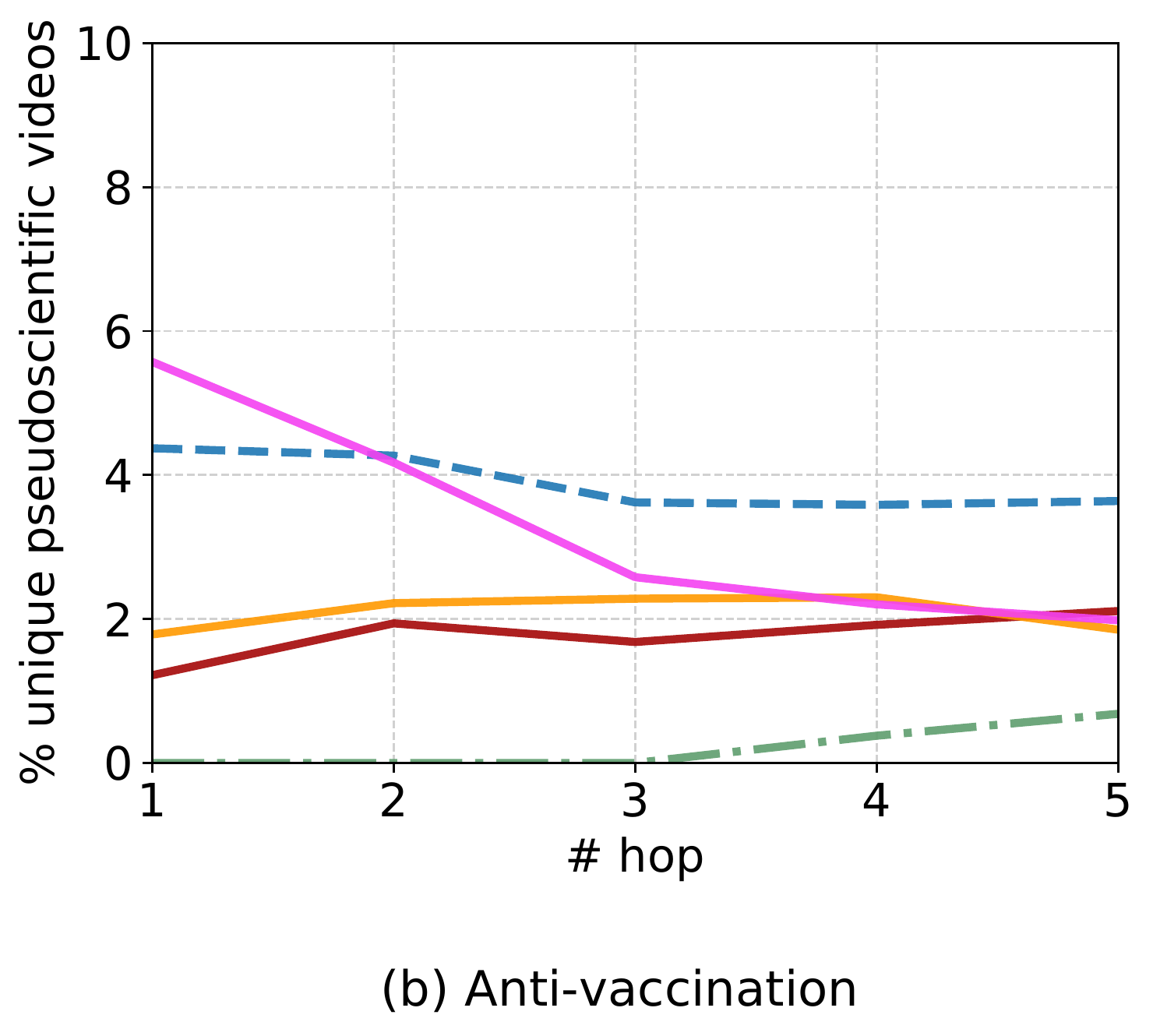}
\includegraphics[width=0.49\linewidth]{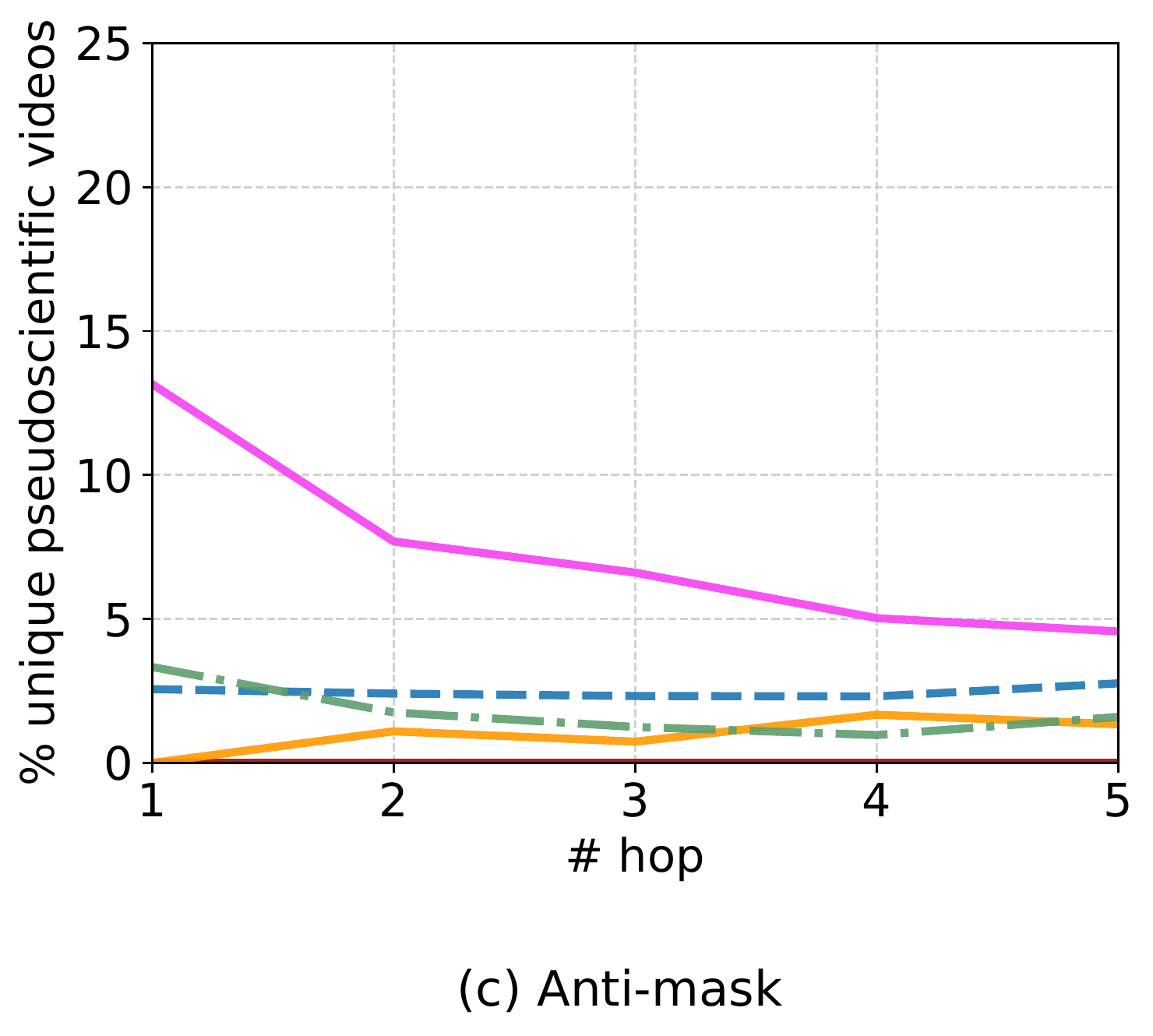}
\includegraphics[width=0.49\linewidth]{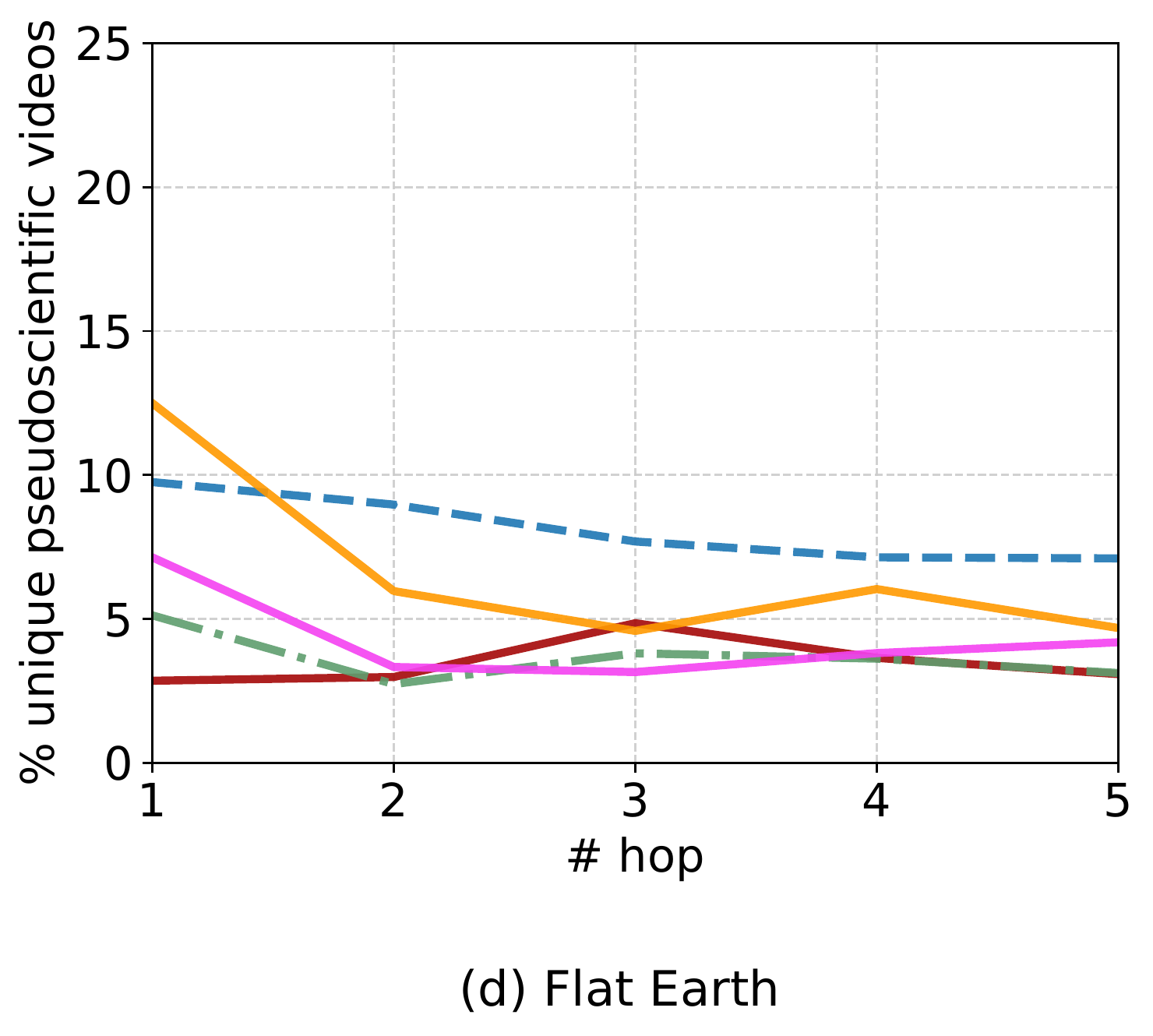}
\caption{Percentage of unique pseudoscientific videos that the random walker encounters at hop $k$ per user profile (December 2020).}
\label{fig:pseudoscience_random_walks_pseudoscience_found}
\end{figure*}

\subsubsection{Video Recommendations}
\label{subsubsec:pseudoscience_random_walks}
Last but not least, we set out to assess YouTube's recommendation algorithm's pseudoscience problem by performing controlled, live random walks on the recommendation graph while again measuring the effect of a user's watch history. This allows us to simulate the behavior of users with varying interests who search the platform for a video and subsequently watch several videos according to recommendations.
Note that videos are nodes in YouTube's recommendation graph, and video recommendations are directed edges connecting a video to its recommended videos.
For example, a YouTube video page can be seen as a snapshot of YouTube's recommendation graph showing a single node (video) and all the directed edges to all its recommended videos in the graph.

For our experiments, we use the four pseudoscientific topics considered for the creation of our ground-truth dataset. We initially perform a search query on YouTube and randomly select one video from the top 20 search results for each topic. 
Then, we watch the selected video, obtain its top ten recommended videos, and randomly select one. 
Again, we watch that selected video and randomly choose one of its top ten recommendations.
This simulates the behavior of a user who watches videos based on recommendations, selecting the next video randomly from among the top ten recommendations until he reaches five hops (i.e., six total videos viewed),  thus ending a single live random walk.
We repeat this process for 50 random walks for each search term related to our pseudoscientific topics while automatically classifying each video we visit. 
\revision{For topics with two search queries (i.e., COVID-19), we perform the experiment twice and average their results. We also ensure that the same video is not selected twice within the same random walk and that all random walks of a user profile performed for the same topic are unique.
We perform this experiment with all three Google accounts (logged-in users), the user with no profile (browser), and the YouTube Data API during December, 2020.
}

\revision{
Note that the recommendations collected using the API differ from the recommendations collected in the experiments performed using a browser.
In fact, the API allows us to collect the "related" videos of a given video, which are basically recommendations provided by YouTube's recommendation algorithm based on video item-to-item similarity, as well as general user engagement and satisfaction metrics.
Second, the API does not provide a functionality to watch YouTube videos. 
Hence, in the experiments we perform using the API we do not watch the videos selected during our random walks.}

For each user profile's random walks, we calculate the percentage of pseudoscientific videos encountered over all the unique videos that the random walker visits up to the $k$-th hop. 
Note that we have already assessed the amount of pseudoscientific content in the search results. 
Hence, in this experiment, we focus on video recommendations and do not consider, in our calculations, the initial video of each random walk selected from the search results.

Figure~\ref{fig:pseudoscience_random_walks_pseudoscience_found} plots this percentage per hop for each of the pseudoscientific topics explored.
Looking at the percentage of pseudoscientific videos encountered by each user profile in all the random walks of each pseudoscientific topic, we highlight some interesting findings.
For all topics, the amount of pseudoscientific content being suggested to the Pseudoscience profile after five hops are higher than the Science profile (see Figure~\ref{fig:pseudoscience_random_walks_pseudoscience_found}). 
In particular, the portion of unique pseudoscientific videos encountered by the Pseudoscience profile after five hops is $2.1\%$, $3.6\%$, $2.8\%$, and $7.1\%$ for COVID-19, Anti-vaccination, Anti-mask, and Flat Earth,  respectively, while for the Science profile, it is $0.8\%$, $1.9\%$, $0.0\%$, and $3.1\%$.
We also validate the statistical significance of the differences in the portion of pseudoscientific content suggested to the Pseudoscience profile compared to the Science profile for Anti-vaccination, Anti-mask, and Flat Earth, via the Fisher's Exact test ($p<0.05$).

Lastly, we find that for more traditional pseudoscientific topics like Flat Earth, YouTube suggests more pseudoscientific content to all types of users, except the YouTube Data API, compared to the other three more recent pseudoscientific topics. 
Using Fisher's exact test, we confirm that this difference between Flat Earth and COVID-19 is statistically significant for all types of users ($p<0.05$), while for Anti-mask this holds for the Science profile and the no profile (browser), and for Anti-vaccination this holds for the Pseudoscience profile and the no profile (browser).
This is another indication that YouTube has taken measures to counter the spread of pseudoscientific misinformation related to important topics like the COVID-19 pandemic. 

Overall, in most cases, the watch history of the user \emph{does} affect user recommendations and the amount of pseudoscientific content suggested by YouTube's algorithm.
This is also evident from the results of the random walks performed on the browser by the user with no profile. 
This profile does not maintain a watch history. 
It is recommended less pseudoscientific content than all the other profiles after five hops when starting from a video related to COVID-19 ($0.7\%$), and mainly to Anti-vaccination ($0.7\%$) and Flat earth ($3.1\%$).

Finally, we find a higher amount of pseudoscientific content in the random walks performed using the YouTube Data API than the random walks performed with the other non-logged-in user on the browser. 
In particular, the amount of unique pseudoscientific videos encountered by the YouTube Data API after five hops is $1.3\%$, $2.0\%$, $4.6\%$, and $4.2\%$ for COVID-19, Anti-vaccination, Anti-mask, and Flat earth, respectively, while, for the no profile (browser), it is $0.7\%$, $0.7\%$, $1.6\%$, and $3.1\%$.
However, this difference is not statistically significant and this indicates that the YouTube Data API results do not account for user personalization, neither the API maintains a watch history.
On the other hand, this difference may indicate that the YouTube Data API is more sensitive to item-to-item mapping~\cite{linden2003amazon} of the videos by the recommendation engine.

\begin{figure*}[t!]
\centering
\includegraphics[width=1.0\linewidth]{figures/pseudoscience_random_walks_legend.pdf}\\
\includegraphics[width=0.49\linewidth]{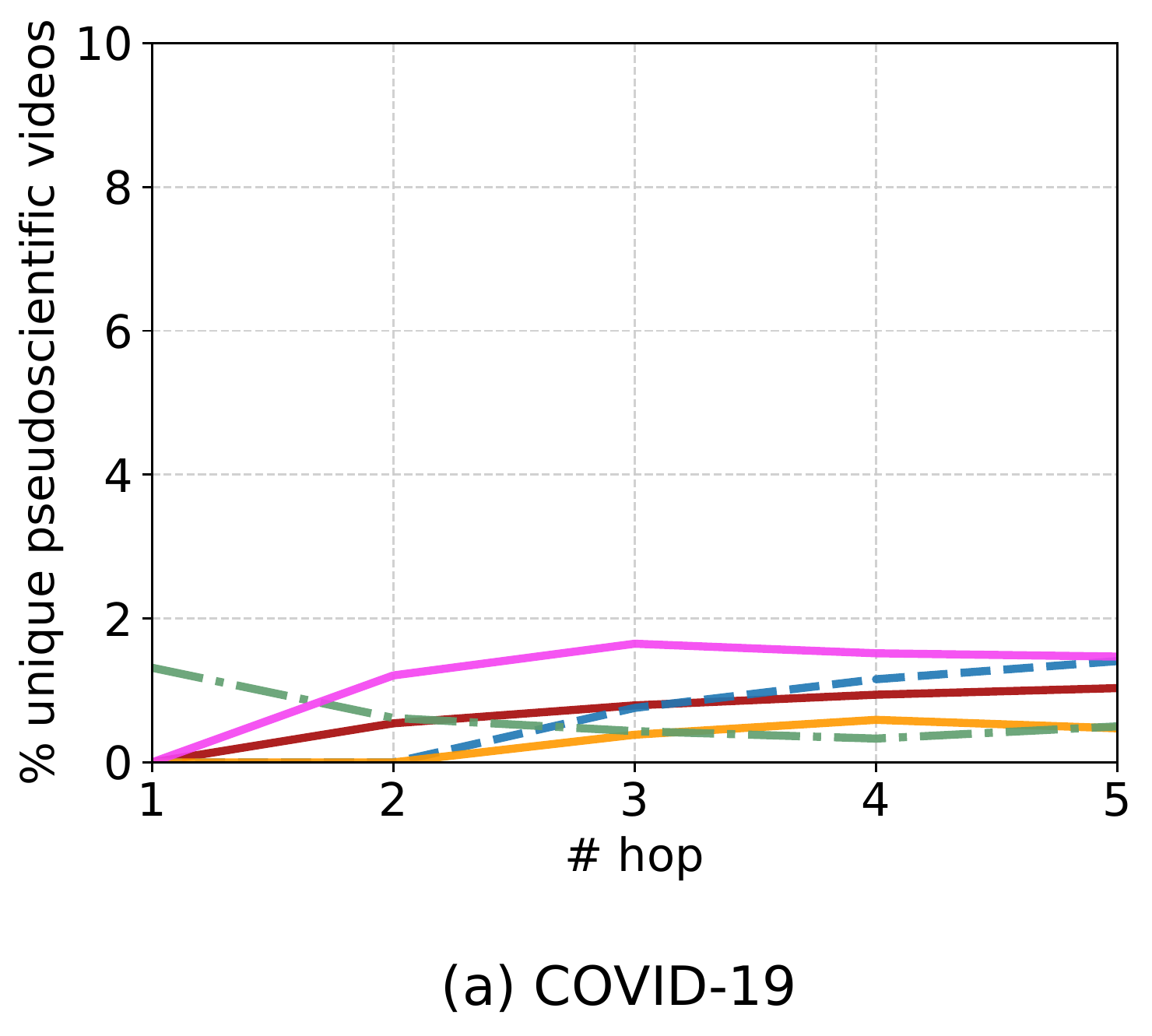}
\includegraphics[width=0.49\linewidth]{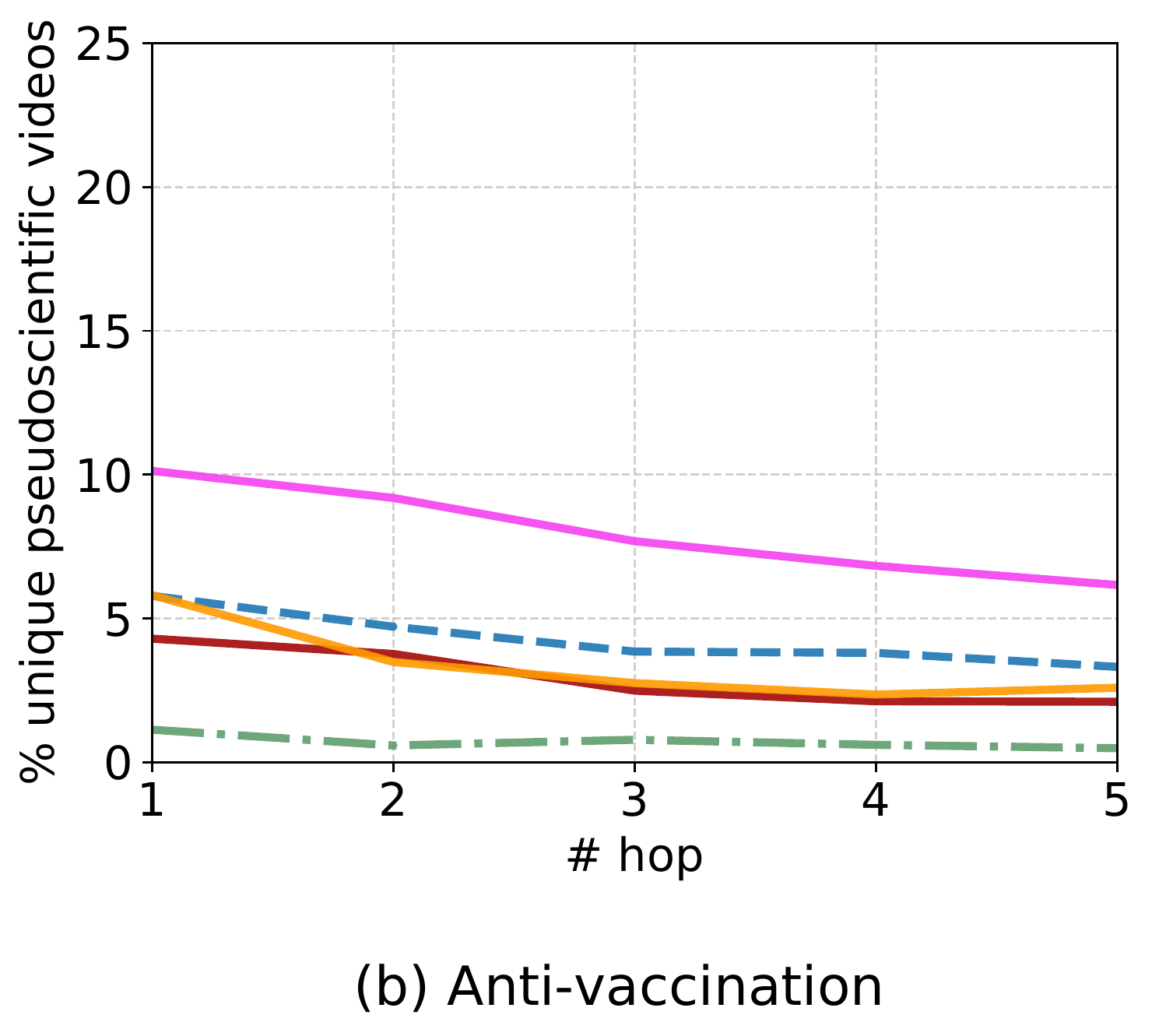}
\includegraphics[width=0.49\linewidth]{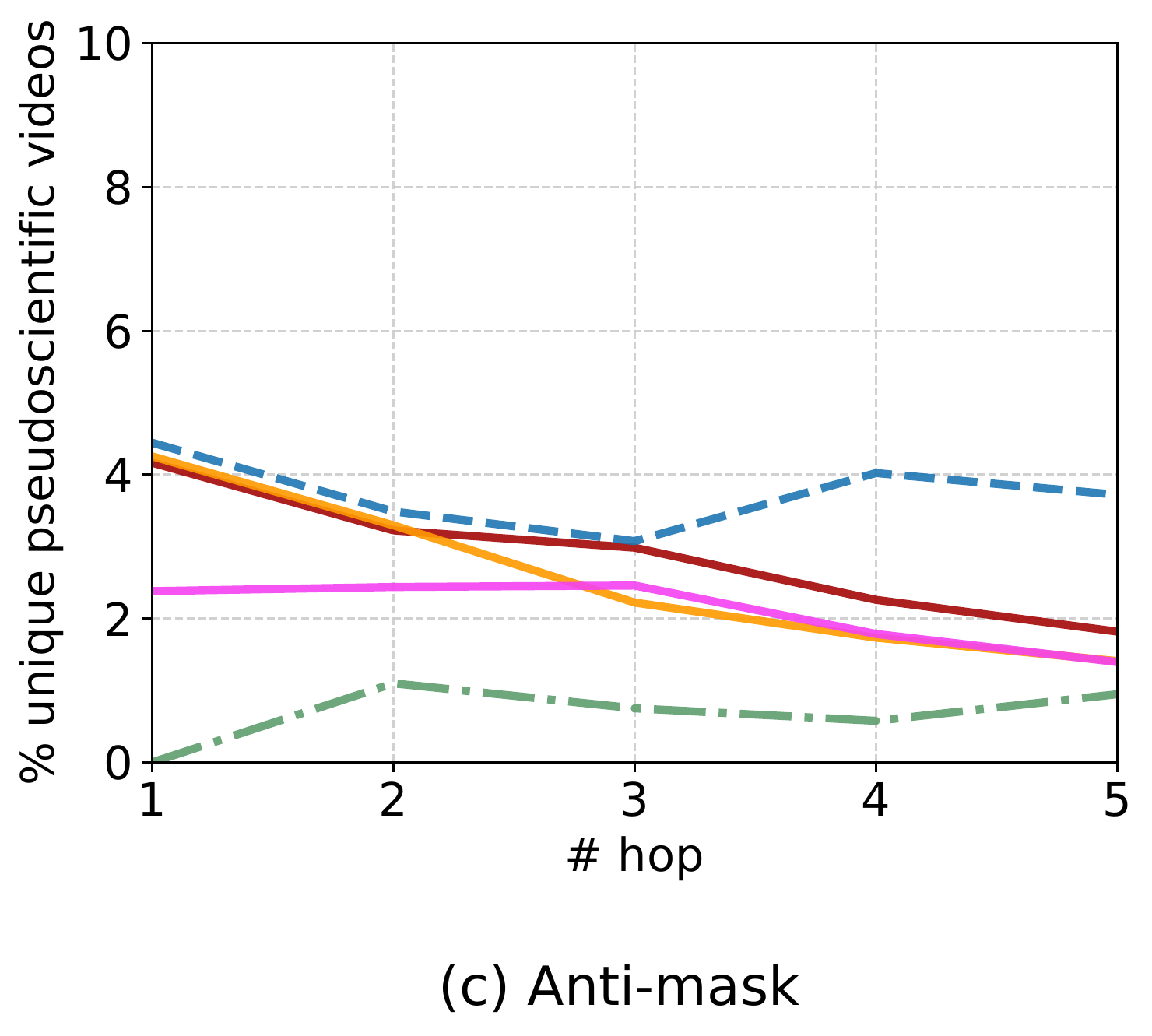}
\includegraphics[width=0.49\linewidth]{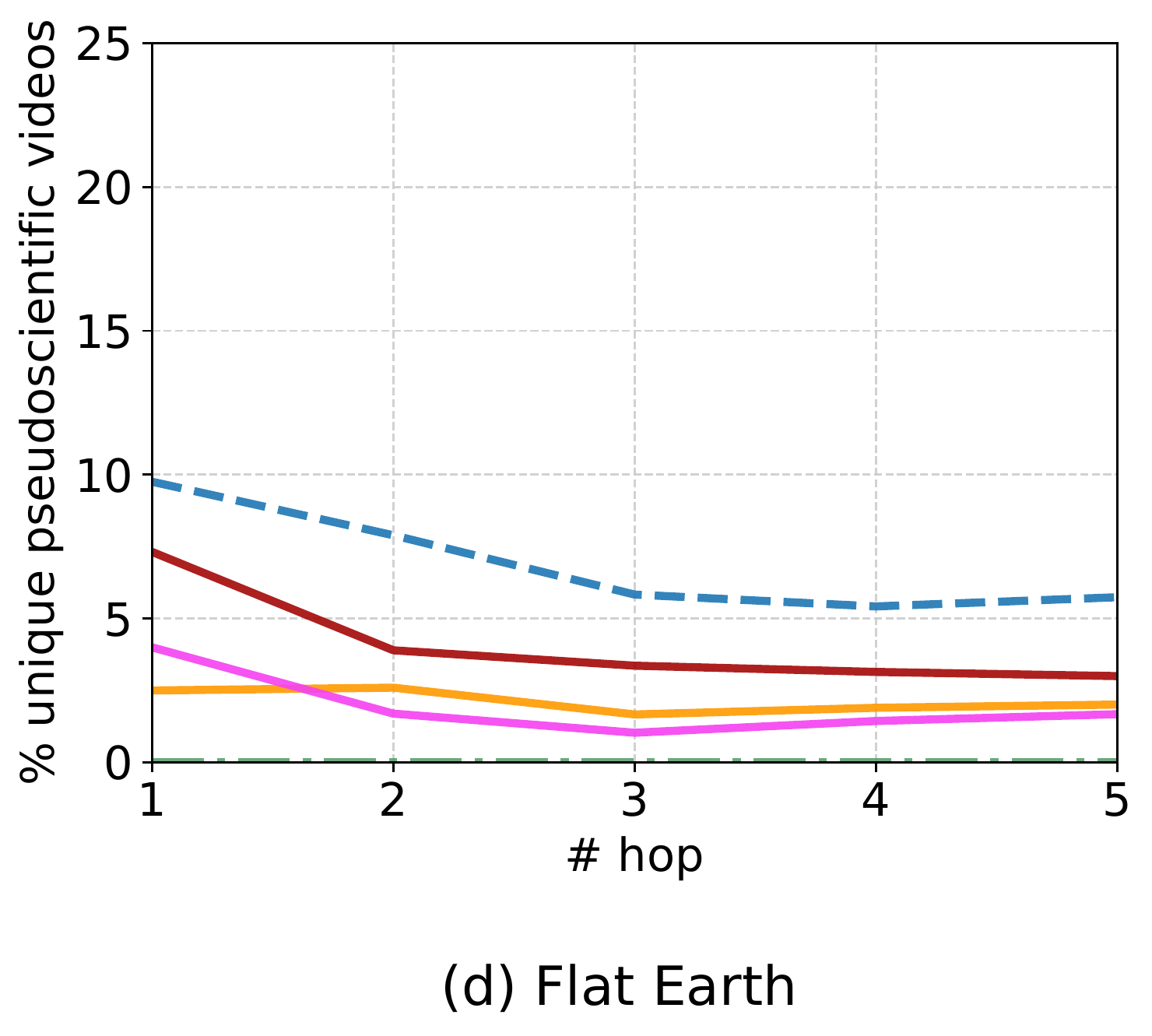}
\caption{Percentage of unique pseudoscientific videos that the random walker encounters at hop $k$ per user profile (April-May 2021).}
\label{fig:pseudoscience_random_walks_pseudoscience_found_second}
\end{figure*}

\subsection{Temporal Sensitivity}
\revision{
Here, we investigate any variations in the results of our experiments over time either due to changes in the recommendation algorithm or to the effectiveness of the moderation strategies employed by YouTube.
Although we are not aware of any significant changes in the recommendation algorithm and YouTube has not officially announced any changes to its system, the company did announce during the COVID-19 pandemic that it will revert to human moderators to effectively tackle misinformation on its platform~\cite{ft2020humanannotators}.
Hence, to investigate the temporal sensitivity of our results, we perform the video recommendations experiment once again between April and May 2021 using the same experiment setup and user profiles.
Figure~\ref{fig:pseudoscience_random_walks_pseudoscience_found_second} plots the percentage of unique pseudoscientific videos that the random walker encounters at hop $k$ for each of the pseudoscientific topics explored.
Importantly, we find that, for all pseudoscientific topics, the Pseudoscience profile receives more pseudoscientific content that the Science/Pseudoscience and the Science profiles.
Overall, we make similar observations as with the results of the video recommendations experiment performed in December, 2020.
}

\revision{
Next, we compare the results of each pseudoscientific topic with the respective results of the identical experiment performed in December. Importantly, we observe a slight decrease in the amount of pseudoscientific content being suggested to all user profiles for Flat Earth, while for the other topics the differences are negligible.
In general, we find that our results are not substantially affected by changes in the algorithm and data that may have occurred over the studied time period.}

\revision{
Nevertheless, we argue that the results of this line of work, which mostly derive from a single point in time are valuable.
This is because we mainly focus on timely pseudoscientific topics pertaining to the COVID-19 pandemic that are increasingly popular and of great societal interest.
The topics we analyze also allow us to:
1) evaluate the effectiveness of “novel” methodologies employed by YouTube to tackle misinformation around specific topics like COVID-19; 
2) investigate how YouTube responded to misinformation against crucial topics and whether the response was timely; and 
3) investigate the effectiveness of the usual mitigation strategies employed by YouTube when compared to other special mitigation strategies employed. 
The latter can be done by comparing the percentage of pseudoscientific content observed for traditional topics (i.e., Flat Earth) to the one observed for timely topics like COVID-19.
}

\begin{table}[t!]
\footnotesize
\centering
\begin{tabular}{llrrrrrrr}
\toprule
\multirow{2}{*}{\textbf{Part of YouTube}} & \textbf{Pseudos.} & \textbf{Science} & \textbf{Pseudoscience} & \textbf{Sci./Pseudos.} & \textbf{No Profile} & \textbf{No Profile}  \\
& \textbf{Topic} & \textbf{Profile} & \textbf{Profile} & \textbf{Profile} & \textbf{(Browser)} & \textbf{(API)}  \\
\toprule
\textbf{Homepage} & \multirow{2}{*}{\begin{tabular}[c]{@{}l@{}}-\end{tabular}} & \multirow{2}{*}{\begin{tabular}[c]{@{}l@{}}$2.4\%$\end{tabular}} & \multirow{2}{*}{\begin{tabular}[c]{@{}r@{}}$9.8\%$\end{tabular}} & \multirow{2}{*}{\begin{tabular}[c]{@{}r@{}}$4.4\%$\end{tabular}} & \multirow{2}{*}{\begin{tabular}[c]{@{}r@{}}$1.9\%$\end{tabular}} & \multirow{2}{*}{\begin{tabular}[c]{@{}r@{}}-\end{tabular}} \\
{\bf (Top 30)}\\
\midrule
\multirow{5}{*}{\begin{tabular}[c]{@{}l@{}}\textbf{Search Results}\\ \textbf{(Top 20)}\end{tabular}} 
& COVID-19 & 0.0\% & 0.0\% & 0.0\% & 0.0\% & 0.0\% \\
 & Anti-vacc & 10.0\% & 18.0\% & 16.1\% & 15.0\% & 13.4\% \\
 & Anti-mask & 4.8\% & 9.5\% & 9.5\% & 9.1\% & 10.0\% \\
 & Flat Earth & 15.0\% & 20.0\% & 20.0\% & 20.0\% & 19.0\% \\
 & All Topics & 6.6\% & 10.9\% & 10.3\% & 9.8\% & 9.2\% \\
\midrule
\multirow{5}{*}{\begin{tabular}[c]{@{}l@{}}\textbf{Video}\\ \textbf{Recommendations}\end{tabular}}
 & COVID-19 & 0.8\% & 2.1\% & 0.7\% & 0.7\% & 1.3\% \\
 & Anti-vacc & 2.1\% & 3.6\% & 1.9\% & 0.7\% & 2.0\% \\
 & Anti-mask & 0.0\% & 2.8\% & 1.3\% & 1.6\% & 4.6\% \\
 & Flat Earth & 3.1\% & 7.1\% & 4.7\% & 3.1\% & 4.2\% \\
 & All Topics & 1.5\% & 3.6\% & 1.9\% & 1.2\% & 2.5\% \\
\bottomrule
\end{tabular}%
\caption{Percentage of unique pseudoscientific videos encountered by each user profile in the three main parts of YouTube.}
\label{tab:pseudoscience_experiments_pseudoscience_videos_found_details}
\end{table}

\subsection{Experiments Summary}
We now summarize the main findings of our experiments.
Table~\ref{tab:pseudoscience_experiments_pseudoscience_videos_found_details} reports the percentage of unique pseudoscientific videos appearing on the YouTube homepage, search results, and the video recommendations section for each user profile out of all the unique videos encountered by each user profile in each experiment.

The highest percentage of pseudoscientific videos occurs in the search results.
That experiment shows that, for all pseudoscientific topics except COVID-19, the Pseudoscience and the Science/Pseudoscience profiles encounter more pseudoscientific content when searching for these topics than the Science profile.
For COVID-19, none of the profiles see any pseudoscientific content. 
When it comes to recommendations, in all the random walks (except Anti-mask), the Pseudoscience profile gets more pseudoscientific content than all the other profiles. 
For Anti-mask, we find a higher proportion of pseudoscientific content using the Data API.

\section{Limitations}
\revision{
Naturally, this line of work has several limitations. 
First, we use crowdworkers who are unlikely to have any expertise in identifying pseudoscientific content. 
Hence, a small percentage of the annotated videos may be misclassified. 
However, we mitigated this issue by not including annotators with low accuracy on a classification task performed on a test dataset and annotating each video based on the majority agreement. 
We also evaluated our crowdsourced annotation’s performance by manually reviewing a randomly selected set of videos from our ground-truth dataset, yielding 0.92 precision, 0.91 recall, and 0.92 F1 score.
To further eliminate this issue, we also collapsed our three labels into two, combining the science with the irrelevant videos into an “Other” category.
}

\revision{
Second, our ground-truth dataset is relatively small for such a subjective classification task. 
Nonetheless, the classifier provides a meaningful signal, which, supported by manual review, allows us to assess YouTube’s recommendation algorithm’s behavior with respect to pseudoscientific content. 
Third, there might be videos in our experiments that are pseudoscientific and have been classified as “Other.”
Hence, to verify our results’ accuracy, we manually reviewed a random sample ($10\%$) of the videos encountered during our experiments and classified them as “Other,” finding that $98.0\%$ of them were correctly classified. 
Finally, as for user personalization, we only work with watch history, which is a fraction of YouTube’s signals for user personalization.
}

\section{Remarks}
In this line of work, we studied pseudoscientific content on the YouTube platform. 
We collected a dataset of 6.6K YouTube videos, and by using crowdsourcing, we annotated them according to whether or not they include pseudoscientific content. 
We then trained a deep learning classifier to detect pseudoscientific videos. We used the classifier to perform experiments assessing the prevalence of pseudoscientific content on various parts of the platform while accounting for the effects of the user's watch history.
To do so, we crafted a set of accounts with different watch histories.

Overall, we found that the user's watch history does substantially affect future user recommendations by YouTube's algorithm.
This should be taken into consideration by research communities aiming to audit the recommendation algorithm and understand how it drives users' content consumption patterns. 
We also found that YouTube search results are more likely to return pseudoscientific content than other parts of the platform like the video recommendations section or a user's homepage.
However, we also observed a non-negligible number of pseudoscientific videos on both the video recommendations section and the users' homepage.
By investigating the differences across multiple pseudoscientific topics, we showed that the recommendation algorithm is more likely to recommend pseudoscientific content from traditional pseudoscience topics, e.g., Flat Earth, compared to more controversial topics like COVID-19.
This likely indicates that YouTube takes measures to counter the spread of harmful information related to critical and emerging topics like the COVID-19 pandemic. 
However, achieving this in a proactive and timely manner across topics remains a challenge.

Our work provides insights on pseudoscientific videos on YouTube and provides a set of resources to the research community (the dataset, the classifier, and all the source code of our experiments). 
In particular, the ability to run this kind of experiment while taking into account users' viewing history will be beneficial to researchers focusing on demystifying YouTube's recommendation algorithm---irrespective of the topic of interest.
In other words, our methodology and codebase are generic and can be used to study other topics besides pseudoscience, e.g., additional conspiracy theories.

\chapter{Discussion \& Conclusions}
\label{chapter:conclusions}

In this thesis, we studied several types of abhorrent, misinformative, and mistargeted content on YouTube.
Specifically, we shed light on three main relevant lines of inquiry:
1) characterizing and detecting inappropriate videos targeting young children;
2) characterizing hateful and misogynistic content on the platform through the lens of the Incel community; and
3) assessing the effect of a user's watch history on pseudoscientific video recommendations.
Below, we provide the main insights and possible future directions for each line of work.

\section{Characterizing and Detecting Inappropriate Videos Targeting Young Children on YouTube}

\descr{Remarks.} 
An increasing number of young children are shifting from broadcast to streaming video consumption, with YouTube providing an endless array of content tailored toward young viewers. 
While much of this content is age-appropriate, there is also an alarming amount of inappropriate material available.
In this line of work, we studied the problem at a large-scale and we provided the first characterization of inappropriate or disturbing videos targeted at young children on YouTube.
By developing a deep learning classifier~\cite{disturbedyoutubeclassifier} we were able to detect disturbing toddler-oriented videos with $0.84$ accuracy.
Using this classifier, we assessed how prominent the problem is on a large-scale and we also performed a live simulation of a toddler who searches the platform for a video and then she watches several videos according to the recommendations. 
The main take-aways from this work are: 
\begin{itemize}
    \item Toddler-oriented disturbing videos have similar characteristics as benign videos, and YouTube’s algorithmic recommendation system is not able to discern them, thus regrettably suggesting them to users.
    
    \item We find that when a toddler watches an Elsagate-related benign video if she randomly follows one of the top ten recommended videos, there is a non-negligible chance that she will end up at a disturbing or restricted video.
    
    \item Young children are likely to encounter inappropriate videos when they randomly browse the platform starting from appropriate videos.
    
    \item As of May 2019, YouTube was still plagued by inappropriate videos targeting young children, and our assessment of YouTube's current mitigatory measures shows that the platform is not able to tackle the problem in a timely manner.
\end{itemize}

Although the scientific debate (and public opinion) on the risks associated with "screen time" for young children is still ongoing, based on our findings, we believe a more pressing concern to be the dangers of crowd-sourced, uncurated content combined with engagement oriented, gameable recommendation systems. 
Considering the advent of algorithmic content creation (e.g., "deep fakes") and the monetization opportunities on sites like YouTube, there is no reason to believe there will be an organic end to this problem. 
Our classifier, and the insights gained from our analysis, can be used as a starting point to gain a deeper understanding and begin mitigating this issue.

Note that in this line of work, we collect and analyze a large number of Elsagate-related, other child-related, as well as random and popular videos available on YouTube. 
Although not representative of the entirety of YouTube, we believe that the set of seed keywords (Elsagate-related and other child-related keywords) cover a wide range of child-related content available on the platform. 
Concerning our sampling process, we believe that by including a wide range of child-related content as well as other types of videos in our ground truth dataset, we aid the proposed model to generalize to different types of videos that are available on YouTube.

\descr{Future Directions.}
Several interesting future directions derive from the finding of this thesis.
First, we propose a deep learning classifier that can detect disturbing videos and we also present a novel methodology that simulates the behavior of users casually browsing YouTube.
This methodology, together with the proposed model architecture, can be used in a lot of different problems of inappropriate content on YouTube to assess how prominent each problem is. 
For example, the proposed model architecture can be used to build classifiers that can detect other types of violent or sexually explicit content, and together with the random walks-based methodology can be used to measure how likely is for users to come across such content on the platform.
With regards to inappropriate or disturbing videos targeted at young children, and considering the non-negligible amount of disturbing videos found in this line of work, an interesting future direction is to study the intent of the uploaders of disturbing videos on the platform.
Another interesting future direction is to further focus on the detection and mitigation of disturbing videos by also considering the video footage, video transcript, and comments posted under toddler-oriented videos.
With regards to user personalization, an interesting future direction is to study the problem by performing random walks using logged-in user profiles with child-related characteristics and watch history.

Moreover, in the meantime, YouTube started taking drastic measures to tackle the problem of disturbing videos targeted at toddlers.
More specifically, as of January 2020, YouTube requires channel owners to specify, during upload, whether their content is directed at children or not.
On videos marked as directed at children, the platform blocks data collection for all viewers, as well as comments and targeted ads.
Hence, it would be interesting to investigate whether these measures can effectively eliminate disturbing videos on the platform.

\section{Characterizing Hateful and Misogynistic Content on You-Tube Through the Lens of the Incel Community}

\descr{Remarks.} 
In this like of work, we performed a characterization of hateful and misogynistic content on YouTube by focusing on the Incel community.
The main goal was to assess how has the Incel community grown on YouTube over the last decade and whether YouTube's recommendation algorithm contributes to steering users towards Incel communities.
We collected YouTube videos shared by users in Incel-related communities within Reddit, we devised a methodology to effectively annotate them as "Incel-related," and used them to understand how Incel ideology spreads on YouTube and study the evolution of the community.
We also analyzed how YouTube’s recommendation algorithm behaves with respect to Incel-related videos. 

The main take-aways from this work are: 
\begin{itemize}
    \item We find an increase in Incel-related activity on YouTube over the past few years and in particular concerning Incel-related videos, as well as comments that include pertinent terms.
    
    \item We find a non-negligible amount of Incel-related videos ($2.9\%$) within YouTube’s recommendation graph being recommended to users.

    \item When a user watches a non-Incel-related video, if they randomly follow one of the top ten recommended videos, there is a $2.8\%$ chance they will end up with an Incel-related video.
    
    \item By performing random walks on YouTube’s recommendation graph, we find that when starting from a random non-Incel-related video, there is a $6.3\%$ probability to encounter at least one Incel-related video within five hops. Simultaneously, Incel-related videos are more likely to be recommended within the first two to four hops than in the subsequent hops.
    
    \item As users choose to watch Incel-related videos, the algorithm recommends other Incel-related videos with increasing frequency.
\end{itemize}

Last, despite YouTube’s attempts to tackle hate~\cite{tacklehate2019youtube}, our results show that the threat is clear and present. 
Also, considering that the Incel ideology is often associated with misogyny, multiple mass murders, and violent offenses~\cite{splc2019malesupremacy,fifthestate2020incels}, we urge that YouTube develops effective content moderation strategies to tackle misogynistic content on the platform.

\descr{Design Implications.}
Prior work has shown apparent user migration to increasingly extreme subcommunities within the Manosphere on Reddit~\cite{ribeiro2020pick}, and indications that YouTube recommendations serve as a pathway to radicalization. 
When taken along with our results, a more complete picture with respect to online extremist communities begins to emerge.

Radicalization and online extremism is clearly a \emph{multi-platform} problem. 
Social media platforms like Reddit, designed to allow organic creation and discovery of subcommunities, play a role, and so do platforms with algorithmic content recommendation systems.  The immediate implication is that while the radicalization process and the spread of extremist content generalize (at least to some extent) across different online extremist communities, the specific mechanism likely does not generalize across different platforms.
However, that does not mean that specific platform oriented-solutions should exist in a vacuum.
For example, an approach that could benefit both platforms involves using Reddit activity to help tune the YouTube recommendation algorithm and using information from the recommendation algorithm to help Reddit perform content moderation. 
In such a hypothetical arrangement, Reddit, whose content moderation team is intimately familiar with the troublesome communities, could help YouTube understand how the content these communities consume fits within the recommendation graph. 
Similarly, Reddit's moderation efforts could be bolstered with information from the YouTube recommendation graph. 
The discovery of emerging dangerous communities could be aided by understanding where the content posted by them fits within the YouTube recommendation graph compared to the content posted by known troublesome communities.

\descr{Future Directions.} 
Several future directions can be derived from the findings of this line of work.
First, misogynistic ideologies and other Manosphere communities on YouTube (e.g., Men Going Their Own Way) are relatively unstudied.
We believe that the scientific community can use and build upon our text-based approach to study other misogynistic ideologies on the platform, which tend to have their particular glossary.
Besides understanding each distinct community, it would also be interesting to study user migration between Manosphere and other reactionary communities on YouTube.
Moreover, our random walks analysis does not consider per-user personalization; the video recommendations we collect represent only some of the recommendation system’s facets.
Hence, an interesting future direction is the implementation of crawlers that will allow the simulations of real users and perform random walks on YouTube analyzing the problem of misogynistic content with personalization.
However, note that this task is not straightforward as it requires understanding and replicating multiple meaningful characteristics of Incels' behavior. 
Finally, we believe that an important future direction is the detection and mitigation of misogynistic content on YouTube.

\section{Assessing the Effect of Watch History on YouTube’s Pseudoscientific Video Recommendations}

\descr{Remarks.} 
In this line of work, we studied pseudoscientific content on YouTube by focusing on four topics, namely, COVID-19, the anti-vaccine movement, the anti-mask movement, and the Flat earth theory.
We first collected videos from YouTube related to these topics, annotate them using crowdsourcing and use them to build a classifier for detecting pseudoscientific videos.
The relatively low accuracy of the classifier and the low agreement score of our crowdsourced annotation highlights the subjective nature of the pseudoscientific vs. scientific content on YouTube.
Then, we devise a novel methodology for assessing the prevalence of pseudoscientific content on various parts of the platform, while also accounting for the effect of the user's watch history.
More specifically, with this methodology, we can simulate the behavior of users with different watch histories who browse YouTube while using our classifier to annotate all the videos we visit.%

The main take-aways of this line of work are:
\begin{itemize}
    \item The watch history of the user substantially affects what videos are suggested to the user.
    
    \item It is more likely to encounter pseudoscientific videos in the search results (i.e., when searching for a specific topic) than in the video recommendations section or the homepage of a user, except in the case of the COVID-19 topic.
    
    \item For ``traditional'' pseudoscience topics (e.g., Flat Earth), there is a higher rate of recommended pseudoscientific content than for more emerging/controversial topics like COVID-19, anti-vaccination, and anti-mask. For COVID-19, we find an even smaller amount of pseudoscientific content being suggested, which may result from measures YouTube took to mitigate misinformation concerning the COVID-19 pandemic.
    
    \item Although YouTube seems to tackle COVID-19 related misinformation in its search results, all profiles used in our experiments still receive recommendations to questionable content related to the pandemic.
    
    \item The difference between the results of the YouTube Data API and the no profile (browser) is statistically insignificant; this indicates that recommendations returned using the API are not subject to personalization. This finding can be helpful to other researchers that use YouTube's Data API for their experiments.
\end{itemize}

\descr{Future Directions.} 
Several interesting future directions derive from the main take-aways of this thesis.
First, we present a novel methodology for assessing the effect of watch history on pseudoscientific recommendations on YouTube.
Our findings indicate that recommendations returned using the YouTube API are not subject to personalization, thus highlighting the importance of accounting for personalization when studying YouTube for more representative analysis.
For this reason, we make our methodology publicly available~\cite{pseudosciencerepository}, and the research community can use it
to assess the effect of the watch history while studying a lot of other topics of interest on the platform.
For instance, this methodology can be used to create user profiles with child-related histories and investigate several types of inappropriate content being recommended to children when browsing YouTube.
Another interesting direction is to build upon our methodology and devise a more comprehensive user personalization methodology to account for factors outside of watch history, such as account characteristics and user engagement.
With regards to the detection of pseudoscientific misinformation, the relatively low agreement score of our crowdsourced annotation points to the difficulty in objectively identifying whether a video is pseudoscientific or not and also confirms that it is not easy to automate the discovery of misinformation.
\revision{
Hence, further research is required to develop classification models that can more effectively detect pseudoscientific misinformation on YouTube considering additional features as input to the classification models (e.g., the sentiment of the comments of the video, audio-related features, etc.).
}
Finally, another future direction is to understand how people share and view pseudoscientific content on other social networks, including Twitter and Facebook, and how people engage with such content.

\section{Challenges and Limitations}
\revision{
Here we summarize the main technical challenges we faced in this thesis, how we addressed them and we highlight the limitations of this thesis.
}

\descr{Challenges.}
\revision{
Our data collection and annotation efforts in this thesis faced several challenges.
First, there are no datasets of YouTube videos available related to the problems we analyzed, neither there are any state-of-the-art classifiers that can detect with an acceptable performance the type of videos we analyzed.
To address this issue and build our datasets, we followed different approaches depending on the peculiarities of the problem under study.
To collect disturbing videos targeting young children, we use Reddit to compile lists of keywords from various sources and use them to search YouTube and collect videos.
To collect Incel-related videos, we again use Reddit and we collect videos shared in communities associated with the Incel community.
In our latest line of work, to collect pseudoscientific videos, we search YouTube using keywords relevant to the topics of interest.
}

\revision{
Second, devising methodologies for annotating the collected videos and for detecting videos with abhorrent and misleading content is not trivial.
To do this, we employed distinct methodologies for each different type of problematic content.
For the detection of disturbing videos, we built a deep learning classifier that analyzes various metadata of a given video, as well as its thumbnail.
Similarly, for the detection of pseudoscientific content, we built another deep learning classifier that considers the comments and the transcript of a given video along with its various metadata (e.g., title, description, etc.).
On the other hand, when analyzing misogynistic communities, we detect misogynistic content based on a lexicon of terms that considers both the video's transcript and its comments. We believe that the scientific community can use and build upon our classifiers and text-based methodologies to study other types of inappropriate, misogynistic, and misinformative content on YouTube.
}

\descr{Limitations.}
\revision{
This thesis is not without limitations.
First, the methodologies and results presented in this thesis heavily rely on annotators who may have no expertise in identifying problematic content.
Hence, a small percentage of the annotated videos in this thesis may be misclassified.
However, we mitigate this issue by employing multiple annotators and by annotating each video based on the majority agreement.
Furthermore, when crowdsourcing is employed, we take additional measures to reduce subjectivity.
That is, we do not include annotators with low accuracy on a classification task performed on a test database, and we exclude from our datasets all the videos where all annotators disagree.
}

\revision{
Second, our video annotation methodologies might flag some benign videos as inappropriate.
However, by considering the various metadata of a video we can achieve an acceptable detection accuracy that uncovers a substantial proportion of the inappropriate videos. To mitigate the issues of misclassification when studying pseudoscientific content, we use the developed classifier to uncover possible pseudoscientific videos that we later manually review to eliminate false positives.
Another limitation of this thesis is that we may miss some inappropriate videos. 
Notwithstanding such limitation, we believe that our methodologies approach a lower bound of the inappropriate videos in our datasets, allowing us to conclude that the implications of YouTube's recommendation algorithm on disseminating abhorrent and misleading content are at least as profound as we observe. 
}
\revision{
Third, we acknowledge that the ground truth datasets we use to train our classifiers are relatively small and the performance of our classifiers is highly affected by this issue. We were unable to provide larger annotated datasets mainly due to the limited resources for the annotation process.
Nonetheless, we argue that the proposed classifiers provide a meaningful signal, which allows us to assess YouTube’s recommendation algorithm’s behavior with respect to inappropriate and pseudoscientific content. 
}

\revision{
Moreover, for some of the problems analyzed in this thesis, we do not consider per-user personalization; the video recommendations we collect represent only some of the recommendation system's facet.
However, we believe that the recommendations we obtain in the first two lines of work in this thesis allow us to understand how YouTube’s recommendation system is behaving in our scenario.  At the same time, in our latest line of work in which we consider user personalization, we only work with watch history, which is a fraction of YouTube's signals for user personalization.
}

\section{Conclusion}
In this thesis, we shed light on the prevalence of several types of abhorrent and mistargeted content on YouTube.
Our work reveals the need to further study and understand the role of YouTube's recommendation algorithm in promoting abhorrent and misleading content.
Also, it prompts the need for YouTube and the research community to develop more effective content moderation strategies to tackle such content on the platform.
Nevertheless, our findings indicate that the most proper way for YouTube to cope with certain types of content (i.e., pseudoscientific misinformation) on the platform effectively is to use deep learning models that signal potential harmful videos to human annotators who examine the videos and make the final decision.
We argue that the aforementioned are of paramount importance. We are confident enough that the findings and the resources provided by this thesis 
will be beneficial to researchers focusing on demystifying YouTube’s recommendation algorithm,
hence shedding additional light on the recommendation algorithm and its potential influence, irrespective of the topic of interest.

\small{\printbibliography}

\end{document}